\documentclass[12pt]{article}
\usepackage{hyperref}
\usepackage{amsmath}
\usepackage{url}
\usepackage{graphicx}
\usepackage{cite}

\newcommand{\ca}[1]{\mathcal{#1}}

\def\sinh{{\mathrm{sinh}}}

\newcommand{\bea}{\begin{eqnarray}}
\newcommand{\eea}{\end{eqnarray}}

\newcommand{\nn}{\nonumber}
\newcommand{\Tr}{\text{Tr}}

\newcommand{\Op}{\mathcal{O}}

\newcommand{\ket}[1]{\left| #1 \right>}
\newcommand{\bra}[1]{\left< #1 \right|}

\newcommand{\de}{\partial}

 \def\ep{{\epsilon}}

 \def\a{{\alpha}}
 
 \def\frac#1#2{{#1\over #2}}

 \def\s{\sqrt}

\def\be{\begin{equation}}
\def\ee{\end{equation}}
\def\ba{\begin{eqnarray}}
\def\ea{\end{eqnarray}}
\numberwithin{equation}{section}

 \def\de{\partial}

 \def\f {\frac}
 \def\ti{\tilde}
 \def\ap{\alpha}

 \def\ddd{\cdot\cdot\cdot}
 \def\no{\nonumber \\}
\def\nn{\nonumber \\}
 \def\la{\langle}
 \def\lb{\rangle}
 \def\ep{\epsilon}

\begin{document}

\begin{titlepage}
\thispagestyle{empty}

\begin{flushright}
YITP-19-41
\\
IPMU19-0077
\\
\end{flushright}

\bigskip

\begin{center}
\noindent{{ \large \textbf{Double Local Quenches in 2D CFTs and Gravitational Force}}\\
\vspace{2cm}
Pawel Caputa$^a$, Tokiro Numasawa$^{b,c}$,  Teppei Shimaji$^a$, \\
Tadashi Takayanagi$^{a,d}$, 
and Zixia Wei$^a$
 \vspace{1cm}\\
{\it
$^{a}$Center for Gravitational Physics, 
Yukawa Institute for Theoretical Physics,\\
Kyoto University, Kyoto 606-8502, Japan\\
$^{b}$
Department of Physics, McGill University, Montr\'eal, Qu\'{e}bec, H3A 2T8, Canada \\
$^{c}$
Department of Physics, Graduate School of Science,
Osaka university, \\
Toyonaka 560-0043, Japan \\
$^{d}$Kavli Institute for the Physics and Mathematics of the Universe (WPI),\\
University of Tokyo, Kashiwa, Chiba 277-8582, Japan
}}
\vskip 2em
\end{center}

%%%%%%%%%%%%%%%
%%%%%%%%%%%%%%%
\begin{abstract}
In this work we extensively study the dynamics of excited states created by instantaneous local quenches at two different points, i.e.~double local quenches. We focus on setups in two dimensional holographic and free Dirac fermion CFTs. We calculate the energy stress tensor and entanglement entropy for double joining and splitting local quenches. In the splitting local quenches we find an interesting oscillating behaviors. Finally, we study the energy stress tensor in double operator local quenches. In all these examples, we find that, in general, there are non-trivial interactions between the two local quenches. Especially, in holographic CFTs, the differences of the above quantities between the double local quench and the simple sum of two local quenches tend to be negative. We interpret this behavior as merely due to gravitational force in their gravity duals.
\end{abstract}
%%%%%%%%%%%%%%%
%%%%%%%%%%%%%%%

\end{titlepage}
\tableofcontents

\newpage

%%%%%%%%%%%%%%%
%%%%%%%%%%%%%%%
\section{Introduction}
%%%%%%%%%%%%%%%
%%%%%%%%%%%%%%%
The AdS/CFT  \cite{Ma,GKPW} provides us with a very beautiful and useful relation between the dynamics of gravity and that of conformal field theories (CFTs).  For example, the information of metric in gravity is essentially captured by that of entanglement entropy (EE)  \cite{BKLS,Sr,HLW,CC,CH} via the holographic calculation \cite{RT,HRT,HEER}. To better understand the correspondence between the dynamics of both theories, we recall that the gravity has its characteristic property of non-linear interactions, which leads to gravitational forces. 

There have been successful progresses which explain a part of non-linear Einstein equation from the properties of entanglement entropy in CFTs based on the perturbative expansions \cite{Faulkner:2017tkh,Sarosi:2017rsq} along the line of \cite{Faulkner:2013ica,Nozaki:2013vta}. To go further, from a different perspective, it will also be helpful to understand how gravitational forces between two heavy objects are interpreted from the viewpoint of the dynamics of CFTs. This requires the full non-perturbative analysis of gravitational interactions. 

\begin{figure}[h!]
  \centering
  \includegraphics[width=7cm]{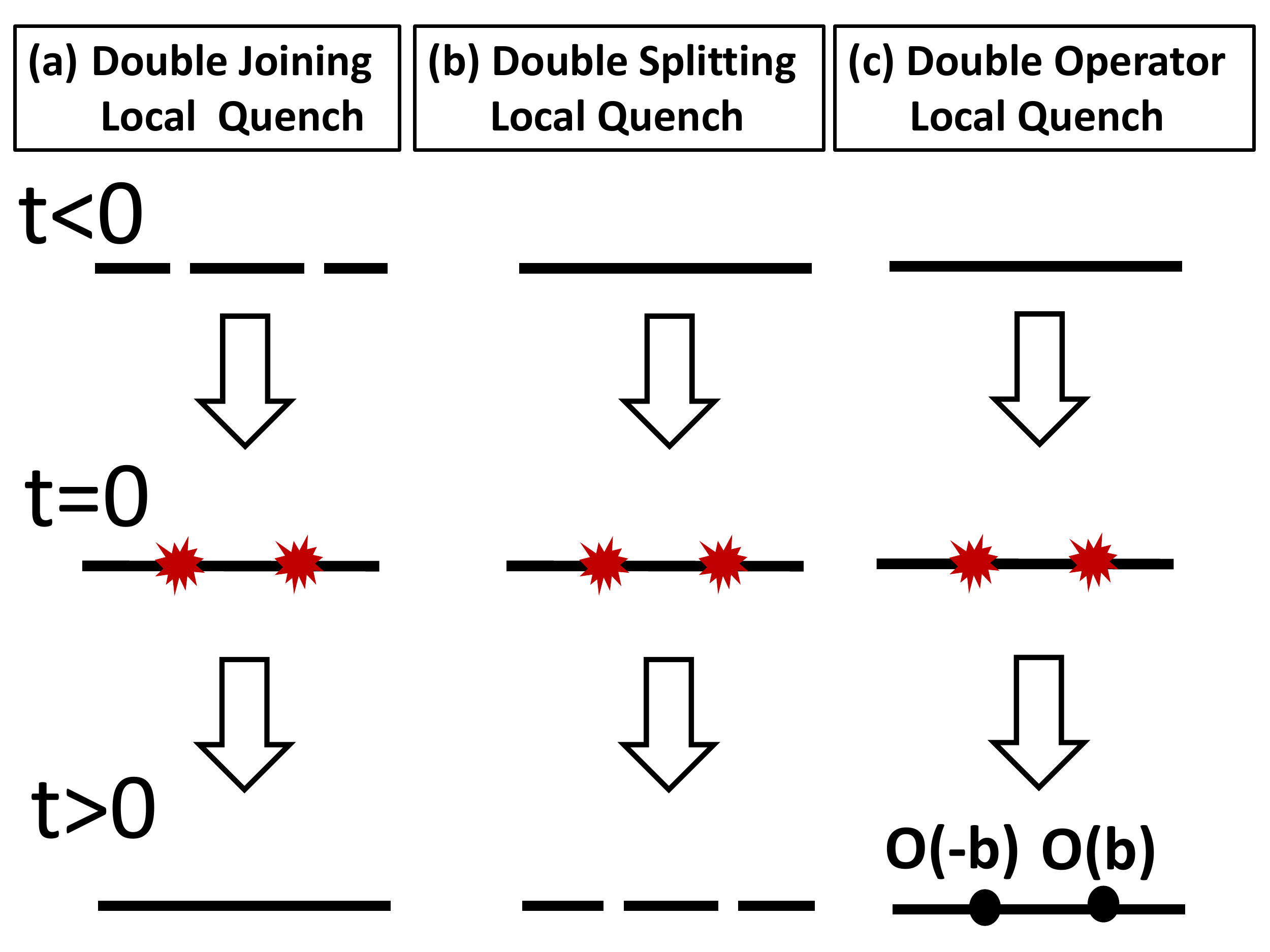}
 \caption{The three different double local quenches are sketched: the
 joining local quench ((a): left), the splitting local quench ((b): middle), 
and the operator local quench ((c): right) in two dimensional CFTs. We choose the two points where the local quench occurs to be $x=\pm b$. }
\label{fig:dqsetup}
\end{figure}

A purpose of this work is to study a class of explicit examples where gravitational force 
between two heavy objects in AdS plays a crucial role. For this we will study the simplest possible setups in AdS/CFT, namely double local quenches in two dimensional CFTs (2d CFTs). Even without thinking of the AdS/CFT, the double local quenches 
are at the same time intriguing non-equilibrium processes, which have not been studies well so far.
We consider three different types of local quenches: (a) Joining local quenches \cite{CCL}, (b) Splitting local quenches \cite{STW}, and (c) Operator local quenches \cite{NNT,Nozaki:2014uaa}, whose double quenches are depicted in figure \ref{fig:dqsetup}. 

A single local quench describes a local excitation at one point and is in general dual to a certain localized massive object via the AdS/CFT.  The gravitational force toward the AdS horizon dictates the motion of the object and its back-reaction leads to intriguing time-dependences of entanglement entropy. The precise holographic descriptions of the above three (single) local quenches (a), (b) and (c) were first worked out in \cite{Ugajin:2013xxa}, \cite{STW}, and \cite{HLQ}, respectively. The gravity duals of (a) and (b) are given by evolutions of spacetime boundary surfaces (or hard walls) in AdS \cite{STW,Ugajin:2013xxa} in the AdS/BCFT construction \cite{AdSBCFT}, while that of (c) is given by a massive particle falling into the AdS horizon \cite{HLQ}.
The results of entanglement entropy from gravity duals can be reproduced from the CFT calculations in the large $c$ limit (i.e. holographic CFTs) \cite{Asplund:2014coa}. Refer to \cite{Raj,NSTW} for other classes of topology changing quantum operations in CFTs such as projections and partial identifications. 
 
If we perform two local quenches at the same time (the locations of the two local quenches are taken to be $x=\pm b$), which we call the double local quench, then its gravity dual corresponds to two heavy objects in AdS.  Therefore, this double quench provides us a basic holographic setup where we can study gravitational force between two heavy objects. Note that, in addition, there is still the gravitational force which pulls the objects into the AdS horizon as in the single quench case. Hence, it is interesting to consider the following difference for a physical quantity $q$, which has positive contributions from excitations, such as energy density and entanglement entropy:
\ba
q^D-q^{S(x=b)}-q^{S(x=-b)}, \label{difQ}
\ea
where $q^D$ is the value of $q$ under the double local quench at $x=\pm b$ and $q^{S(x=b)}$ (or  $q^{S(x=-b)}$) is the quantity under a single local quenches at $x=b$ (or $x=-b$). If there is no interaction between two local quenches in the double quench, dual to the gravitational force between two objects, then the difference (\ref{difQ}) should vanish. Therefore, we believe that, such a quantity should a good probe of gravitational force in the holographic dual.

For double local quenches, we can again consider the three different setups 
(a) Joining local quenches, (b) Splitting local quenches, and (c) Operator local quenches, which are sketched in figure \ref{fig:dqsetup}. Only few results have been known so far for the double local quenches.
In \cite{Caputa:2016yzn,Numasawa:2016kmo}, the behavior of EE has been analyzed for (c) operator local quenches in 2d rational CFTs (RCFTs), such as free CFTs and minimal models. In this special case, the time evolution of EE is so simple that the EE is just a sum of two single operator local quenches i.e. (\ref{difQ}) does vanish. There have been no known results for more interacting CFTs, including holographic CFTs.
Also there have been no computations done for (a) joining and (b) splitting double local quenches of CFTs.

In this work, we will provide extensive investigations of double local quenches for (a) joining and  (b) splitting setups. It is intriguing to note that even though the path-integral description of single 
joining/single splitting quench can for both be conformally mapped into an upper half plane, the situation is different in double local quenches. The double joining quench is still described by an upper half plane. However, the double splitting quench is now transformed into an annulus and we will have a phase transition depending on the values of quench parameters. 

We will study the behaviors of the energy stress tensor and the entanglement entropy (EE). In the case of double splitting quenches, we will see characteristic oscillating behaviors, absent in the double joining quenches. 
We will also analyze the energy stress tensors for (c) double operator local quenches. Then we will analyze their differences (\ref{difQ}) and probe the gravitational forces from CFTs. To obtain the exact results, we will mainly work with two choices of CFTs: holographic CFTs and integrable CFTs such as the free fermion CFT and Ising model.

This work is organized as follows: In section two, we present a brief review of single local quenches as well as our strategy to calculate the EE. In section three, we give an outline of this work. We explain the quantities that we are interested in and our main results. In section four, we present our results of energy stress tensor and EE for (a) double joining local quenches. In section five, we show our results of energy stress tensor and EE for (b) double splitting local quenches. In section six, we study the behavior of energy stress tensor for (c) double operator local quenches. In section seven, we will summarize our conclusions and discuss future problems. In appendix A, we present explicit analytical expressions of 
single joining/splitting quenches. In appendix B, we explain the geometric picture of the connected geodesic length in the calculation of HEE for single joining quenches. 
In appendix C, we present the analytical calculations of  EE for the Dirac fermion CFT in the limit where the subsystem is far away from the quench points.

We became aware of a parallel work \cite{KuMiP}, where the evolution of entanglement entropy under double operator local quenches is studied for holographic CFTs. It is complementary to the present work.

%%%%%%%%%%%%%%%%%%%%%%%%%%%%%%
%%%%%%%%%%%%%%%%%%%%%%%%%%%%%%
\section{Brief Review of Single Local Quenches}
%%%%%%%%%%%%%%%%%%%%%%%%%%%%%%
%%%%%%%%%%%%%%%%%%%%%%%%%%%%%%
In this section, we give a brief review on the descriptions and known results of single local quenches (refer to \cite{STW} for more details). We will start with computations of entanglement entropy based on both the field theoretic and holographic analysis. We will consider three different types of local quenches: (a) Joining local quenches \cite{CCL}, (b) Splitting local quenches \cite{STW}, and (c) Operator local quenches \cite{NNT,Nozaki:2014uaa}. We will consider the descriptions of local quenches in the holographic CFT and Dirac free fermion CFT in two dimensions, so that we can have analytical control.  Then we will review the main features of entanglement entropy in three kinds of local quantum quenches.\footnote{A quantum quench is to prepare an initial state with a Hamiltonian $H_{0}$ and then see its time evolution under another Hamiltonian $H_{1}$. Usually, the ground state of $H_{0}$ is chosen to be the initial state. If $H_{0}-H_{1}$ has its supports only on a local space region, then we call it a local quench. ``Local operator quench'' below is not included in this definition. However we still call it a local quench by generalizing the idea of local quenches to localized excitations in field theories.} 
%%%%%%%%%%%%%%%%%%%%%%%%%%%%%%
\subsection{Entanglement entropy in 2d CFTs}\label{sec:EE}
%%%%%%%%%%%%%%%%%%%%%%%%%%%%%%
Let us start with the computation of entanglement entropy in 2d CFTs \cite{HLW,CC}.
For a density matrix $\rho$ defined on Hilbert space $\mathcal{H}_{tot}=\mathcal{H}_{A}\otimes\mathcal{H}_{A^c}$, the entanglement entropy (EE for short) of subsystem $A$ is defined by the von Neumann entropy:
\ba
S_A=-\mbox{Tr}[\rho_A\log\rho_A].
\label{EEdef}
\ea
Here, $A$ is a subsystem of the whole physical system, and $A^c$ is its complementary system. 
In this work we always choose $A$ to be an interval.
$\rho_A$ is the reduced density matrix defined by tracing out $A^c$: $\rho_A = \mbox{Tr}_{A^c} [\rho]$. Besides this, for a natural number $n\geq2$, the $n$-th R\'{e}nyi entropy of $\rho_A$ is defined by:
\ba
S_{A}^{(n)}=\frac{1}{1-n}\log\mbox{Tr}(\rho_A)^n,
\ea
and the von Neumann entropy is given by the $n\rightarrow1$ limit of the R\'{e}nyi entropy:
\ba
S_{A}=\lim_{n\rightarrow1}S_{A}^{(n)}.
\ea
When we want to compute EE in a quantum field theory, we often rely on its the replica trick definition
\ba
S_A = -\left.\frac{\partial}{\partial n} \operatorname{Tr} (\rho_{A})^{n}\right|_{n=1} = -\left.\frac{\partial}{\partial n}\log\Big(\operatorname{Tr} (\rho_{A})^{n}\Big)\right|_{n=1}. 
\ea

Let us then consider a 2d CFT on a plane $R^2$ and use complex coordinate $(w,\bar{w})$ to describe it. The Euclidean time and space coordinate $(\tau,x)$ are defined as
\be
w=x+i\tau,\ \ \ \bar{w}=x-i\tau.
\ee
To get the real time we can perform analytic continuation
\be
\tau=it.
\ee
For a primary operator $\Op$ in a 2d CFT, its two point function is given by
\be
\la \Op(w_1,\bar{w}_1)\Op(w_2,\bar{w}_2)\lb=\frac{1}{|w_1-w_2|^{2(h+\bar{h})}},
\label{twopp}
\ee
where $(h,\bar{h})$ is the chiral/anti-chiral conformal dimension of $O$.
For a subsystem $A=\{x|x\in(x_1,x_2)\}$ at Euclidean time $\tau$, 
\ba
\mbox{Tr}(\rho^n_A) \propto \la \sigma_n(x_1+i\tau,x_1-i\tau)\bar{\sigma}_n(x_2+i\tau,x_2-i\tau)\lb,
\ea
where $\sigma_n$ is the twist operator which has the conformal dimension $h=\bar{h}=\frac{c}{24}(n-1/n)$. Here, $c$ is the central charge of the CFT. Therefore the EE of subsystem $A$ for the CFT vacuum is given by
\be
S_A=-\frac{\de}{\de n}\log \la \sigma_n(x_1,x_1)\bar{\sigma}_n(x_2,x_2)\lb\Bigr|_{n=1}=\frac{c}{3}\log \frac{x_2-x_1}{\ep},
\label{gree}
\ee
where $\ep$ is the UV cut off corresponding to the lattice spacing. The R\'{e}nyi entanglement entropy can also be computed similarly. For the vacuum state, 
\be
S^{(n)}_A=\frac{c}{6}\left(1+\frac{1}{n}\right)\log \frac{x_2-x_1}{\ep}.
\ee
Also note that we can use a conformal transformation
\ba
\xi = f(w)
\ea
to map $(w,\bar{w})$ to a new coordinate $(\xi, \bar{\xi})$ and do the analysis on it. In this case, UV cutoff $\ep$ introduced in $(w,\bar{w})$ is mapped to $\ti{\ep}(\xi)$ in $(\xi, \bar{\xi})$, which is related to $\ep$ by\footnote{In this work, we always use $(w,\bar{w})$ to denote the physical system we are thinking about and other notations including $(\xi,\bar{\xi})$ to denote an artificial frame on which calculations are easier. So the physical UV cutoff $\ep$ is introduced in $(w,\bar{w})$ frame as a constant and corresponding UV cutoff $\ti{\ep}(\xi)$ turns out to be different at different spacetime points.}
\ba
\ti{\ep}(\xi) = |f'(w)|\ep.
\ea

If a CFT is defined on a manifold $M$ with boundaries $\de M$, and a linear combination of conformal symmetry is preserved on $\de M$, we call it a boundary conformal field theory (BCFT). Indeed, to describe (a) joining and (b) splitting local quenches we need to introduce such a conformal boundary.
Two point functions in a BCFT are essentially the same as four point functions (or even higher order correlation functions) in a CFT without boundaries. Therefore we cannot analytically compute them in general. However, in some special CFTs, such as Dirac free fermion CFT and holographic CFTs, the calculation of EE can be analytically performed even with boundaries as we will explain below. 

EE in the Dirac free fermion CFT can be explicitly computed on several different $M$. 
This is for example, when $M$ parameterized by $(w,\bar{w})$ has a single connected boundary, we can map it to an upper half plane with a conformal map\footnote{From now on, we always use $f$ to denote a conformal map from $(w, \bar{w})$ to some artificial coordinates, and use $g$ to denote its inverse.} $w = g(\xi)$. The EE is given by
\ba
S_A=\frac{1}{6}\log\left(\frac{|\xi_1-\xi_2|^2|\xi_1-\bar{\xi}_1||\xi_2-\bar{\xi}_2||g'(\xi_1)||g'(\xi_2)|}{\ep^2(\xi_1-\bar{\xi}_2)(\xi_2-\bar{\xi}_1)}\right), \label{saf}
\ea
(see \cite{NSTW,CH,ANT,UT} for details). 

%%%%%%%%%%%%%%%%%%%%%%%%%%%%%%%%%
\subsection{Holographic Entanglement Entropy}\label{sec:HEE}
%%%%%%%%%%%%%%%%%%%%%%%%%%%%%%%%%
A 2d holographic CFT has a 3d AdS dual. The holographic entanglement entropy (HEE)
\cite{RT,HRT,HEER} is given by 
\ba
S_A = \frac{L}{4G_N}
\label{HEE}
\ea
where $L$ is the length of the geodesic which connects the two boundary points of the subsystem $A$. Moreover we impose a homology condition which requires that the geodesic is homologous to the subsystem $A$ in the AdS geometry. In this work we always choose $A$ to be an interval.
The Newton constant $G_N$ is related to the central charge $c$ in the CFT by $1/(4G_N)=c/6$. In this work we set the AdS radius to be $1$. For example, the vacuum state of the CFT defined on $(w,\bar{w})$ is dual to an AdS geometry given by the Poincar\'{e} metric:
\ba
ds^2=\frac{dz^2+dwd\bar{w}}{z^2}.
\ea 
The length of the minimal geodesic which connects $(w,\bar{w})=(x_1+i\tau,x_1-i\tau),\ (x_2+i\tau,x_2-i\tau)$ is given by
\ba
L = \log\frac{(x_2-x_1)^2}{\ep^2}
\ea
and thus (\ref{HEE}) matches (\ref{gree}).
\begin{figure}
  \centering
  \includegraphics[width=7.5cm]{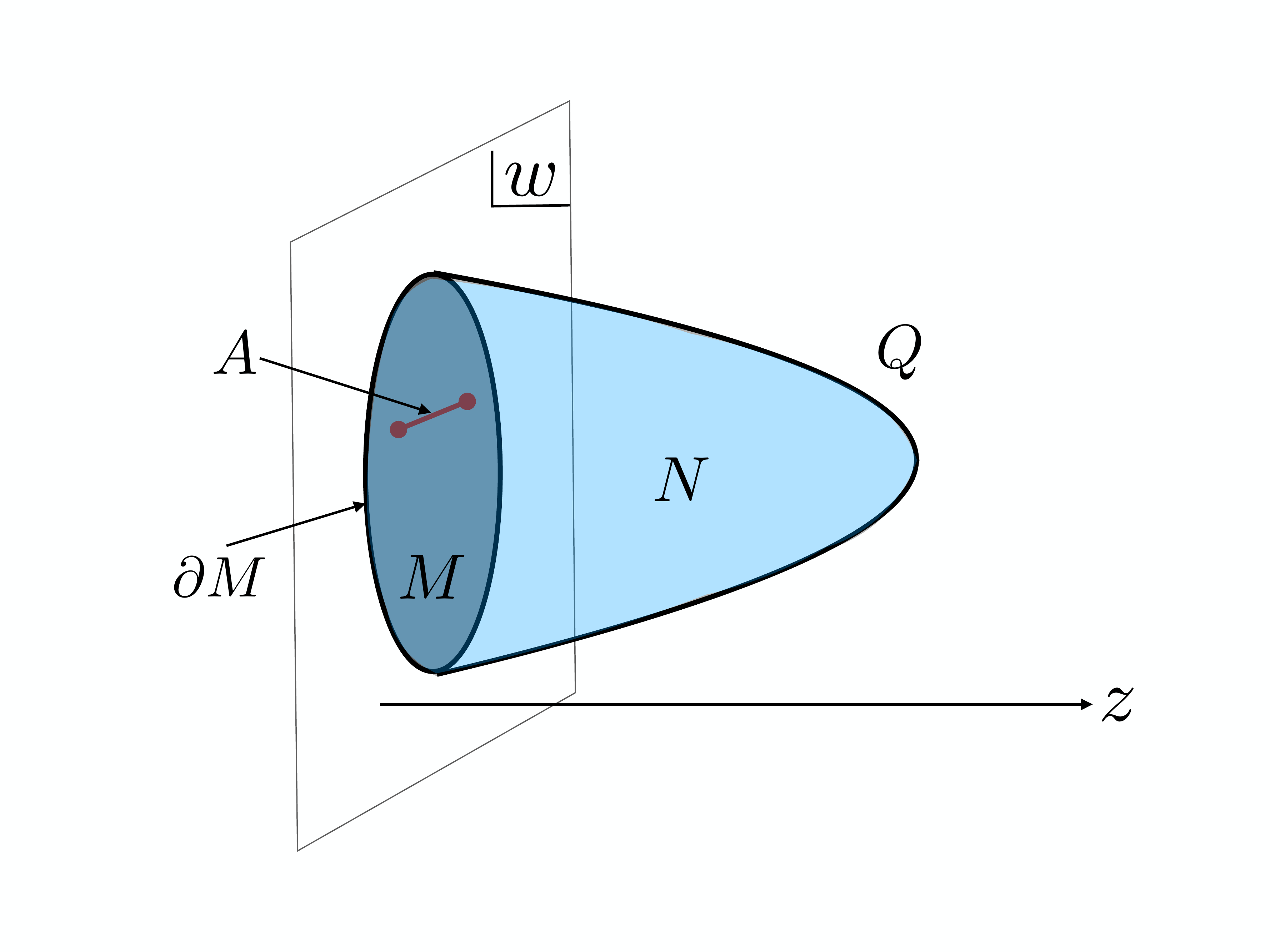}
   \includegraphics[width=7.5cm]{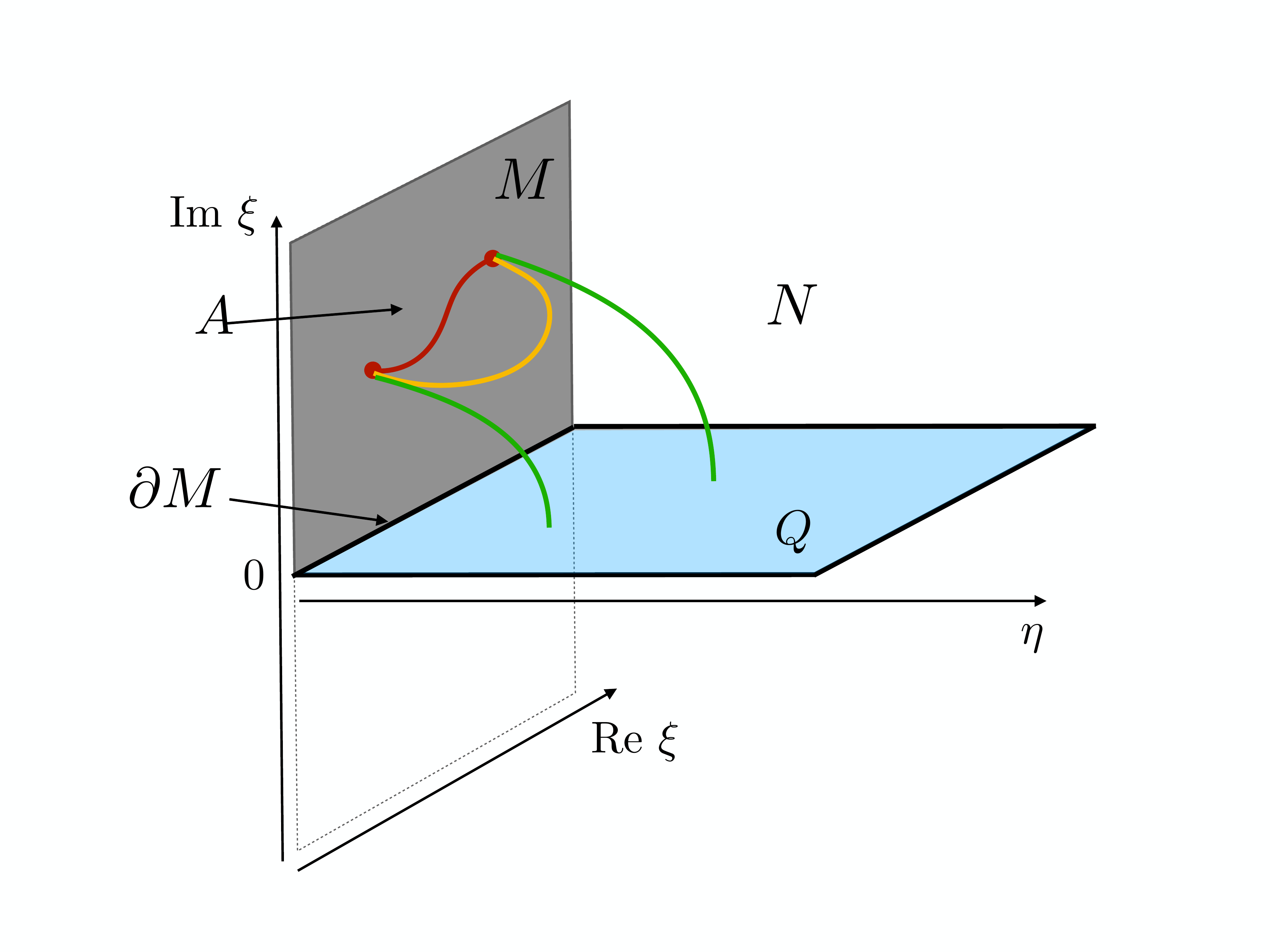}
 \caption{A sketch of AdS/BCFT setups for AdS$_3$. A holographic CFT on $M$ (with the boundary $\de M$) is dual to gravity on $N$. We have $\de N=M\cup Q$ and $\de Q = \de M$. The left picture shows a CFT defined on $M$ with coordinate $(w,\bar{w})$ where $M$ has a single connected boundary, and its gravity dual. The right picture shows how it looks like in $(\xi,\bar{\xi})$, where $M$ is a upper half plane, and its gravity dual. The red curve is the subsystem $A$. The yellow curve and the green curve are connected geodesic and disconnected geodesic, respectively.}
\label{fig:AdSBCFT}
\end{figure}

Let us then consider a conformal field theory defined on a manifold $M$ with boundaries $\de M$, where 
a half of the full conformal symmetries are preserved. Then this is a BCFT \cite{Cardy:1989ir} as we introduced before. A gravity dual of a BCFT can be described by following the AdS/BCFT construction \cite{AdSBCFT} (see also earlier work \cite{KR})\footnote{Though we focus on the three dimensional gravity dual, we can discuss higher dimensional setups in the same way (refer to e.g. \cite{Tonni,KNSW} for calculations of HEE in higher dimensional examples)}. The left picture in figure \ref{fig:AdSBCFT} sketches a typical AdS/BCFT setup. We call the manifold where the gravity dual is defined $N$. We introduce a boundary surface $Q$ in the bulk which satisfies $\de N=M\cup Q$ and $\de Q = \de M$. Also we impose the following boundary condition on $Q$
\be
K_{\mu\nu}-Kh_{\mu\nu}=-T_{BCFT}\cdot h_{\mu\nu}, \label{KT}
\ee
where $h_{\mu\nu}$ is the induced metric on $Q$ and $K_{\mu\nu}$ is the extrinsic curvature on $Q$;
$K$ is the trace $h^{\mu\nu}K_{\mu\nu}$. The constant $T_{BCFT}$ describes the tension of the ``brane'' or ``wall'' $Q$
and can take both positive and negative values in general. Since $Q$ in the bulk is dual to the boundary in the BCFT, it is natural to expect that the boundary conformal symmetry should be preserved. In fact, we can confirm that this boundary condition (\ref{KT})  preserves the boundary conformal symmetry in explicit examples \cite{AdSBCFT}. 

To find the metric in the bulk we need to solve the Einstein equation with the boundary condition (\ref{KT}), where the presence of $Q$ gives back-reactions and modifies the bulk metric \cite{AdSBCFT,NTU,Chu}.

The size of $N$ increases as the tension $T_{BCFT}$ gets larger, which implies that $T_{BCFT}$ estimates the degrees of freedom on the boundary $\de M$.
Indeed, as found in \cite{AdSBCFT}, the tension is monotonically related to the boundary entropy $S_{bdy}$ \cite{AFL}:
\be
S_{bdy}=\frac{c}{6}~\mbox{arctanh}(T_{BCFT}). \label{tenrs}
\ee

Holographic entanglement entropy in the AdS/BCFT setup is still written as (\ref{HEE}), where we regard the boundary surface $Q$ as just a point when we impose the homology condition. In practice, we divide the geodesics into two types: connected geodesics which connects two points given by $\de A$ on $M$, and disconnected geodesics each of which connects either of the two points $\de A$ and a point on the boundary surface $Q$ (as in the right picture in figure \ref{fig:AdSBCFT}). We call the quantities calculated from these two kinds of geodesics with (\ref{HEE}), connected EE $S_A^{con}$ and disconnected EE $S_A^{dis}$, respectively. The correct HEE is given by the minimum among them:
\ba
S_A = \min\{S_A^{con}, S_A^{dis}\}.
\ea
In a class of AdS/BCFT setups, we can firstly map it to a well-studied setup, and perform the calculation in the latter one. Figure \ref{fig:AdSBCFT} shows an example. The left figure shows a BCFT defined on a manifold $M$ with a connected boundary. As shown in the right figure, this can be mapped to an upper half plane with a conformal map $\xi = f(w)$. The geometry of the latter gravity dual is given by the Poincar\'{e} metric
\ba
ds^2=\frac{d\eta^2+d\xi d\bar{\xi}}{\eta^2},  \label{pool}
\ea
which is easy to work with. In this case, for a subsystem $A$ with boundary points $(w,\bar{w}) = (w_1,\bar{w}_1), (w_2,\bar{w}_2)$, connected HEE and disconnected HEE are given by
\ba
S_A^{con} &=& \frac{c}{6}\log \frac{|\xi_1-\xi_2|^2}{\ti{\ep}_1\ti{\ep}_2}=\frac{c}{6}\log\left[\frac{|f(w_1)-f(w_2)|^2}{\ep^2|f'(w_1)||f'(w_2)|}\right], \no
S_A^{dis} &=& \frac{c}{6}\log \frac{2\mbox{Im}\xi_1}{\ti{\ep}_1}+\frac{c}{6}\log \frac{2\mbox{Im}\xi_2}{\ti{\ep}_2} + 2S_{bdy} \\
&=&\frac{c}{6}\log\left(\frac{4(\mbox{Im}f(w_1))(\mbox{Im}f(w_2))}{\ep^2|f'(w_1)||f'(w_2)|}\right)+2S_{bdy}. \nonumber
\ea
The single joining/splitting and double joining local quench, are classified as this class of setups, i.e. can be mapped into the gravity dual of upper half-plane. On the other hand, as we will see later, the double splitting local quench is mapped into a gravity dual of a cylinder. However we can still apply, the basic rule of the AdS/BCFT formulation and the analysis of HEE remains the same. Refer also to
\cite{MRTW,VanRaamsdonk:2018zws,Cooper:2018cmb,Numa} for other quantum information theoretical understandings of AdS/BCFT.

%%%%%%%%%%%%%%%%%%%%%%%%%%%%%%%%%
\subsection{Single Joining Local Quench } \label{sec:SJQ}
\begin{figure}[t!]
  \centering
  \includegraphics[width=10cm]{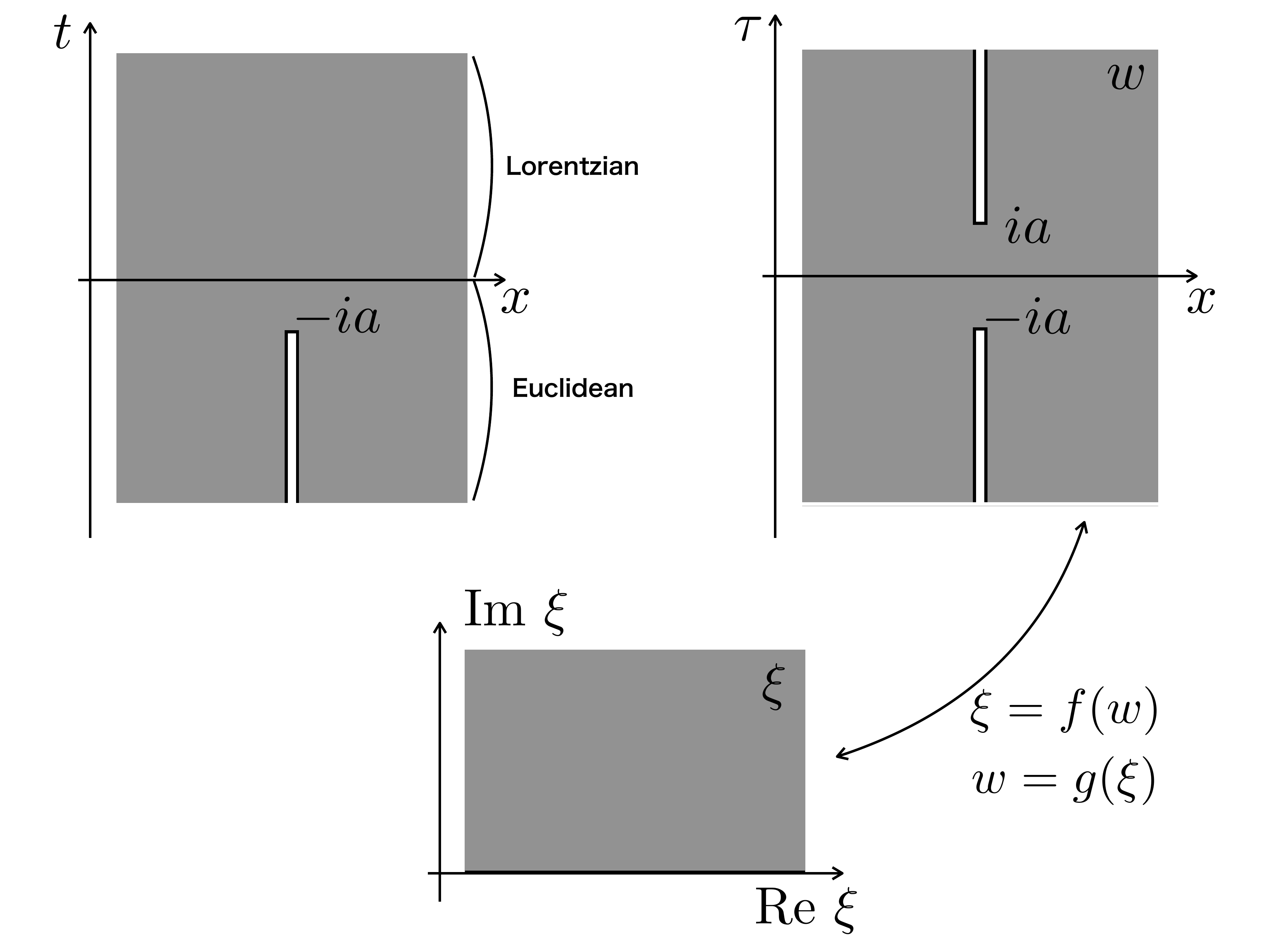}
 \caption{The left figure shows how to realize a single joining quench using path integral in a (1+1)d CFT. The right figure shows the corresponding Euclidean setup. It can be mapped into an upper half plane as showed in the lower figure using (\ref{SJQmap}) or (\ref{SJQmaprev}).}
\label{fig:SJQ}
\end{figure}

%%%%%%%%%%%%%%%%%%%%%%%%%%%%%%%%%
In a 2d physical system, let us prepare the initial state separately on $x<0$ and $x>0$ and then turn on the interaction at the neighborhood of $x=0$ at time $t=0$. We call it a single joining quench, because in this process we join two initially separated systems together, and we have exactly one joining point. We can use the path integral showed in figure \ref{fig:SJQ} to realize a single joining quench in 2d CFT. The Euclidean setup can be mapped into an upper half plane using the conformal map:\footnote{The map used here is related to that used in \cite{STW} by a simple coordinate transformation $w\rightarrow-w$. This map can be reduced from the map for the double joining quench (\ref{xcmap}) as we will see later.}
\ba
\xi=i\s{\f{w+ia}{ia-w}}\equiv f(w),  \label{pastef}
\label{SJQmap}
\ea
or equivalently 
\ba
w = ia\frac{\xi^2+1}{\xi^2-1}\equiv g(\xi).
\label{SJQmaprev}
\ea
Note that the $a$ here arises from the regularization of the local quench and has nothing to do with $\epsilon$, which is the physical cutoff corresponding to the lattice spacing. 

For a general subsystem $A=[x_1,x_2]$ where $0<|x_1|<x_2$, we can get the analytical results of the EE. However let us leave the details to appendix \ref{EESQ} or our previous work \cite{STW}. For numerical plots, we choose the subsystem as $A=[50,100]$ or $A=[0.1,1000]$ here as examples to summarize the main features of the EE after a single joining quench.

%%%%%%%%%%%%%%%%%%%%%%%%%%%%%%%%%
\subsubsection{EE in holographic CFT}
%%%%%%%%%%%%%%%%%%%%%%%%%%%%%%%%%

\begin{figure}[h!]
  \centering
  \includegraphics[width=7cm]{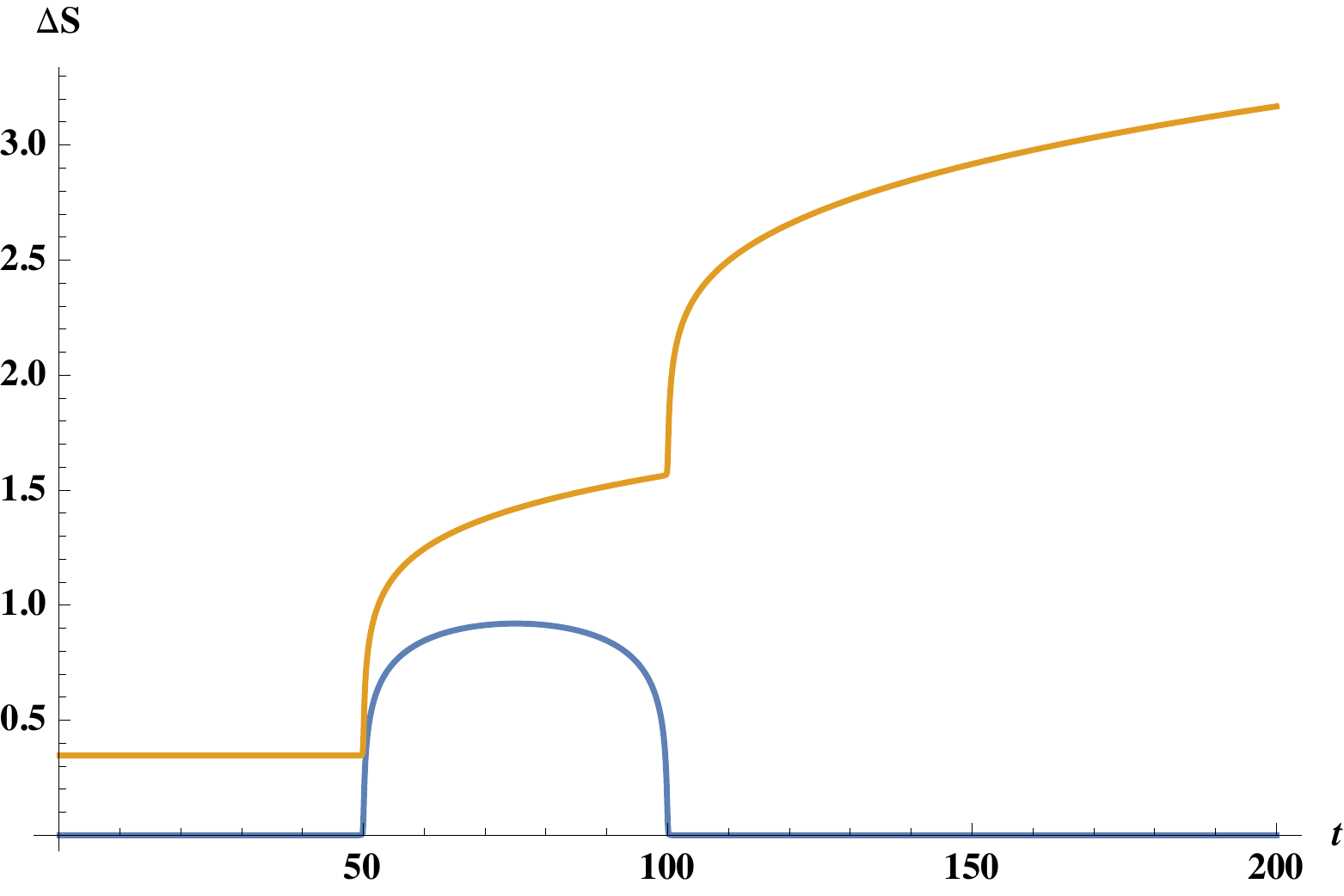}
  \includegraphics[width=7cm]{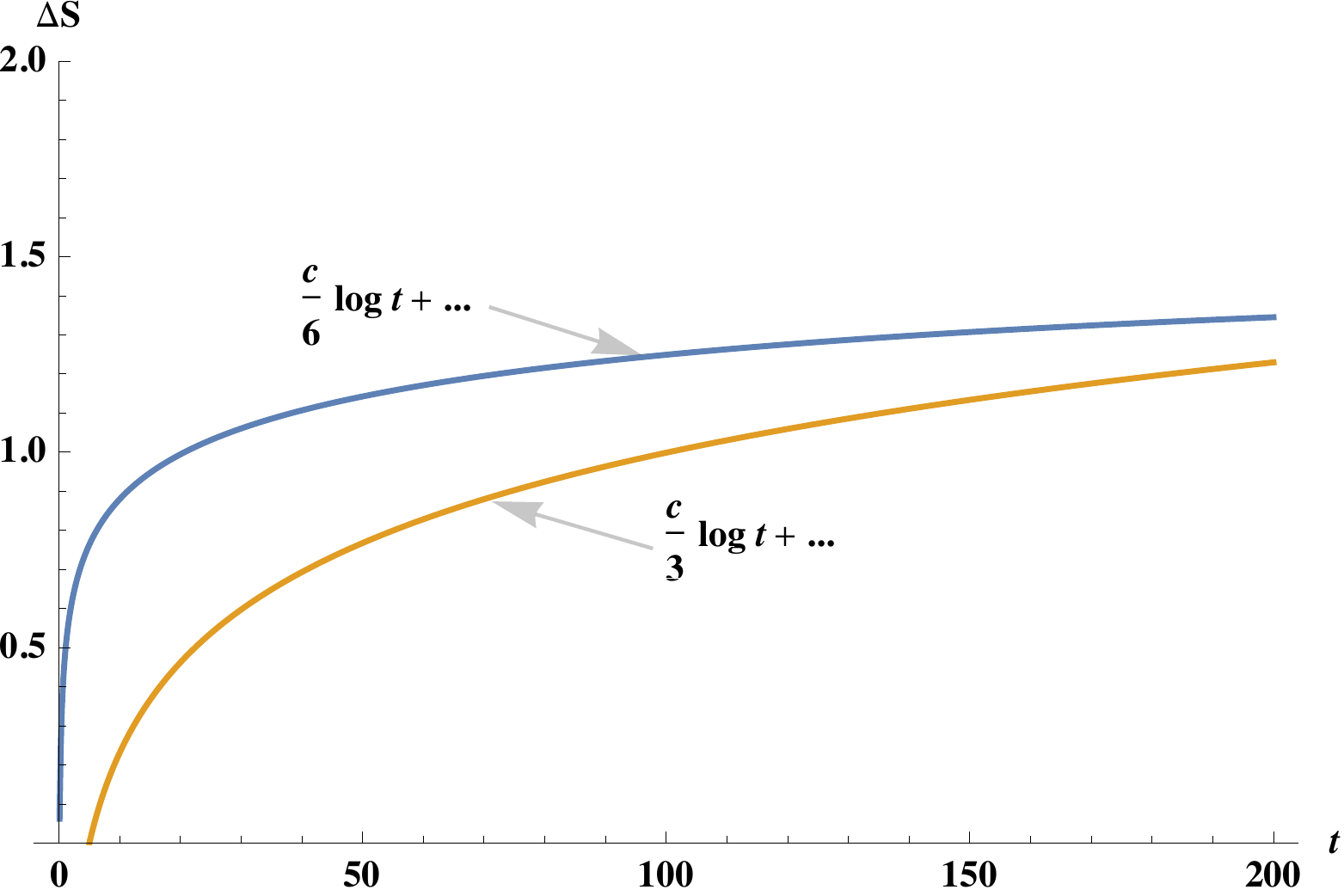}
 \caption{$\Delta S^{con}_A=S^{con}_A -S_A^{(0)}$ (blue lines) and $\Delta S^{dis}=S^{dis}_A -S_A^{(0)}$ (orange lines) after a single joining quench in a holographic CFT, where $S_A^{(0)}$ is the EE of the vacuum state. $A=[50,100]$ in the left figure and $A=[0.1,1000]$ in the right figure. We set $a=0.1$ and $c=1$. The boundary entropy $S_{bdy}$ is set to be zero.}
\label{fig:HolSJQ}
\end{figure}

The gravity dual can be constructed based on the AdS/BCFT \cite{Ugajin:2013xxa,STW} 
(see also \cite{AM}). The gravity counterpart of the map (\ref{pastef}) maps the gravity dual to just a half of Poincar\'{e} AdS.  The calculation of HEE follows from this construction.
Figure \ref{fig:HolSJQ} shows the connected EE and the disconnected EE after a single joining quench at $t=0$. We can see the following features in this figure.
First, both the connected EE and the disconnected EE have a discontinuity in their time derivative at $t=|x_1|$ and $t=x_2$. Next, at $|x_1|\ll t\ll x_2$, the connected EE and the disconnected EE has a logarithmic growth as below:
\ba
S_A^{con}&=&\frac{c}{6}\log \frac{t}{a}+... \ , \label{SJEEcon}\\
S_A^{dis}&=&\frac{c}{6}\log \frac{t}{a}+\frac{c}{6}\log \frac{t}{\epsilon}+... \ =\ \frac{c}{3} \log t+... \ . \label{SJEEdis}
\ea
These behaviors have clear geometric interpretations \cite{STW}, for which we will give a brief review in section \ref{sec:singleBS}. When the subsystem $A$ is semi-infinite, the disconnected one always dominates the HEE at $x_1\ll t\ll x_2$.
%%%%%%%%%%%%%%%%%%%%%%%%%%%%%%%%%
\subsubsection{EE in Dirac free fermion CFT}
\begin{figure}[h]
  \centering
  \includegraphics[width=7cm]{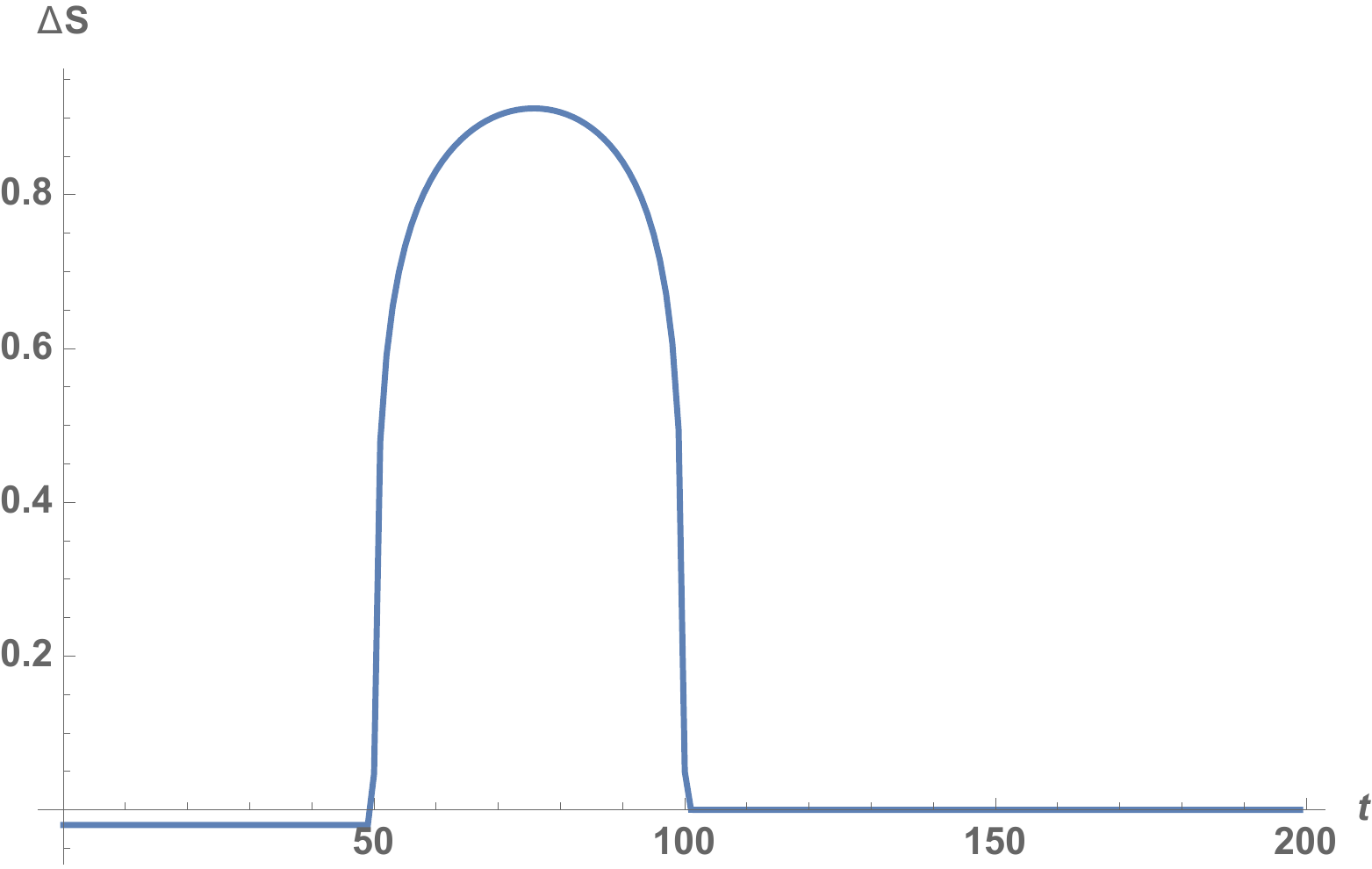}
  \includegraphics[width=7cm]{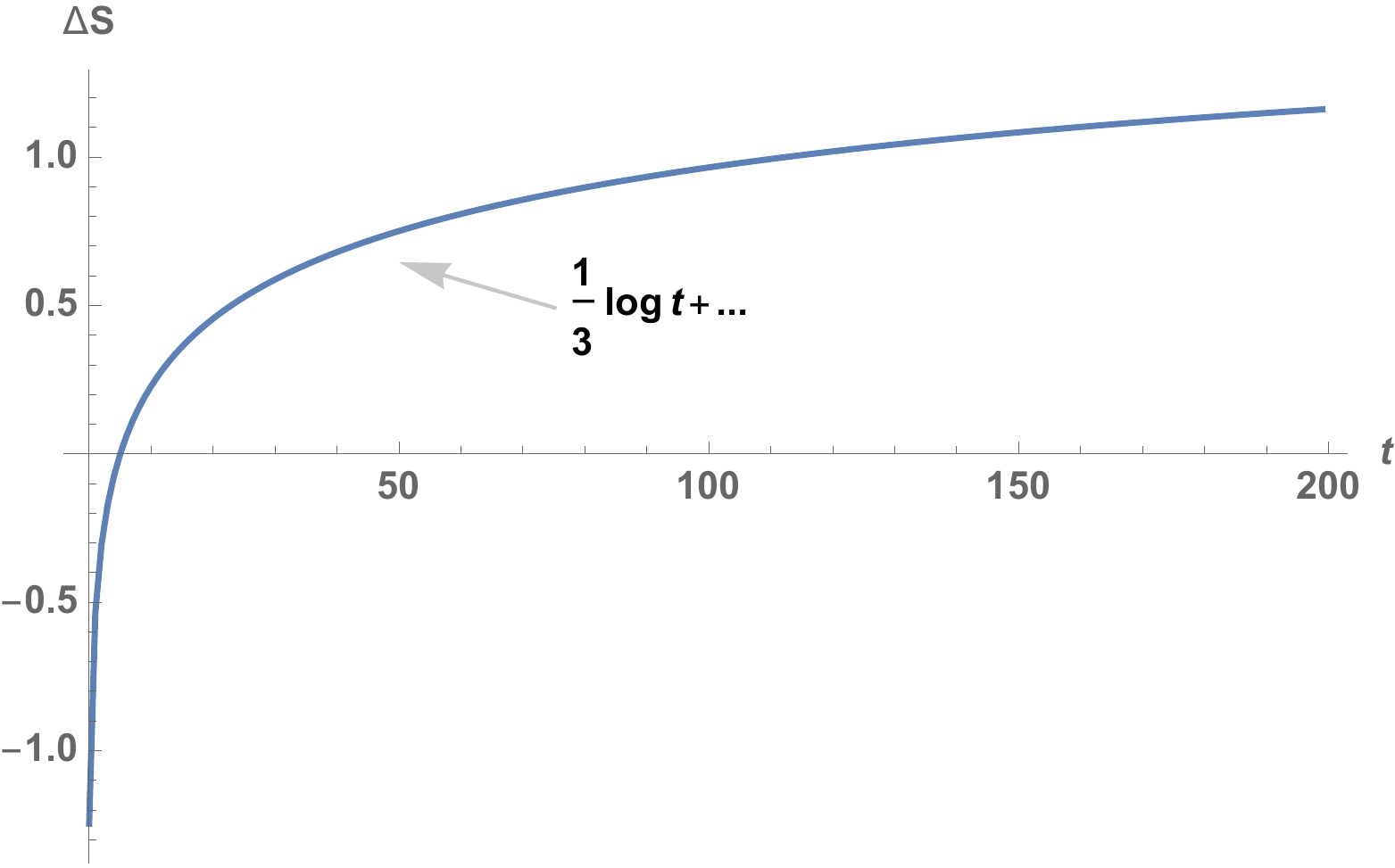}
 \caption{$\Delta S_A=S_A -S_A^{(0)}$ after a single joining quench in a Dirac free fermion CFT, where $S_A^{(0)}$ is the EE of the vacuum state. $A=[50,100]$ in the left figure and $A=[0.1,1000]$ in the right figure. We set $a=0.1$.}
\label{fig:DiracSJQ}
\end{figure}

%%%%%%%%%%%%%%%%%%%%%%%%%%%%%%%%%
Figure \ref{fig:DiracSJQ} shows the EE after a single joining quench at $t=0$. 
The EE has a discontinuity in its time derivative at $t=|x_1|$ and $t=x_2$.
At $|x_1|\ll t\ll x_2$, the EE has a logarithmic growth as below \cite{CCL,STW}:
\ba
S_A&=&\frac{1}{3}\log t+... \ .
\ea

%%%%%%%%%%%%%%%%%%%%%%%%%%%%%%%%%
\subsection{Single Splitting Local Quench}\label{sec:SSQ}
\begin{figure}
  \centering
  \includegraphics[width=10cm]{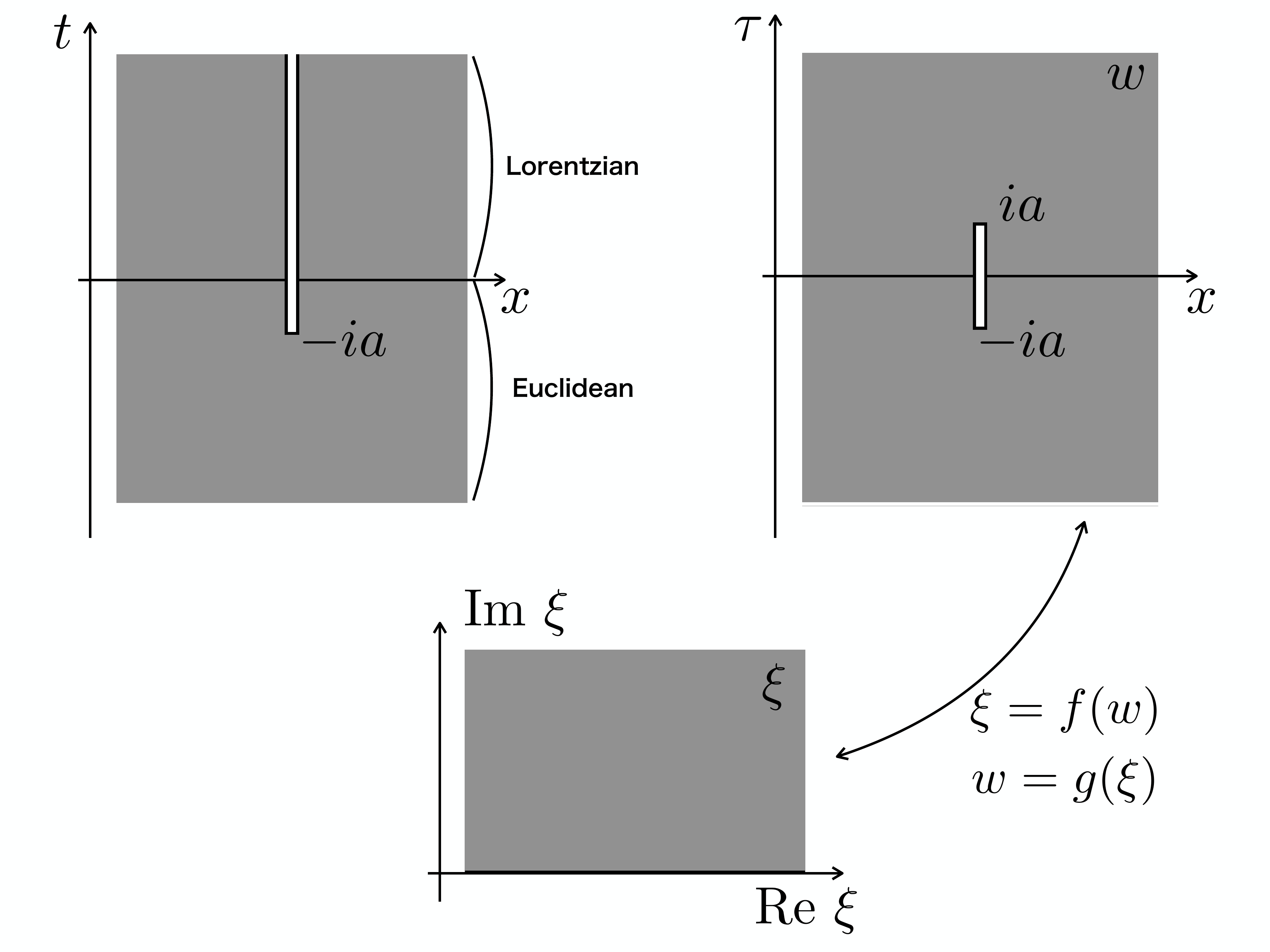}
 \caption{The left figure shows how to realize a single splitting quench using path integral in a (1+1)d CFT. The right figure shows the corresponding Euclidean setup. It can be mapped into an upper half plane as showed in the lower figure using (\ref{SSQmap}) or (\ref{SSQmaprev}).}
\label{fig:SSQ}
\end{figure}

%%%%%%%%%%%%%%%%%%%%%%%%%%%%%%%%%
In a 2d physical system, let us use a Hamiltonian which has a support on the whole space region to prepare the initial state and then turn off the interaction at the neighborhood of $x=0$ at time $t=0$. We call it a single splitting quench, because in this process we split an originally connected system into two parts, and we have exactly one splitting point. We can use the path integral showed in figure \ref{fig:SSQ} to realize a single splitting quench in 2d CFT. The Euclidean setup can be mapped into an upper half plane using the conformal map:
\ba
\xi=i\s{\f{w+ia}{w-ia}}\equiv f(w),  \label{pasteff}
\label{SSQmap}
\ea
or equivalently 
\ba
w = ia\frac{\xi^2-1}{\xi^2+1}\equiv g(\xi).
\label{SSQmaprev}
\ea
Again note that the $a$ here arises from the regularization of the local quench and has nothing to do with $\epsilon$, which is the physical cutoff corresponding to the lattice spacing. 

For a general subsystem $A=[x_1,x_2]$ where $0<|x_1|<x_2$, we can get the analytical results of the EE. However let us again leave the details to appendix \ref{EESQ}  and  \cite{STW}.
For numerical calculations we choose the subsystem $A=[50,100]$ and $A=[0.1,1000]$ here as examples to summarize the main features of the EE after a single splitting quench.

%%%%%%%%%%%%%%%%%%%%%%%%%%%%%%%%%
\subsubsection{EE in holographic CFT}
\begin{figure}[h]
  \centering
  \includegraphics[width=7cm]{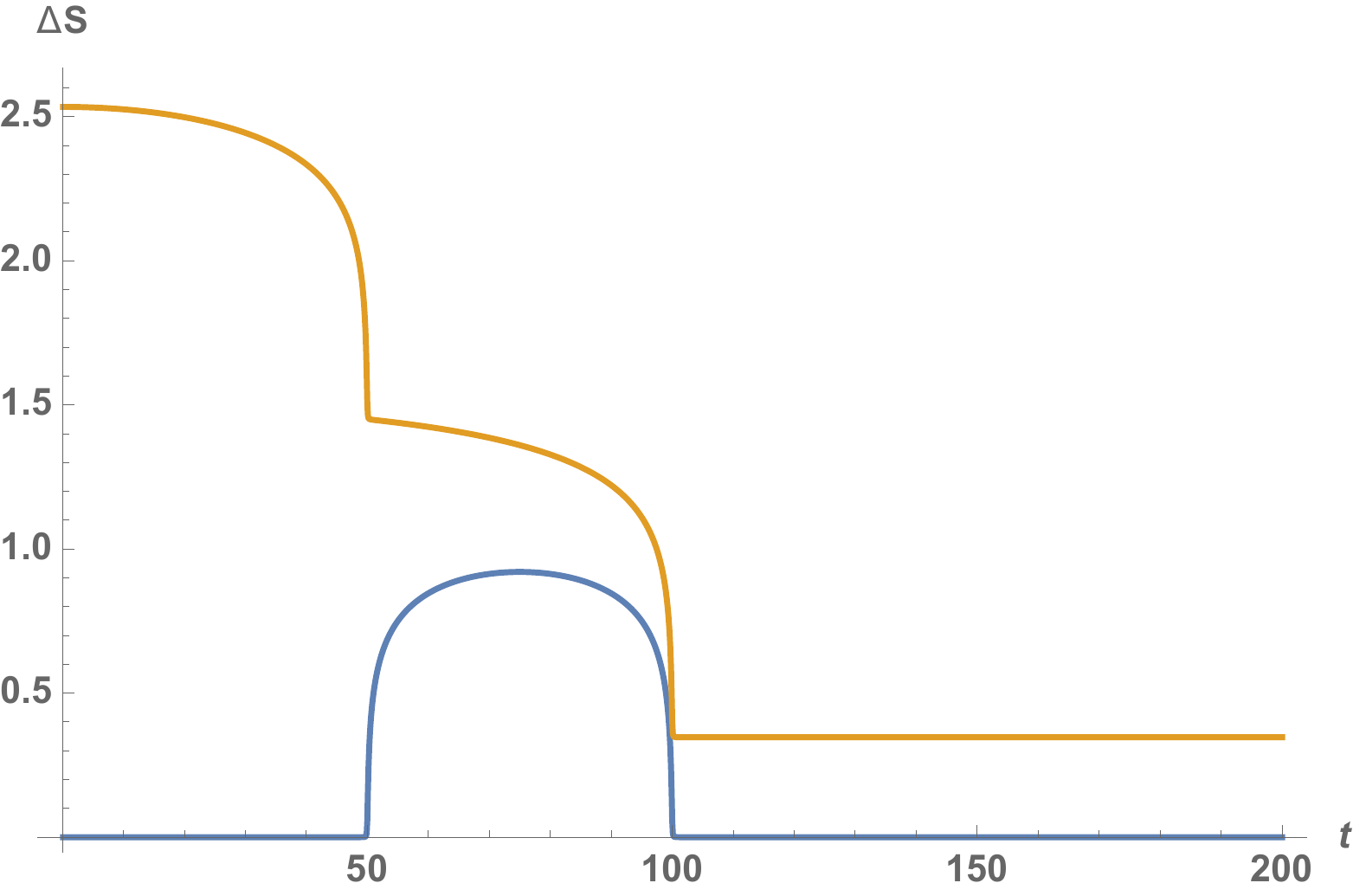}
  \includegraphics[width=7cm]{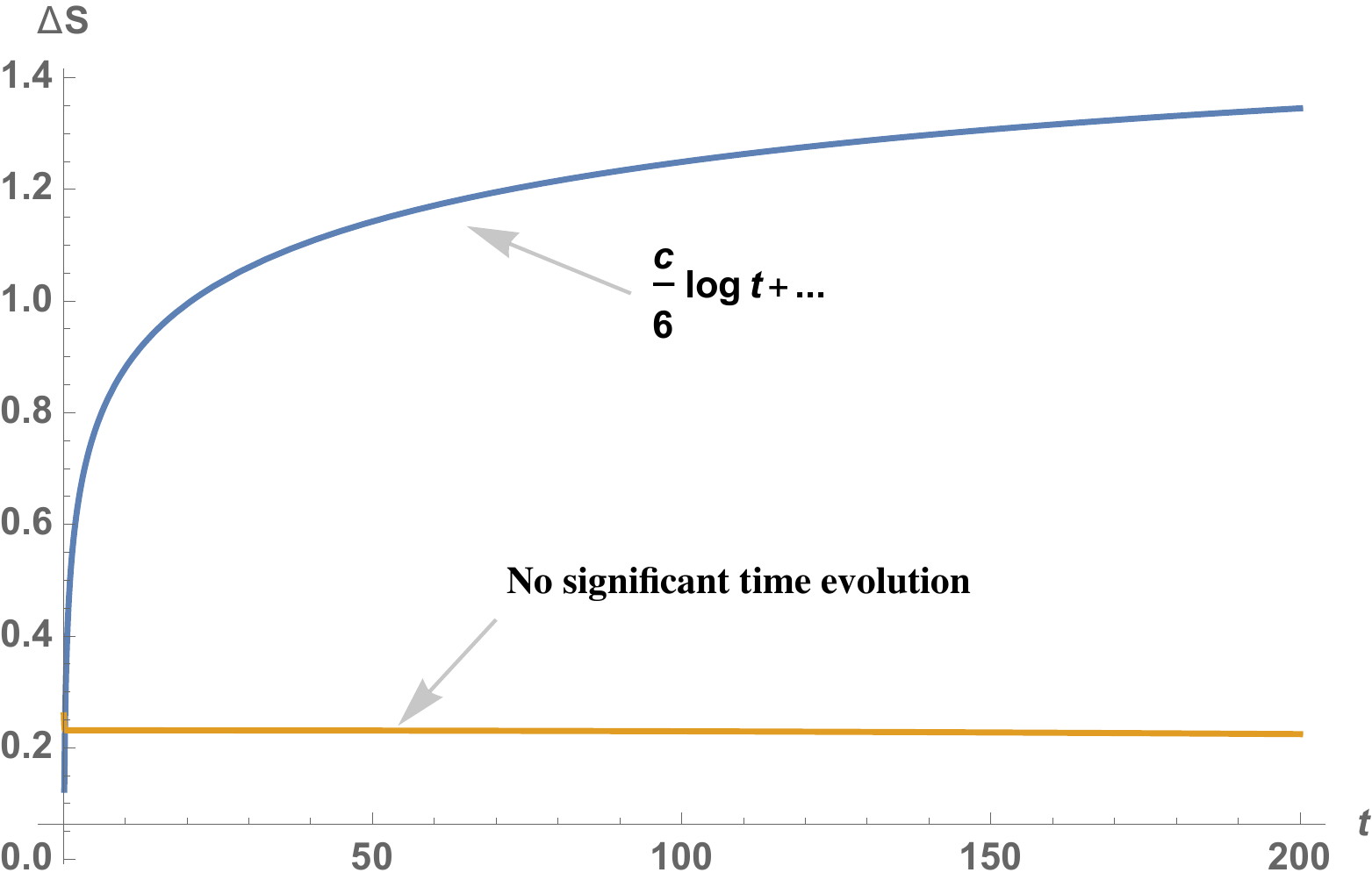}
 \caption{$\Delta S_A^{con}=S^{con}_A -S_A^{(0)}$ (blue lines) and $\Delta S_A^{dis}=S^{dis}_A -S_A^{(0)}$ (orange lines) after a single splitting quench in a holographic CFT, where $S_A^{(0)}$ is the EE of the vacuum state. $A=[50,100]$ in the left figure and $A=[0.1,1000]$ in the right figure. We set $a=0.1$ and $c=1$. The boundary entropy $S_{bdy}$ is set to be zero.}
\label{fig:HolSSQ}
\end{figure}

%%%%%%%%%%%%%%%%%%%%%%%%%%%%%%%%%
Again the gravity dual and its HEE can be found based on the AdS/BCFT \cite{STW}. 
Figure \ref{fig:HolSSQ} shows the connected EE and the disconnected EE after a single splitting quench at $t=0$. We can see the following features in this figure:
Both the connected EE and the disconnected EE have a discontinuity in their time derivative at $t=|x_1|$ and $t=x_2$. At $|x_1|\ll t\ll x_2$, the connected EE has a logarithmic growth. On the other hand, the disconnected EE has no significant time evolution:
\ba
S_A^{con}&=&\frac{c}{6}\log \frac{t}{a}+... \ ,\\
S_A^{dis}&=&\mbox{const.}+....
\ea
These behaviors also have clear geometric interpretations in the gravity dual \cite{STW}.
At late time, the disconnected one dominates the HEE.

\subsubsection{EE in Dirac free fermion CFT}
\begin{figure}[h]
  \centering
  \includegraphics[width=7cm]{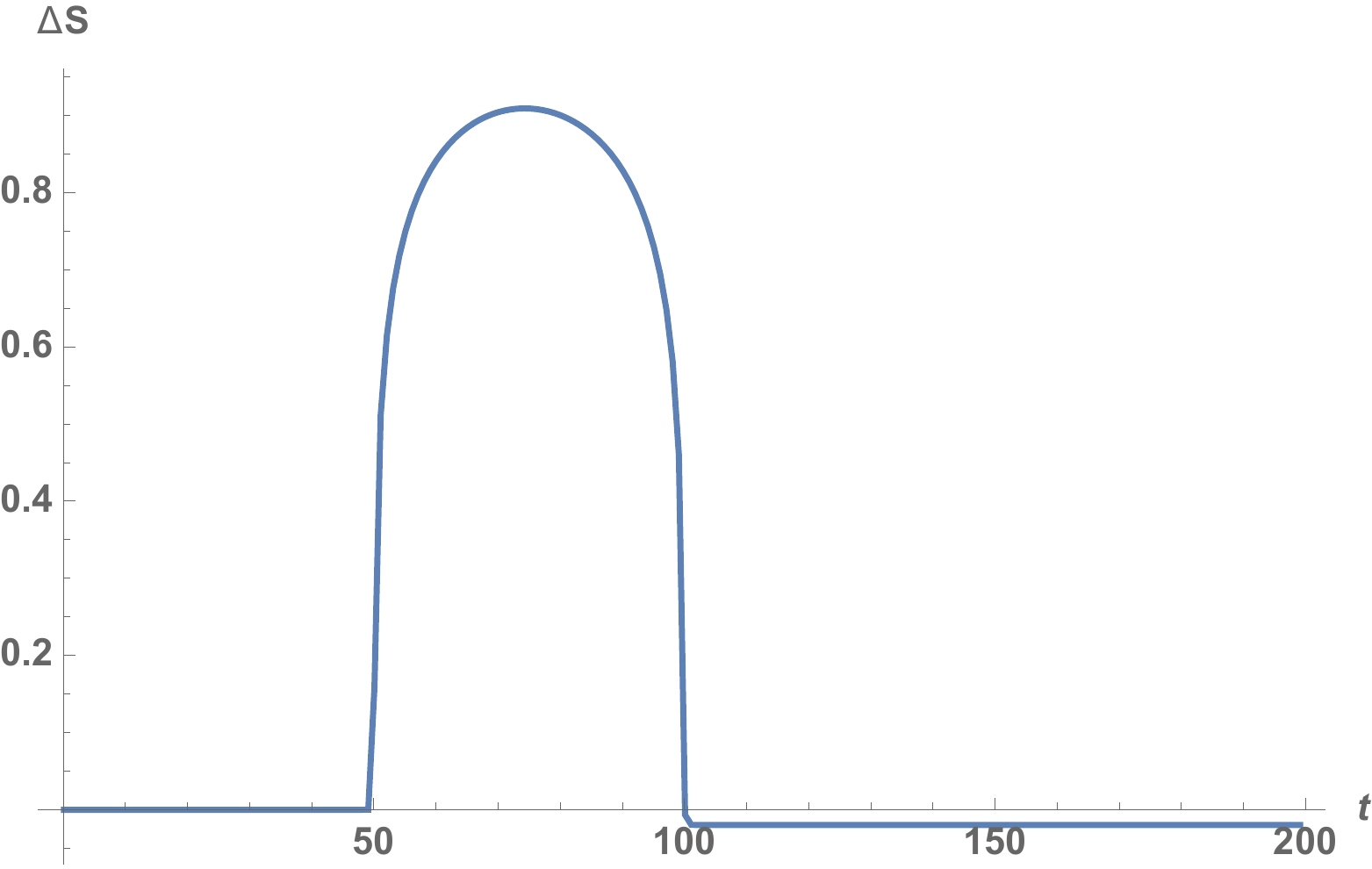}
  \includegraphics[width=7cm]{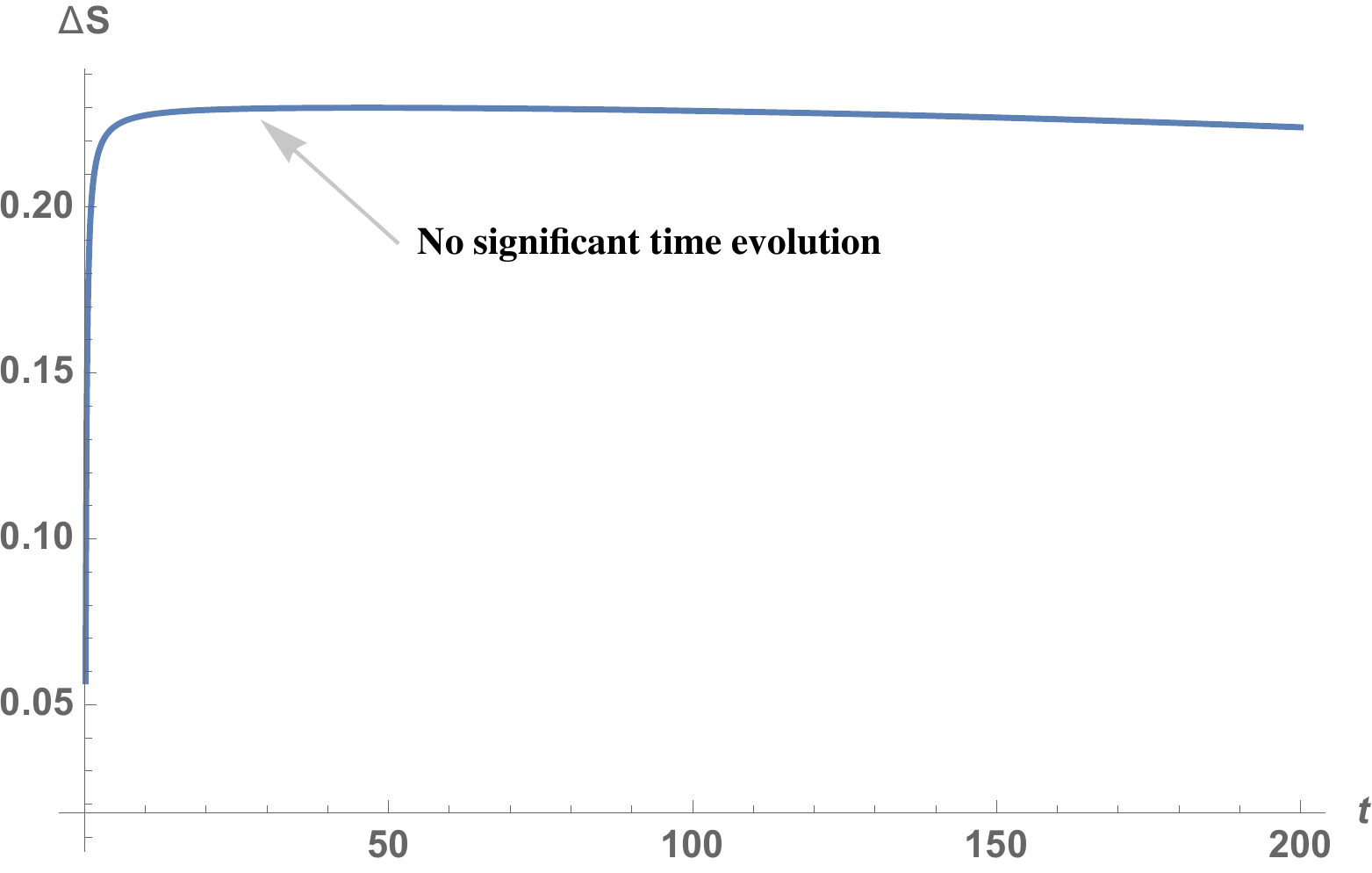}
 \caption{$\Delta S_A=S_A -S_A^{(0)}$ after a single splitting quench in a Dirac free fermion CFT, where $S_A^{(0)}$ is the EE of the vacuum state. $A=[50,100]$ in the left figure and $A=[0.1,1000]$ in the right figure. We set $a=0.1$.}
\label{fig:DiracSSQ}
\end{figure}

Figure \ref{fig:DiracSSQ} shows the EE after a single splitting quench at $t=0$. 
The EE has a discontinuity in its time derivative at $t=|x_1|$ and $t=x_2$.
At $|x_1|\ll t\ll x_2$, the EE has no significant time evolution:
\ba
S_A&=&\mbox{const.}+\ddd  \ .
\ea

%%%%%%%%%%%%%%%%%%%%%%%%%%%%%%%%%
\subsection{Single Operator Local Quench}\label{sec:SOQ}
%%%%%%%%%%%%%%%%%%%%%%%%%%%%%%%%%

Finally, we moved on to (c) single operator local quench \cite{NNT,Nozaki:2014uaa}.
The quenched state is produced by inserting a primary operator $\mathcal{O}(x)$ at a point $x=0$ and time $t=0$. Its time evolved state is expressed as
\ba
|\Psi(t)\lb={\ca N_{\Op}} e^{-iHt}\cdot e^{-a H}\Op(0)|0\lb,  \label{lopqs}
\ea
where $a$ is introduced as a regularization parameter which is infinitesimally small. $\ca N_{\Op}$ is introduced as a normalization factor to preserve the unit norm of the state. Note that $a$ is a regulator for local quench and has nothing to do with $\ep$ i.e. the UV cut off of the field theory itself. 

In this work, we focus on an operator quench in 2d CFTs. 
It is straightforward to calculate the energy stress tensor for this single operator quench. In the Lorentzian signature we obtain the value of left-moving component of energy stress tensor  
$T_{ww}(=T_{++})$ \cite{Caputa:2014vaa} at the spacetime point $(t,x)$ as follows 
\ba
T_{ww}=\frac{2\Delta_{\Op}a^2}{((x-t)^2+a^2)^2},
\ea
where $\Delta_{\Op}$ is the total conformal dimension (i.e. the sum of chiral and anti-chiral dimension) of the primary operator $\Op(x)$. Note that this result is universal and is true for any 2d CFTs. However, if we consider double operator local quenches, we lose this universal behavior and the expectation value of the energy stress tensor depends on which 2d CFT we consider as we will see 
later.

For the calculations of EE, we choose the subsystem $A$ to be an interval $[x_1,x_2]$. (we can take
$0<x_1<x_2$). In free CFTs or more generally rational CFTs (RCFTs), the time evolution of entanglement entropy $S_A$ is very simple \cite{NNT,Nozaki:2014uaa,He:2014mwa}. The growth of EE $\Delta S_A=S_A-S^{(0)}_A$ (here $S^{(0)}_A$ is the EE for the ground state) gets positive only during the time $x_1<t<x_2$, otherwise we have $\Delta S_A=0$.  This means that we can explain the behavior of EE by a simple relativistic particle propagation \cite{NNT,Nozaki:2014uaa,He:2014mwa}. The local quench creates an entangled pair at $x=0$ and 
each of the pair propagates at the speed of light in the opposite directions. Moreover, the amount of 
the growth $\Delta S_A>0$ for  $x_1<t<x_2$ is given by the logarithm of a quantity called quantum dimension \cite{He:2014mwa}.

On the other hand, in holographic CFTs, we find different behavior  of 
entanglement entropy \cite{HLQ,Caputa:2014vaa}. The single operator local quench at $x=0$ in a holographic CFT corresponds to a heavy particle falling towards the Poincar\'{e} horizon along $x=0$ in the AdS. In this description, (\ref{oplog}) comes from the back reaction of the heavy particle. In particular, when  $x_1\ll t \ll x_2$
the HEE shows a logarithmic time evolution
\ba
\Delta S_A\simeq \frac{c}{6}\log \frac{t}{a}, \label{oplog}
\ea
which is missing in the previous RCFT results. The behavior (\ref{oplog}) can be reproduced from field theoretic analysis for large central charge CFTs \cite{Asplund:2014coa}. Refer to \cite{AB,Caputa:2014eta,deBoer:2014sna,Guo:2015uwa,Chen:2015usa,Nozaki:2015mca,Caputa:2015qbk,Caputa:2015tua,Caputa:2016tgt,Caputa:2015waa,Stikonas:2018ane,Rangamani:2015agy,Sivaramakrishnan:2016qdv,Caputa:2016yzn,Numasawa:2016kmo,Nozaki:2016mcy,David:2016pzn,Caputa:2017tju,Nozaki:2017hby,JaTa,AKT,He:2017lrg,KuTa,Guo:2018lqq,Ku,KuMi} for further developments of operator local quenches.

%%%%%%%%%%%%%%%%%%%%%%%%%%%%%%%%%%%%%%%%%%%%%%%%%%%%%
\section{Outline of Our Analysis for  Double Local Quenches}\label{sec:Out}
%%%%%%%%%%%%%%%%%%%%%%%%%%%%%%%%%%%%%%%%%%%%%%%%%%%%%

Before we get into detailed computations for three types of double local quenches  (a) Joining, (b) Splitting, and (c) Operator, as depicted in figure \ref{fig:dqsetup}, we would like to give an outline of our following analysis.
In particular, here, we would like to explain quantities we will study during double local quenches and what kind of behavior we expect from their gravity duals for holographic CFTs. 

The most interesting aspect of double local quenches which is missing in single ones is the interactions
between two local excitations. If we consider holographic CFTs, these interactions correspond to gravitational forces between two heavy objects in AdS, which are dual to the two excitations 
via the AdS/CFT. The objects are massive particles for (a) operator local quench \cite{HLQ},
while they are extended objects in (b) joining and (c) splitting local quenches \cite{STW}. 
Therefore, the difference between a double local quench and two single ones is expected to be related to the gravitational forces between the two objects. 

In this work, we trigger the double quench at the two points $x=\pm b$. Therefore we should choose the two single quenches to occur at $x=b$ and $x=-b$, respectively, so that the difference is well-defined. More explicitly, let us chose a physical quantity $q$ which vanishes at the CFT vacuum and increases in the presence of local excitations, and consider the difference $q^{D(x=\pm b)}-q^{S(x=b)}-q^{S(x=-b)}$, where $q^{D(x=\pm b)}$ and $q^{S(x=\pm b)}$ denotes the value of the quantity under a double local quench and single one quenched at the specified points, respectively.

We would like to argue that for right choices of $q$, we have the inequality 
\ba
q^{D(x=\pm b)}-q^{S(x=b)}-q^{S(x=-b)}\leq 0,   \label{ineq}
\ea
and that this negative value is a manifestation of the fact that the gravitational force is attractive. 

A natural candidate of $q$ is the energy density or expectation value of the energy stress tensor because it vanishes at the ground state and increases under the local quenches. Another candidate of $q$ is the growth of entanglement entropy $\Delta S_A=S_A-S^{(0)}_A$ ($S^{(0)}_A$ is the EE for the ground state). 

First of all, it is clear that the difference vanishes when the distance between two quenches gets larger i.e. $b/a\to\infty$, where we remember that $a$ is the quench cut off parameter:
\ba
\lim_{b/a\to\infty} \left[q^{D(x=\pm b)}-q^{S(x=b)}-q^{S(x=-b)}\right]=0. \ \ \  (\mbox{for (a),(b), and (c)}).  \label{abc}
\ea 
This is simply because we can neglect the correlations between two objects when the distance gets larger. 
Moreover, if we focus on (a) joining and (b) splitting local quench, it is possible to see that the opposite limit $b/a\to 0$ is equivalent to a single quench at $x=0$ (this is also obvious from figure \ref{fig:dqsetup})
, which leads to 
\ba
\lim_{b/a\to 0} q^{D(x=\pm b)}=q^{S(x=0)}. \ \ \ (\mbox{for (a) and (b)}).  \label{bc}
\ea
Clearly these relations (\ref{abc}) and (\ref{bc}) are consistent with the argued inequality (\ref{ineq}).

The purpose of our analysis in this work is then to examine the inequality (\ref{ineq}) for the energy stress tensor $T_{ww}$ and the growth of EE $\Delta S_A=S_A-S^{(0)}_A$ for generic values of $b/a$.  For the energy stress tensor, we will be able to confirm the inequality (\ref{ineq}) for the three types of local quenches (a),(b),(c). This result turns out to be true for any 2d CFTs in the case of (a) and (b). We can show it is true for all CFT we studied (holographic CFTs, Ising CFT and free CFTs) in the case of (c).

For the growth of EE, as we will find from numerical analysis for (a) joining and (b) splitting local quenches, the inequality (\ref{ineq}) is true in holographic CFTs as long as the subsystem $A$ is away from the quench points $|x|=b$ by a certain finite distance.\footnote{If we take the limit $A$ to be infinitely far away, then the first law of EE \cite{Bhattacharya:2012mi}
tells us that the EE is proportional to the energy density and thus this property is just reduced to that of the energy stress tensor.} However, we will observe that the inequality (\ref{ineq}) is not true in general 
for the free Dirac fermion CFT in the case of (a) and (b) even when the subsystem is far away from the quench points. On the other hand, in the case of (c), for the Dirac fermion CFT or more generally rational CFTs,  it was already known that the inequality (\ref{ineq}) for EE is saturated \cite{Numasawa:2016kmo}.

\begin{figure}
  \centering
  \includegraphics[width=7cm]{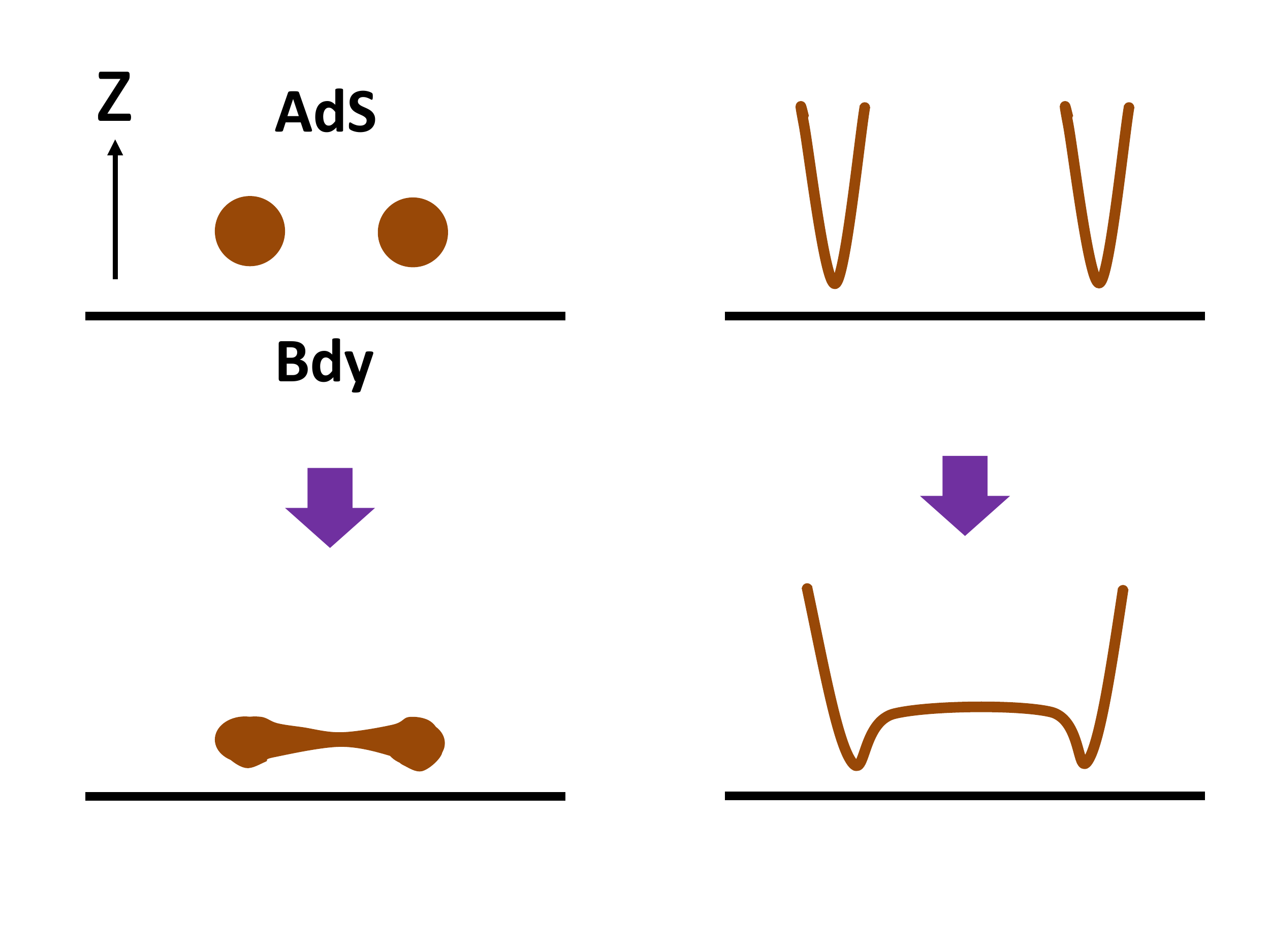}
 \caption{Sketches of gravity duals of double local quenches. The upper left (or right) picture describes the gravity dual of a simple superposition of two single operator (or joining) local quenches, where no gravitational force between two heavy objects (brown colored ones) is taken into account. The lower left (or right) depicts a gravity dual of double operator (or joining) local quench where the back reactions are incorporated. 
In the left, a local operator is dual to a massive particle. In the right, the heavy string is dual to the boundary surface in the joining local quench. In both examples, gravitational forces bring the two objects closer and their back-reaction appears weaker to a distant observer.}
\label{fig:gravityd}
\end{figure}

In the gravity dual picture we can understand the inequality (\ref{ineq}) in an intuitive way as explained 
in figure \ref{fig:gravityd}. Consider two heavy objects placed in AdS. Due to their gravitational force, they will tend to attract each other. This makes their configuration squeezed towards the center in the bulk AdS and thus an outside observer feels that the back-reaction due to these objects is getting reduced. This attractive nature of gravitational force makes the value of $q$ smaller for double quenches and in this way we expect the non-positivity of the difference i.e. (\ref{ineq}). This feature common to the three different double quenches can be explicitly observed, for example, in the behaviors of energy stress  tensor, which are depicted in  the middle picture 
of  Fig.\ref{fig:ratio} for the double joining quench (a), in the right picture of Fig.\ref{EMsp} for the double splitting quench (b), and in the Fig.\ref{fig:RatioTTf}
for the double operator local quench (c).

After this intuitive outline, from the next section, we will begin more technical analysis of the double local quenches.

%%%%%%%%%%%%%%%%%%%%%%%%%%%%%%%%%
%%%%%%%%%%%%%%%%%%%%%%%%%%%%%%%%%
\section{Double Joining Local Quenches}\label{sec:DJLQ}
%%%%%%%%%%%%%%%%%%%%%%%%%%%%%%%%%
%%%%%%%%%%%%%%%%%%%%%%%%%%%%%%%%%

\begin{figure}
  \centering
  \includegraphics[width=9cm]{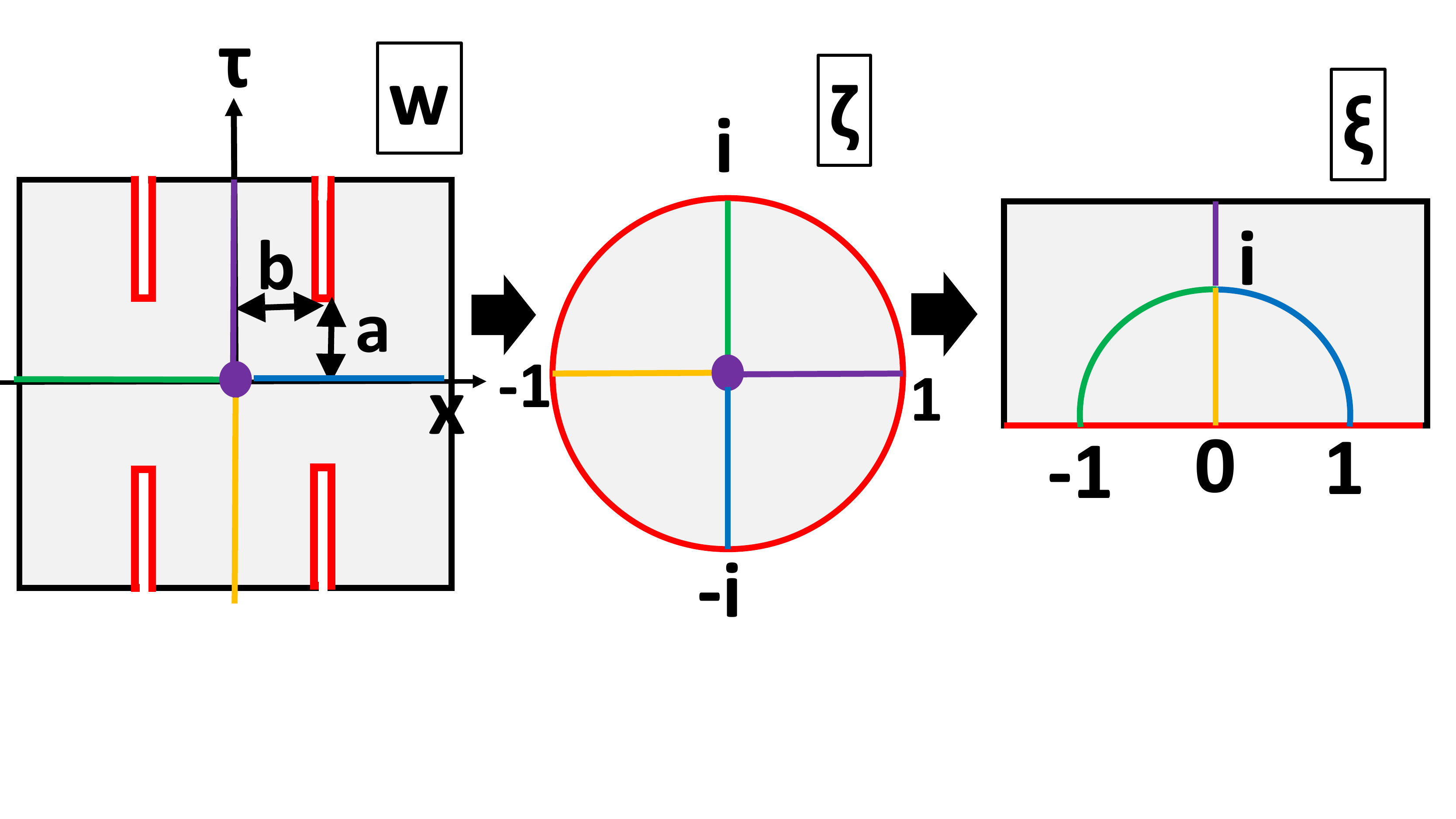}
 \caption{The conformal map (\ref{xcmap}).}
\label{fig:cmap}
\end{figure}

To describe a double joining local quench, let us consider the map \cite{Nehari} (refer to figure \ref{fig:cmap})
\ba
&& w=i\left(\frac{\sin^2\ap}{2}\log\left(\frac{1+\zeta}{1-\zeta}\right)+\cos^2\ap\left(\frac{\zeta}{1+\zeta^2}\right)\right)\equiv g(\xi), \no
&& \zeta=\frac{\xi-i}{\xi+i}. \label{xcmap}
\ea
This combined map $w=g(\xi)$ transforms a complex plane $(w,\bar{w})$ with four vertical slits into an upper half plane $(\xi,\bar{\xi})$ with Im$\,\xi>0$. Also note that the middle coordinate $\zeta$ describes the unit radius disk $|\zeta|\leq 1$.

\begin{figure}
  \centering
  \includegraphics[width=7cm]{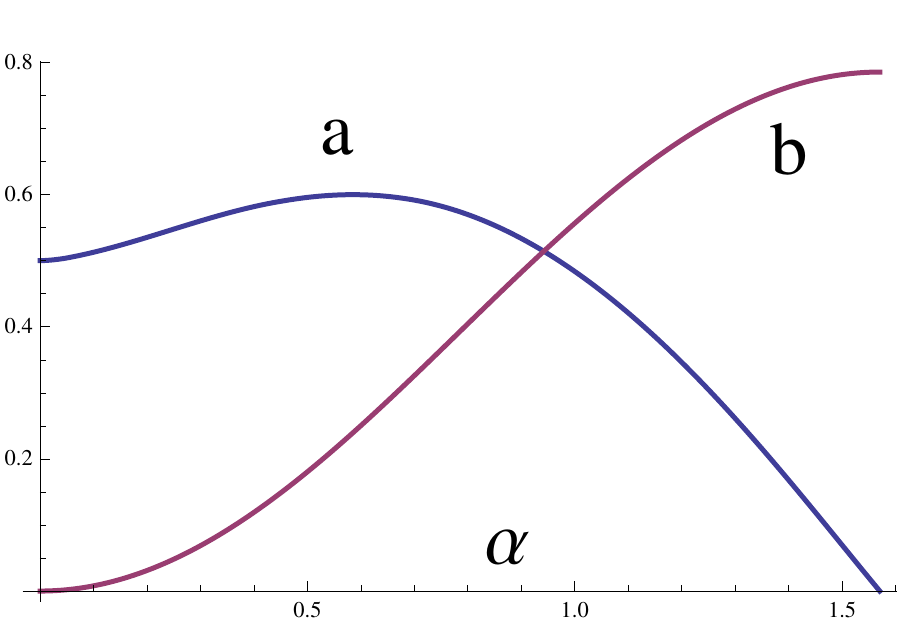}
 \caption{The plot of $a_0$ (blue) and $b_0$ (red) as a function of $\ap$.}
\label{fig:ab}
\end{figure}

The end points of the four splits, which extend to infinity, in the former are given by $\pm b_0\pm ia_0$,
where 
\ba
&& a_0(\ap)=\frac{\sin^2\ap}{2}\log\left(\cot\frac{\ap}{2}\right)+\frac{1}{2}\cos\ap,\no
&& b_0(\ap)=\frac{\pi}{4}\sin^2\ap. \label{qqq}
\ea
The behavior of $a_0$ and $b_0$ as functions of $\ap$ is depicted in figure \ref{fig:ab}.
It is useful to note that if we set $\ap=0$, then we find $w=i\frac{\zeta}{1+\zeta^2}$, which leads to the map
\ba
\xi=i\s{\frac{\frac{i}{2}+w}{\frac{i}{2}-w}}.
\ea
This coincides with the single joining quench with $a=\frac{1}{2}$.

It is straightforward to analyze more general values of $a$ and $b$ by simply rescaling them 
$(a,b,w,\bar{w})\to \lambda(a,b,w,\bar{w})$, where $\lambda$ is an arbitrary positive constant. 
Therefore below we treat $a$ and $b$ as independent parameters of the double quench.

We introduce the $R^2$ coordinate $(\tau,x)$ as before: 
\be
w=x+i\tau,
\ee
where $\tau$ is the Euclidean time and we can consider its Lorentzian continuation as
$\tau=it$,
The coordinate $(w,\bar{w})$ describes a double joining local quench where the points 
$x=b$ and $x=-b$ are joined at the same time as in figure \ref{fig:dqsetup}.

\begin{figure}[h!]
  \centering
  \includegraphics[width=7cm]{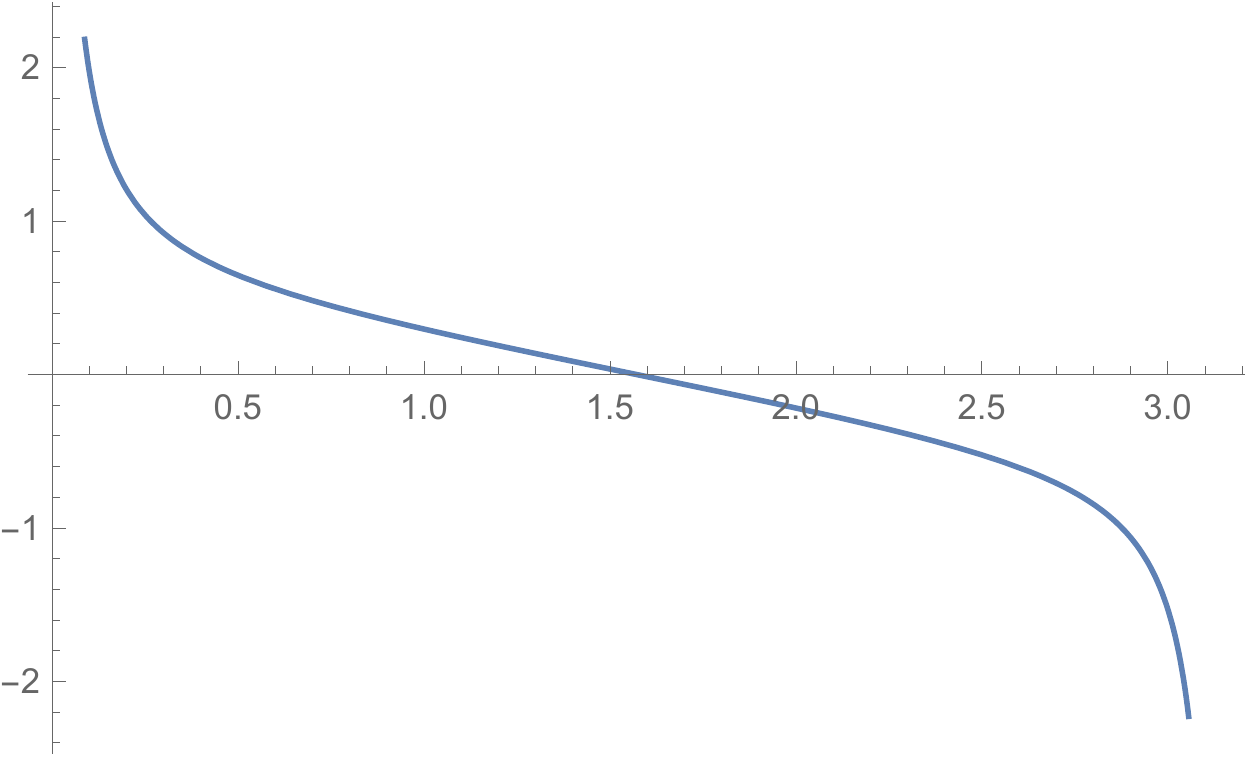}
 \caption{The graph of the function  $g(e^{i\theta})$ defined by (\ref{gt}) at $\alpha=1$.}
\label{fig:invmap}
\end{figure}

In the Lorentzian time evolution, the coordinate 
$(w,\bar{w})=(x-t,x+t)$ takes real values.
 Then we introduce real valued functions $(\theta,\bar{\theta})$ such that 
\ba
w=x-t=w(e^{i\theta}),\ \ \ \ \bar{w}=x+t=w(e^{i\bar{\theta}}),  \label{wmapa}
\ea
where we defined
\be
w(e^{i\theta})=\frac{1}{2}\cot\theta\cdot\cos^2\ap+\frac{1}{2}\left(\frac{\pi}{2}-\theta\right)\sin^2\ap,
\label{gt}
\ee
as depicted in figure \ref{fig:invmap}. This function $w$ takes the values: 
\be
w(\theta=0)=\infty,\ \ \  w(\theta=\pi/2)=0,\ \ \  w(\theta=\pi)=-\infty.
\ee
Note that the real valued coordinates $\theta$ and $\bar{\theta}$ are independent.
Only when $t=0$, we have $\theta=\bar{\theta}$.

Finally, we can numerically find the inverse functions
\ba
\theta=\theta(x-t),\ \ \ \ \bar{\theta}=\bar{\theta}(x+t),   \label{wmapaa}
\ea
when we would like to compute time evolutions of various quantities.

%%%%%%%%%%%%%%%%%%%%%%%%%%%%%%%%%%%
\subsection{Energy Stress Tensor for Double Joining Local Quenches}
%%%%%%%%%%%%%%%%%%%%%%%%%%%%%%%%%%%

Since the energy stress tensor vanishes in the coordinate $(\xi,\bar{\xi})$ we find the energy stress tensor in the original coordinate $(w,\bar{w})$ from the standard conformal transformation rule as 
\ba
T_{ww}=\frac{c}{24}\cdot \frac{2g'(\xi)g'''(\xi)-3g''(\xi)^2}{g'(\xi)^4}.
\ea

For example, for a single joining local quench at $x=0$ with the regularization parameter $a$, we find the energy stress tensor 
\be
T^{S(x=0)}_{ww}=\frac{c}{6}\cdot \frac{3a^2}{4(w^2+a^2)^2}.
\ee

Now let us turn to the double joining local quench which we are interested in.
In particular, we can explicitly confirm that at $\ap=0$ (i.e. $a=1/2$ and $b=0$), 
the above energy stress tensor for the double local quench coincides with that for a single local quench
$T^{S(x=0)}_{ww}$.

\begin{figure}
  \centering
 \includegraphics[width=4cm]{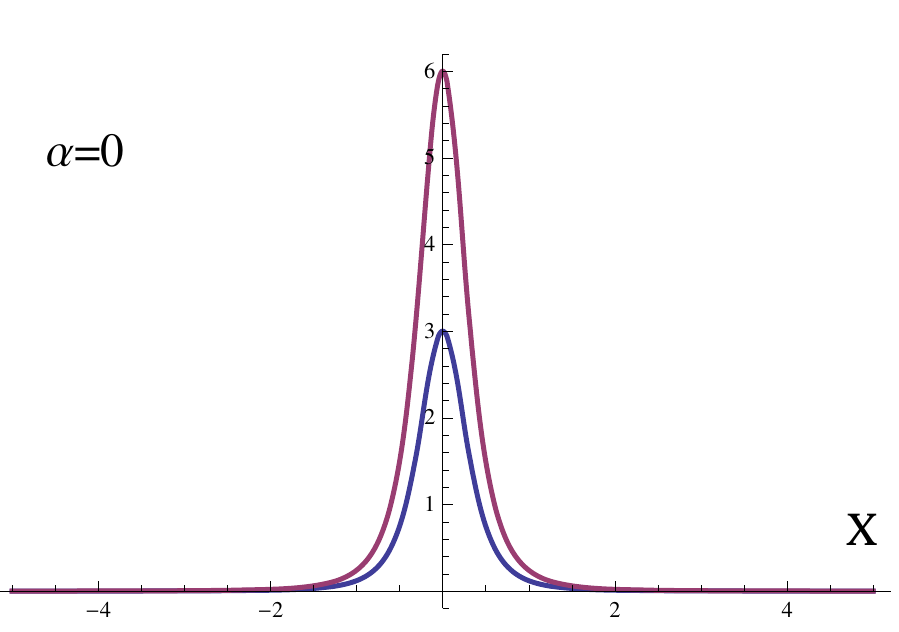}
\hspace{3mm}
  \includegraphics[width=4cm]{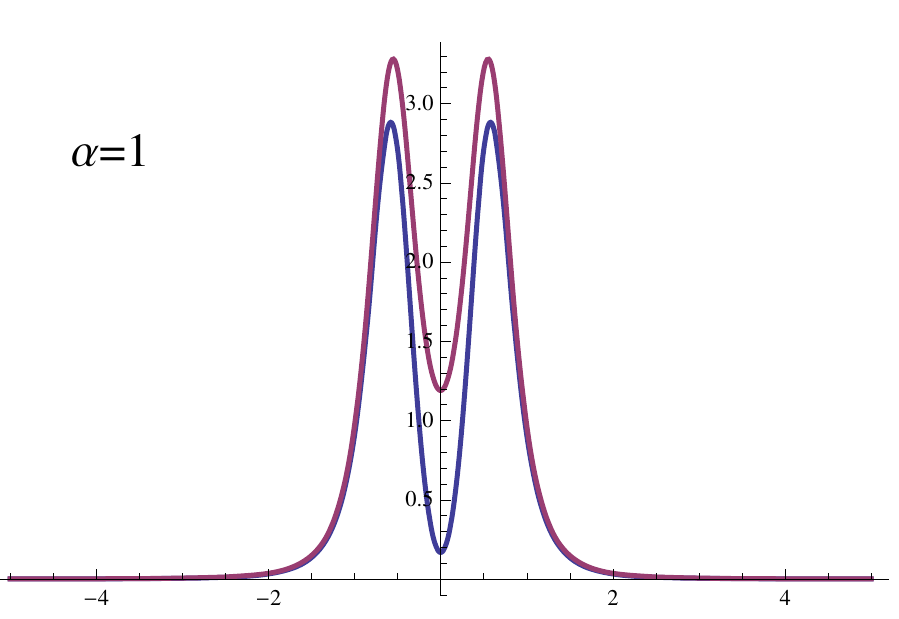}
\hspace{3mm}
  \includegraphics[width=4cm]{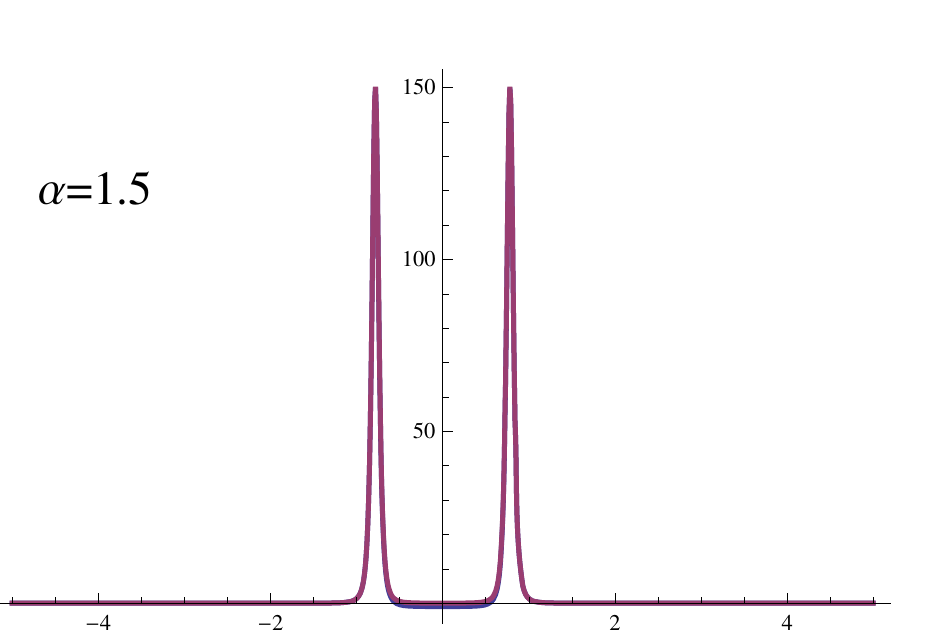}
\hspace{3mm}
 \includegraphics[width=4cm]{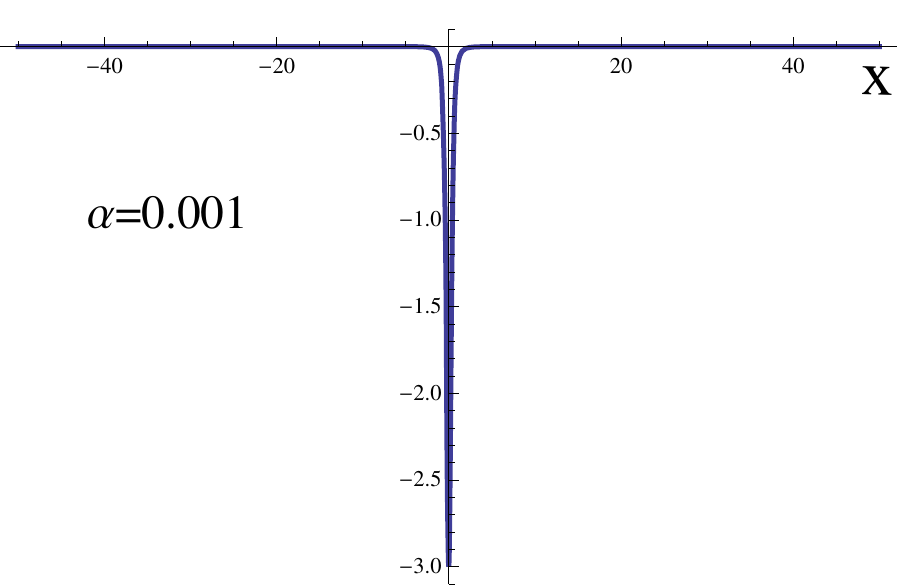}
\hspace{3mm}
  \includegraphics[width=4cm]{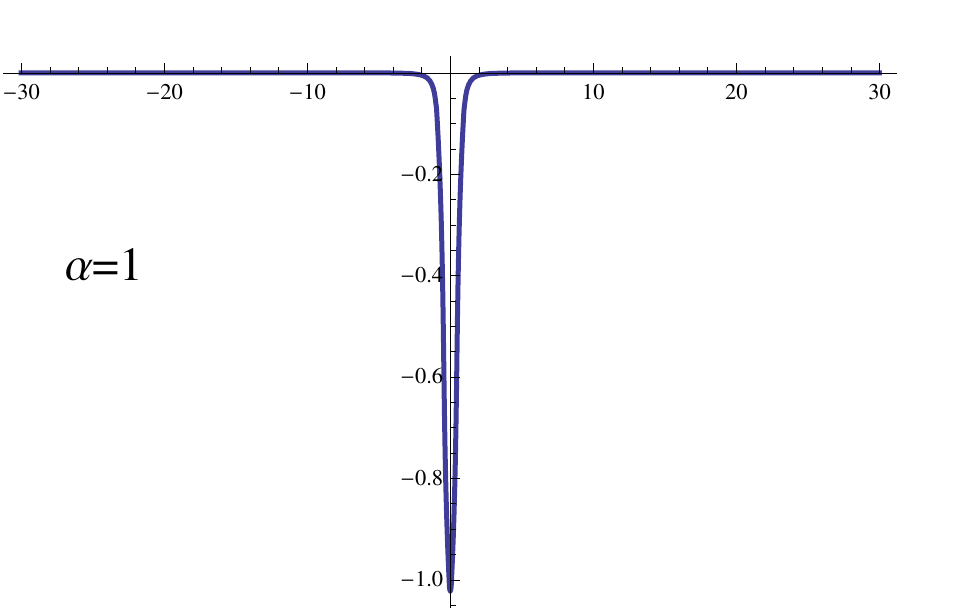}
\hspace{3mm}
  \includegraphics[width=4cm]{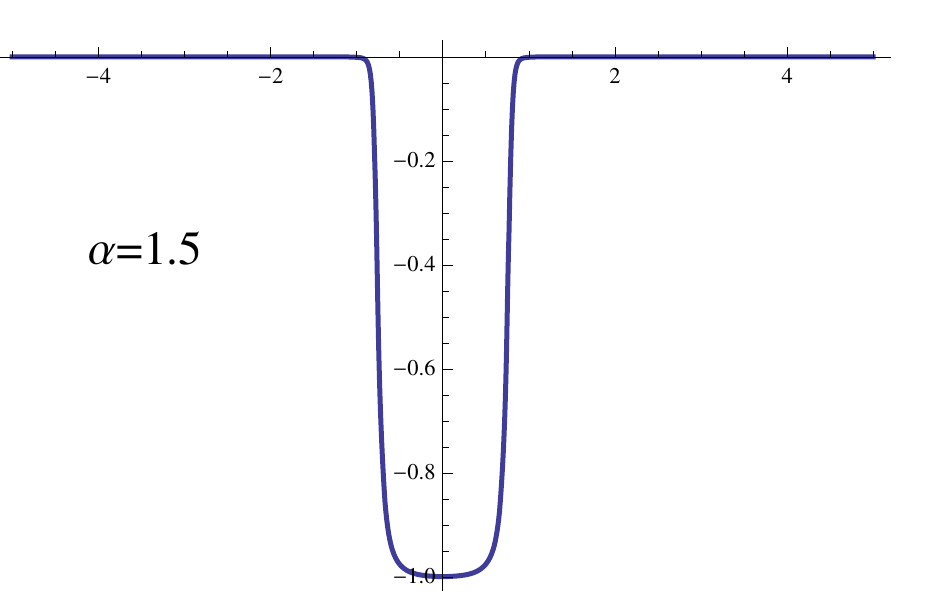}
 \caption{The above three pictures are the plots of energy stress tensor $T_{ww}$ at $t=0$ for $\ap=0.001$ (left), $\ap=1$ (middle) and $\ap=1.5$ (right).  
The red graph describes the sum of two single local quenches, while the blue one does the double local quench.
In the right graph ($\ap=1.5$), the blue and red graph almost coincide. The lower three graphs are the plots of the difference $T^{D}_{ww}-(T^{S(x=b)}_{ww}+T^{S(x=-b)}_{ww})$, which turns out 
to be non-positive.}
\label{fig:EM}
\end{figure}

When $\ap>0$, we find that the result for the double quench deviates from that of the single quench as depicted in the above three pictures of figure \ref{fig:EM}.
As $\ap$ approaches to $\pi/2$, the result for the double quench gets close to that for the two single quenches. 
Moreover, for any values of $0<\ap<\pi/2$, we can confirm the inequality (\ref{ineq}) by setting $q=T_{ww}$ as we can also find from figure \ref{fig:EM}.
However, note that this calculation for energy stress tensor is universal in that the result does not depend on 
the types of 2d CFTs we consider. As we can see from the lower three graphs of figure \ref{fig:EM}, we
can confirm that the difference of (\ref{ineq}) is always non-positive:
\ba
T^{D}_{ww}-(T^{S(x=b)}_{ww}+T^{S(x=-b)}_{ww}) \leq 0.  \label{ineqt}
\ea

\begin{figure}
  \centering
  \includegraphics[width=4cm]{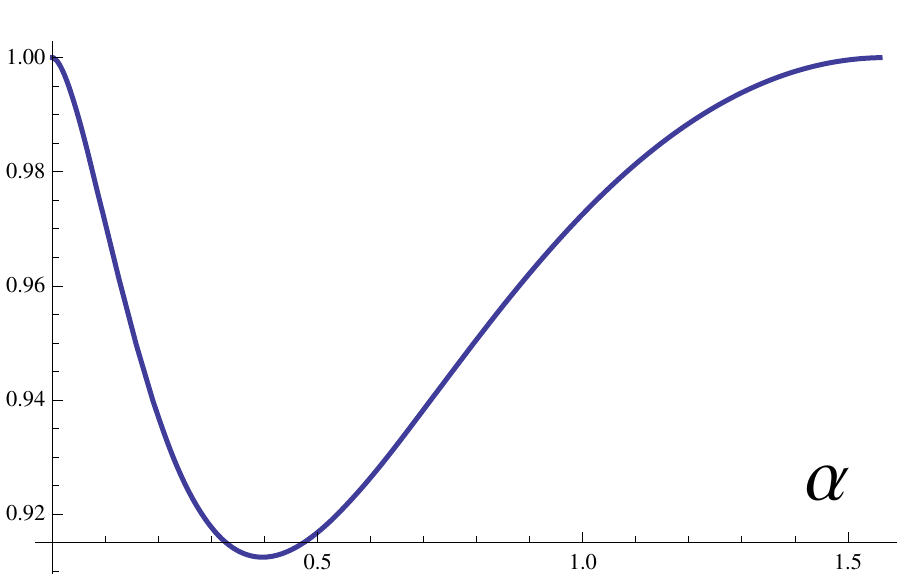}
\hspace{3mm}
 \includegraphics[width=4cm]{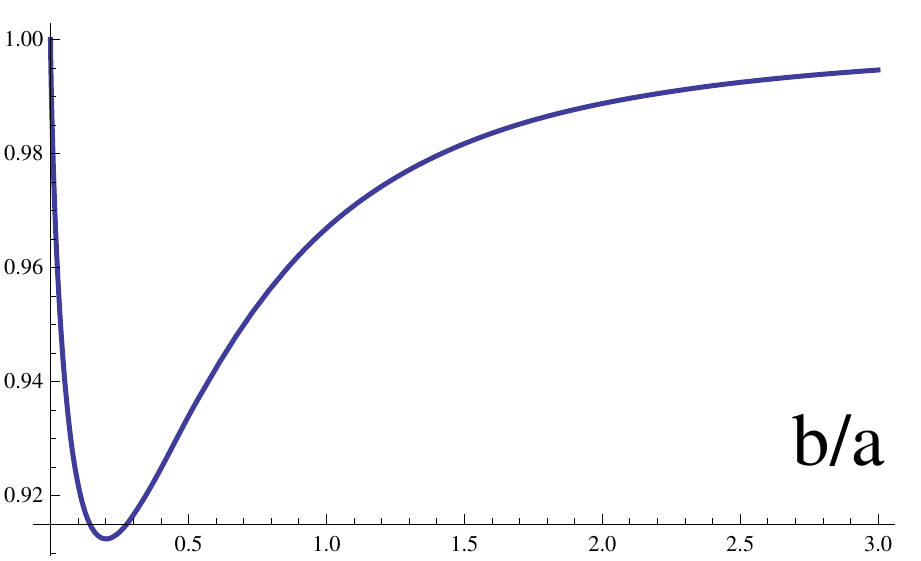}
\hspace{3mm}
 \includegraphics[width=4cm]{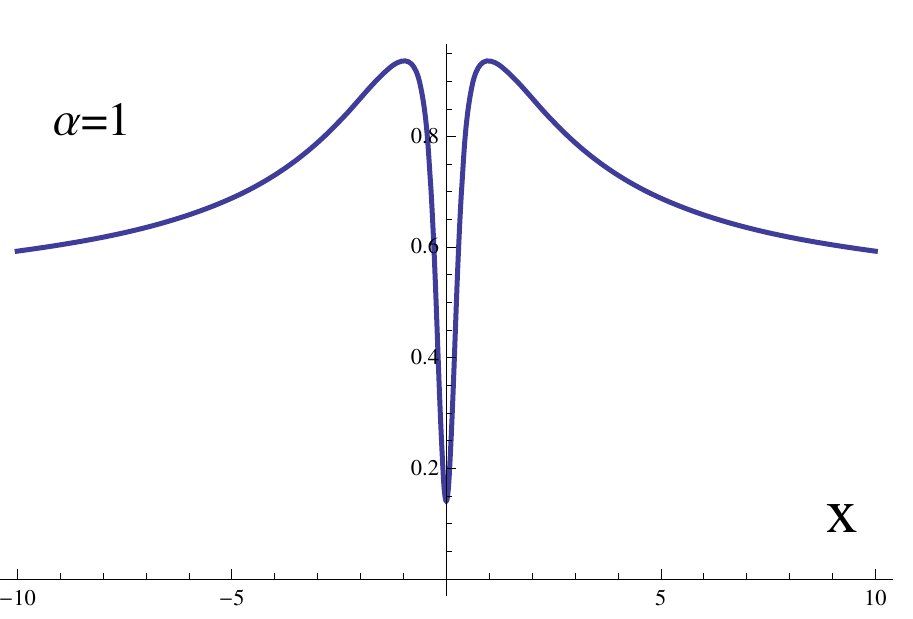}
 \caption{The ratio $\lim_{x\to\infty}\frac{T^{D}_{ww}(x)}{T^{S}_{ww}(x)}=F(b/a)$ 
as a function of $\ap$ (left) and 
as a function of $b/a$ (middle) at the spacial infinity limit $x\gg 1$ at $t=0$.
The right graph shows the ratio $\frac{T^{D}_{ww}(x)}{T^{S(x=b)}_{ww}(x)+T^{S(x=-b)}_{ww}(x)}$
as a function of $x$ at $t=0$ and $\ap=1$.}
\label{fig:ratio}
\end{figure}

Consider the limit $x\to\infty$ such that the subsystem $A$ is far away
 from the quench points $x=\pm b$ at $t=0$. In this limit we can analytically estimate 
$T_{ww}$ as follows
\be
T^{D}_{ww}\simeq \frac{c}{256x^4}\cdot \left(7+4\cos(2\ap)-3\cos(4\ap)\right).
\ee
Note that this gets vanishing at $\ap=\pi/2$ (i.e. $a=0$), where the energy density is delta functionally 
localized at $x=\pm b$. We can confirm $T^{D}_{ww}\simeq T^{S}_{ww}$ at $\ap=0$ and 
$\ap=\pi/2$, while in general we have $T^{D}_{ww}<T^{S}_{ww}$ for $0<\ap<\pi/2$. This is plotted in the left graph of 
figure \ref{fig:ratio}.

By a simple scale transformation,  we can also find for generic values of 
$(a,b)$, the energy stress tensor at $x$ and $t=0$, is given by 
\ba
T^{D}_{ww}(x,a,b)\simeq \frac{ca^2}{8x^4}\cdot F(b/a)=T^{S}_{ww}(x,a)\cdot F(b/a),\label{Q1}
\ea
where we introduced the function
\ba
F(b/a)=\frac{7+4\cos(2\ap)-3\cos(4\ap)}{32 a_0(\ap)^2}, \label{frat}
\ea 
with $\ap$ is determined by  $\frac{b}{a}$ via $\frac{b}{a}=\frac{b_0(\ap)}{a_0(\ap)}$.
The ratio $T^{D}_{ww}/T^{S}_{ww}$ in the limit $x\to \infty$, which coincides with 
$F(a/b)$ is plotted in the middle graph of 
figure \ref{fig:ratio}. We also plotted the ratio $\frac{T^{D}_{ww}}{T^{S(x=b)}_{ww}+T^{S(x=-b)}_{ww}}$ at each point $x$ for $\ap=1$ in the right graph of figure \ref{fig:ratio}. Indeed this ratio is always less than $1$ which confirms the inequality (\ref{ineqt}). Note that this ratio approaches to $F(a/b)$ in the limit
$x\to\infty$.

We would also like to comment on the time evolutions of the energy stress tensor.
Actually, $T_{ww}$ and $T_{\bar{w}\bar{w}}$ only depend on $x-t$ and $x+t$, respectively. 
Therefore the time evolution of each of them is just a simple shift $x\to x-t$ and the above observations are true at any time 
$t$ in a straightforward way.

%%%%%%%%%%%%%%%%%%%%%%%%%%%%%%%%%%%%
\subsection{Entanglement Entropy for Double Joining Local Quenches}
%%%%%%%%%%%%%%%%%%%%%%%%%%%%%%%%%%%%

As in the energy stress tensor analysis, we are interested in properties of entanglement entropy.
We would like to compare the results for the double joining quench with those for the single joining quenches. First we would like to note the following two basic observations. At $b=0$ (or $\ap=0$)  the entanglement entropy growth $\Delta S_A=S_A-S^{(0)}_A$ ($S^{(0)}$ is the ground state entanglement entropy) for the double joining quench coincides with that for the single joining quench, where the quench occurs at the origin $x=0$. Also in the opposite  limit $a\to 0$ (or $\ap\to \pi/2$), the entanglement entropy for the double joining quench gets identical to that for the sum of two single joining quenches each of which we join the points $x=b$ or $x=-b$, respectively. Therefore we have
\ba
&& \lim_{b/a\to 0}\Delta S^{double(x=\pm b)}_A=\Delta S^{single(x=0)}_A,\no
&& \lim_{a/b\to 0}\Delta S^{double(x=\pm b)}_A=\Delta S^{single(x=b)}_A+\Delta S^{single(x=-b)}_A.
\label{eqjoin}
\ea

Below we will study two explicit examples: holographic CFTs and massless Dirac fermion CFT.

%%%%%%%%%%%%%%%%%%%%%%%%%%%%%%%%%%%%
\subsection{Entanglement Entropy in Holographic CFTs}
%%%%%%%%%%%%%%%%%%%%%%%%%%%%%%%%%%%%

Consider the entanglement entropy $S_A(x_1,x_2,t)$ for a subsystem $A$ given by the interval $[x_1,x_2]$ at time $t$. In holographic CFTs, as we briefly reviewed in section 2.2, we can calculate $S_A$ by the following formula when the dual geodesic $\Gamma_A$ is connected:
\ba
S^{con}_A=\frac{c}{6}\log\left(\frac{|f(w_1)-f(w_2)|^2}{\ep^2|f'(w_1)||f'(w_2)|}\right)=\frac{c}{6}\log\left(\frac{|\xi_1-\xi_2|^2|g'(\xi_1)||g'(\xi_2)|}{\ep^2}\right),
\label{HEEcon}
\ea
where $\xi=f(w)$ and its inverse $w=g(\xi)$ are the conformal map (\ref{xcmap}) for the double local quenches.

In the presence of conformal boundaries, we can apply the AdS/BCFT formulation, as is so in single local quenches. 
In that case, $\Gamma_A$ can be disconnected and we have the following formula when the dual geodesic $\Gamma_A$ is disconnected:
\be
S^{dis}_A=\frac{c}{6}\log\left(\frac{|\xi_1-\bar{\xi}_1||\xi_2-\bar{\xi}_2||g'(\xi_1)||g'(\xi_2)|}{\ep^2}\right)+2S_{bdy},
\ee
where $S_{bdy}$ is the boundary entropy.
In the end, the HEE is given by the smaller one among $S^{con}_A$ and $S^{dis}_A$.

We are interested in an inequality of the form (\ref{ineq}). We would like to argue the following inequality 
for the HEE of connected geodesic, is always true in our holographic double joining quench when the subsystem $A$ is an arbitrary interval:
\ba
\Delta S^{con,D(x=\pm b)}_A-\left(\Delta S^{con, S(x=b)}_A+\Delta S^{con, S(x=-b)}_A\right)\leq 0.
\label{ineqJ}
\ea
Note that we do not expect that such an inequality is satisfied for the disconnected geodesic in general. This is partly 
because the boundary entropy contributions $S_{bdy}$ does not cancel in the above difference. 

We can find that $S^{con}_A$ is smaller than $S^{dis}_A$ when a finite size  subsystem $A$ is away from the quench points $x=\pm b$ by a certain distance. Therefore, our result (\ref{ineqJ}) leads to 
the inequality (\ref{ineq}) for the genuine HEE 
\ba
\Delta S^{D(x=\pm b)}_A-\left(\Delta S^{S(x=b)}_A+\Delta S^{S(x=-b)}_A\right)\leq 0.
\label{ineqJJ}
\ea
when the subsystem $A$ is enough separated from the quench points.

Below we will show explicit results of $\Delta S_A$, which support the above inequalities.
 We will choose $b=50$ throughout this work.  We will set $S_{bdy}=0$ below for the disconnected 
geodesic contribution.

%%%%%%%%%%%%%%%%%%%%%%%%%%%%%%%%%%%%
\subsubsection{Holographic Entanglement Entropy at $t=0$}
%%%%%%%%%%%%%%%%%%%%%%%%%%%%%%%%%%%%

\begin{figure}
\centering
\begin{minipage}{0.32\hsize}
\includegraphics[width=5cm]{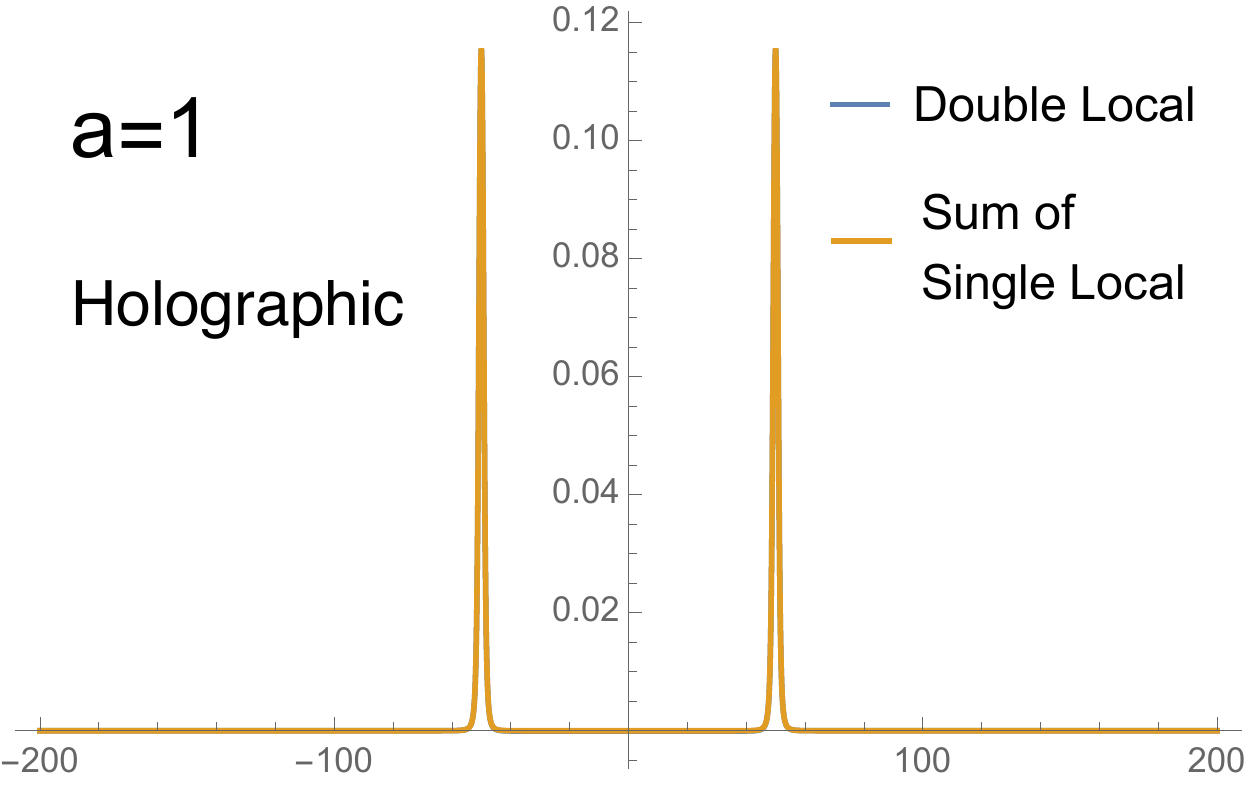}
\end{minipage}
\begin{minipage}{0.32\hsize}
\includegraphics[width=5cm]{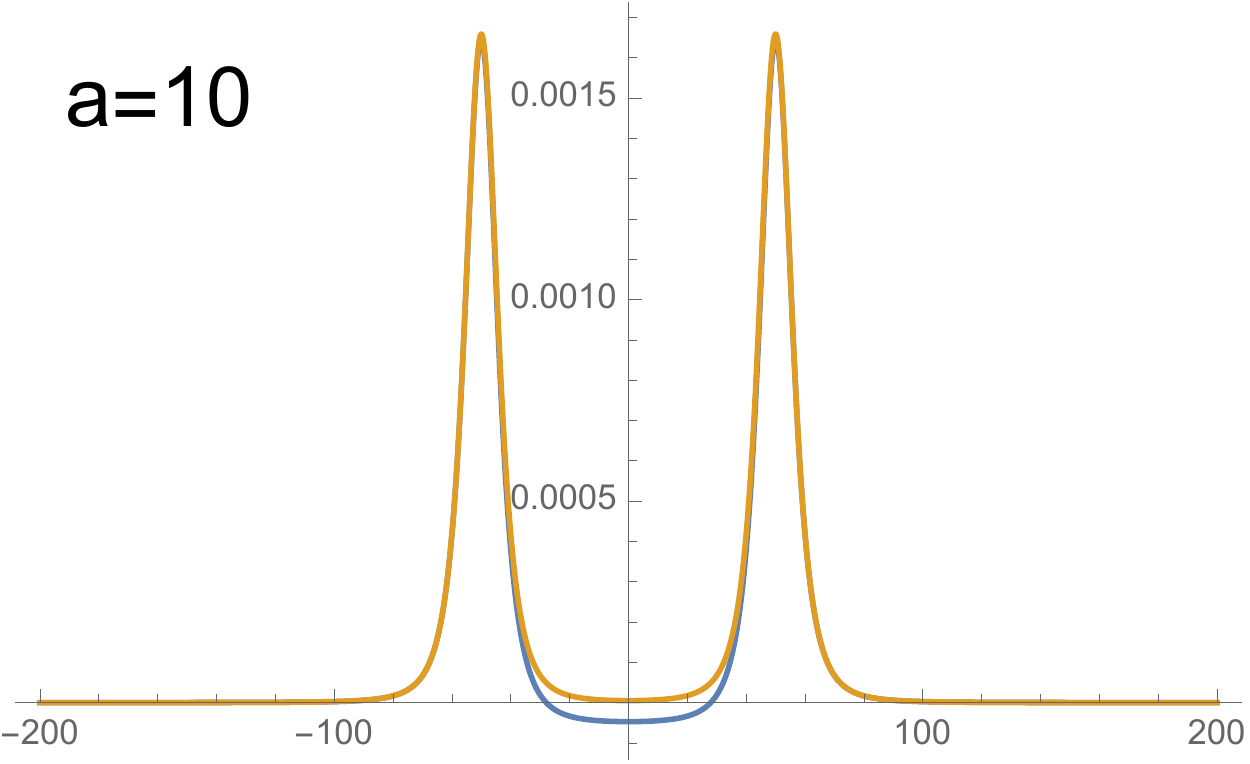}
\end{minipage}
\begin{minipage}{0.32\hsize}
\includegraphics[width=5cm]{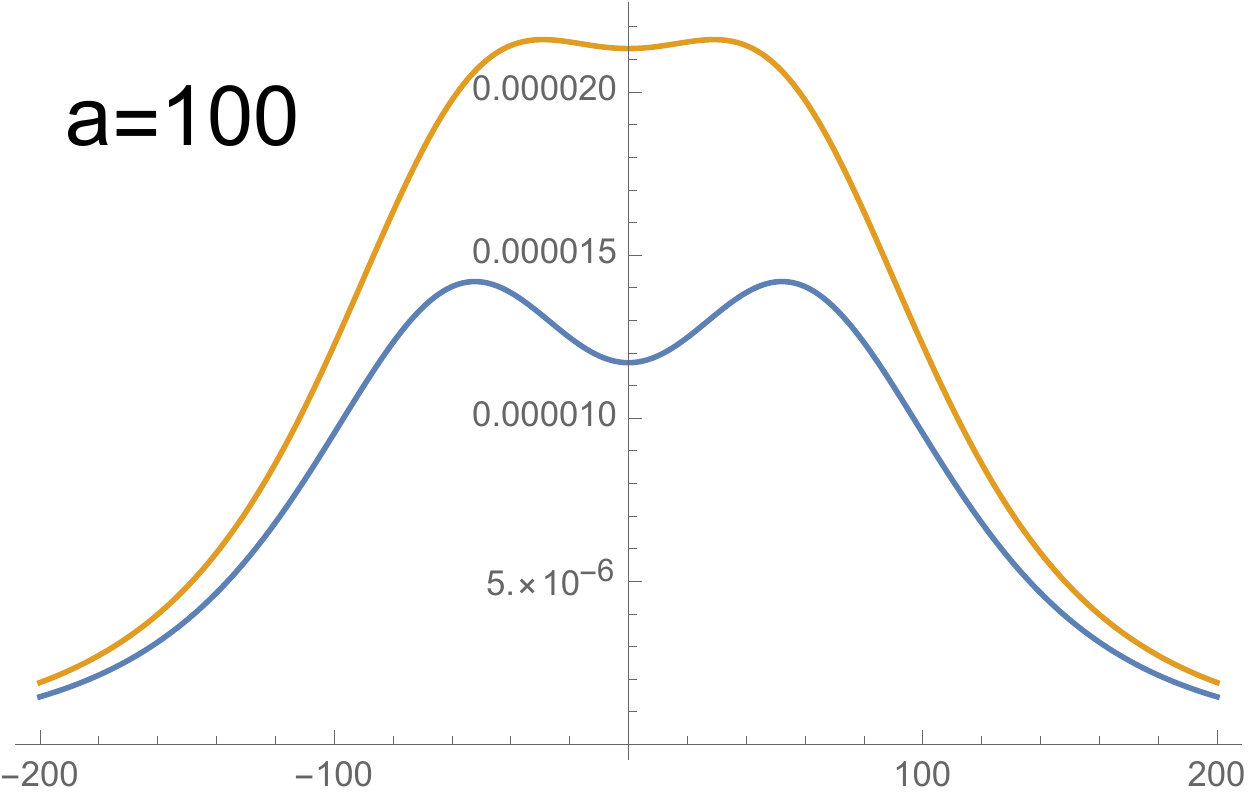}
\end{minipage}
\begin{minipage}{0.06\hsize}
        \vspace{10mm}
      \end{minipage} \\
\begin{minipage}{0.32\hsize}
\includegraphics[width=5cm]{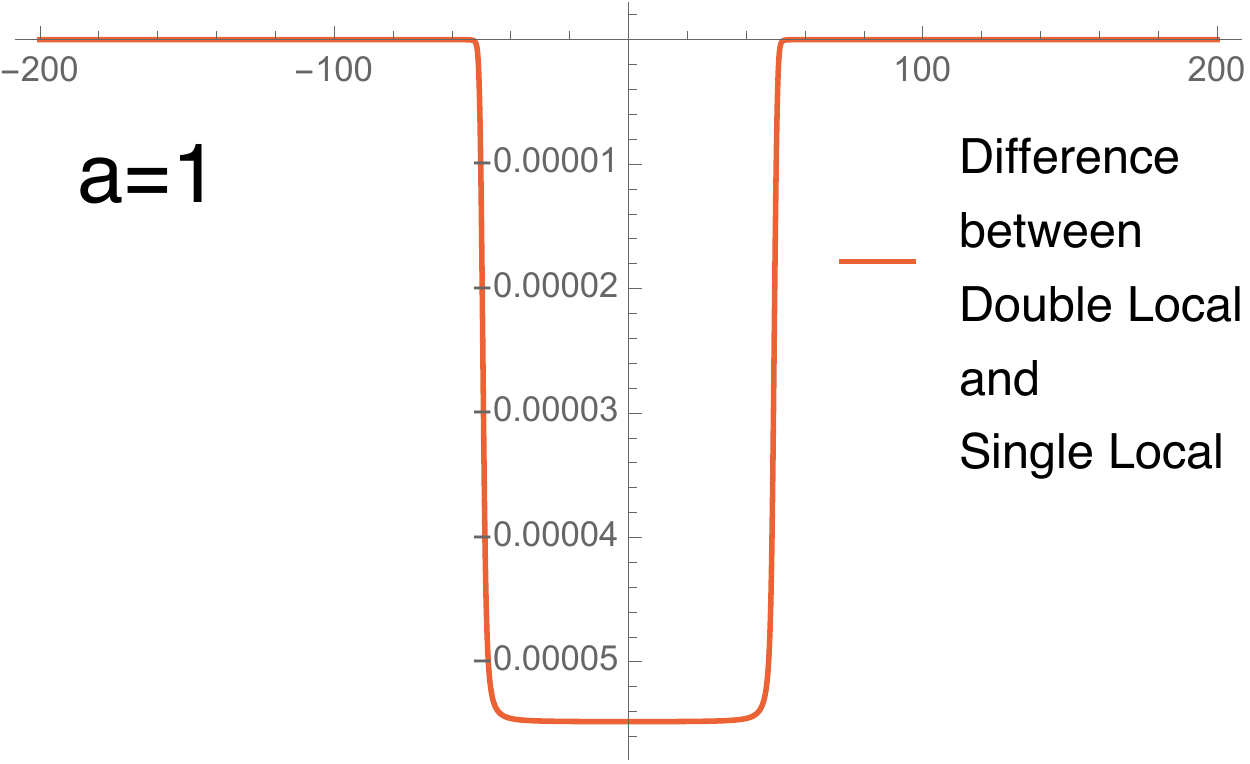}
\end{minipage}
\begin{minipage}{0.32\hsize}
\includegraphics[width=5cm]{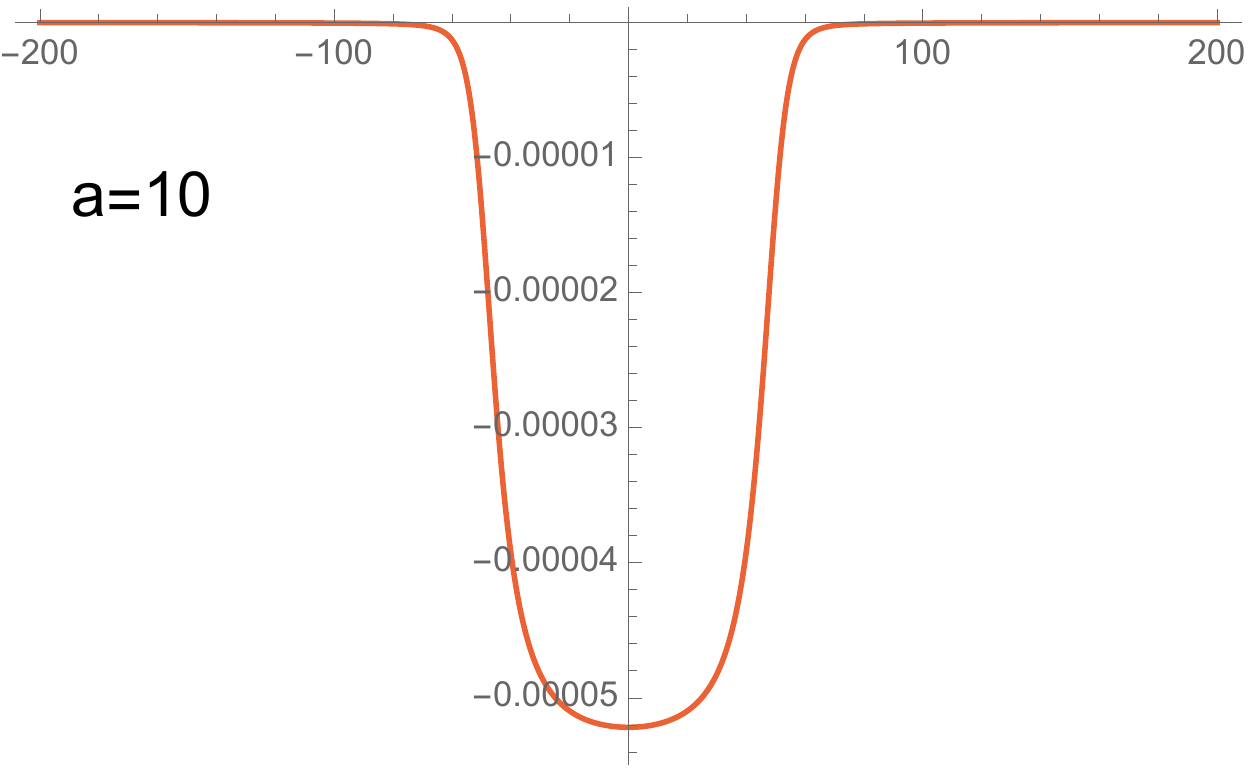}
\end{minipage}
\begin{minipage}{0.32\hsize}
\includegraphics[width=5cm]{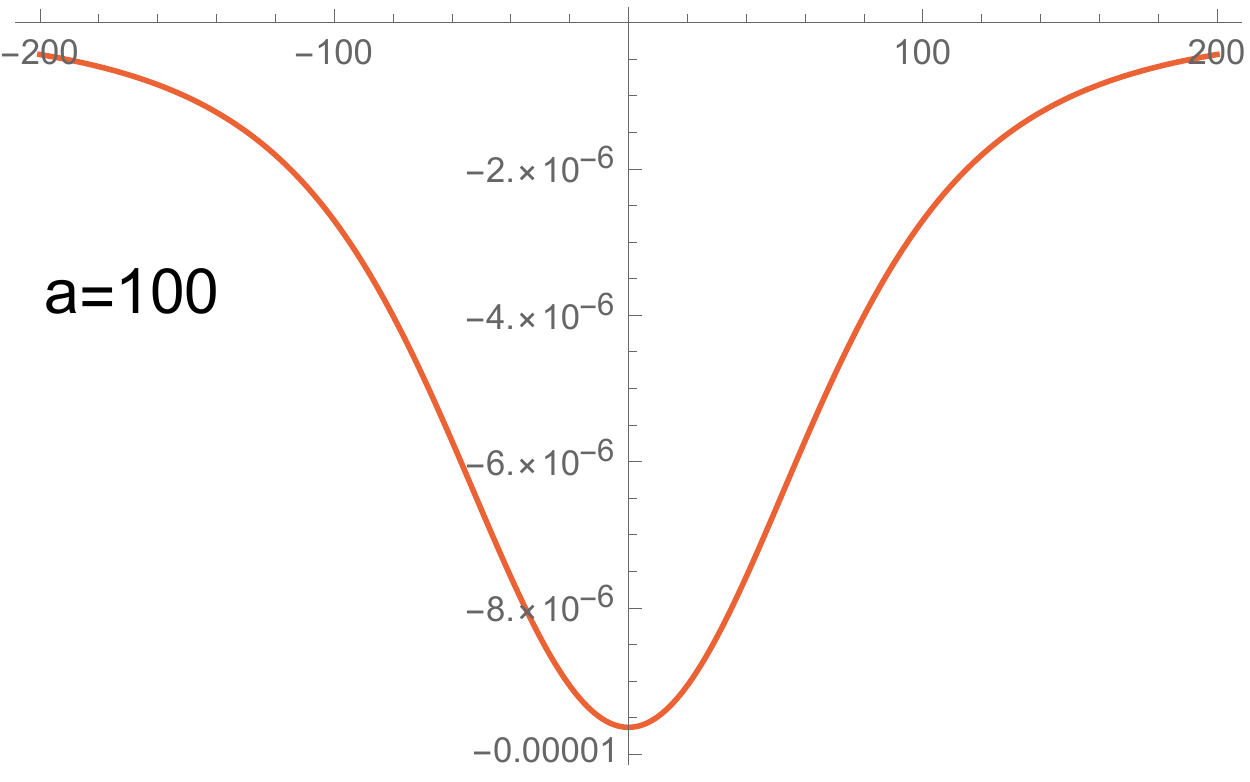}
\end{minipage}
\begin{minipage}{0.06\hsize}
        \vspace{10mm}
      \end{minipage} \\
\caption{{\bf Top:} The behaviors of connected holographic entanglement entropy $S^{con}_A$ at $t=0$ and $b=50$ for the double local quench (blue) and the sum of two single local quenches (orange) 
at $a=1$ (left), $a=10$ (middle) and $a=100$ (right). We chose the subsystem $A$ to be 
$A=[x-1,x+1]$ and plotted $\Delta S^{con}_A$ as a function $x$. In the left graph at $a=1$, the blue and orange graphs almost coincide. {\bf Bottom:} The difference $ \Delta S^{D}_A-\Delta S^{S(x=b)}_A-\Delta S^{S(x=-b)}_A$ for connected geodesics.}
\label{fig:JoinHEEstatic}
\end{figure}

Here we consider the HEE under the double joining quenches at $t=0$.
We presented numerical plots of $\Delta S^{con}_A$ as a function of $x$ in figure \ref{fig:JoinHEEstatic}, where we chose the subsystem $A$ as $[x-1,x+1]$. For such a small subsystem, we can confirm that always the connected geodesics are favored for the calculations of HEE. We can confirm the behaviors (\ref{eqjoin}) and the inequality (\ref{ineqJ}).

In particular, let us consider the distant limit of the subsystem $A=[x_1,x_2]$: 
\be
l\equiv x_{1}-x_2 \to 0, \ \  \mbox{and}\ \   x_{1}\simeq x_2(=x)\gg b,a.
\ee
In this limit, the entanglement entropy $\Delta S^{con}_A$ can be analytically estimated as 
\ba
\Delta S^{con,D}_A\simeq F(b/a)\cdot \frac{ca^2l^2}{24x^4}= F(b/a)\cdot \Delta S^{con, S}_A,\label{Q2}
\ea
where $F$ was defined in (\ref{frat}). This behavior follows from the first law of EE \cite{Bhattacharya:2012mi}
\be
\Delta S^{con}_A\simeq \frac{l^2}{3}T_{ww}.  \label{firstla}
\ee 
Thus in this limit, the inequality (\ref{ineqJJ}) is equivalent to that for the energy stress tensor
(\ref{ineqt}).

%%%%%%%%%%%%%%%%%%%%%%%%%%%%%%%%%%%%
\subsubsection{Time Evolutions of Holographic Entanglement Entropy}
%%%%%%%%%%%%%%%%%%%%%%%%%%%%%%%%%%%%

\begin{figure}
  \centering
 \includegraphics[width=8cm]{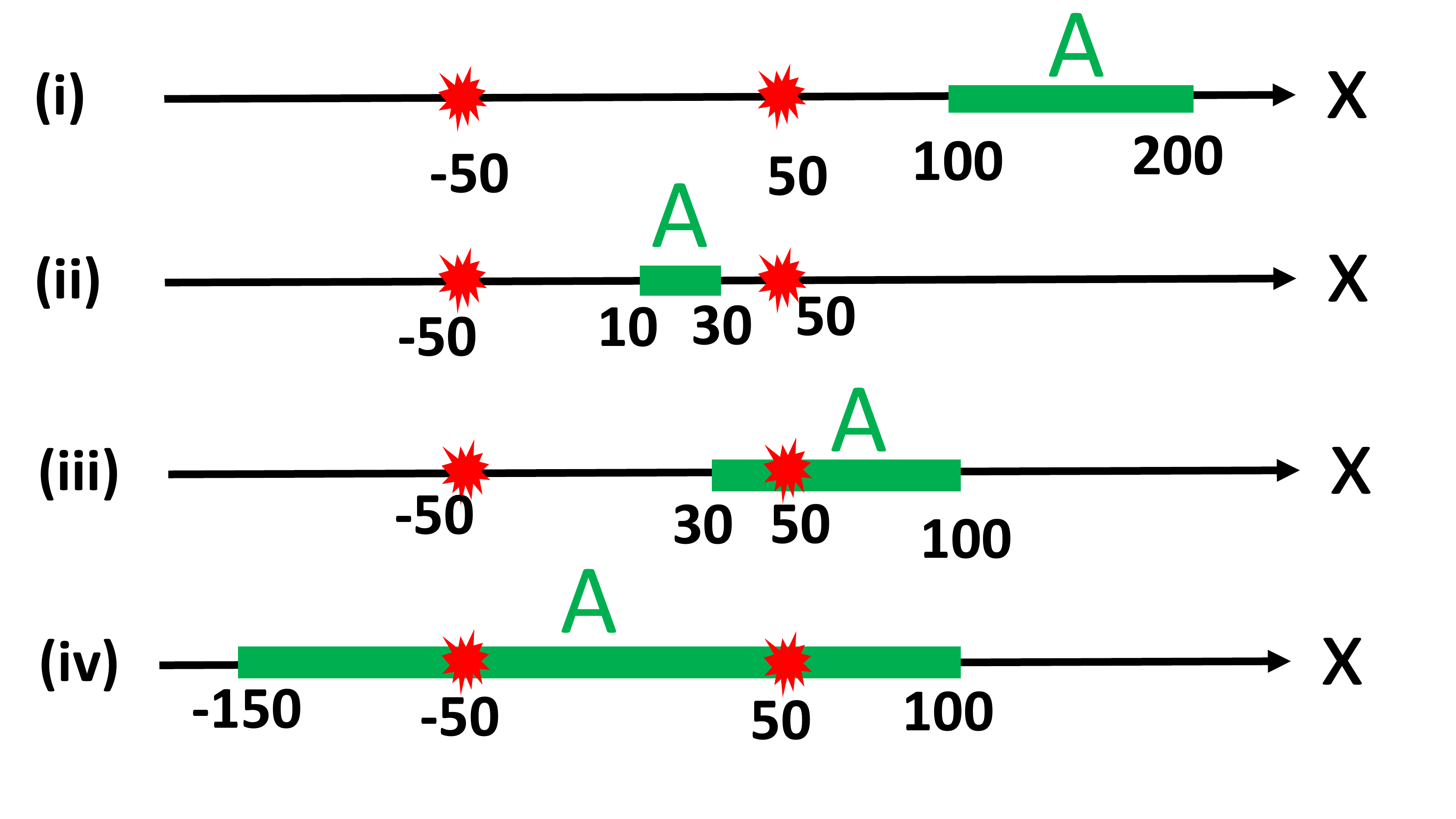}
 \caption{The four different choices of subsystem $A$. We always choose $b=50$.}
\label{fig:eesetup}
\end{figure}

Next we present numerical results for the time evolutions of HEE under the double joining quenches.
We consider $\Delta S_A$ for the four different choices of the subsystem $A$: (i), (ii), (iii), (iv) sketched in figure \ref{fig:eesetup}.

\begin{figure}
\centering
\begin{minipage}{0.49\hsize}
\includegraphics[width=7cm]{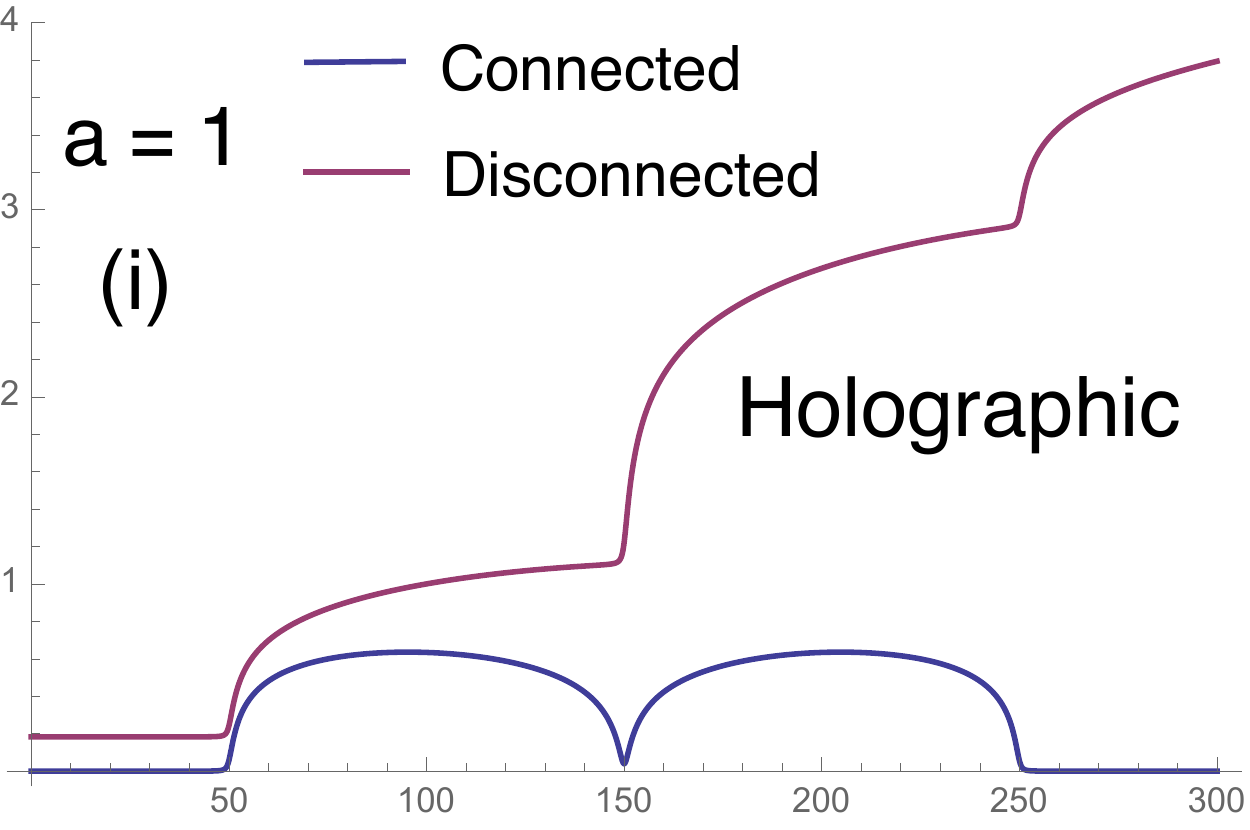}
\end{minipage}
\begin{minipage}{0.49\hsize}
\includegraphics[width=7cm]{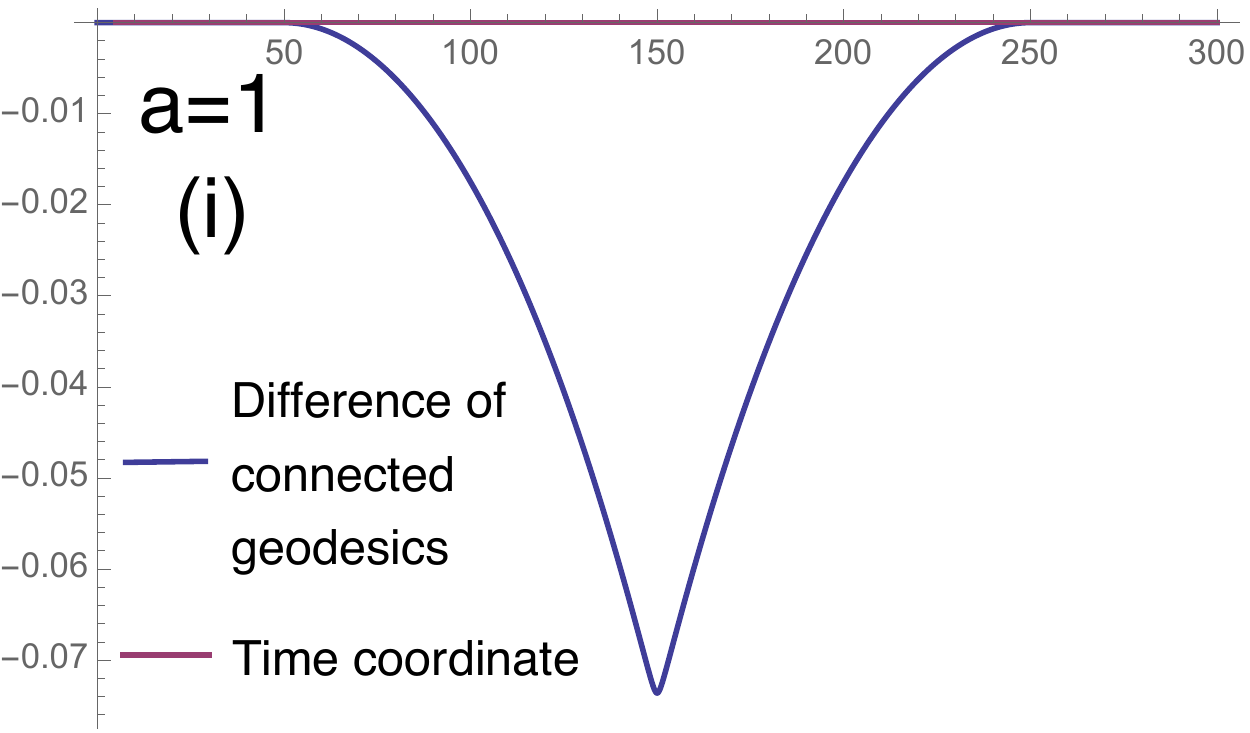}
\end{minipage}
\begin{minipage}{0.02\hsize}
        \vspace{10mm}
      \end{minipage} \\
\begin{minipage}{0.49\hsize}
\includegraphics[width=7cm]{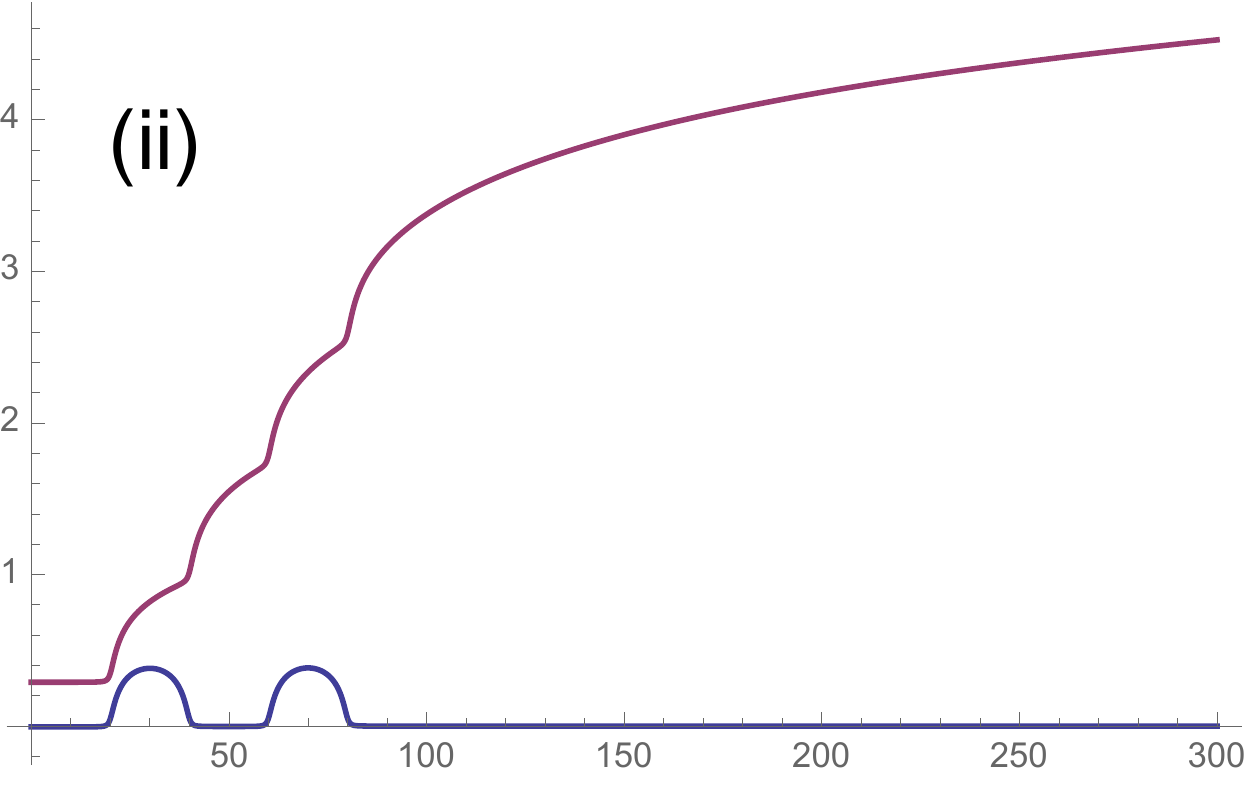}
\end{minipage}
\begin{minipage}{0.49\hsize}
\includegraphics[width=7cm]{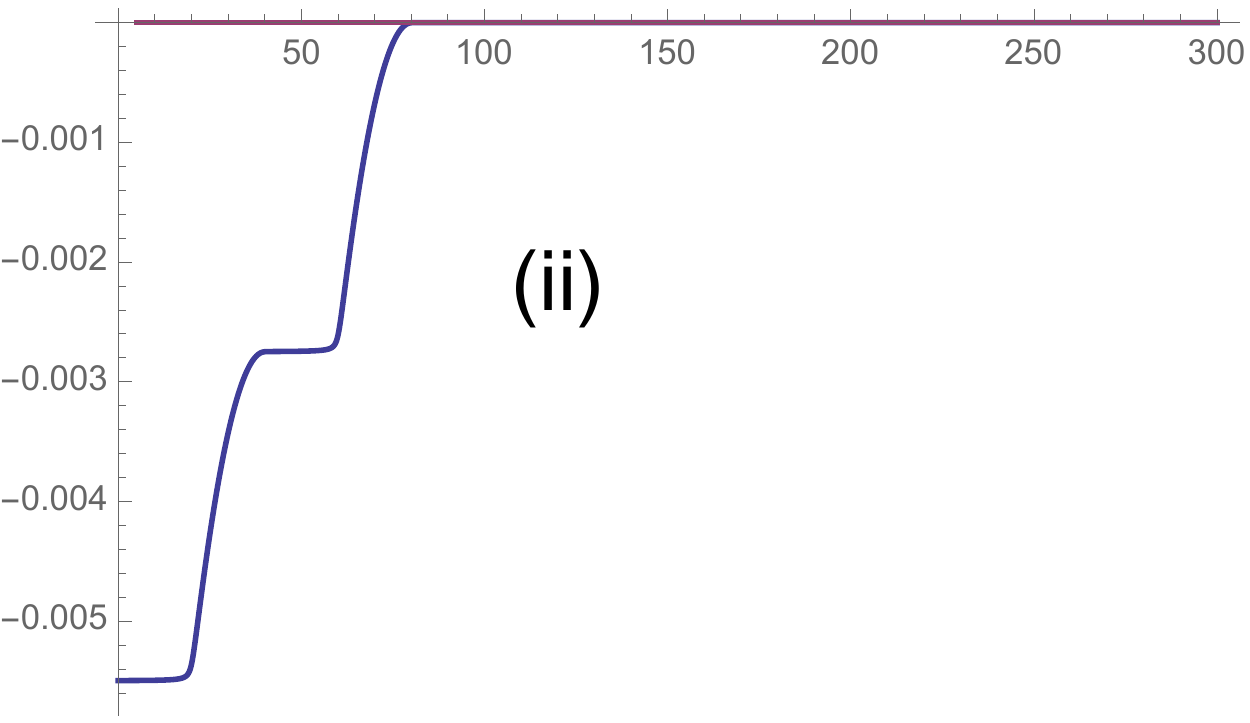}
\end{minipage}
\begin{minipage}{0.02\hsize}
        \vspace{10mm}
      \end{minipage} \\
\begin{minipage}{0.49\hsize}
\includegraphics[width=7cm]{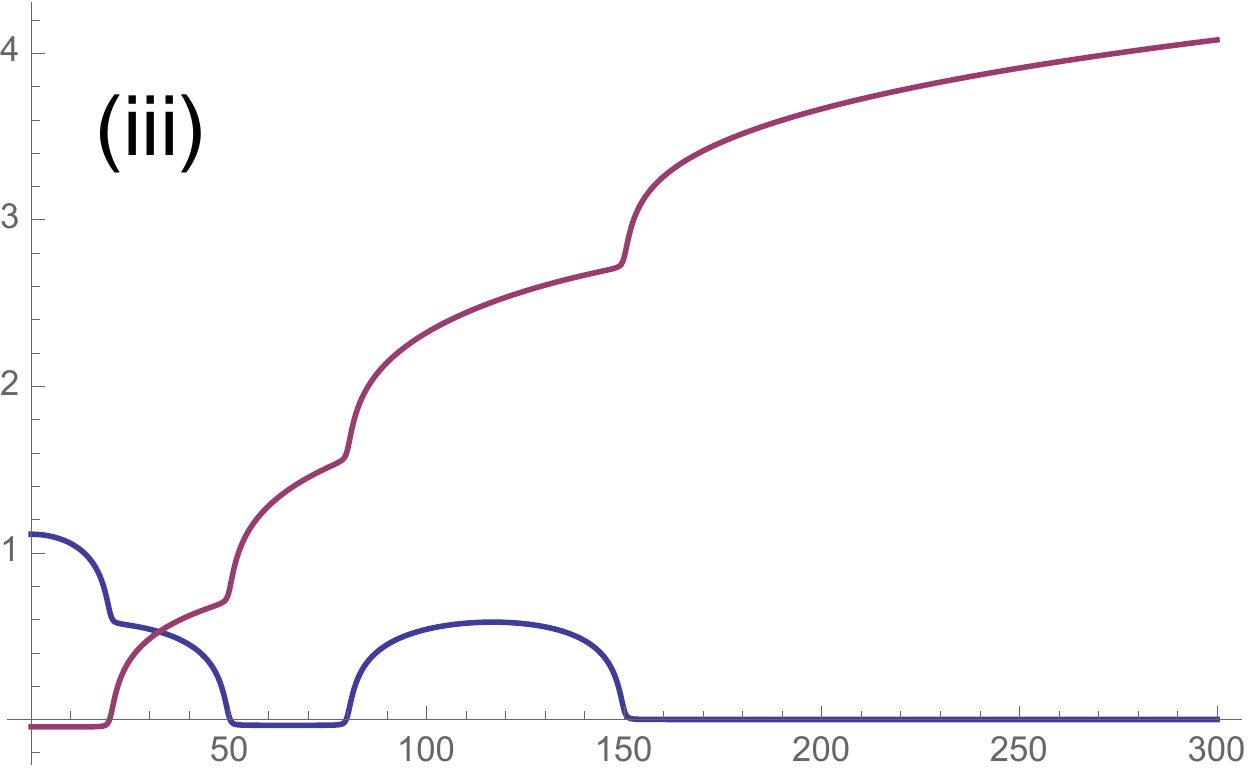}
\end{minipage}
\begin{minipage}{0.49\hsize}
\includegraphics[width=7cm]{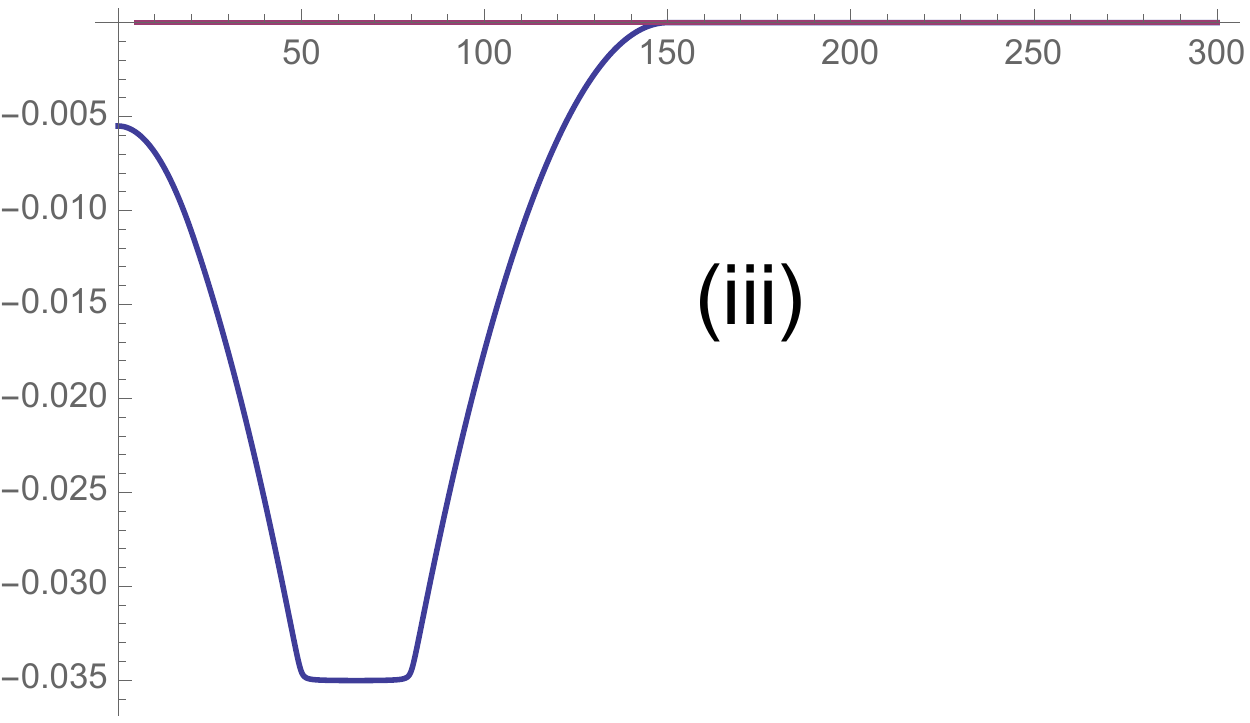}
\end{minipage}
\begin{minipage}{0.02\hsize}
        \vspace{10mm}
      \end{minipage} \\
\begin{minipage}{0.49\hsize}
\includegraphics[width=7cm]{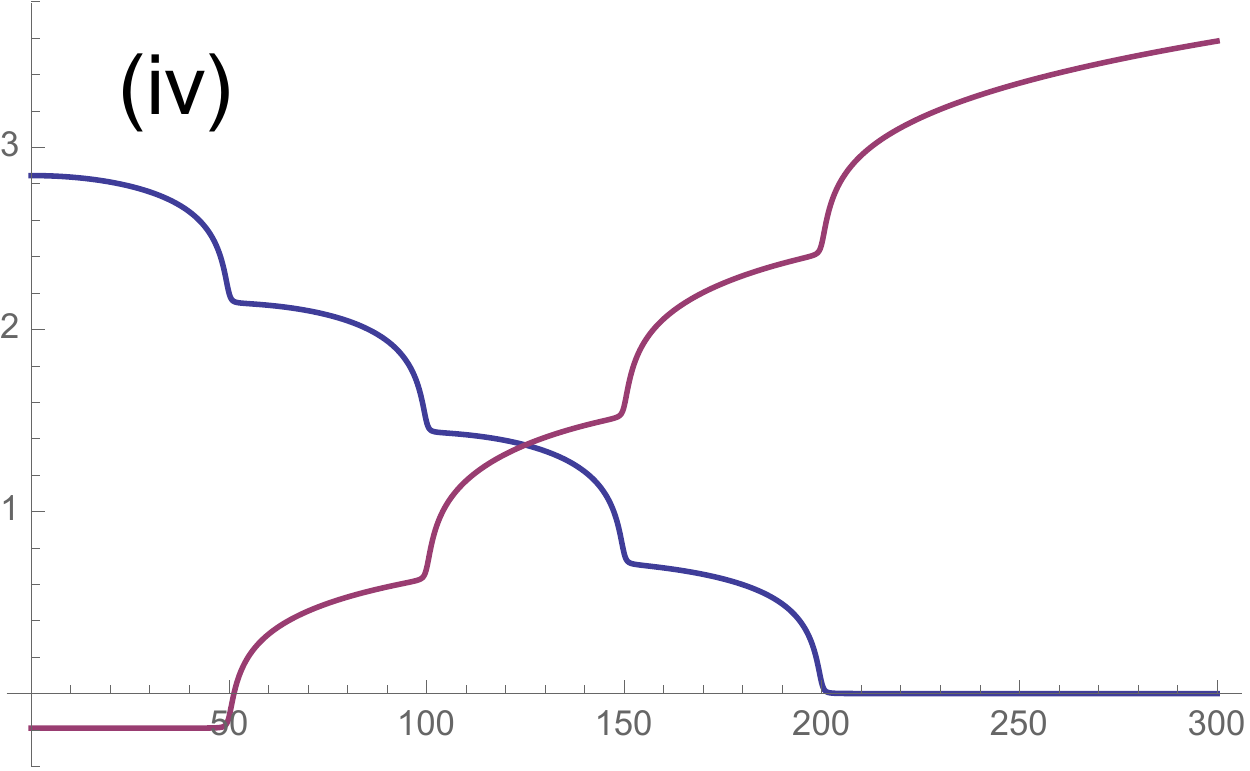}
\end{minipage}
\begin{minipage}{0.49\hsize}
\includegraphics[width=7cm]{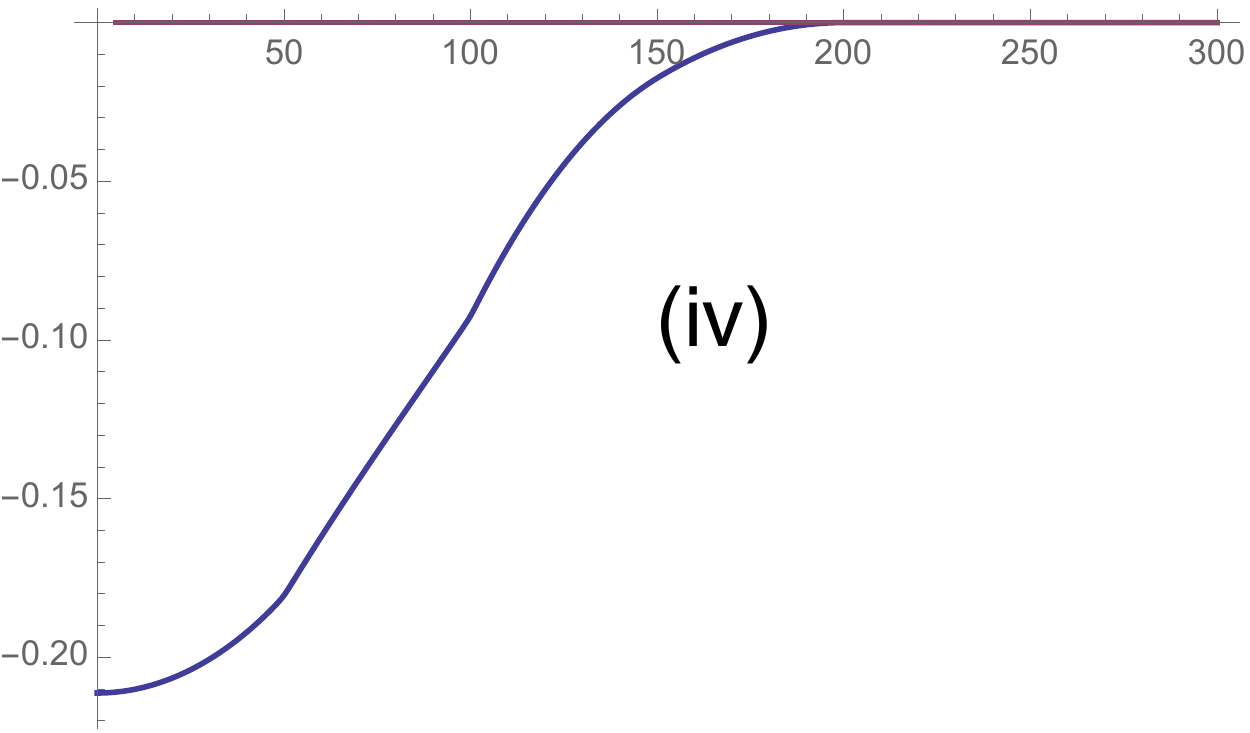}
\end{minipage}
\caption{The plots of $\Delta S^{con/dis, D}_A$ for the double joining local quench (left) and
the difference $\Delta S^{con,D}_A-\Delta S^{con, S(x=b)}-\Delta S^{con, S(x=-b)}$ 
between the double and single quench  (right) as a function of $t$. We set $b=50$ and $a=1$.
From the top to bottom we presented results for the four different setups (i), (ii), (iii) and (iv).
{\bf Left: } In the left plots, the blue and red graph describes $\Delta S^{con}_A$ and $\Delta S^{dis}_A$, 
respectively. {\bf Right: }In the right plots, the blue graphs describe the difference $\Delta S^{D}_A-\Delta S^{S(x=b)}_A-\Delta S^{S(x=-b)}_A$ for the connected geodesic, while the red line just shows the axis of time coordinate.}
\label{fig:JoinHEEtime}
\end{figure}

Our numerical results for these four different cases are presented in figure \ref{fig:JoinHEEtime}.
Qualitative features for the time evolutions of HEE under double quenches are very similar to those for the sum of two single quenches. However the difference between them (i.e. blue graphs in the right pictures of figure \ref{fig:JoinHEEtime}) shows non-trivial time-dependence. The difference is always non-positive and this confirms the inequality  (\ref{ineqJ}). Also, the difference gets larger when 
signals from both of the two quenches arrive in the subsystem $A$.

In the case (i), there are two lumps in the regions $t\in [x_1-b,x_2-b]$ and $[x_1+b,x_2+b]$ and they are due to the two local quenches at $x=\pm b$, which create entangled pairs each propagating in the left or right direction at the speed of light. Also in the other cases (ii), (iii) and (iv), similarly we can understand the presence of lumps from the viewpoint of entangled pair creations at the quench points $x=\pm b$.

At late time $t\gg b,a$, the HEE is dominated by the connected geodesic contribution and we obtain the following behavior 
\ba
&& \Delta S^{(con)D}_A\simeq F(b/a)\cdot \frac{a^2l^2}{24t^4}= F(b/a)\cdot 
\Delta S^{(con)S}_A. \label{Q3}
\ea
Indeed these are obtained from the first law relation (\ref{firstla}).
Note that this clearly shows the inequality  (\ref{ineqJ}) at late time.

\begin{figure}
\centering
\includegraphics[width=5cm]{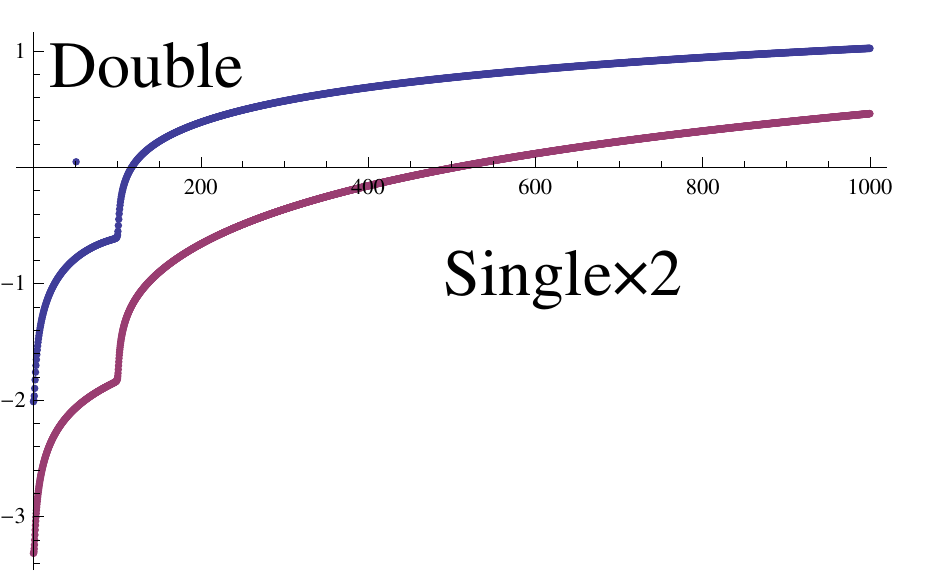}
\includegraphics[width=5cm]{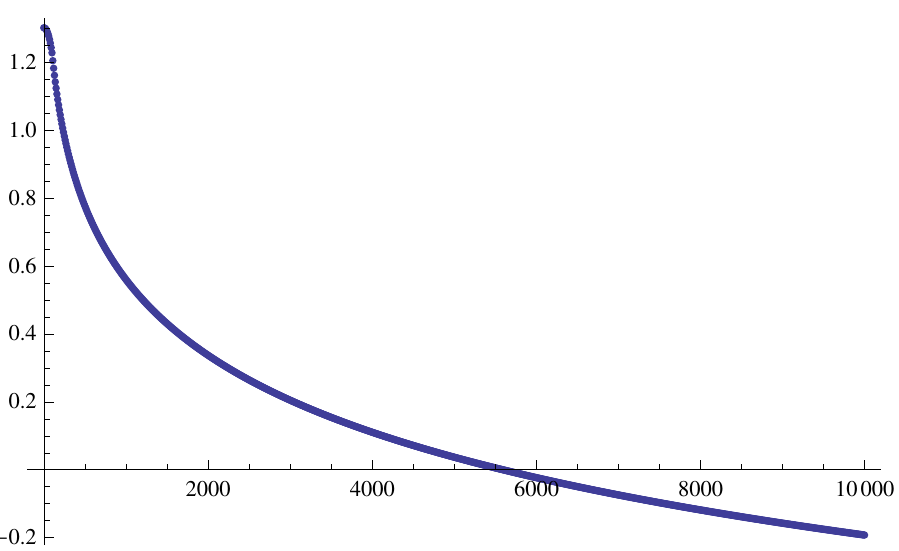}
\includegraphics[width=5cm]{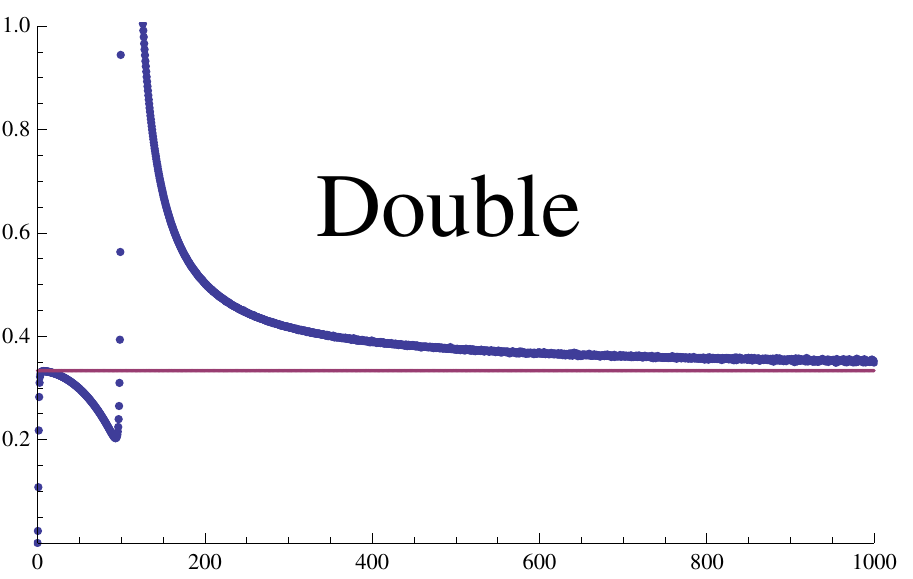}
\caption{The behaviors of holographic entanglement entropy as a function of the time 
$t$ when the subsystem $A$ is almost semi-infinite. We chose $A=[51,10^6]$ and $a=1, b=50$. In the left picture, we plotted $\Delta S^{dis}_A$ for the double quench (blue) and the sum of two singles quenches (red) as a function of time.  In the middle 
picture we plotted the difference $\Delta S^{dis,D}_A-\Delta S^{dis,S(x=b)}_A
-\Delta S^{S(dis,x=-b)}_A$ as a function of time.
In the right graph, we plotted $t\frac{d\Delta S^{dis}_A}{dt}$ (the coefficient of $\log t$ for the disconnected geodesic contribution) for the double joining quench as a function of time.}
\label{fig:JoinHEEhalf}
\end{figure}

A special feature of EE for holographic CFTs can be found when the subsystem $A$ is semi-infinite.
The numerical behaviors are plotted in figure \ref{fig:JoinHEEhalf}. Note that for such a large subsystem with one of the end points of $A$ close to $b$, the disconnected geodesic is favored for the HEE computation. Thus at late time we found 
\ba
&& \Delta S^{dis,D}_A\simeq \frac{c}{3}\log t,  \no
&& \Delta S^{dis,S(x=b)}+ \Delta S^{dis,S(x=-b)}\simeq \frac{2c}{3}\log t. 
\label{timehalf}
\ea
This result can be understood as follows. At late time, the excitations which were 
created at the quench points $x=\pm b$, already propagate a long distance. Therefore the differences 
of the two quench points are negligible and the situation is very similar to $b=0$ case 
(see \ref{eqjoin}). Again this confirms the inequality  (\ref{ineqJ}) at late time.

\begin{figure}
\begin{minipage}{0.49\hsize}
\includegraphics[width=70mm]{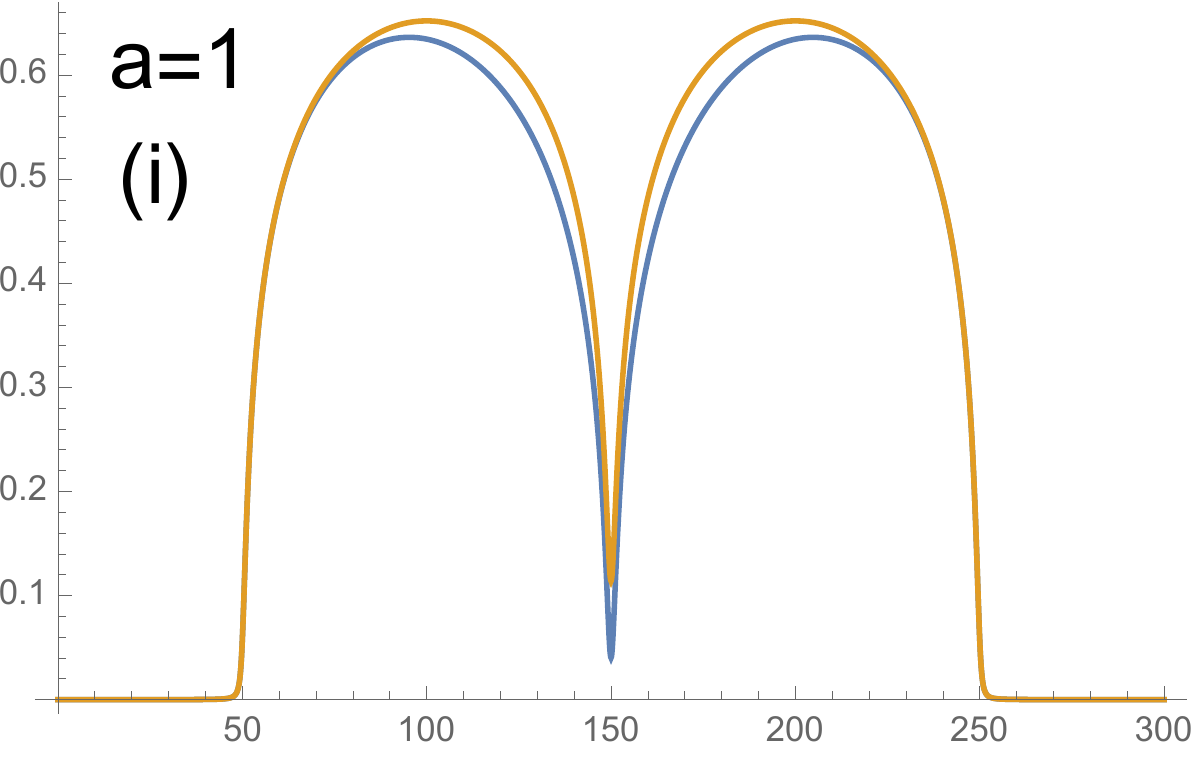}
\end{minipage}
\begin{minipage}{0.49\hsize}
\includegraphics[width=70mm]{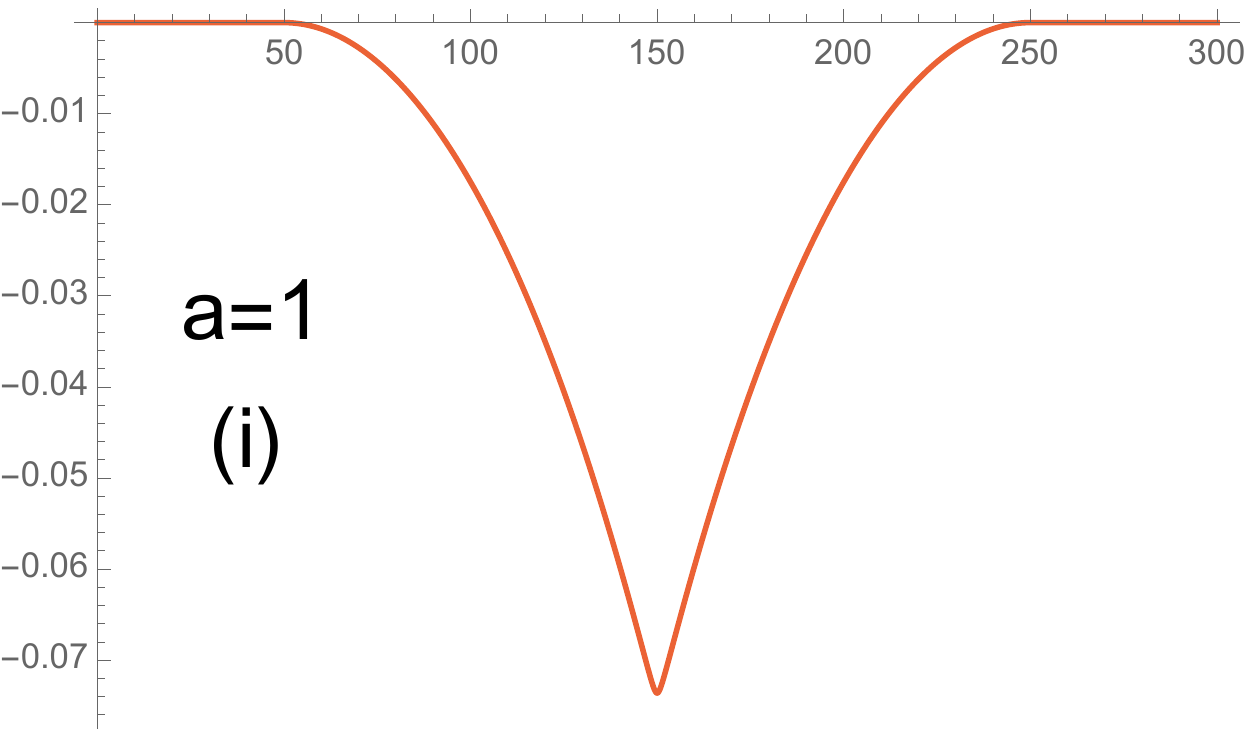}
\end{minipage}
\begin{minipage}{0.2\hsize}
        \vspace{10mm}
      \end{minipage} \\
\begin{minipage}{0.49\hsize}
\includegraphics[width=70mm]{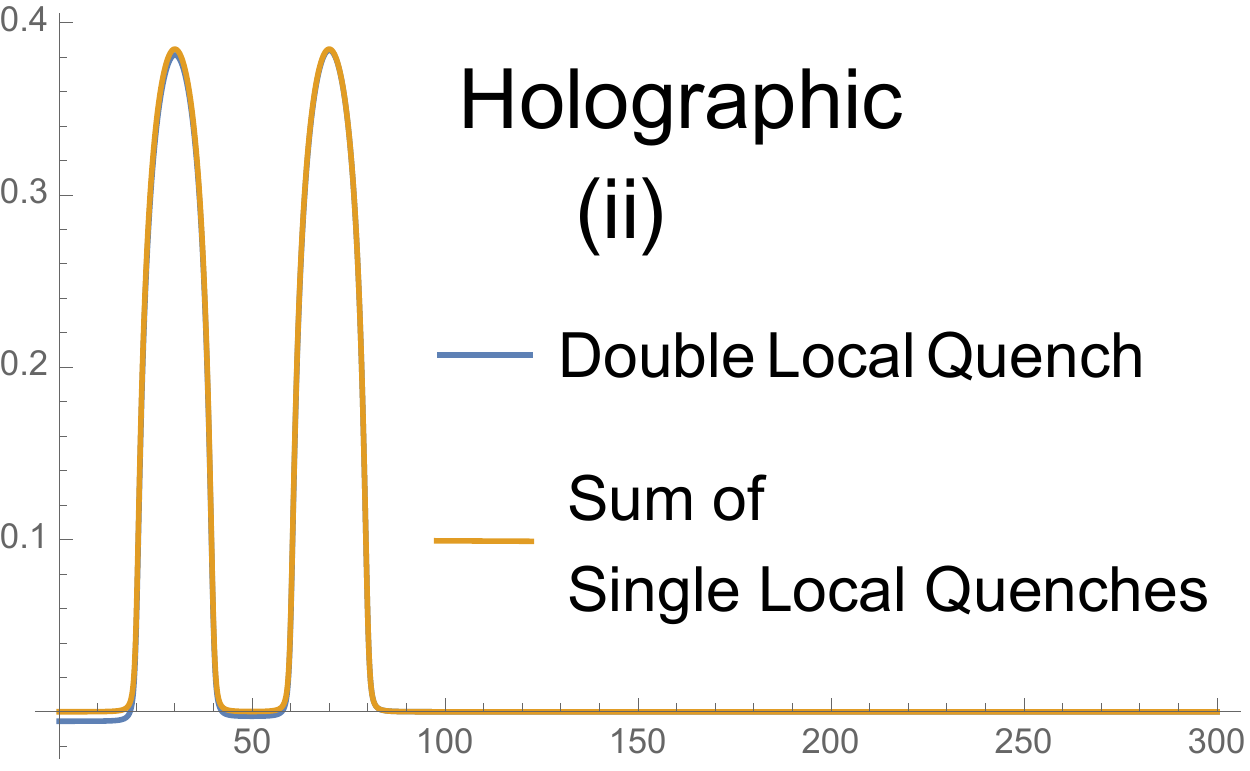}
\end{minipage}
\begin{minipage}{0.49\hsize}
\includegraphics[width=70mm]{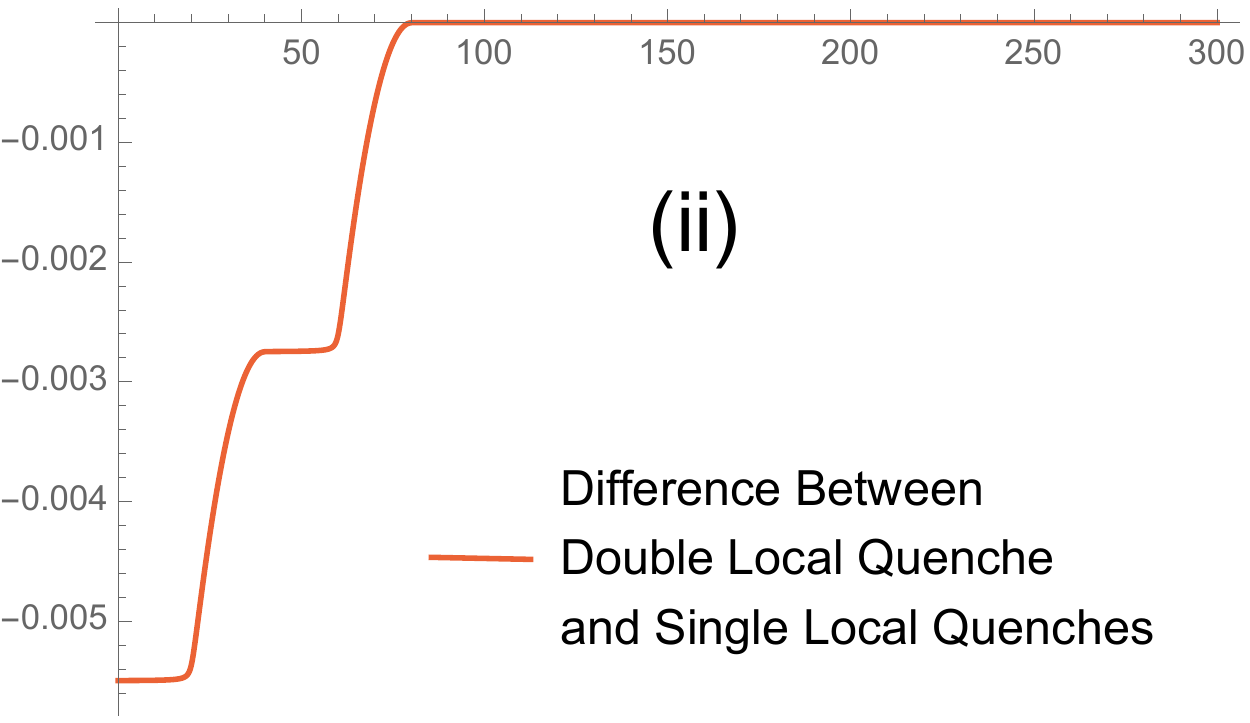}
\end{minipage}
\begin{minipage}{0.2\hsize}
        \vspace{10mm}
      \end{minipage} \\
\begin{minipage}{0.49\hsize}
\includegraphics[width=70mm]{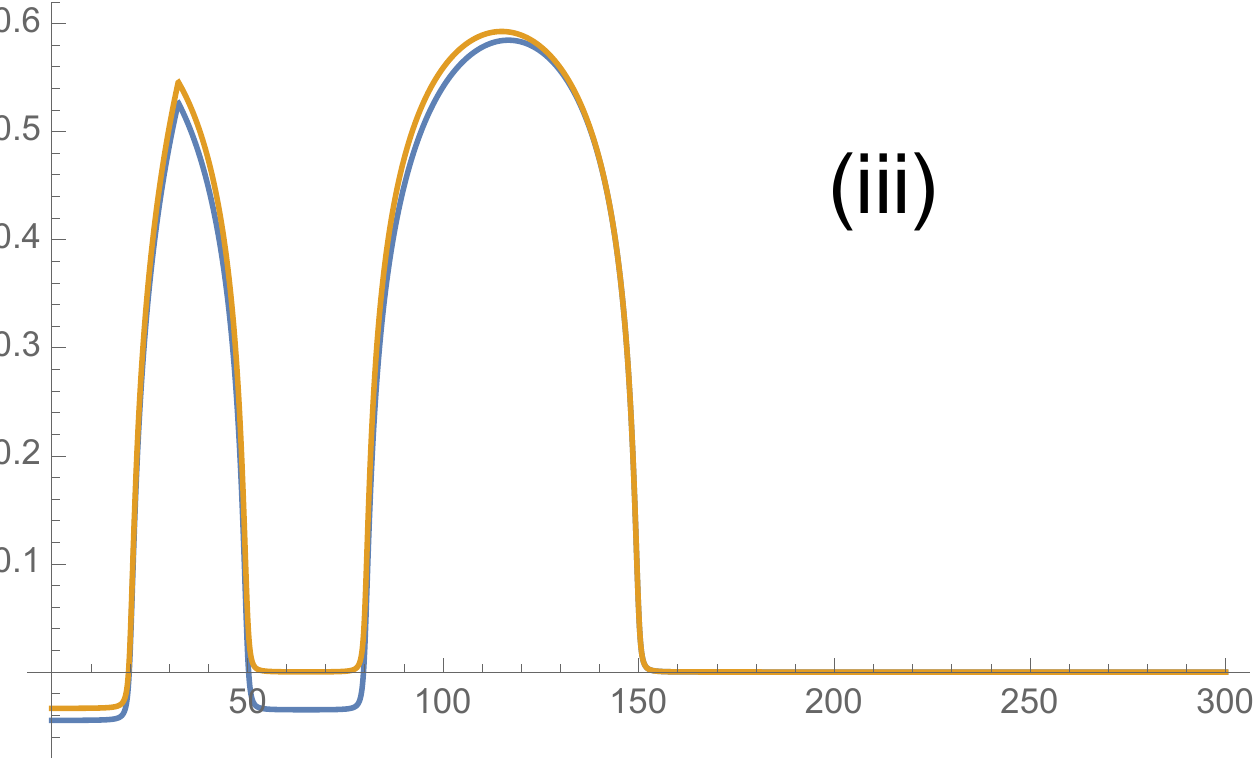}
\end{minipage}
\begin{minipage}{0.49\hsize}
\includegraphics[width=70mm]{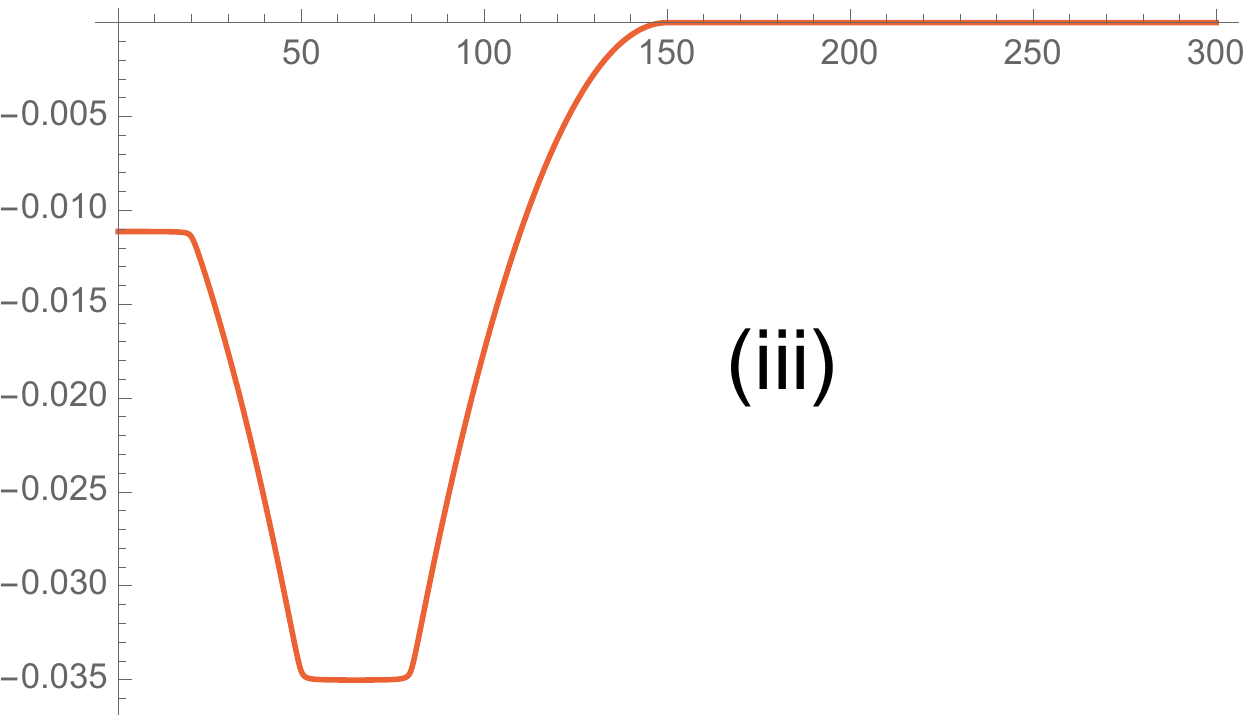}
\end{minipage}
\begin{minipage}{0.2\hsize}
        \vspace{10mm}
      \end{minipage} \\
\begin{minipage}{0.49\hsize}
\includegraphics[width=70mm]{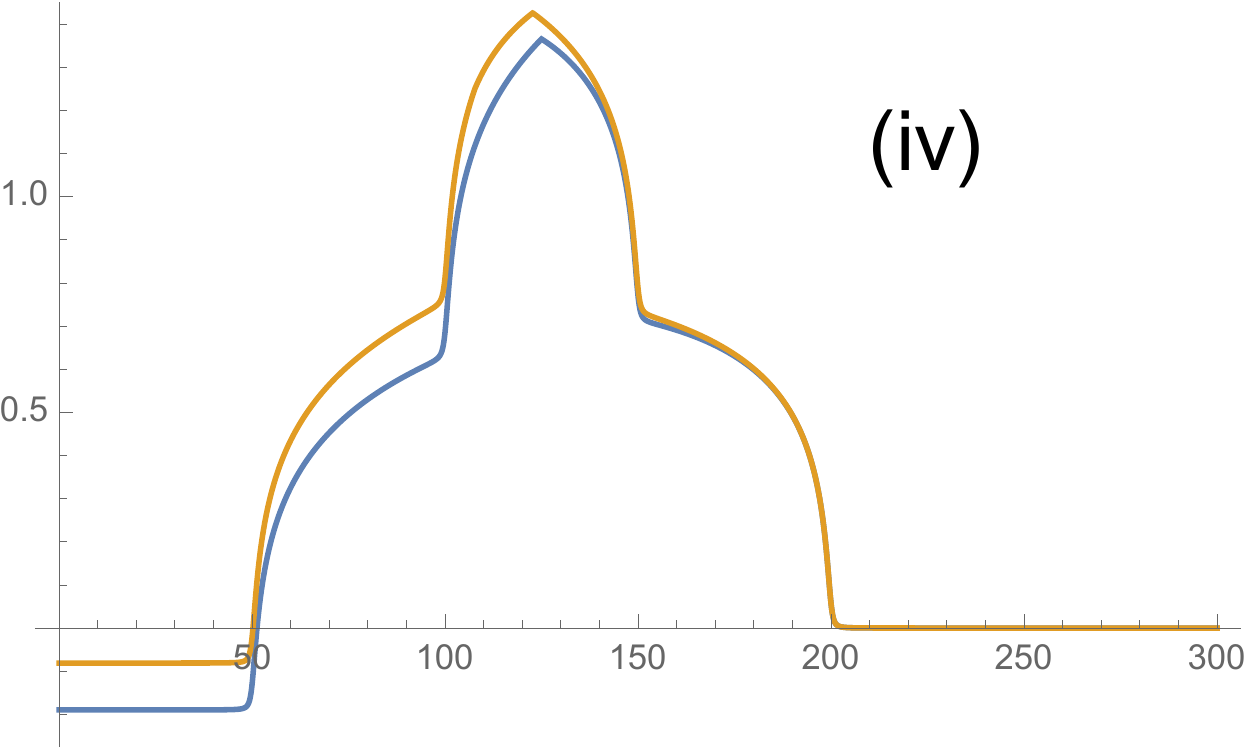}
\end{minipage}
\begin{minipage}{0.49\hsize}
\includegraphics[width=70mm]{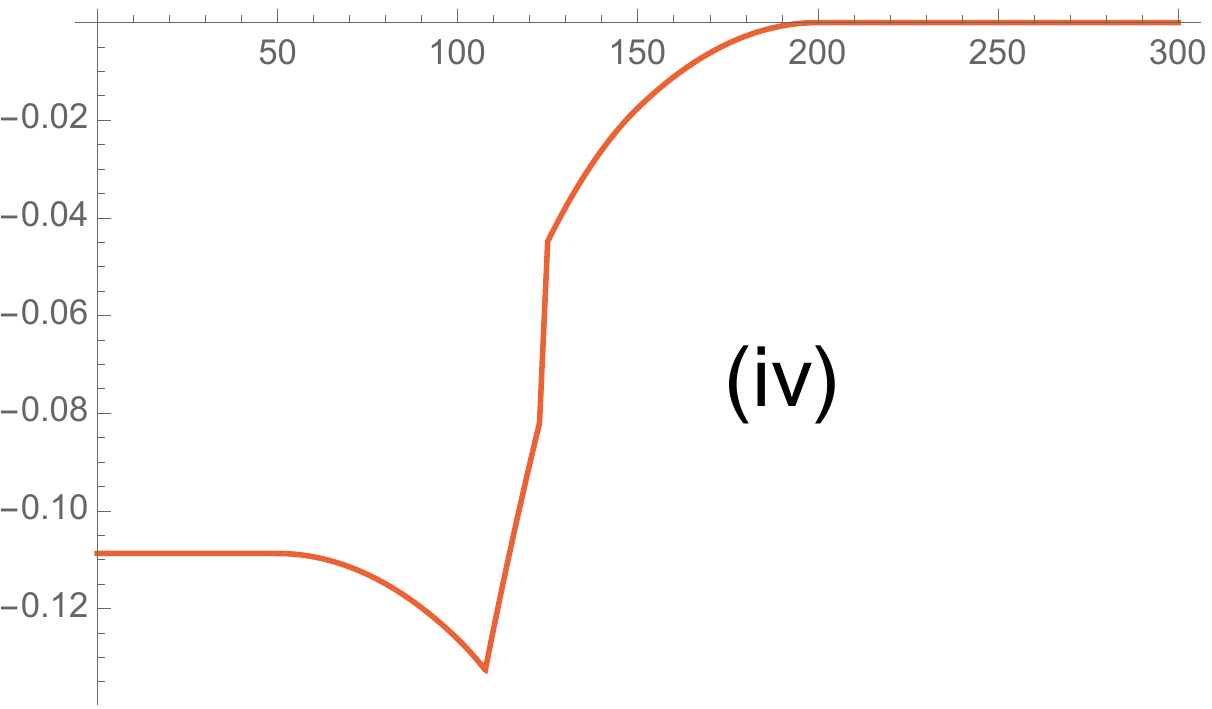}
\end{minipage}
\caption{ The plots of the full HEE $\Delta S^{D}_A$ for the double joining local quench and
the difference $\Delta S^{D}_A-\Delta S^{S(x=b)}-\Delta S^{S(x=-b)}$ 
between the double and single quench  as a function of $t$. We set $b=50$ and $a=1$.
From the top to bottom we presented results for the four different setups (i), (ii), (iii) and (iv). 
{\bf Left:}The plot of the full HEE $\Delta S_A$  for the double local quench.
{\bf  Right:} The plot of the difference $\Delta S_A^{D} - \Delta S_A^{S(x=b)} - \Delta S_A^{S(x=-b)}$ between double local quench and single local quenches.
From the top to bottom we presented results for the four different setups (i), (ii), (iii) and (iv).
  \label{fig:DLholographicTime}}
\end{figure}

Holographic entanglement entropy that is physically realized is the smaller one between $S_A^{con}$ and $S_A^{dis}$. It is also interesting to study the full holographic entanglement entropy in a double joining quench and see the difference from the sum of the full holographic entanglement entropy in single local quenches.
In figure \ref{fig:DLholographicTime}, we plot the holographic entanglement entropy, which is the smaller one among connected geodesics and disconnected geodesics, for both of double local quenches and single local quenches. In this setup (the same as that of figure \ref{fig:JoinHEEtime}), the difference of holographic entanglement entropy $\Delta S^{D}_A-\Delta S^{S(x=b)}_A-\Delta S^{S(x=-b)}_A$ becomes negative even after taking the minimal one between $S_A^{con}$ and $S_A^{dis}$.

\begin{figure}
\centering
\includegraphics[width=7cm]{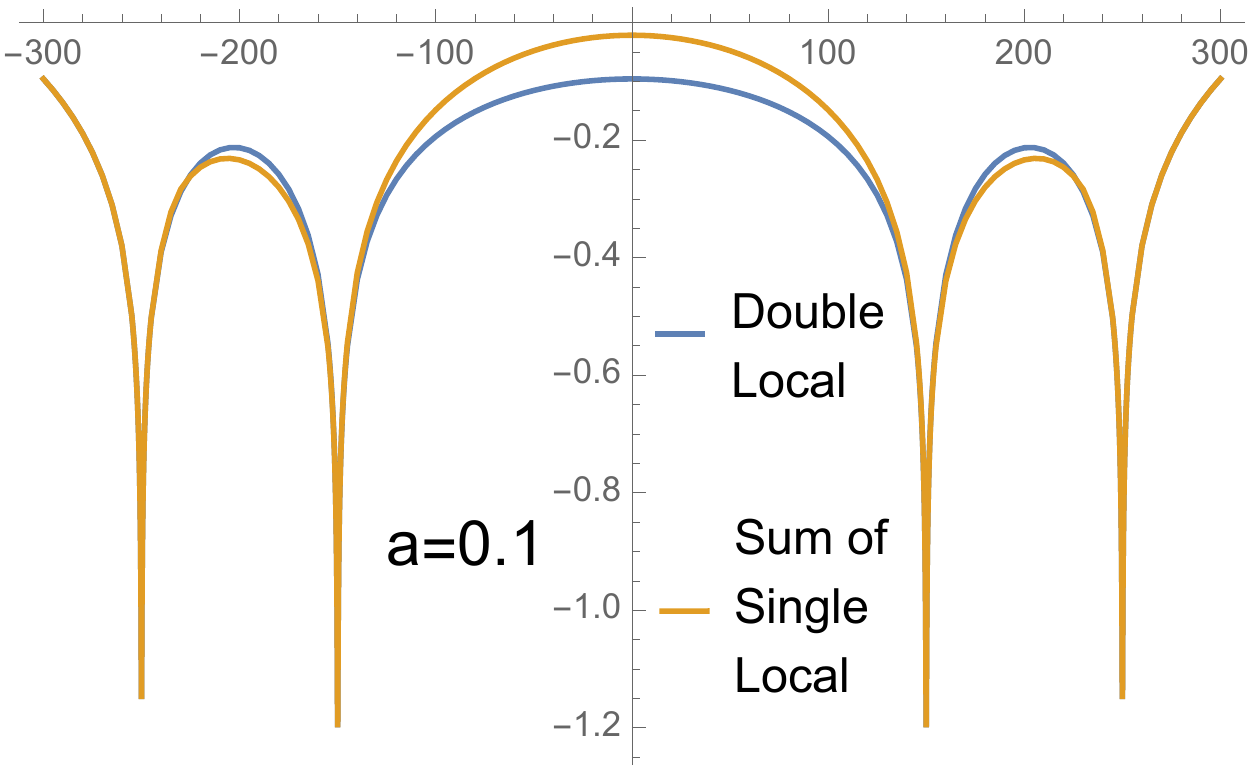}
\includegraphics[width=7cm]{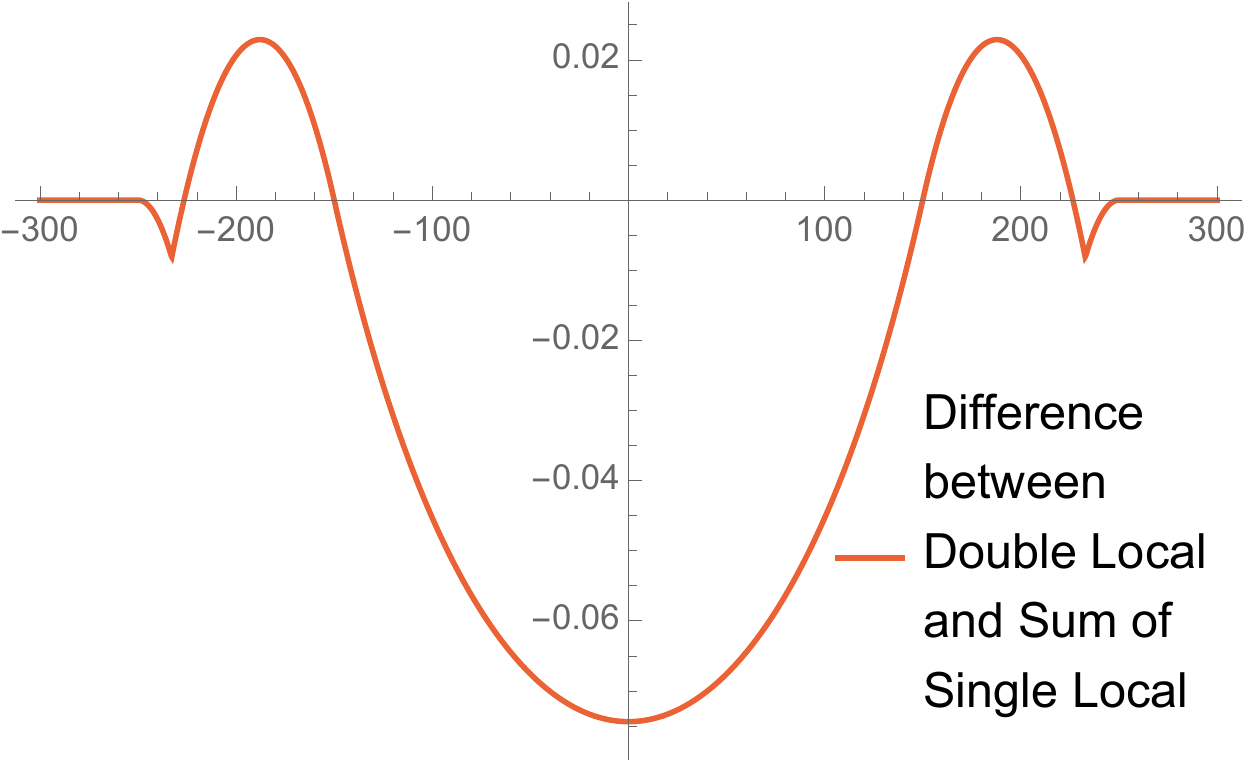}
\caption{The plot of the full HEE $\Delta S^{D}_A$ for the double joining local quench and
the difference $\Delta S^{D}_A-\Delta S^{S(x=b)}-\Delta S^{S(x=-b)}$ 
between the double and single quench at $t=0$ as a function of $x$. 
We now set $b=50$ and $a=0.1$.
We choose $A$ to be $[x-200,x+200]$ and we plot holographic entanglement entropy as functions of $x$.
{\bf Left:} The plot of $\Delta S_A$ in holographic CFT for double local quench as well as the sum $\Delta S^{S(x=b)}+\Delta S^{S(x=-b)}$  of holographic entanglement entropy for single local quenches.
{\bf  Right:} The plot of the difference $\Delta S_A^{D} - \Delta S_A^{S(x=b)} - \Delta S_A^{S(x=-b)}$ between double local quench and single local quenches.
\label{fig:JoinHEEstaticv}} 
\end{figure}

\begin{figure}
\centering
\includegraphics[width=5cm]{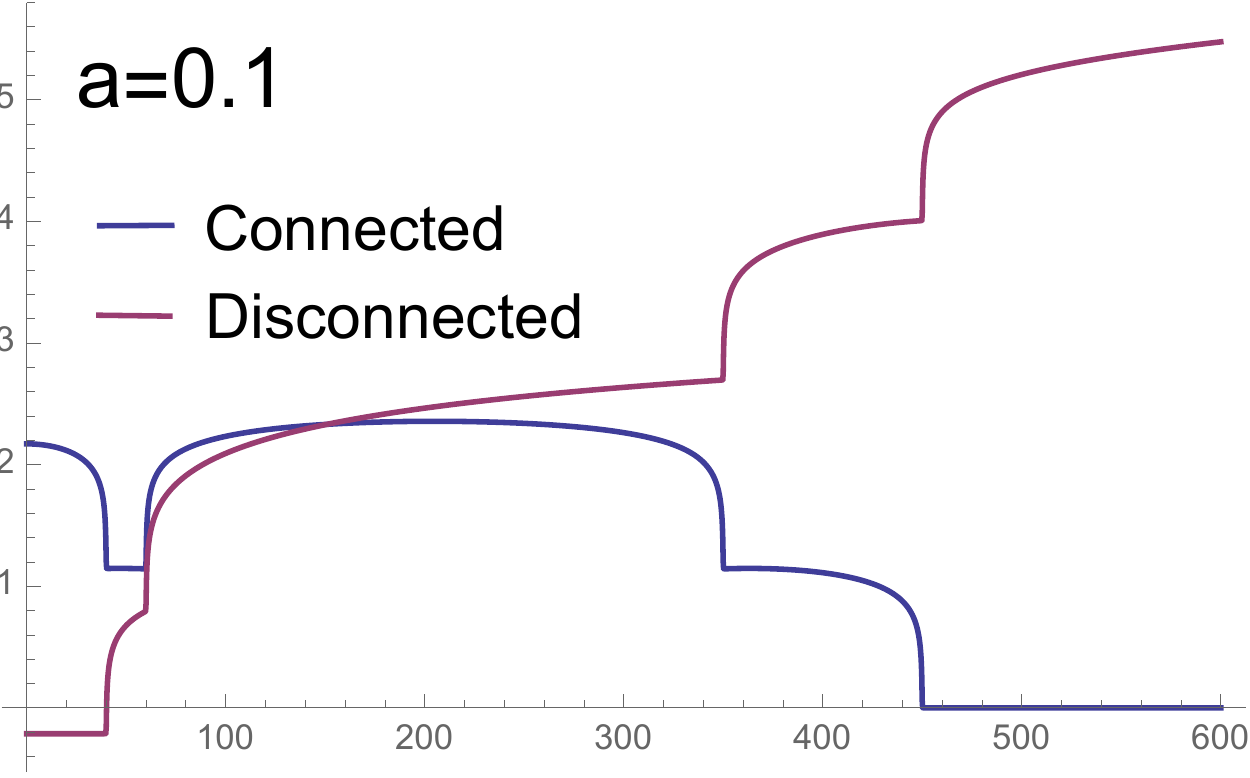}
\includegraphics[width=5cm]{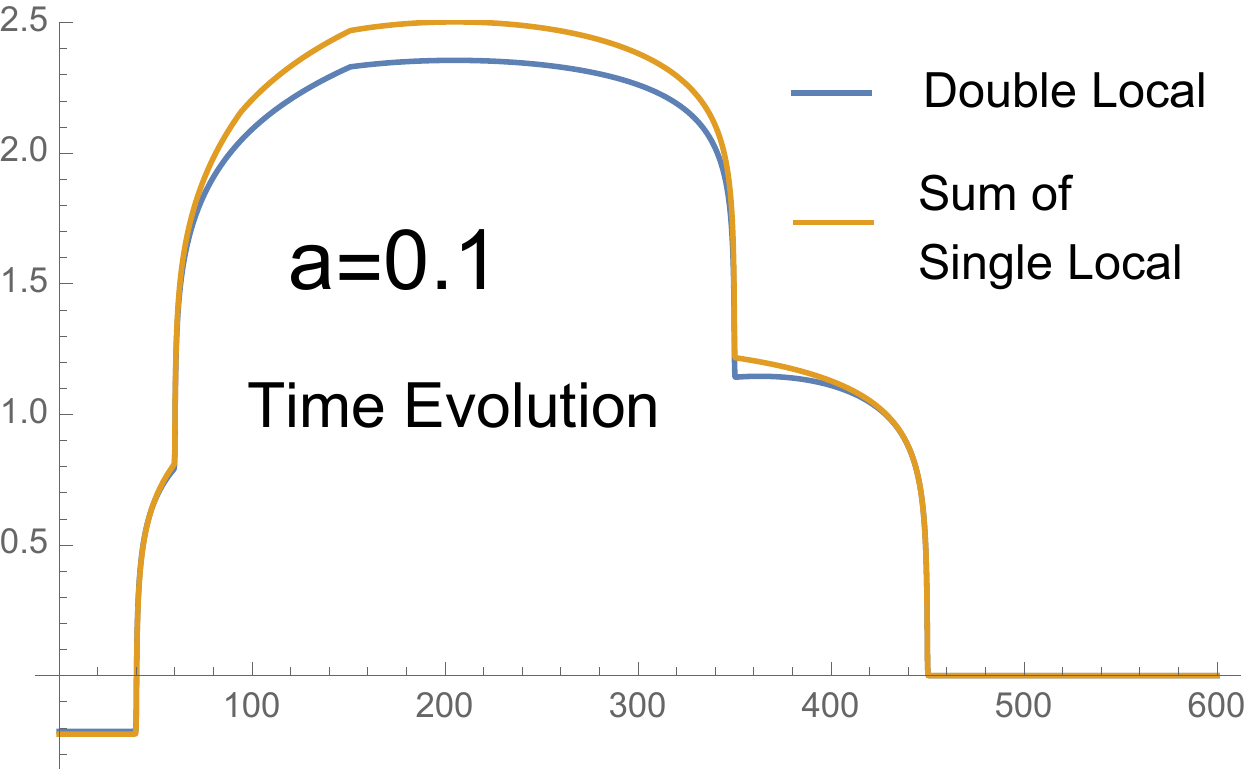}
\includegraphics[width=5cm]{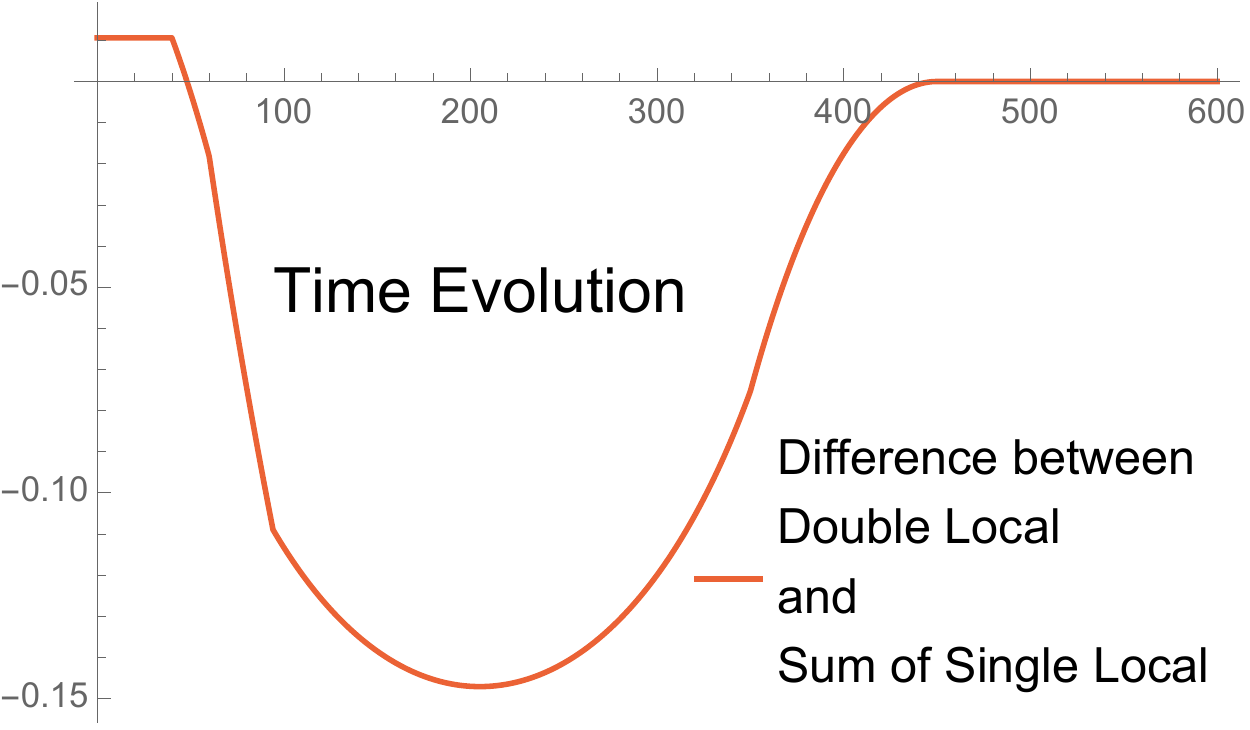}
\caption{The plot of the full HEE $\Delta S^{D}_A$ for the double joining local quench and
the difference $\Delta S^{D}_A-\Delta S^{S(x=b)}-\Delta S^{S(x=-b)}$  
as a function of $t$. We set $b=50$ and $a=0.1$.
We choose $A$ to be $[10,400]$.
{\bf Left:} In the left plots, the blue and red graph describe $\Delta S^{con}_A$ and $\Delta S^{dis}_A$, 
respectively.
{\bf Middle:} The plot of $\Delta S_A$ for double local quench as well as the sum $\Delta S^{S(x=b)}_A+\Delta S^{S(x=-b)}_A$  for single local quenches.
{\bf  Right:} The plot of the difference $\Delta S_A^{D} - \Delta S_A^{S(x=b)} - \Delta S_A^{S(x=-b)}$ between double local quench and single local quenches.
\label{fig:JoinHEEtimev}} 
\end{figure}

However, $\Delta S^{D}_A-\Delta S^{S(x=b)}-\Delta S^{S(x=-b)} \le 0$ is not always satisfied even in holographic theory and can be violated when the regions are close to the excitation points. We plot the full holographic entanglement entropy for double joining quench with $a=0.1, b=50$ in (\ref{xcmap}) at $t=0$ with $A = [x-200,x+200]$ in figure \ref{fig:JoinHEEstaticv}.
The plots show that the difference $\Delta S_A^{D} - \Delta S_A^{S(x=b)} - \Delta S_A^{S(x=-b)}$ becomes positive when the endpoint of $A$ is close ($x \sim 200 $ or $x \sim -200$) to the quench points.
We can also see that $\Delta S_A^{D} - \Delta S_A^{S(x=b)} - \Delta S_A^{S(x=-b)}$ becomes positive.
We plot the full holographic entanglement entropy for double joining quench with $a=0.1$, $b=50$ in (\ref{xcmap}) at $t=0$ with $A = [x-200,x+200]$ in figure \ref{fig:JoinHEEtimev}.
The plots show that the difference $\Delta S_A^{D} - \Delta S_A^{S(x=b)} - \Delta S_A^{S(x=-b)}$ 
is positive at early time. 

%%%%%%%%%%%%%%%%%%%%%%%%%%%%%%%%%%%%%%%%
\subsection{Boundary Surface in Holographic Double Joining Local Quenches}\label{section:DJBS}
%%%%%%%%%%%%%%%%%%%%%%%%%%%%%%%%%%%%%%%%

An essential ingredient of holographic description based on the AdS/BCFT is the presence of 
the boundary surface $Q$ which is a heavy object moving in the bulk. Here we would like to study how the surface $Q$ looks like 
in the gravity dual of double joining local quenches. This will clarify the behavior of HEE which we calculated 
just before using the conformal map into the upper half plane.

%%%%%%%%%%%%%%%%%%%%
\subsubsection{Mapping in the Bulk}
%%%%%%%%%%%%%%%%%%%%
Consider a 2 dimensional CFT on $(w,\bar{w})$ plane, and map it to $(\xi,\bar{\xi})$ plane with conformal transformation:
\ba
&&\xi = f(w) ,\no
&&\bar{\xi}=\bar{f}(\bar{w}). \label{CF}
\ea
%This is the story of the boundary CFT. 
Then let us consider the bulk AdS corresponding to it. Using $(\xi,\bar{\xi}, \eta)$ to 
denote the coordinate of 3 dimensional AdS corresponding to the CFT 
on $(\xi,\bar{\xi})$ plane, let us focus on the case when the metirc is given by Poincar\'{e} metric:
\ba
ds^2=\frac{d\eta^2+d\xi d\bar{\xi}}{\eta^2},  \label{pol}
\ea
where we set the AdS radius to $1$ for simplicity.
In this case, the gravity dual of the conformal transformation (\ref{CF}) is given by 
the following coordinate transformation in AdS$_3$ (see e.g.\cite{Ro}):
\ba
&& \xi=f(w)-\frac{2z^2(f')^2(\bar{f}'')}{4|f'|^2+z^2|f''|^2},\no
&& \bar{\xi}=\bar{f}(\bar{w})-\frac{2z^2(\bar{f}')^2(f'')}{4|f'|^2+z^2|f''|^2},\no
&& \eta=\frac{4z(f'\bar{f}')^{3/2}}{4|f'|^2+z^2|f''|^2}. \label{coradss}
\ea
The metric in the coordinate $(w,\bar{w},z)$ looks like
\ba
ds^2=\frac{dz^2}{z^2}+T_{ww}(w)(dw)^2+\bar{T}_{\bar{w}\bar{w}}(\bar{w})(d\bar{w})^2+\left(\frac{1}{z^2}
+z^2T_{ww}(w)\bar{T}_{\bar{w}\bar{w}}(\bar{w})\right)dwd\bar{w}, \label{metads}
\ea
where
\ba
T_{ww}(w)=\frac{3(f'')^2-2f'f'''}{4f'^2},\ \ \bar{T}_{\bar{w}\bar{w}}(\bar{w})=\frac{3(\bar{f}'')^2-2\bar{f}'\bar{f}'''}{4\bar{f}'^2},
\ea
are the chiral and anti-chiral energy stress tensor, respectively. These together give the gravity dual in the $(w,\bar{w},z)$ coordinate.

%%%%%%%%%%%%%%%%%%%%%%%%%%%%%%%%%%%%%%%%
\subsubsection{Boundary Surface in Euclidean Setup}
%%%%%%%%%%%%%%%%%%%%%%%%%%%%%%%%%%%%%%%%

Here, we denote $g \equiv f^{-1}$, and introduce two new parameters  $F$ and $\bar{F}$ which satisfy
\ba
&&w = g(F), \no
&&\bar{w}=\bar{g}(\bar{F}). \label{TM}
\ea
Then we have
\ba
&& \xi=F + \frac{2z^2g'(F)\bar{g}''(\bar{F})}{4(g'(F)\bar{g}'(\bar{F}))^2+z^2g''(F)\bar{g}''(\bar{F})},\no
&& \bar{\xi}=\bar{F} + \frac{2z^2\bar{g}'(\bar{F})g''(F)}{4(g'(F)\bar{g}'(\bar{F}))^2+z^2g''(F)\bar{g}''(\bar{F})},\no
&& \eta=\frac{4z(g'(F)\bar{g}'(\bar{F}))^{3/2}}{4(g'(F)\bar{g}'(\bar{F}))^2+z^2g''(F)\bar{g}''(\bar{F})}. \label{corads}
\ea

In the case of the double joining quench, the metric of the bulk AdS in $(\xi,\bar{\xi}, \eta)$ coordinate is given 
by (\ref{pol}). When boundary tension $T_{BCFT} = 0$, the boundary surface $Q$ is given by $\xi -\bar{\xi} = 0$ and
the gravity dual is given by $\xi -\bar{\xi} >0$. This means, on the boundary surface, we have\footnote{This is valid only at $F\neq\bar{F}$. In $w$ coordinate, the part of the boundary surface corresponding to $F=\bar{F}$ extends from the BCFT boundary to the bulk. This part does not influence Lorentzian time evolution of joining quenches. However, it is crucial in splitting quenches.}
\ba
z(F,\bar{F}) = \bigg( \frac{4(F-\bar{F})g'^2\bar{g}'^2}{2(g''\bar{g}'-\bar{g}''g')-(F-\bar{F})g''\bar{g}''} \Bigg)^{1/2}.
\label{DJBSpara}
\ea
Then we can let $F$ run in the region $F-\bar{F} > 0$ to find out the boundary surface
given by $(w,\bar{w},z)=(g(F),g(\bar{F}), z(F,\bar{F}) )$.

%%%%%%%%%%%%%%%%%%%%%%%%%%%%%%%%%%%%%%%%
\subsubsection{Single Joining Quench Revisited}\label{sec:singleBS}
%%%%%%%%%%%%%%%%%%%%%%%%%%%%%%%%%%%%%%%%

Discussions so far can be applied to any BCFT setup which can be mapped to an upper half plane using a conformal map. Before we go further into details of double joining quench, let us revisit the single joining quench case. 

\begin{figure}
  \centering
   \includegraphics[width=10cm]{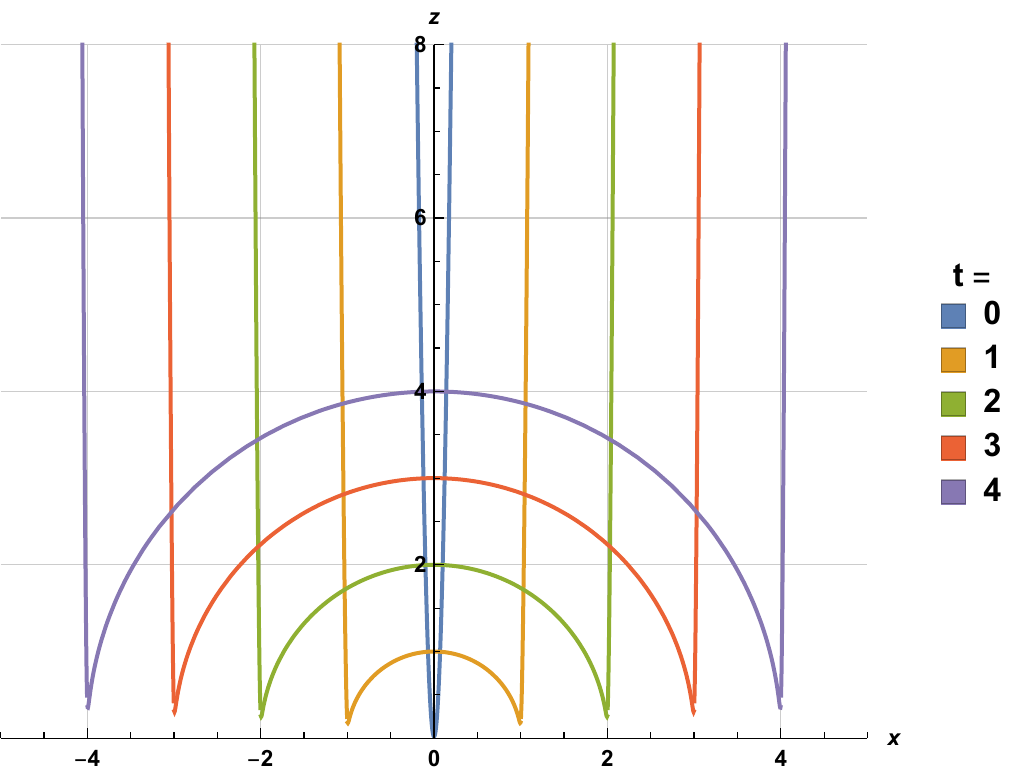}
 \caption{Boundary surfaces of single joining quench at different $t$. $a=0.01$. Note that the boundary surface at $x<t$ is given by $z^2 = (t^2-x^2)\Big[1+\mathcal{O}\Big(\frac{t^2a^2}{(t^2-x^2)^2}\Big)\Big]$ and the $x>t$ part is given by $z^2=\frac{1}{a^2}\frac{4(x^2-t^2)^3}{(x^2+3t^2)}\Big[1+\mathcal{O}\Big(\frac{t^2a^2}{(t^2-x^2)^2}\Big)\Big]$. Initially at $t=0$, there is a sharp boundary surface localized at $x\sim0$. Then it moves to $\pm x$ and falls towards $+z$ direction in the bulk. When $t$ is sufficiently large, the boundary surface can be roughly regarded as two vertical lines and a semicircle between them.}
\label{fig:SJBSTE}
\end{figure}

A single joining quench with cutoff $a$ defined on $w$ plane can be mapped to an upper half plane $\xi$ by the map\footnote{The sign of this map is opposite to that used in \cite{STW}. This map here can be reduced from the map (\ref{xcmap}) by set $\a=0$.}
\ba
w = g\left(\xi\right) = ia\frac{\xi^2+1}{\xi^2-1}.
\ea
In order to figure out bulk properties, we introduce a parameter $F$ which satisfies $w=g\left(F\right)$ as explained before. In this case, it is straightforward to write $F$ as a function of $w=x+i\tau$, and we can get the boundary surface by simply plugging it into (\ref{DJBSpara}). The result is given by
\ba
z = \frac{2\sqrt{x^2+(a+\tau)^2}\sqrt{x^2+(a-\tau)^2}}{\sqrt{a^2-\Big(\sqrt{x^2+(a+\tau)^2}-\sqrt{x^2+(a-\tau)^2}\Big)^2}} \equiv G_E(x,\tau).
\ea
Then applying analytic continuation, we have
\ba
z = \frac{2\sqrt{x^2+(a+it)^2}\sqrt{x^2+(a-it)^2}}{\sqrt{a^2-\Big(\sqrt{x^2+(a+it)^2}-\sqrt{x^2+(a-it)^2}\Big)^2}} \equiv G_L(x,t).
\label{eq:SJBS}
\ea
Figure \ref{fig:SJBSTE} shows how the boundary surface looks like on different time slices. Initially at $t=0$, there is a sharp boundary surface localized at $x\sim0$. Then it moves to $\pm x$ and falls towards $+z$ direction in the bulk. When $t$ is sufficiently large, the boundary surface can be roughly regarded as two vertical lines and a semicircle between them. 

Features appeared in the time evolution of both connected EE and disconnected EE can be understood from the behavior of geodesics in the presence of the boundary surface. Refer to section \ref{sec:SJQ} for a brief review of single joining quenches.

\begin{figure}[h]
  \centering
   \includegraphics[width=7cm]{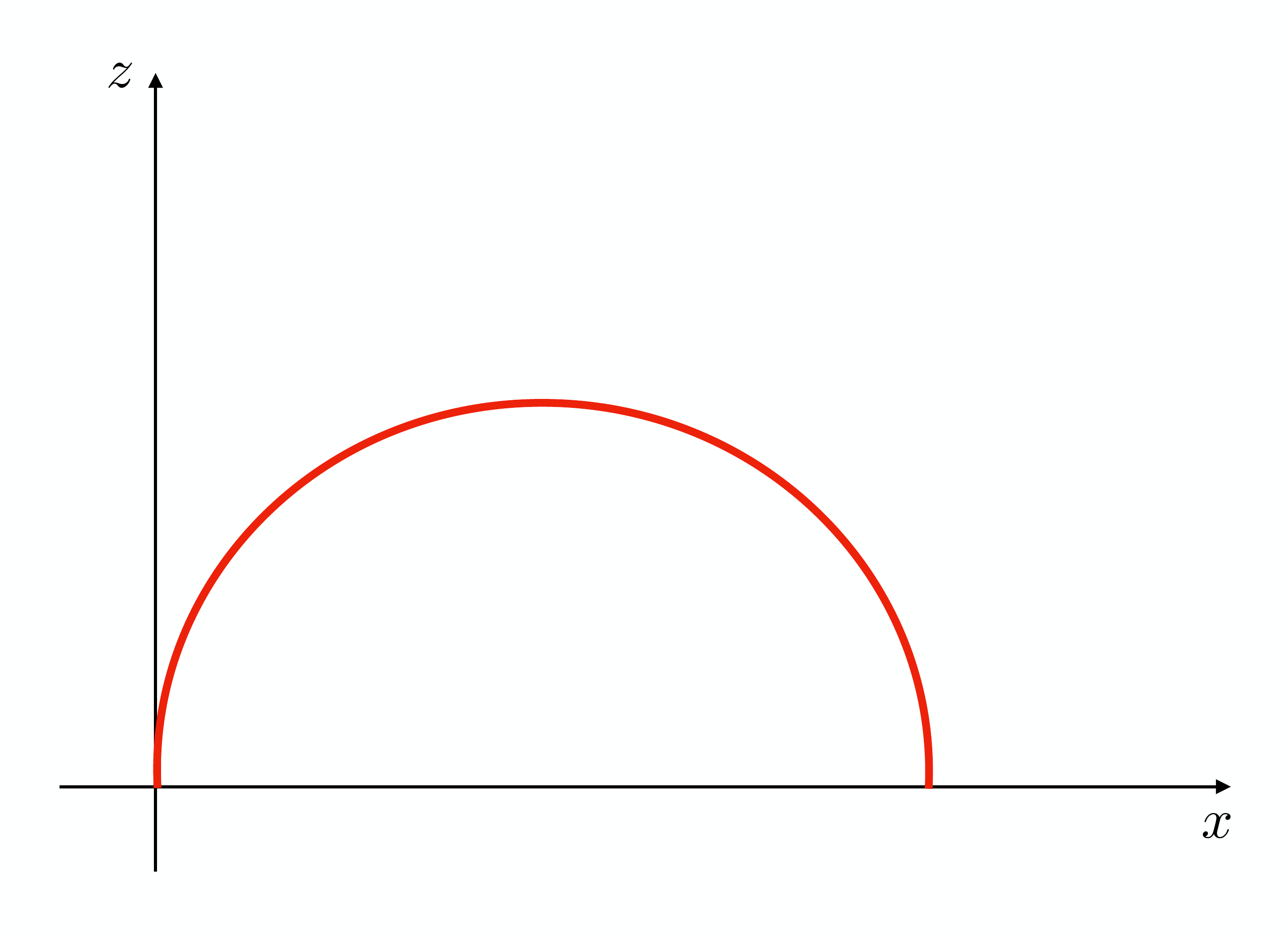}
   \includegraphics[width=7cm]{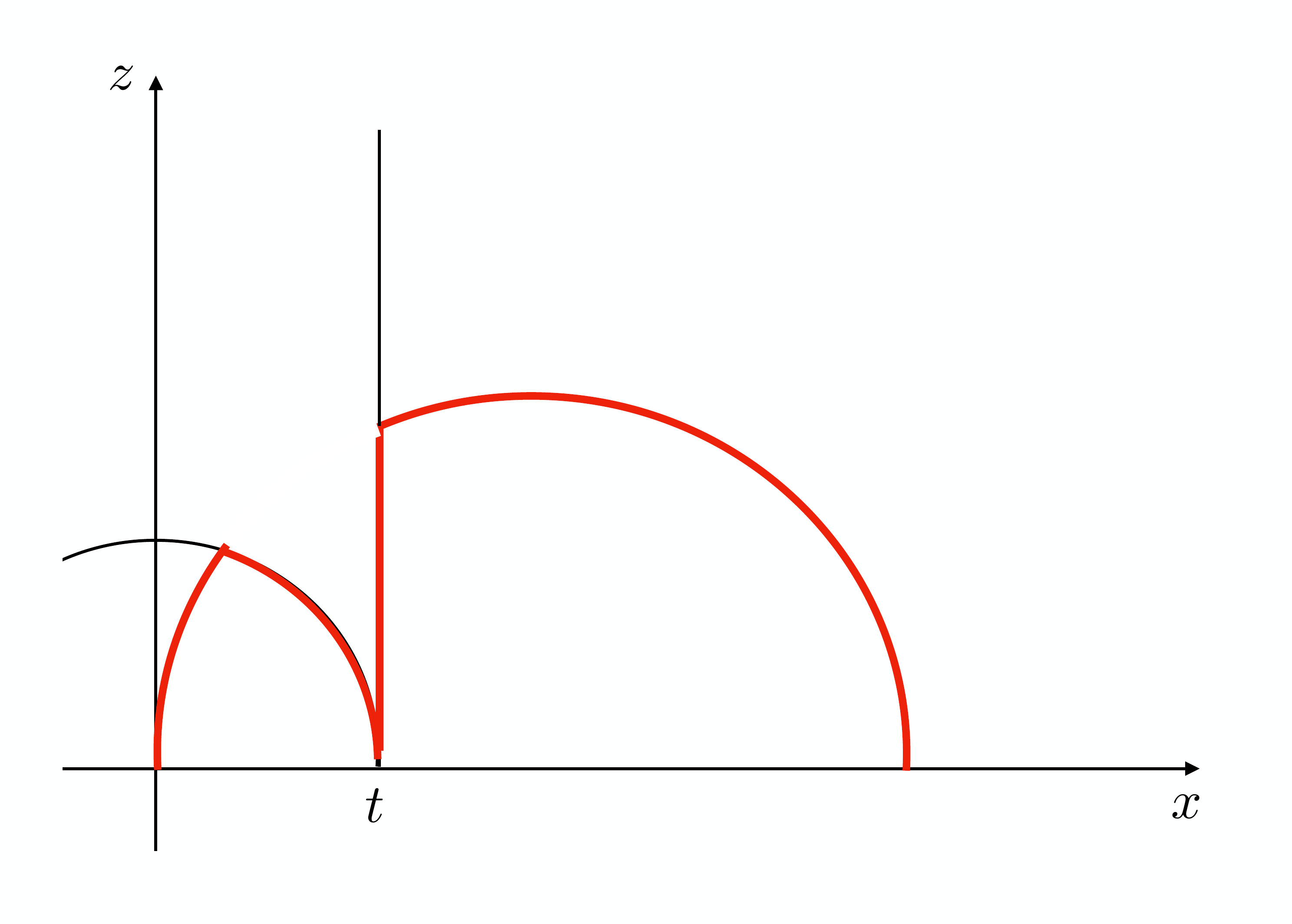}
 \caption{A schematic drawing of connected geodesic (shaded red) on a time slice, without and with the existence of a boundary surface. This detour at the sharp corner of the boundary gives an increase of connected entanglement entropy. Note that the real geodesic does not exactly lie on a time slice.}
\label{fig:SJdetour}
\end{figure}

One of the main features of connected EE is that there are hills (increase followed by decrease) in the graph of time evolution.  
This can be understood by connected geodesic making a detour at the sharp corner of the boundary surface. Figure \ref{fig:SJdetour} shows a schematic drawing of such a detour on a time slice. Though a general geodesic should not exactly lie on a time slice, we can see that, actually for this case, even an estimation restricted on a time slice can give the leading order of connected EE. (See appendix \ref{SJCEE} for details.)

\begin{figure}
  \centering
   \includegraphics[width=7cm]{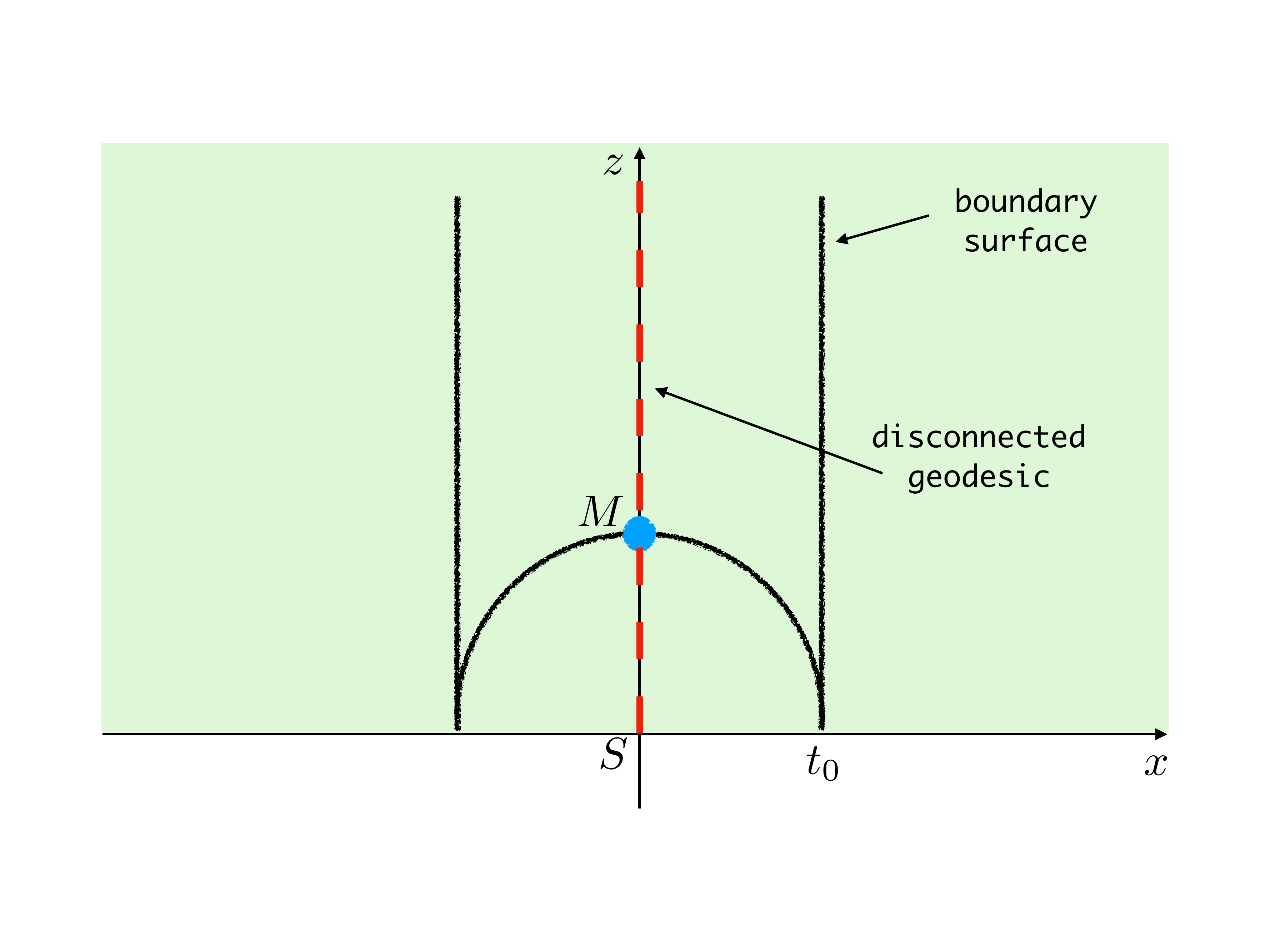}
   \includegraphics[width=7cm]{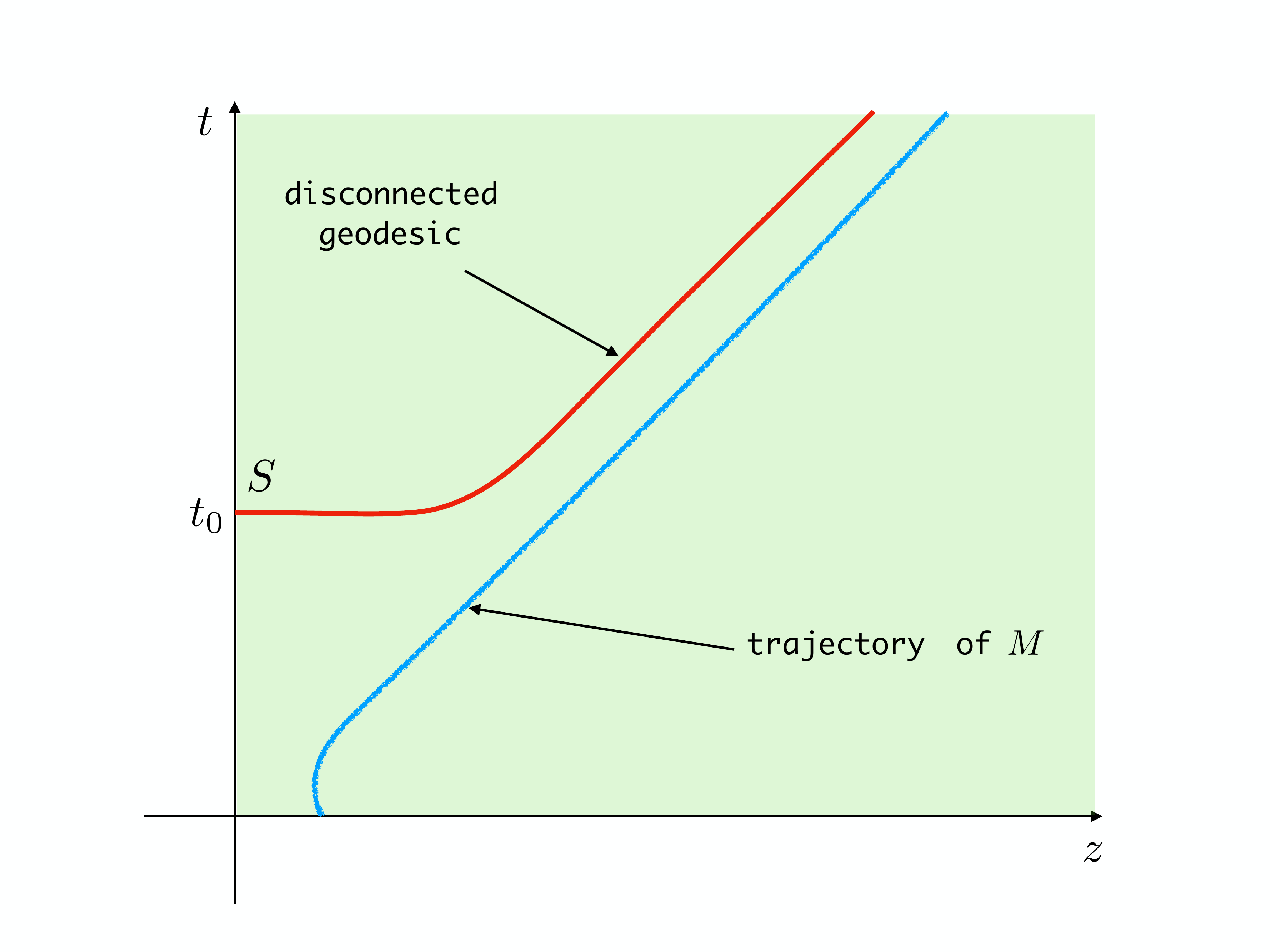}
   \includegraphics[width=8cm]{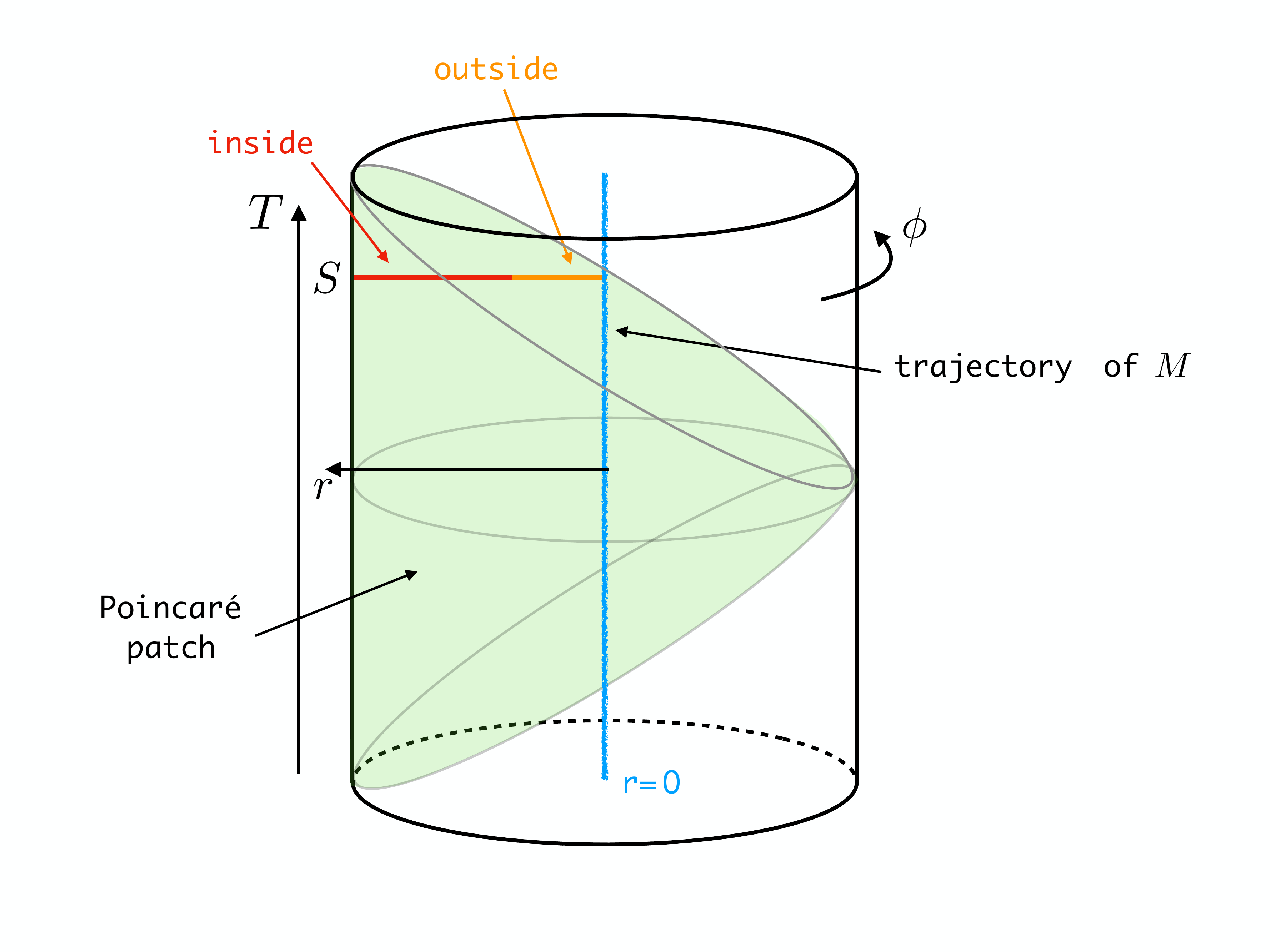}
 \caption{The left figure shows a time slice in Poincar\'{e} coordinate $(t,x,z)$. For simplicity, let us set the start point $S$ of the disconnected geodesic to be $(t_0, 0, 0)$. The red dashed line shows how the geodesic extending from $S$ looks like when projected on the time slice. Note that this geodesic does not lie on the time slice. The blue point $M$ is the middle point of the boundary surface. The right figure shows how these look like on $x=0$. The lower figure shows how these looks like in a global coordinate $(T, r, \phi)$. This coordinate is taken to satisfy that $M$ stays at $r=0$. The Poincar\'{e} patch is shaded green in all three figures. The geodesic has an inside part (red) and an outside part (yellow). Its length is given by $\log (t/\ep)+\log(t/a)$}.
\label{fig:SJdisgeo}
\end{figure}

One of the main features of disconnected EE is that there is an increase in the graph of time evolution.  This can be understood by disconnected geodesic extended towards $z$ direction at late time. Figure \ref{fig:SJdisgeo} shows 
how a disconnected geodesic extended from $(x, t)$ to the boundary surface looks like at $t\gg x$. This geodesic does not end inside the Poincar\'{e} patch. The length of this geodesic inside and outside the Poincar\'{e} patch is given by $\log(t/\epsilon)$ and $\log(t/a)$, respectively. They together contribute to the $(c/3)\log t$ time evolution (\ref{SJEEdis}) in disconnected EE. (See section 6 of \cite{STW} for details.)

%%%%%%%%%%%%%%%%%%%%%%%%%%%%%%%%%%%%%%%%
\subsubsection{Boundary Surface in Double Joining Quench}
%%%%%%%%%%%%%%%%%%%%%%%%%%%%%%%%%%%%%%%%

\begin{figure}
  \centering
   \includegraphics[width=7cm]{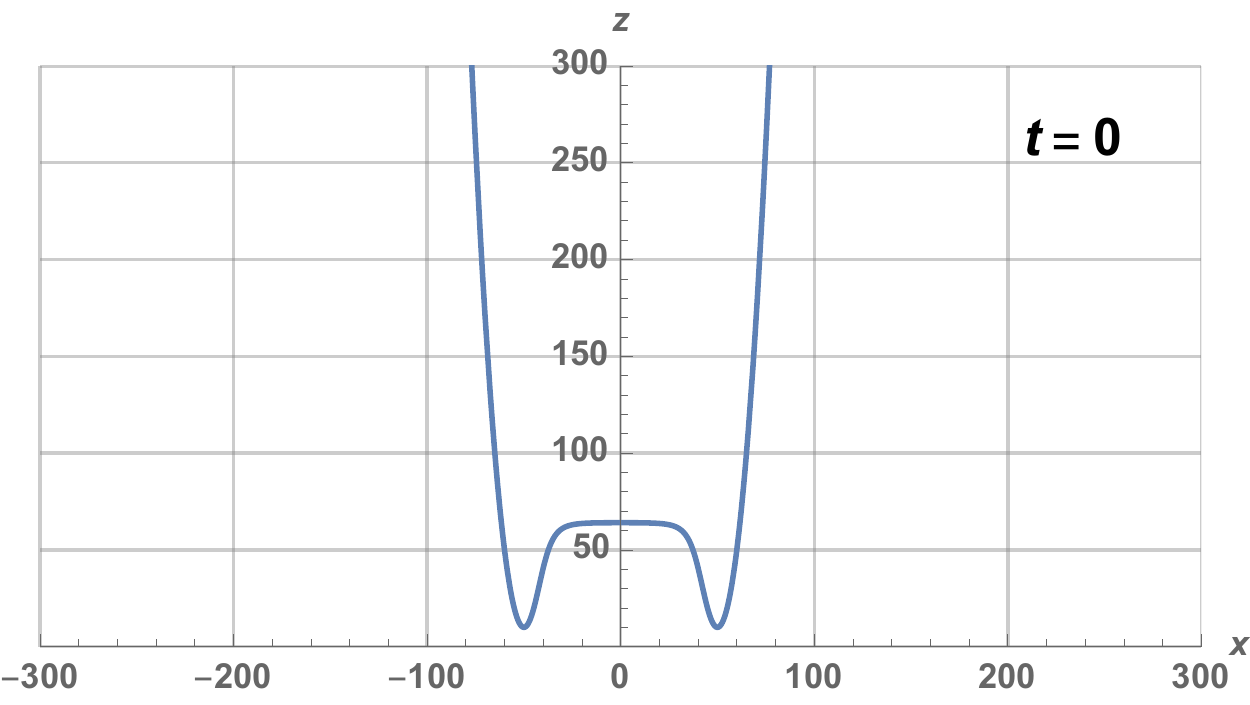}
   \includegraphics[width=7cm]{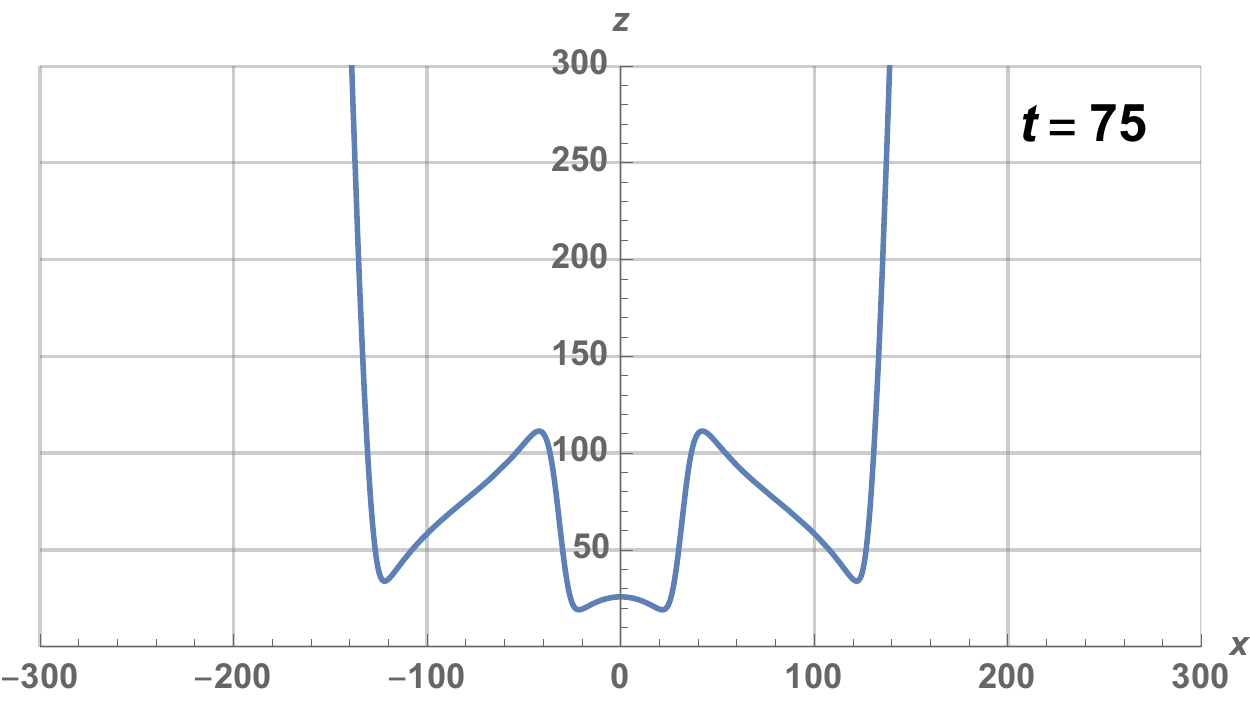}
   \includegraphics[width=7cm]{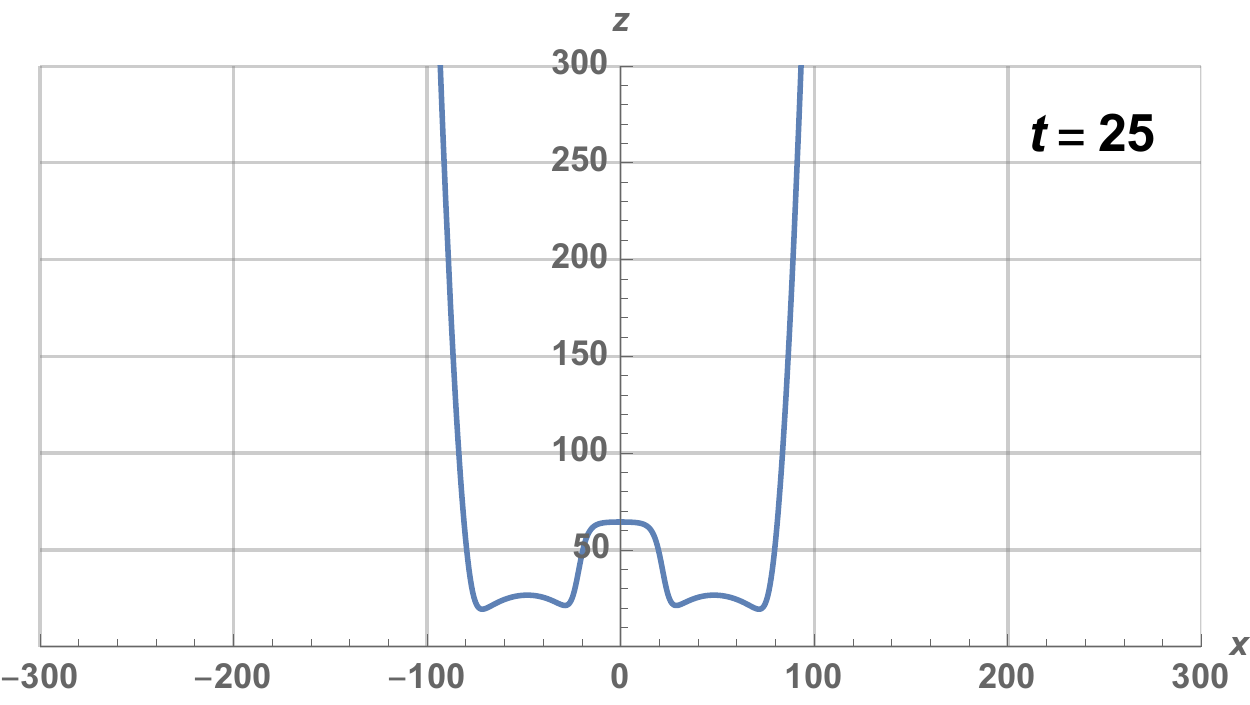}
   \includegraphics[width=7cm]{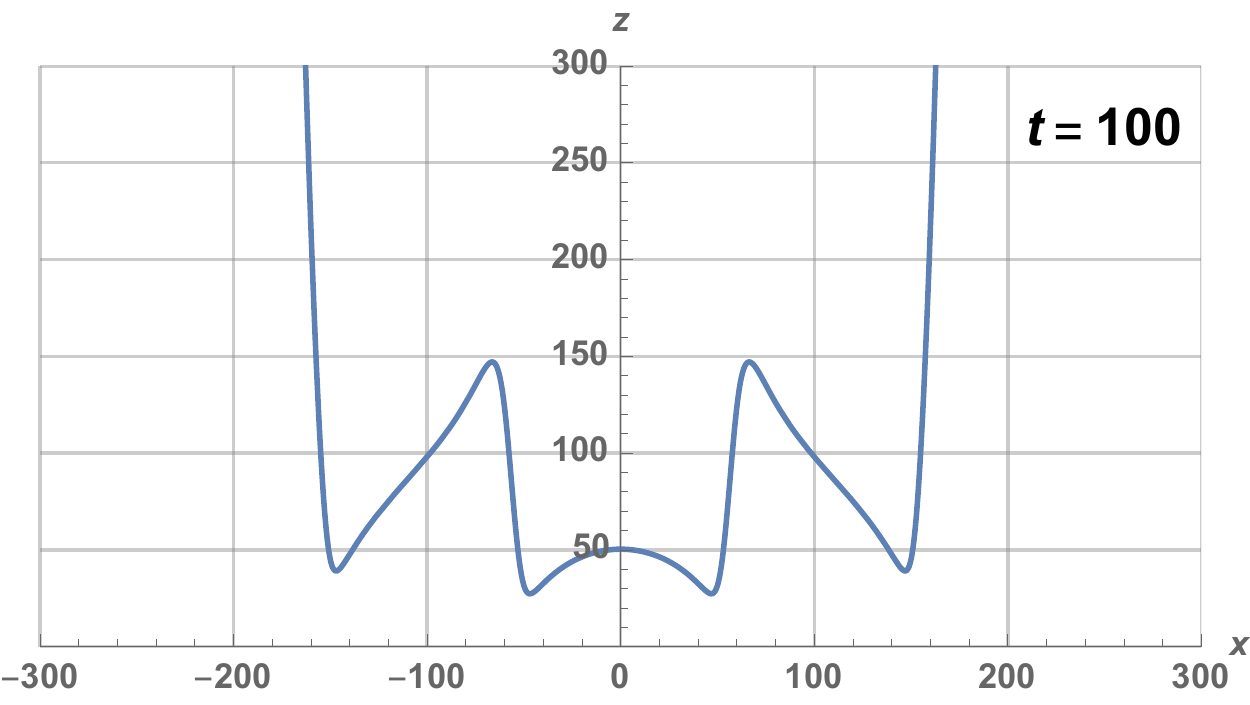}
   \includegraphics[width=7cm]{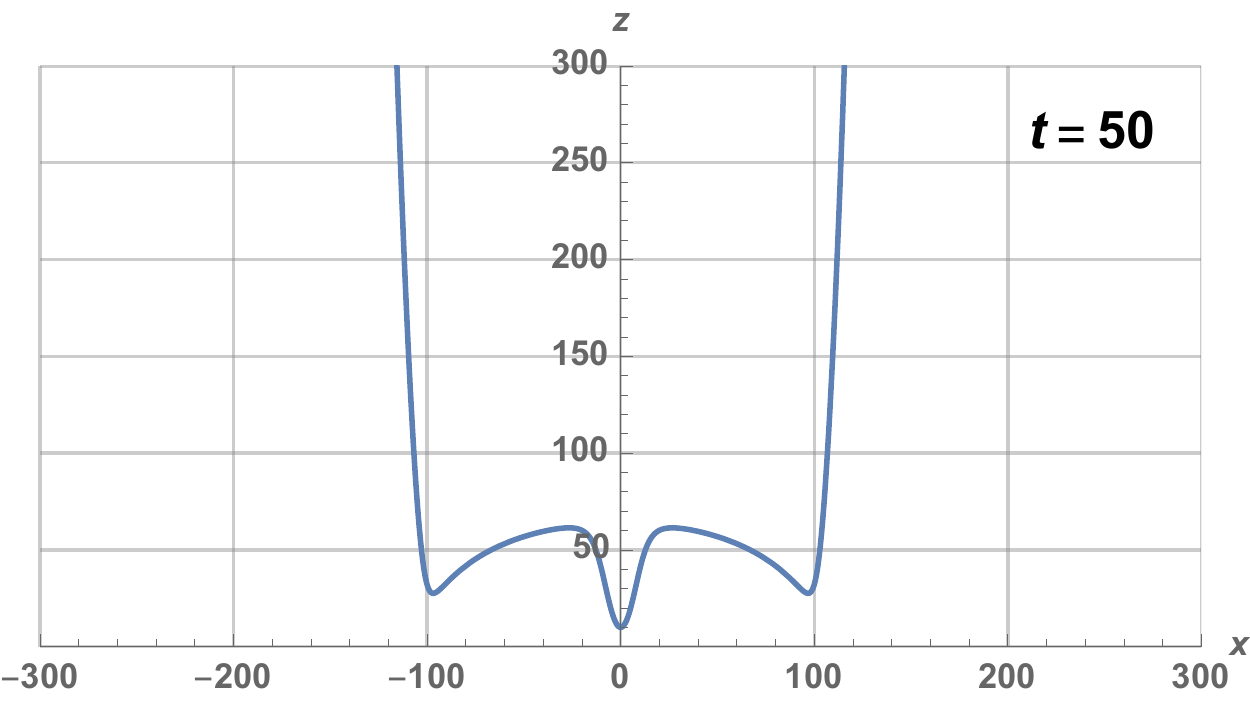}
   \includegraphics[width=7cm]{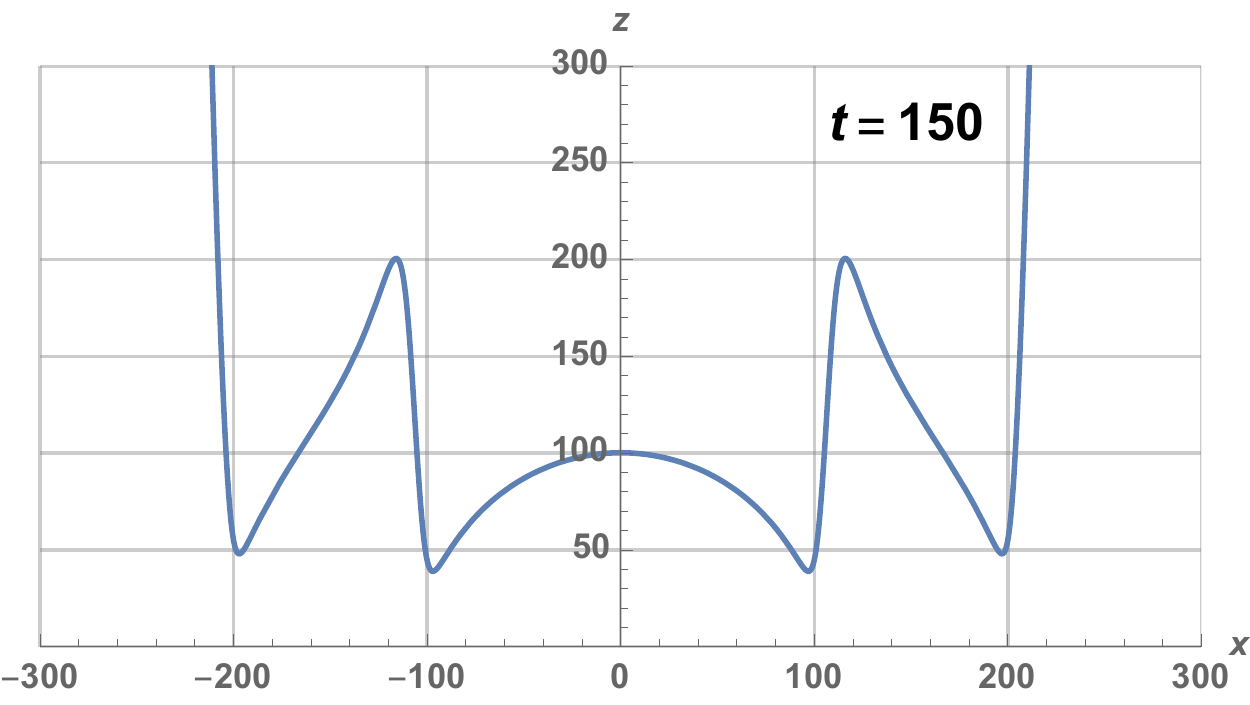}
 \caption{Boundary surface in double joining quench on different time slices with sufficiently small $a$. (Here, we have $b=50$ and $a=5$.) At $t=0$, the boundary has two sharp angles localized at $x=\pm b$, and then both of them moves to $\pm x$ direction. At large $t$, the boundary surface can be roughly regarded as two vertical lines $x=\pm b$, a semicircle $x^2+z^2=(t-b)^2$ and two sharp angles at each side between them. The boundary is falling towards $+z$ direction as a whole.}
\label{fig:DJBSTE}
\end{figure}

In a general double joining quench, the $z$-coordinate of a boundary surface cannot be explicitly represented as a function of $x$ and $t$. However, we can use  the maps (\ref{wmapa}) and 
 (\ref{wmapaa}) to numerically compute it. Figure \ref{fig:DJBSTE} shows the boundary surface of a double joining quench at different time slices. Initially at $t=0$, the boundary has two sharp angles localized at $x=\pm b$, and then both of them moves to $\pm x$ direction. At large $t$, the boundary surface can be roughly regarded as two vertical lines $x=\pm b$, a semicircle $x^2+z^2=(t-b)^2$ and two sharp angles on each side between them. The boundary is falling towards $+z$ direction as a whole. 
 
Let us see how this behavior of boundary surface in a double joining quench is consistent with many results computed in a holographic CFT at $b/a\gg1$. 
\begin{itemize}
\item Energy stress tensor for a double joining quench at $x\gg1$ is almost the same as that for a single quench. The same statement is also true for connected entanglement entropy of a small subsystem at $x\gg1$ due to the first law. From a boundary surface point of view, this is because the heavy boundary surface for a double joining quench looks almost the same as that for a single joining quench in this case. This is shown in the left figure of figure \ref{fig:cat}.

\item It is also true by changing $x\gg1$ in the former statement with $t\gg1$. This is shown in the right figure of figure \ref{fig:cat}.

\item As (\ref{timehalf}) shows, EE for a semi-infinite subsystem has a $(c/3)\log t$ time evolution. Indeed we can confirm that the disconnected geodesic in the double joining
quench is essentially the same as that of the disconnected geodesic in a single joining quench.

\item Qualitative features in figure \ref{fig:JoinHEEtime} can also be understood by considering detour of connected geodesics and contributions from disconnected geodesics.\\
\begin{figure}[h!]
  \centering
   \includegraphics[width=7cm]{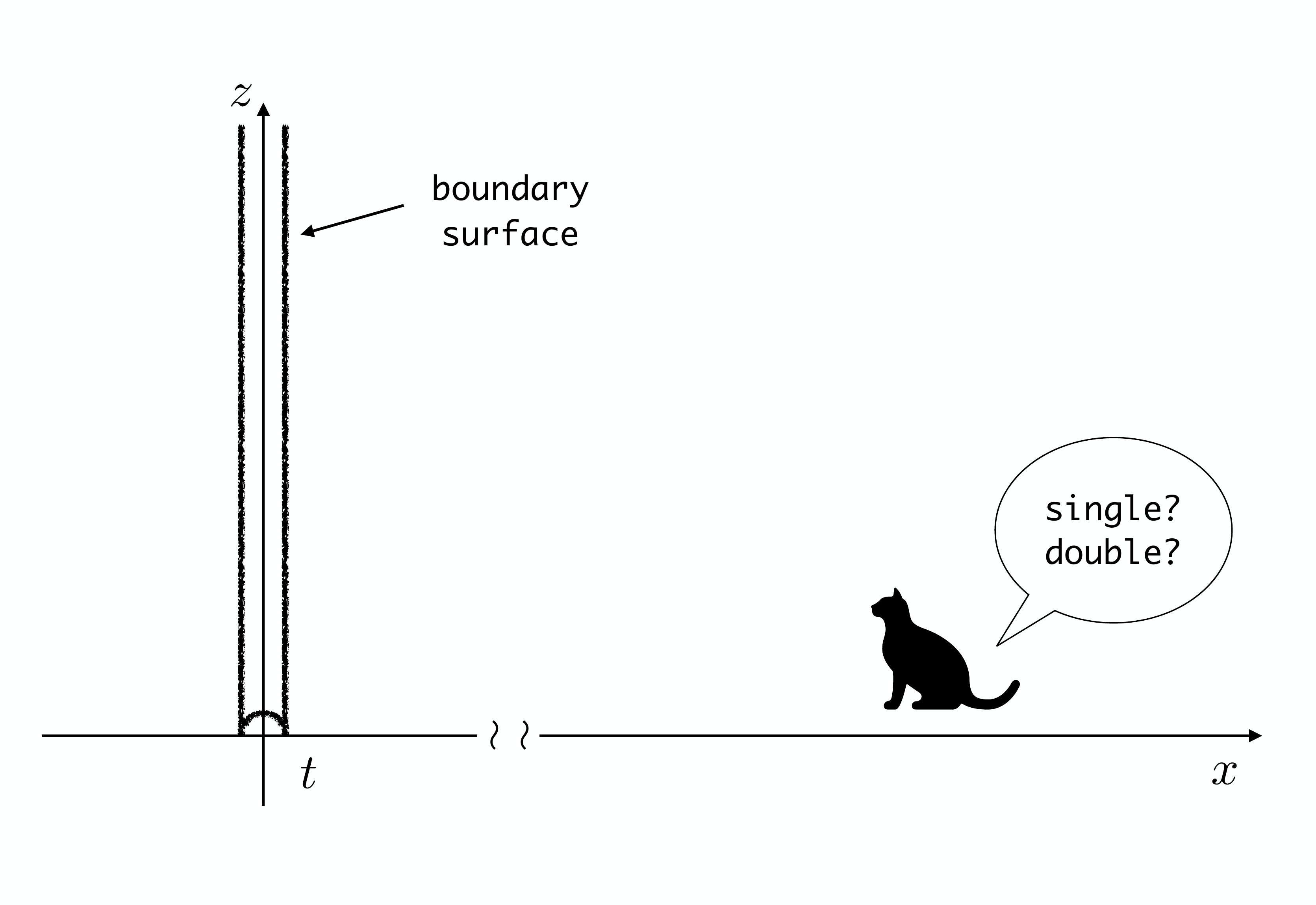}
   \includegraphics[width=7cm]{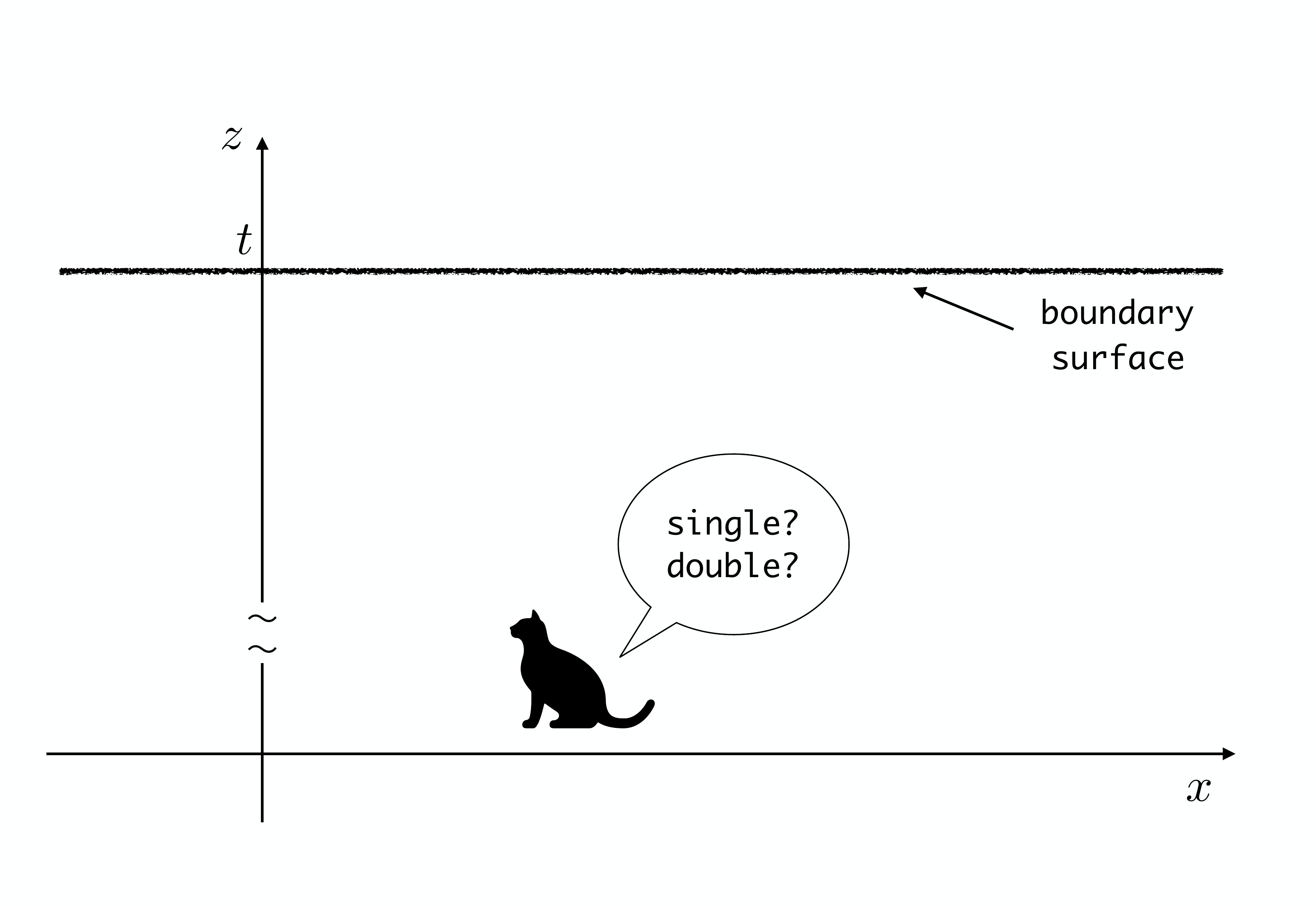}
 \caption{An observer at large $x$ (left figure) or at large $t$ (right figure) can hardly distinguish between a single joining quench and a double joining quench.}
\label{fig:cat}
\end{figure}
\end{itemize}

%%%%%%%%%%%%%%%%%%%%%%%%%%%%%%%%%%%%%%%%
\subsection{Entanglement Entropy in Dirac Fermion CFTs\label{Sec:DiracEE}}
%%%%%%%%%%%%%%%%%%%%%%%%%%%%%%%%%%%%%%%%

Now we would like to consider a free Dirac massless fermion CFTs as the second example.  For the conformal map $w=g(\xi)$, the entanglement entropy for an interval $A$ is computed as follows 
(refer to \cite{NSTW,STW}):
\ba
S^{Dirac}_A=\frac{1}{6}\log\left(\frac{|\xi_1-\xi_2|^2|\xi_1-\bar{\xi}_1||\xi_2-\bar{\xi}_2||g'(\xi_1)||g'(\xi_2)|}{\ep^2(\xi_1-\bar{\xi}_2)(\xi_2-\bar{\xi}_1)}\right).
\ea

\begin{figure}
\begin{minipage}{0.32\hsize}
\includegraphics[width=5cm]{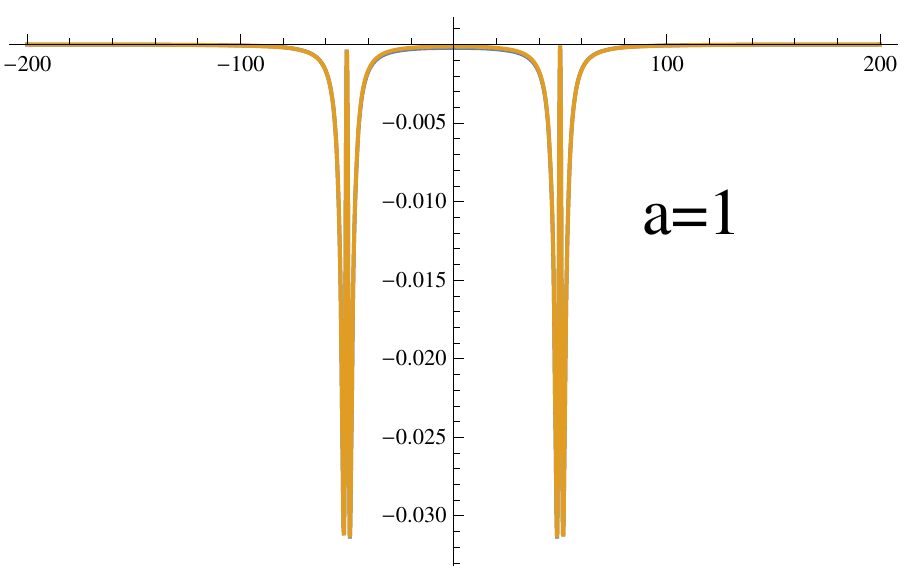}
\end{minipage}
\begin{minipage}{0.32\hsize}
\includegraphics[width=5cm]{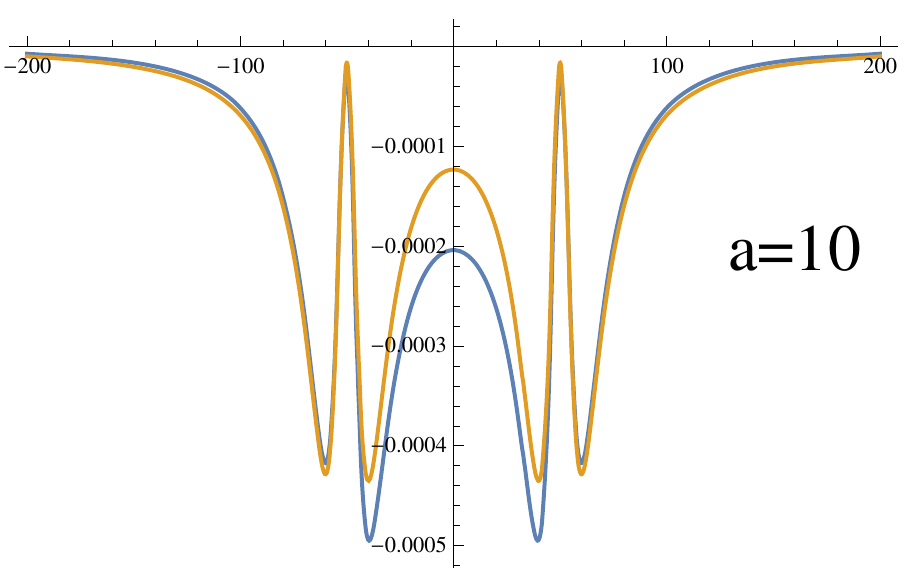}
\end{minipage}
\begin{minipage}{0.32\hsize}
\includegraphics[width=5cm]{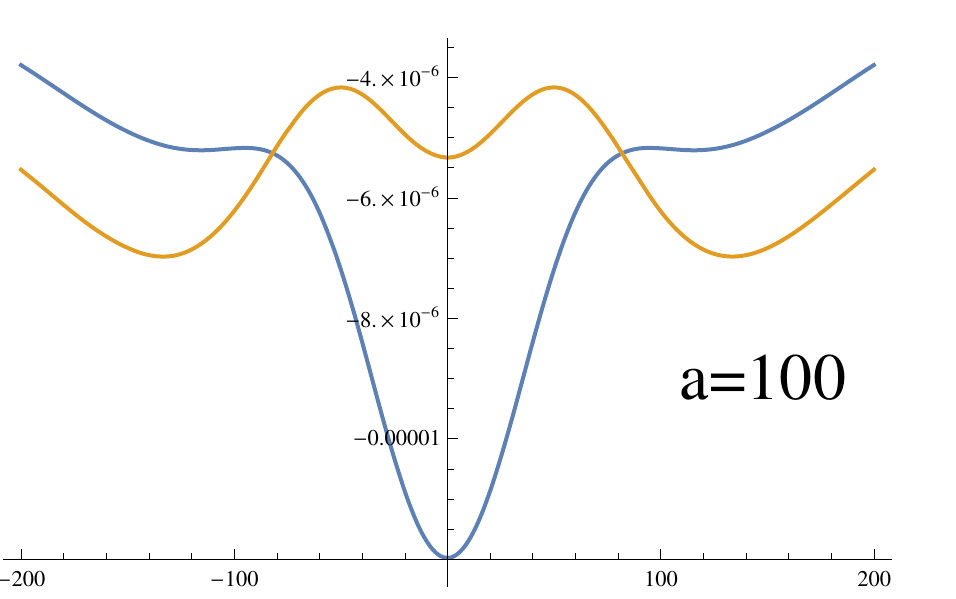}
\end{minipage}
\begin{minipage}{0.06\hsize}
        \vspace{10mm}
      \end{minipage} \\
\begin{minipage}{0.32\hsize}
\includegraphics[width=5cm]{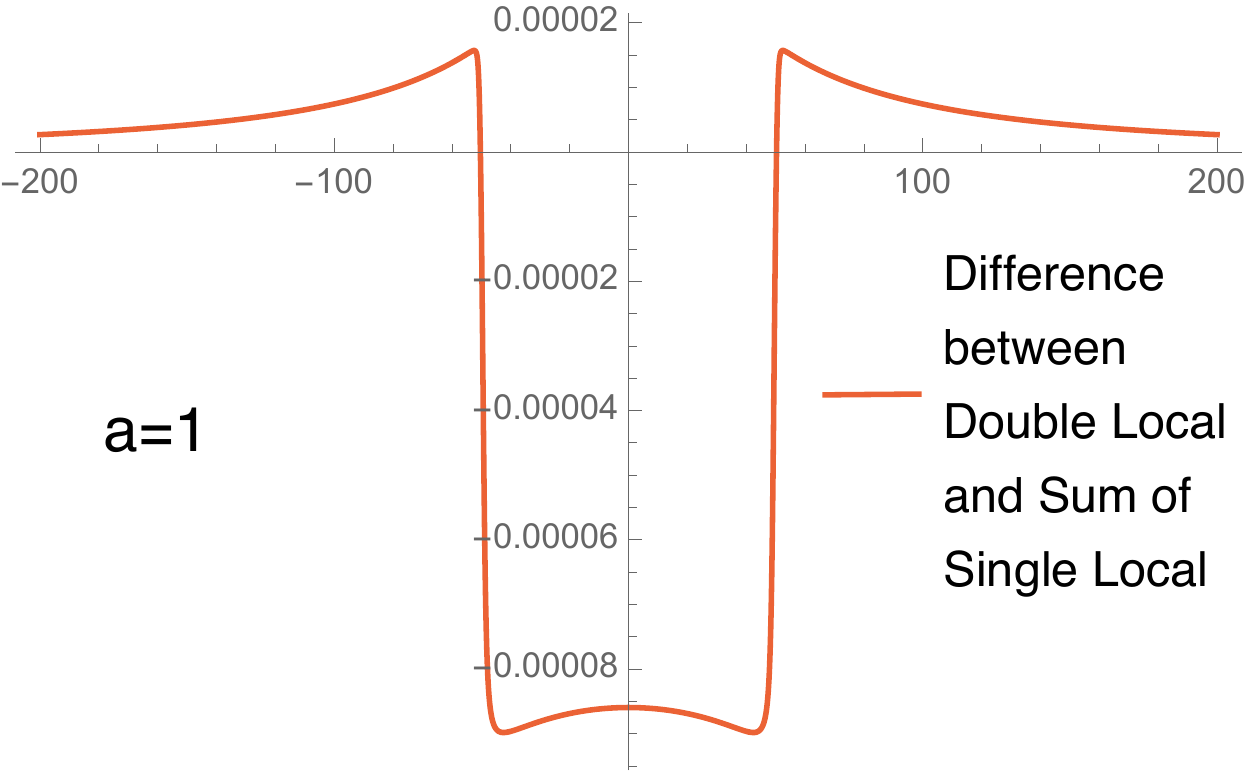}
\end{minipage}
\begin{minipage}{0.32\hsize}
\includegraphics[width=5cm]{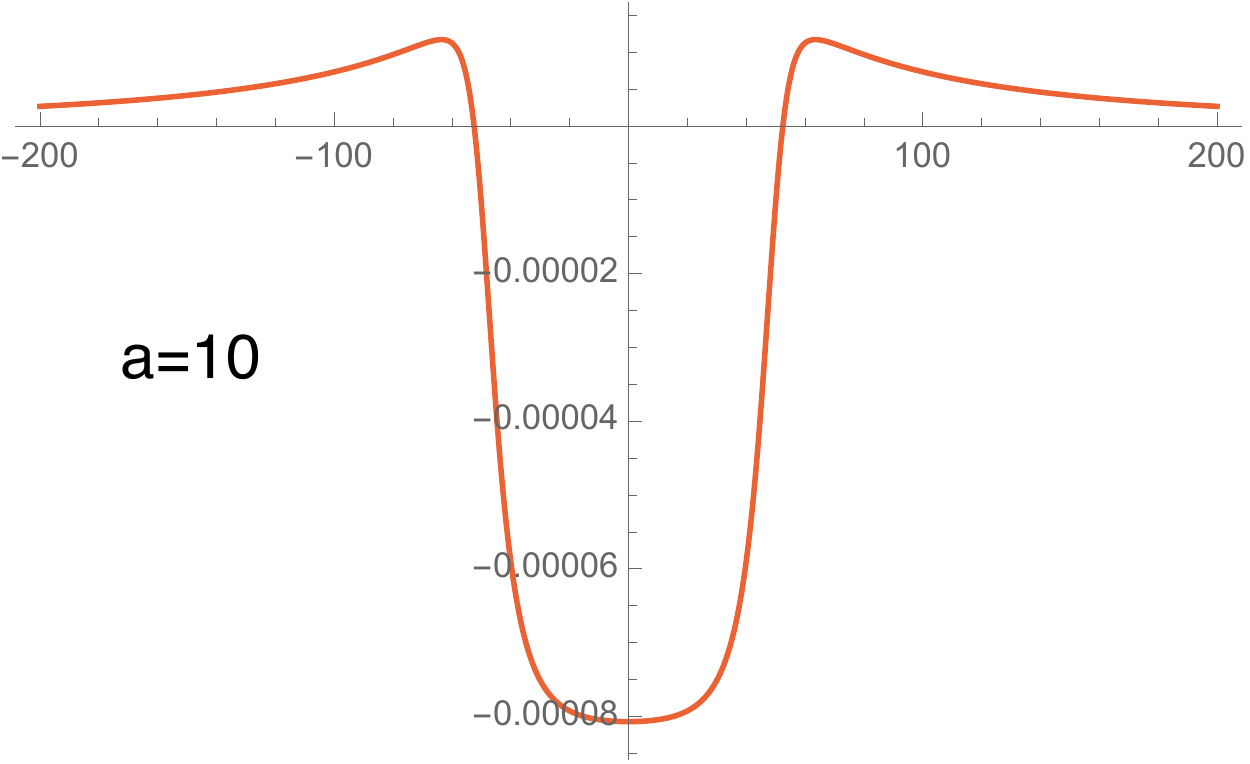}
\end{minipage}
\begin{minipage}{0.32\hsize}
\includegraphics[width=5cm]{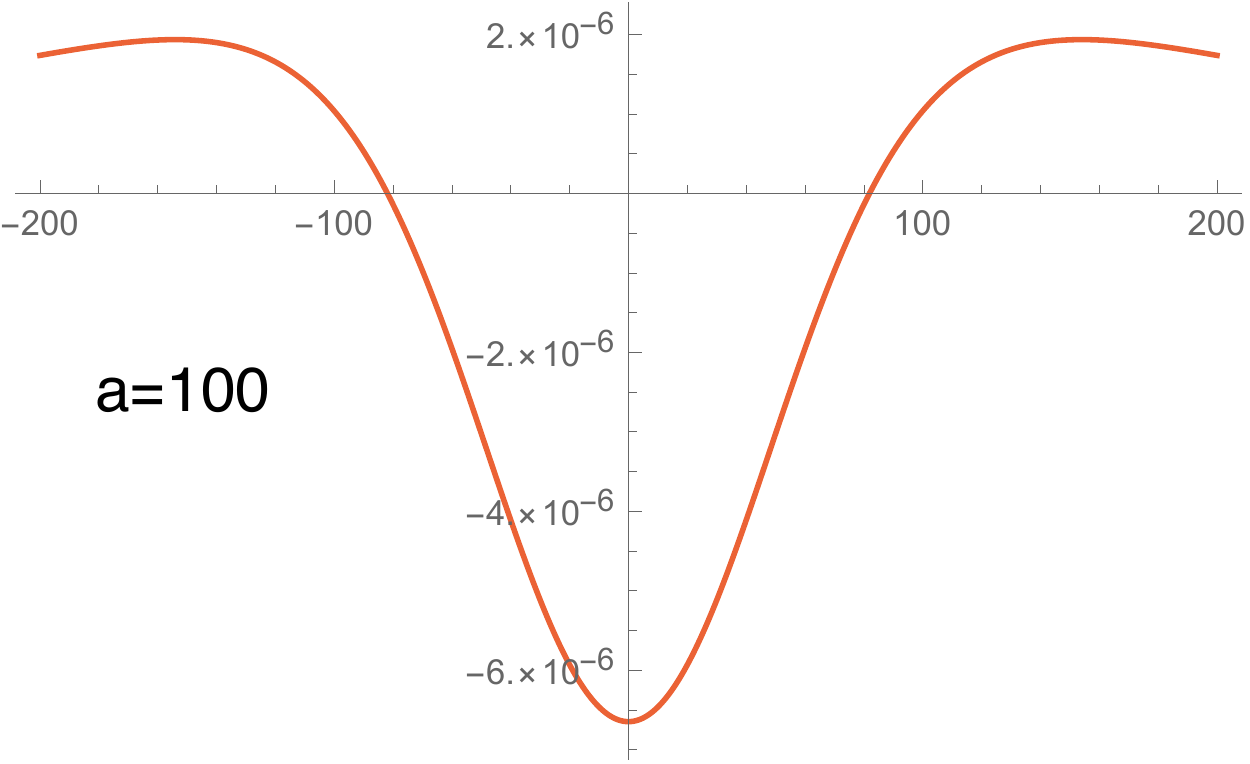}
\end{minipage}
\begin{minipage}{0.06\hsize}
        \vspace{10mm}
      \end{minipage} \\
\caption{ {\bf Top:} The behaviors of entanglement entropy at $t=0$ 
in Dirac Fermion CFT for the double joining local quench (blue) 
and the sum of two single local quenches (orange) 
at $a=1$ (left), $a=10$ (middle) and $a=100$ (right). We chose the subsystem $A$ to be 
$A=[x-1,x+1]$ and plotted $\Delta S_A$ as a function of $x$. In the left graph at $a=1$, the blue and orange graphs almost coincide.
{\bf Bottom:} The behaviors of the difference of entanglement entropy $\Delta S^{D}_A-\Delta S^{S(x=b)}-\Delta S^{S(x=-b)}$ at $t=0$. \label{fig:JoinDiracstatic}}
\end{figure}

We plotted the behavior of $\Delta S_A$ at $t=0$ in figure \ref{fig:JoinDiracstatic}.
In this case of the Dirac fermion, actually we find from the numerical analysis that the inequality  (\ref{ineq}) is violated.  

\begin{figure}
\centering
\includegraphics[width=7cm]{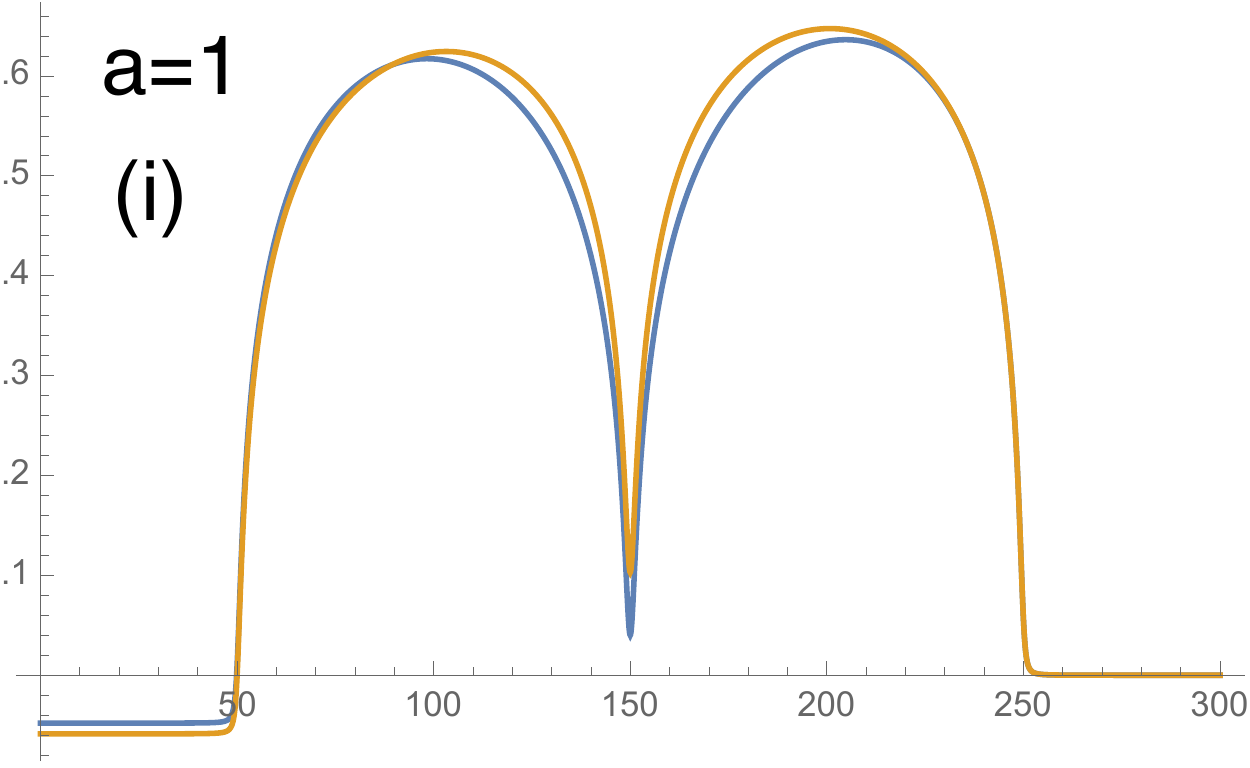}
\includegraphics[width=7cm]{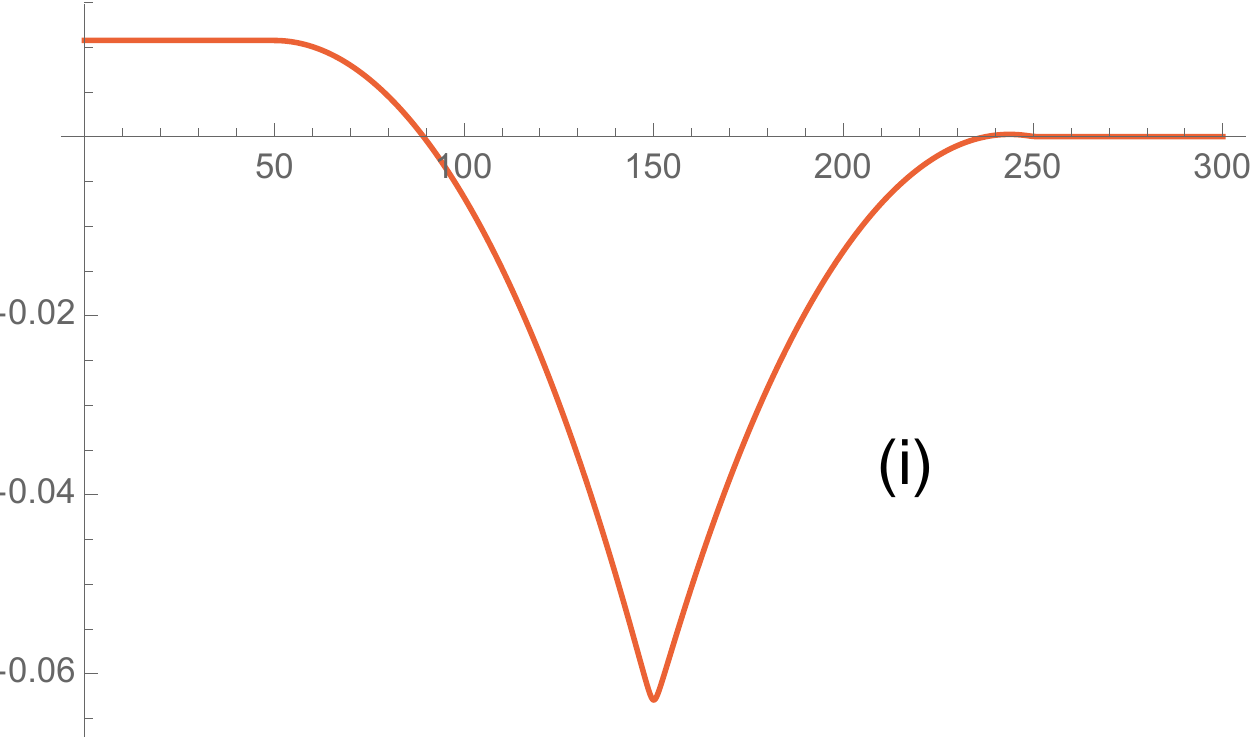}
\includegraphics[width=7cm]{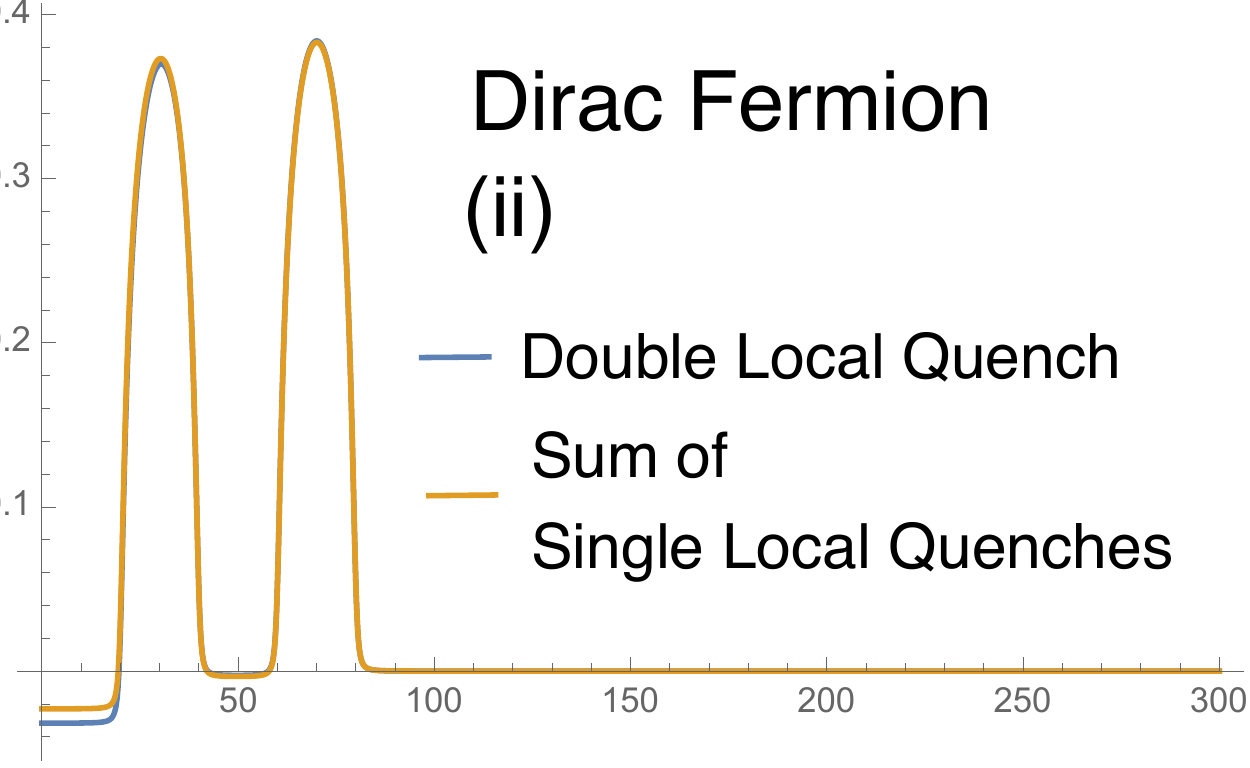}
\includegraphics[width=7cm]{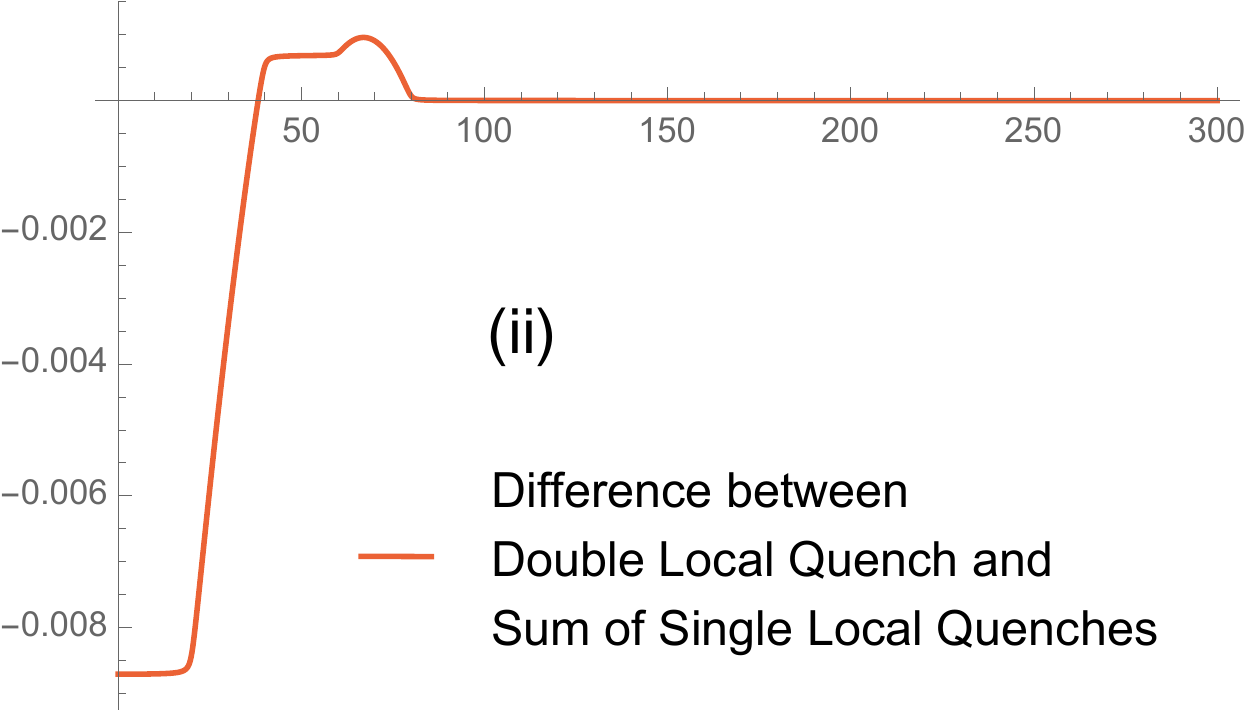}
\includegraphics[width=7cm]{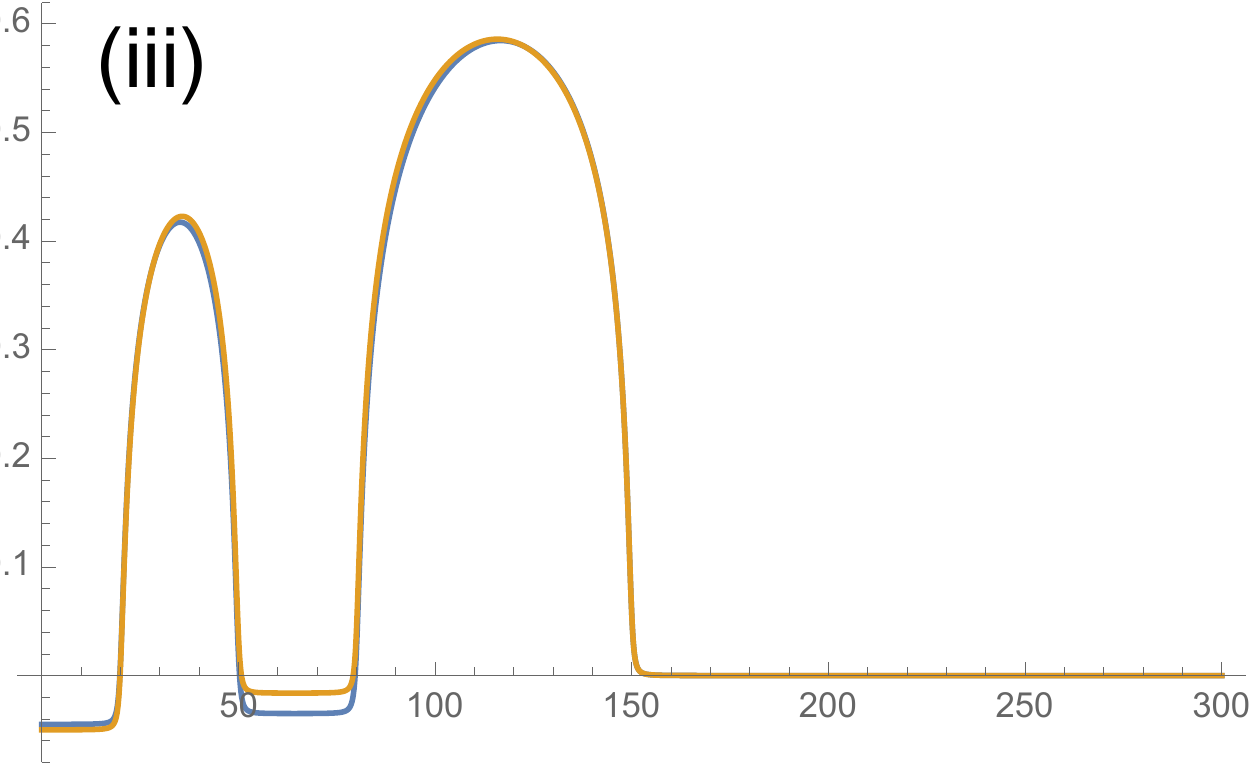}
\includegraphics[width=7cm]{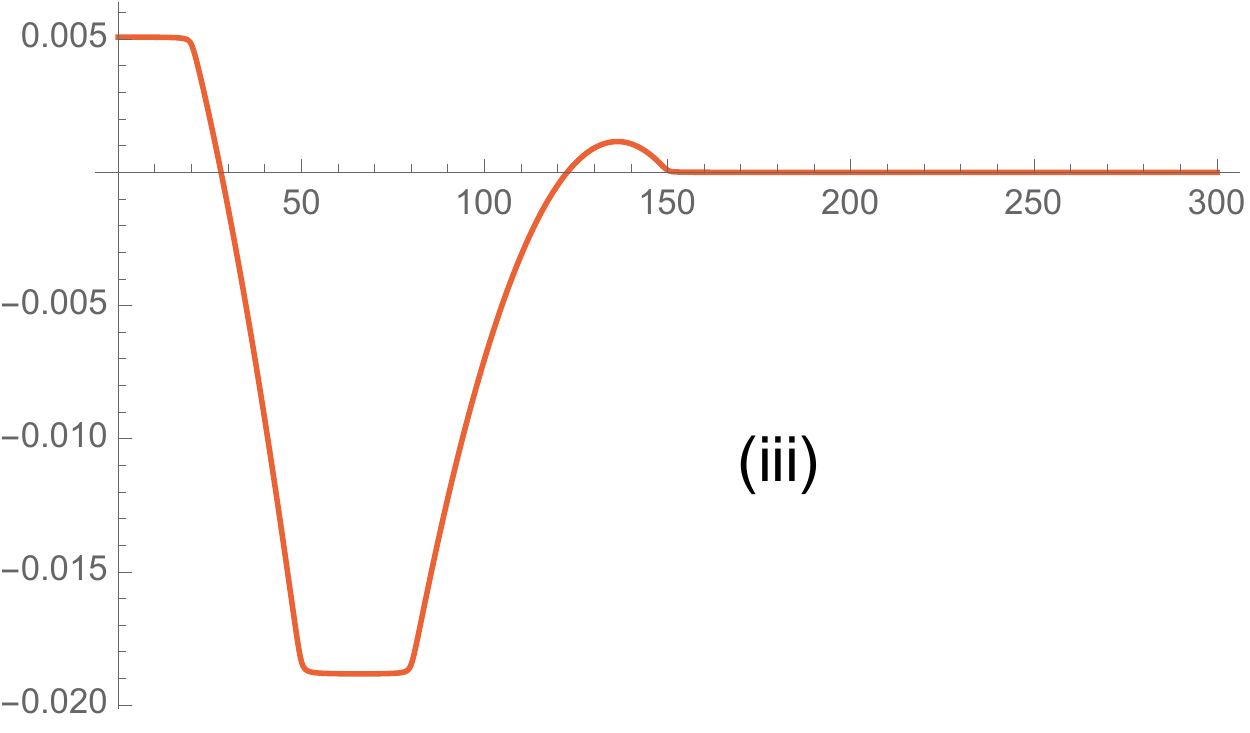}
\includegraphics[width=7cm]{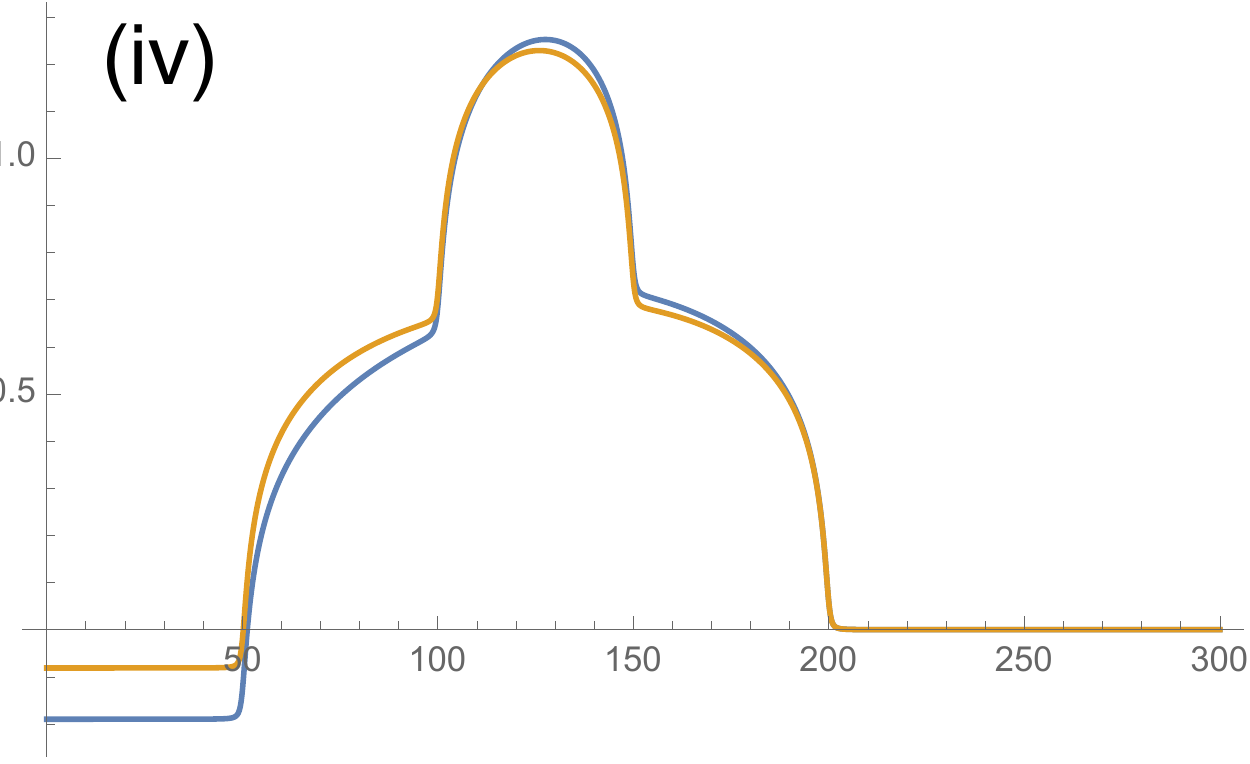}
\includegraphics[width=7cm]{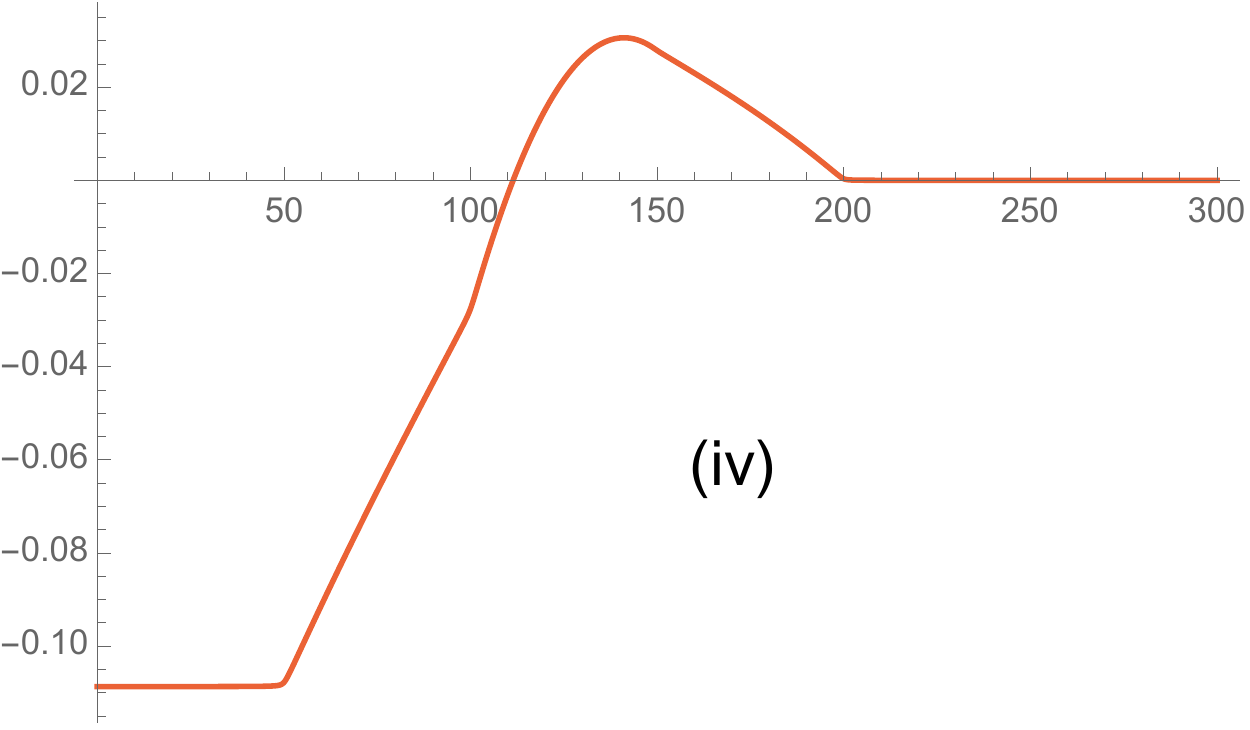}
\caption{The plots of $\Delta S^{double}_A$ of the Dirac Fermion CFT for the double joining local quench (left) and
the difference $\Delta S^{D}_A-\Delta S^{S(x=b)}-\Delta S^{S(x=-b)}$ 
between the double and single quench  (right) as a function of $t$. We set $b=50$ and $a=1$.
From the top to bottom we presented results for the four different setups (i), (ii), (iii) and (iv).
In the left plots, the blue and orange graph describes $\Delta S^{D}_A$ and 
$\Delta S^{S(x=b)}-\Delta S^{S(x=-b)}$, respectively. In the right plots, the red graphs describe the difference 
$\Delta S^{D}_A-\Delta S^{S(x=b)}_A-\Delta S^{S(x=-b)}_A$.}
\label{fig:JoinDiractime}
\end{figure}

Actually, even when we focus on the case where the subsystem $A$ is far away from the quench points  (i.e. $x\gg b$), this violation occurs.
In this limit we have
\be
\Delta S^{D}_A\simeq \Delta S^{S}_A \simeq -\frac{l^2}{24x^2},  \label{Q4}
\ee
both for a single and a double joining quench, where $l$ is the size of $A$. Refer to the appendix \ref{derdirac} for the derivation.
This behavior is different from that in the holographic CFTs as the first law of the form (\ref{firstla}) cannot be applied to free fermion CFTs.
The time evolutions of $\Delta S_A$ for the four different subsystems (i), (ii), (iii), (iv) sketched in figure \ref{fig:eesetup},  are presented in figure \ref{fig:JoinDiractime}. Again we find that the inequality  (\ref{ineqJJ}) is violated in general.

\begin{figure}
\centering
\includegraphics[width=7cm]{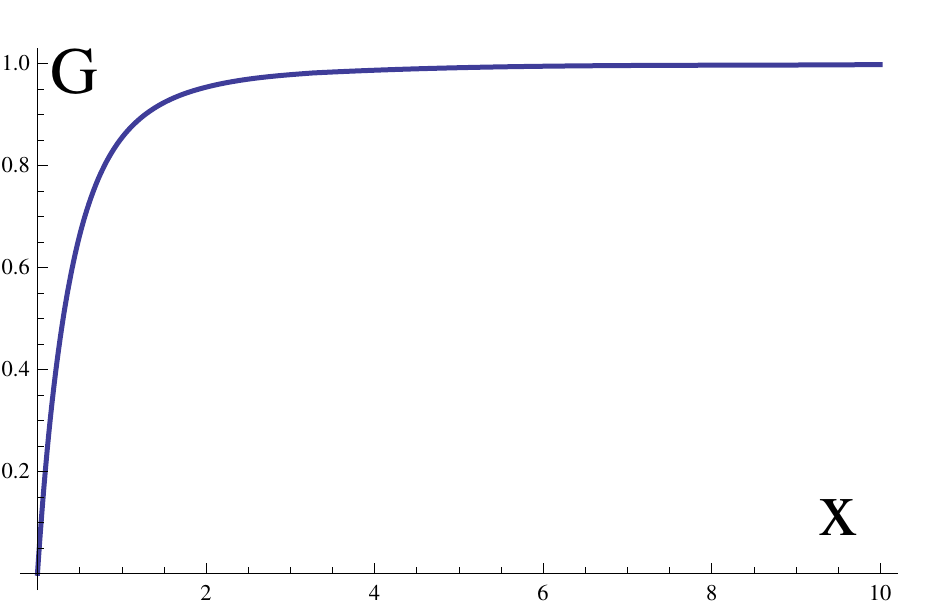}
\caption{A plot of function $G(x)$.}
\label{fig:JoinDiracAS}
\end{figure}

In the late time limit $t\to\infty$,  $\Delta S_A$ in the Dirac fermion CFT behaves as follows 
 ($x$ is the center of the interval $A$ and $l$ is the length of $A$)
\ba
&& \Delta S^{S}_A\simeq \frac{a^2l^2x^2}{3t^6},\no
&& \Delta S^{D}_A\simeq G(b/a)\cdot \frac{a^2l^2}{24t^4},  \label{Q5}
\ea
where the function $G(b/a)$ is plotted in figure \ref{fig:JoinDiracAS}. The exact expression is $G(b/a) = (\cos^2 \alpha \sin ^2 \alpha)/ a_0(\alpha)^2$, which is calculated in Appendix \ref{derdirac}, (\ref{derGba}).
Since $G(0)=0$,  the above result for double quench indeed is reduced to that for the single quench 
at $b=0$. When $b/a\gg 1$, we have $G\simeq1$.  Since we always find $G(x)\leq 1$, the inequality 
(\ref{ineqJJ}) is satisfied.

\
When the subsystem $A$ is semi-infinite $l\gg t$, we found that the behavior of $\Delta S_A$ in the Dirac fermion CFT is very similar to the one for the holographic CFTs. We can again confirm the late time behavior 
(\ref{timehalf}) in this case.

%\newpage
%%%%%%%%%%%%%%%%%%%%%%
%%%%%%%%%%%%%%%%%%%%%%
\section{Double Splitting Local Quenches}\label{sec:DSLQ}
%%%%%%%%%%%%%%%%%%%%%%
%%%%%%%%%%%%%%%%%%%%%%

\begin{figure}[h!]
	\centering
	\includegraphics[width=13cm]{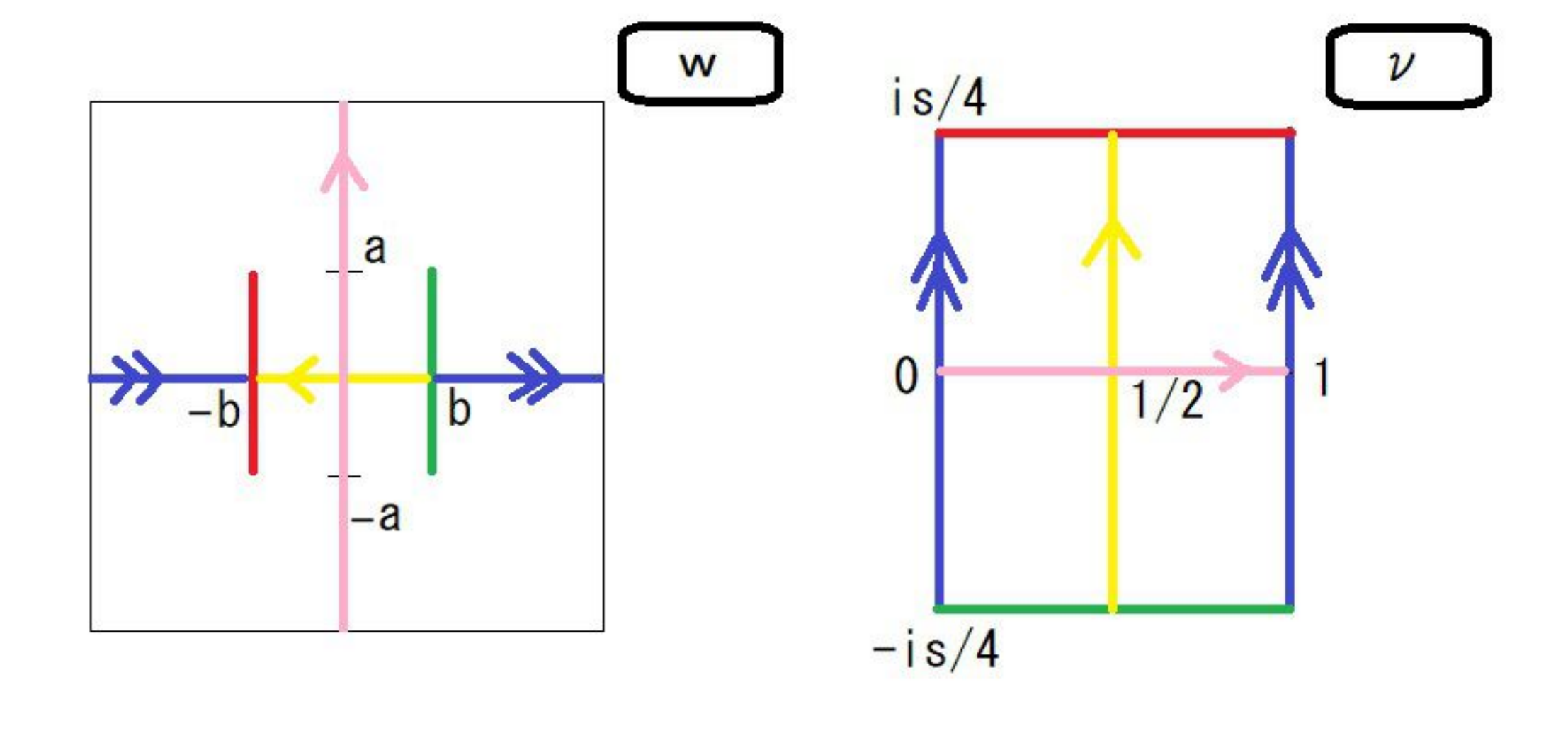}
	\caption{The left figure describes Euclidean spacetime for double splitting quench. There are two vertical boundaries at $\pm b+i\tau\  (-a<\tau<a)$, which means splitting quench at $\pm b$. These vertical boundaries correspond to splitting 1 dimensional system. $a$ is cutoff for quench.
	The right figure is the annulus $\nu$ coordinate which is defined by equation (\ref{DSQ_map}).}
\label{DSQ_mapfig}
\end{figure}

In this section we study double splitting local quenches in two dimensional CFTs.
Our description of a double splitting quench is shown in figure \ref{DSQ_mapfig}. Two vertical boundaries correspond to two quenches by splitting the system. We impose a conformal boundary condition to these boundaries. This boundary condition will be extended to the bulk boundary condition by the AdS/BCFT prescription as in the case of joining quenches.

We prepare a state at $\tau=0$ by the Euclidean path integral from $\tau=-\infty$ to $\tau=0$, and consider the time evolution by Lorentzian path integral via the Wick rotation $\tau= it$. We write the complex coordinate as $w=x+i\tau$ and its complex conjugate as $\overline{w}=x-i\tau$ in the Euclidean space. The Lorentzian spacetime after the Wick rotation is described by the 
coordinate $w_-=x-t,\ w_+=x+t$.

First we consider the Euclidean time picture. The $w$ plane coordinate can be mapped into an annulus coordinate $\nu$ by the following transformation (see \cite{NSTW} for another application of the same 
map\footnote{The parameter $\rho$ in  \cite{NSTW} is related to $s$ in this work via 
$\rho=e^{-\pi s}$.}),
\begin{equation}\label{DSQ_map}
w(\nu) = b\left[ K(\nu)+K\left(\nu+\frac{is}{2}\right)+1 \right],
\end{equation}
where the function $K$ is defined by,
\begin{align}
K(\nu)& =\frac{1}{\pi i}\partial_{\nu'}\log\theta_1(\nu',is)|_{\nu'=\nu}\\[5pt]
\theta_1(\nu,\tau), &=2e^{\pi i \tau/4}\sin \pi\nu \prod_{k=1}^{\infty}(1-e^{2\pi i k\tau})(1-e^{2\pi i \nu}e^{2\pi i k\tau})(1-e^{-2\pi i \nu}e^{2\pi i k\tau}).
\end{align}

Let us make some remarks about this conformal mapping. This maps $C-([-b-ia,-b+ia]\cup [b-ia,b+ia])$ to $[0,1]\times(-is/4,is/4)$ bijectively. Since $\textnormal{Re}(\nu)=0$ line and $\textnormal{Re}(\nu)=1$ line are identified, the $\nu$ coordinate is an annulus coordinate.  $\textnormal{Im}(\nu)=\pm s/4$ lines correspond to the two vertical boundaries in $w$-coordinate. 

As a nontrivial feature of this conformal mapping, $\textnormal{Re}(\nu)$ direction corresponds to $\tau$ direction in $w$-coordinate, and $\textnormal{Im}(\nu)$ direction corresponds to $x$ direction in $w$-coordinate. Because of this, we write $\textnormal{Re}(\nu)$ and $\textnormal{Im}(\nu)$ as $T$ and $X$ in the next subsection.

\begin{figure}[h!]
	\centering
	\includegraphics[width=8cm]{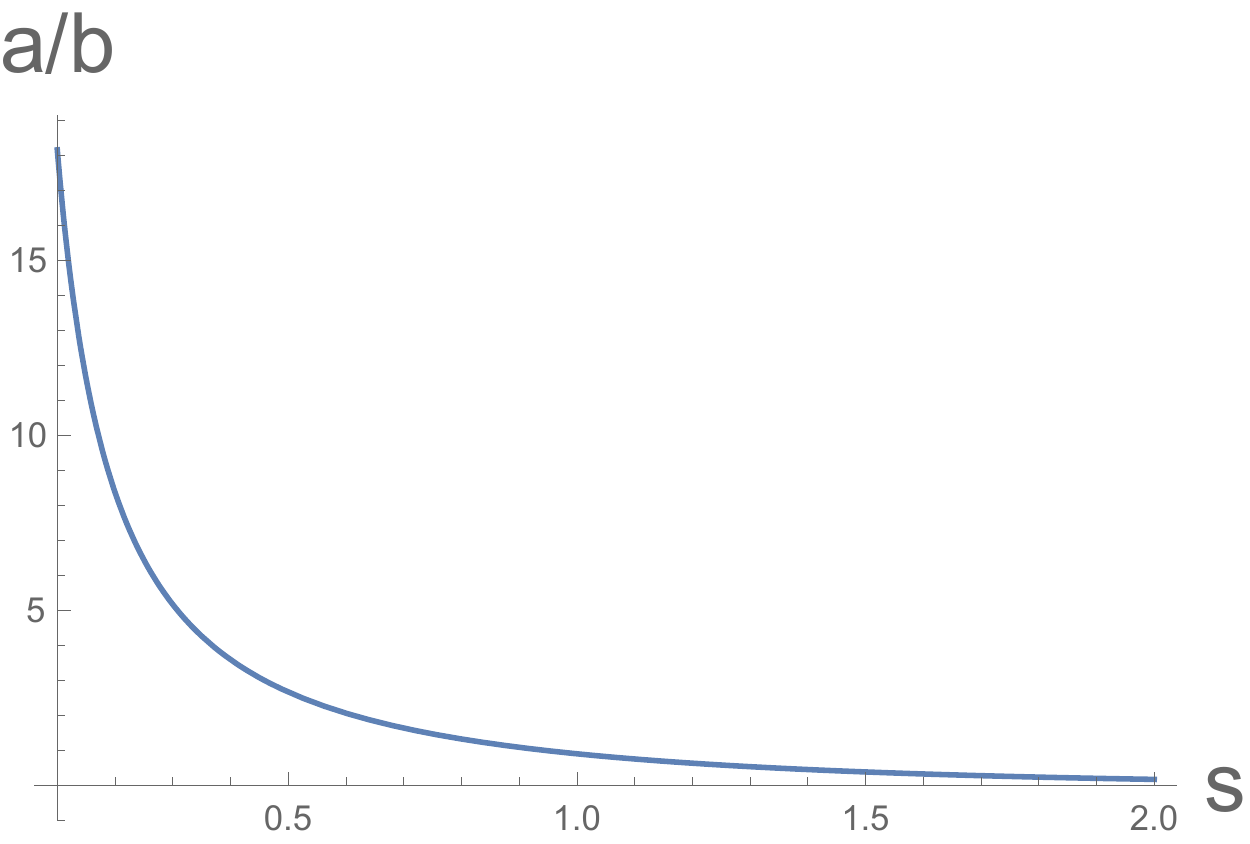}
	\caption{The relation between $s$ and $a/b$. }\label{ModParam_rel}
\end{figure}

Since the $\nu$-annulus which has two boundaries is equivalent to the half of torus, $s$ is the moduli parameter of the torus which is the double of $\nu$-annulus. The moduli parameter $s$ depends only on the ratio $b/a$. We plot the relation between them in figure \ref{ModParam_rel}.
\\
In $s\to 0$ or $s\to \infty$ limit, the relation $s=s(b/a)$ approximates,
\ba
s \sim \begin{cases}
\dfrac{2b}{a}&(s\to 0)\\[9pt]
\dfrac{2}{\pi}\log\dfrac{4b}{a}&(s\to\infty).  
\end{cases} \label{limsab}
\ea
To get above relations, we used the asymptotic behavior of elliptic theta function.
\begin{equation}
\theta_1(\nu,is) \sim \begin{cases}
\dfrac{2}{\sqrt{s}}\exp\left(-\dfrac{\pi}{s}(\nu^2+\dfrac{1}{4})\right)\sinh \dfrac{\pi\nu}{s}&(s\to 0)\\[9pt]
2\exp\left(-\dfrac{\pi s}{4}\right) \sin \pi\nu&(s\to\infty).
\end{cases}
\end{equation}

%%%%%%%%%%%%%%%%%%%%%%
\subsection{Holographic Dual Geometry}
%%%%%%%%%%%%%%%%%%%%%%

In the previous subsection, we considered the annulus coordinate $w,\nu$ with two boundaries and imposed the conformal boundary conditions to these boundaries. Now we consider the AdS spacetime with boundary which is dual to the BCFT. We impose the Neumann boundary condition on the boundary surface $Q$ by following the AdS/BCFT construction \cite{AdSBCFT}. In particular we choose the vanishing tension $T_{BCFT}=0$ of $Q$ or equally the vanishing boundary entropy $S_{bdy}=0$ for simplicity. 

\begin{figure}[h!]
	\centering
	\includegraphics[width=13cm]{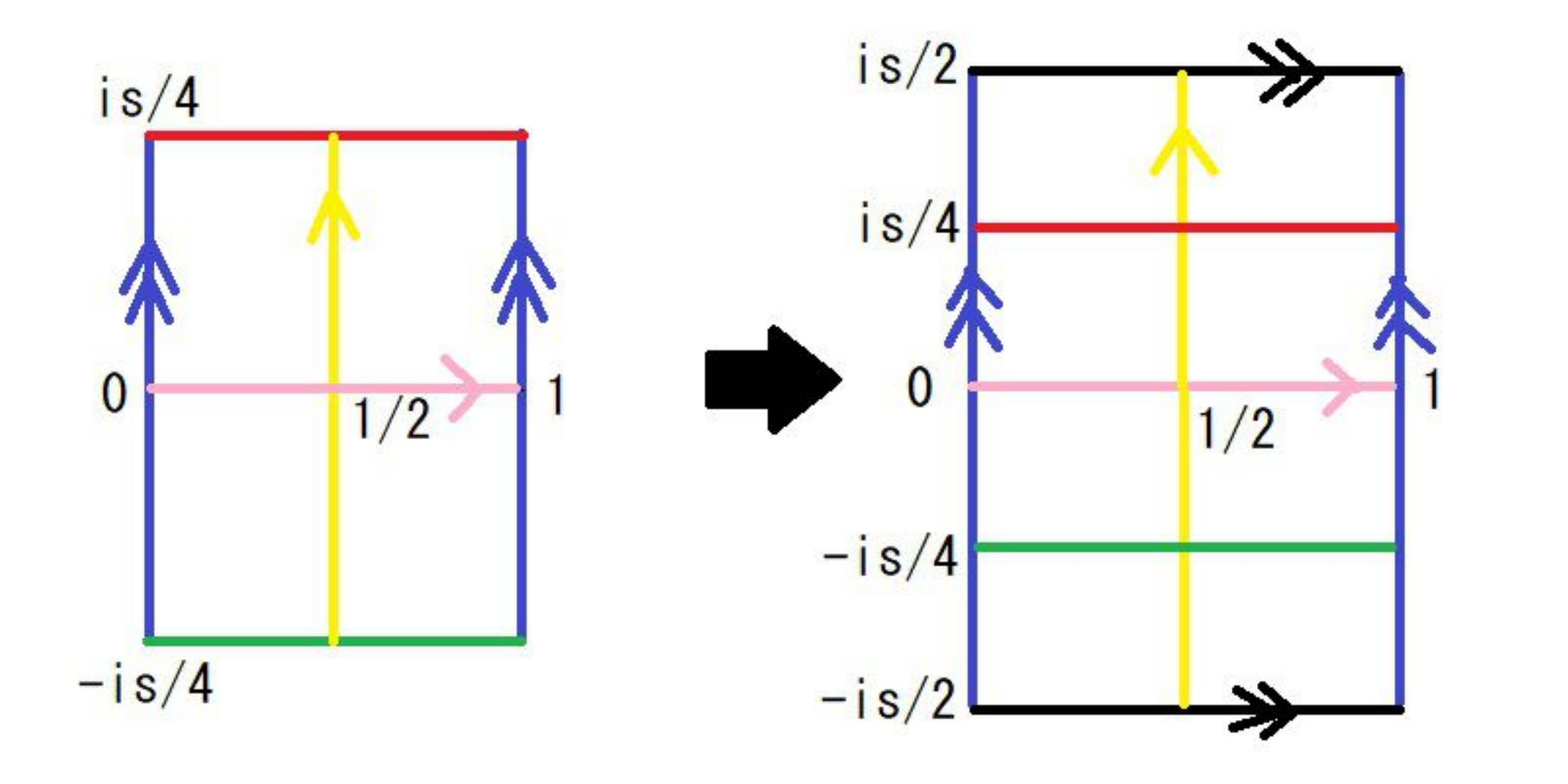}
	\caption{The right figure describes ''doubling'' of $\nu$ annulus coordinate. $\textnormal{Im}(\nu)=\pm s/2$ line are identified, and also $\textnormal{Re}(\nu)=0,1$ line are identified, which corresponds to the whole torus.}\label{DSQ_mirror}
\end{figure}

The gravity dual of a holographic CFT on the $\nu$-annulus is precisely a half of that of a holographic CFT on the torus, which is given by pasting two copies of the $\nu$-annulus. Namely we add  the ``mirror'' region to the original $\nu$-annulus coordinate as depicted in figure \ref{DSQ_mirror}.

The standard known fact in AdS/CFT tells us that there is the Hawking-Page transition of the gravity dual geometry between the thermal AdS phase $s<1$ and BTZ black hole phase $s>1$, depending on the torus moduli parameter $s$. Therefore the gravity dual of the double splitting quench is also classified into the two phases:
\begin{align*}
	({\rm I}) s<1\ \ (b/a<r_*\simeq 1.1064)&\longrightarrow \textnormal{Thermal AdS phase}\\
	({\rm II}) s>1\ \  (b/a>r_*\simeq 1.1064)&\longrightarrow \textnormal{BTZ BH phase}.
\end{align*}
Below we will study their geometries individually.
 
%%%%%%%%%%%%%%%%%%%%%%%%%
\subsubsection{Thermal AdS Geometry ($s<1$)}
\begin{figure}
	\centering
	\includegraphics[width=14cm]{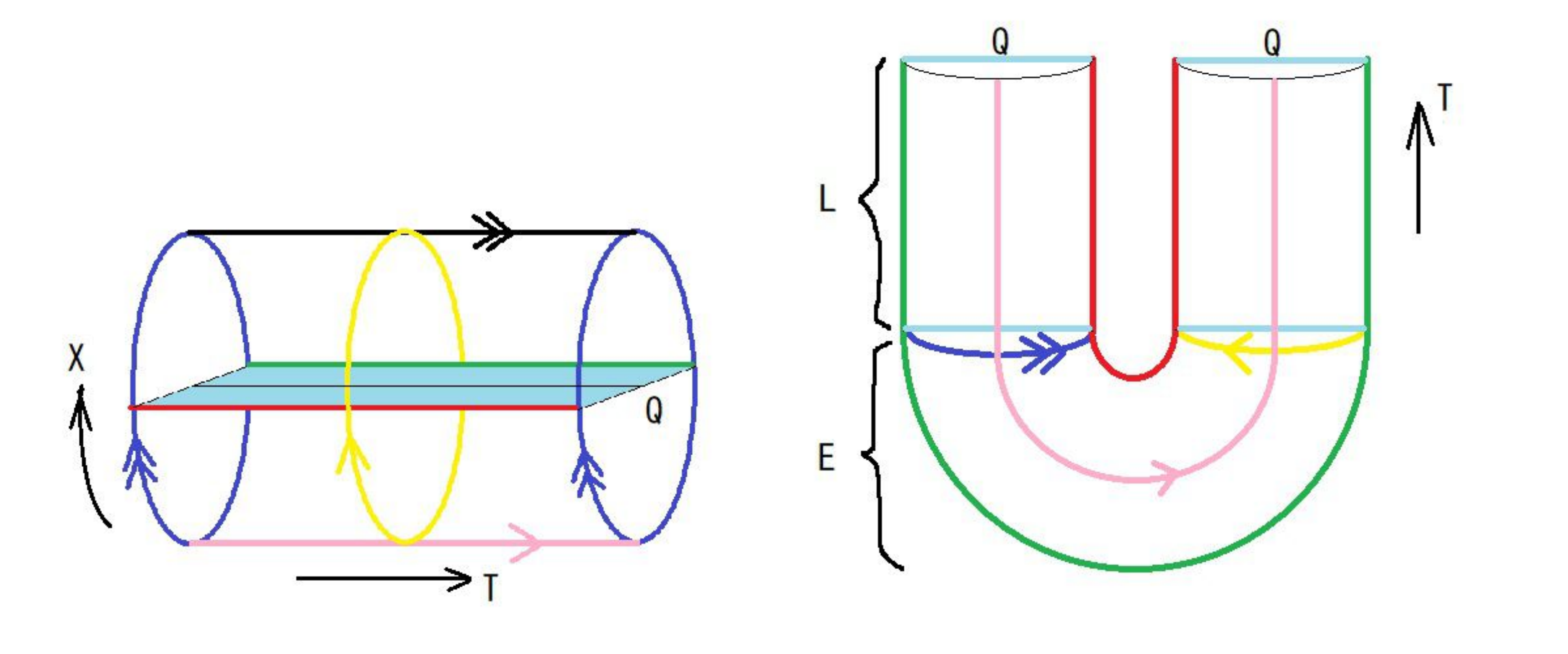}
	\caption{The picture of the dual geometry in the thermal AdS phase. The left figure describes Euclidean geometry, and the right figure describes Lorentzian time evolution. In this case we impose the dual boundary surface $Q$ has one connected component.}\label{TADSfig}
\end{figure}

%%%%%%%%%%%%%%%%%%%%%%%%%
Here we write $-2\pi i\nu = X_T+iT_T$. The periodicity of the torus coordinate becomes,
\begin{equation}
T_T\sim T_T+2\pi,\ \ X_T\sim X_T+2\pi s.
\end{equation}
In $s<1$ case, the $X_T$-direction becomes the smaller circle of the torus and thus will be contractible 
cycle in the gravity dual.  Thus the gravity dual of our double quench is given by a half of the thermal AdS geometry (figure \ref{TADSfig}):
\begin{equation}
ds_T^2=\frac{1}{z_T^2}\left[ \frac{dz_T^2}{1-z_T^2/s^2}+\left(1-z_T^2/s^2\right)dX_T^2+dT_T^2 \right],
\end{equation}
where $(T_T,X_T$) takes the values $0\leq T_T<2\pi$ and $0\leq X_T<\pi s$. In this case the boundary surface $Q$ (which is dual to the boundary of BCFT) has one connected component.
The Lorentzian time evolution is obtained from the Wick rotation $T_T=iT_{L,T}$ as in figure \ref{TADSfig}.

%%%%%%%%%%%%%%%%%%%%%%%%
\subsubsection{BTZ geometry ($s>1$)}
\begin{figure}
	\centering
	\includegraphics[width=14cm]{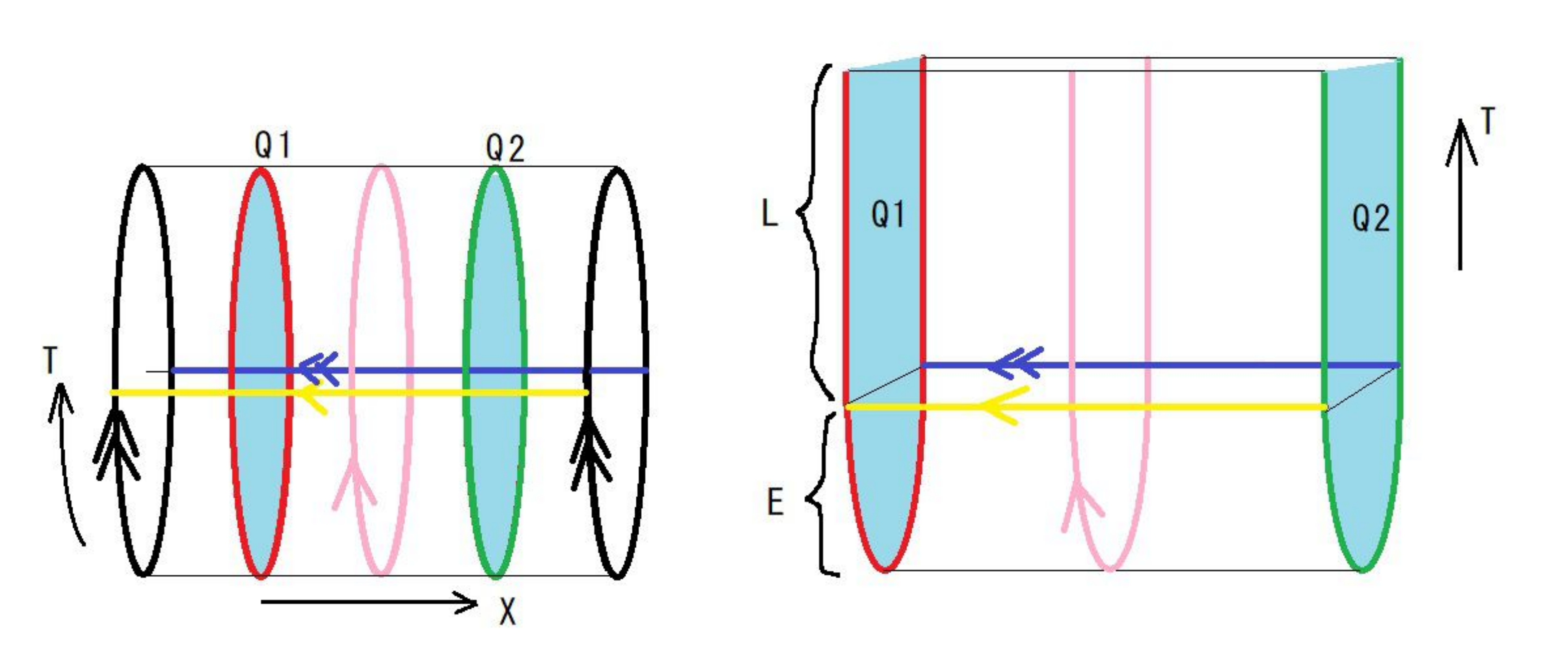}
	\caption{The picture of the dual geometry in the BTZ phase. The left figure describes Euclidean geometry, and the right figure describes Lorentzian time evolution. In this case we impose the dual boundary surface has two connected components $Q_1,Q_2$.}\label{BTZfig}
\end{figure}

%%%%%%%%%%%%%%%%%%%%%%%%
Here we write $-2\pi i\nu/s = X_B+iT_B$. The periodicity of torus coordinate becomes,
\begin{equation}
T_B\sim T_B+2\pi/s,\ \ X_B\sim X_B+2\pi.
\end{equation}
In the $s>1$ case, the $T$-direction becomes the smaller circle of the torus and thus will be contractible 
cycle in the gravity dual.  Thus the gravity dual of our double quench is given by a half of 
the BTZ black hole geometry (figure \ref{BTZfig}),
\begin{equation}
ds_B^2=\frac{1}{z_B^2}\left[ \frac{dz_B^2}{1-s^2z_B^2}+\left(1-s^2z_B^2\right)dT_B^2+dX_B^2 \right],
\end{equation}
where $(T_B,X_B$) takes the values $0\leq T_T<2\pi/s$ and $0\leq X_T<\pi$.
In this case the boundary surface $Q$ (which is dual to the boundary of BCFT) has two connected components. The Lorentzian time evolution is obtained from the Wick rotation 
$T_B=iT_{L,B}$ as in figure \ref{BTZfig}, which can be regarded as a half of the setup in \cite{HaMa}.

It is also useful to remember that the half BTZ geometry has black hole entropy given by
\begin{equation}
S_{BH} = \frac{c}{6}\pi s.  \label{btzentp}
\end{equation}
If we consider the limit $b/a\gg 1$, we find from (\ref{limsab})
\ba
S_{BH}\simeq \frac{c}{3}\log\frac{4b}{a}.  \label{eesplitv}
\ea
Indeed we can identify this with the entanglement entropy between 
two regions $[-b,b]$ and its complement, which are separated by the double splitting quench.
By setting the length of the interval to be $l=2b$ and the cut off scale $\ep$ to be of order 
$a$, the entanglement entropy (\ref{eesplitv}) reproduces the well-known formula of \cite{HLW}.

%%%%%%%%%%%%%%%%%%%%%%%%
\subsection{Energy Stress Tensor}
%%%%%%%%%%%%%%%%%%%%%%%%

The energy stress tensor in our double splitting quench can be computed by 
the conformal transformation (\ref{DSQ_map}), which leads to the following transformation of  the
energy stress tensor:
\ba
T_{ww}=(w')^{-2}\cdot \left(T_{\nu\nu}+\frac{c}{24}\cdot 
\frac{2w'w'''-3(w'')^2}{(w')^2}\right),
\ea
where $w' =\frac{dw}{d\nu}$.

\begin{figure}
	\includegraphics[width=5cm]{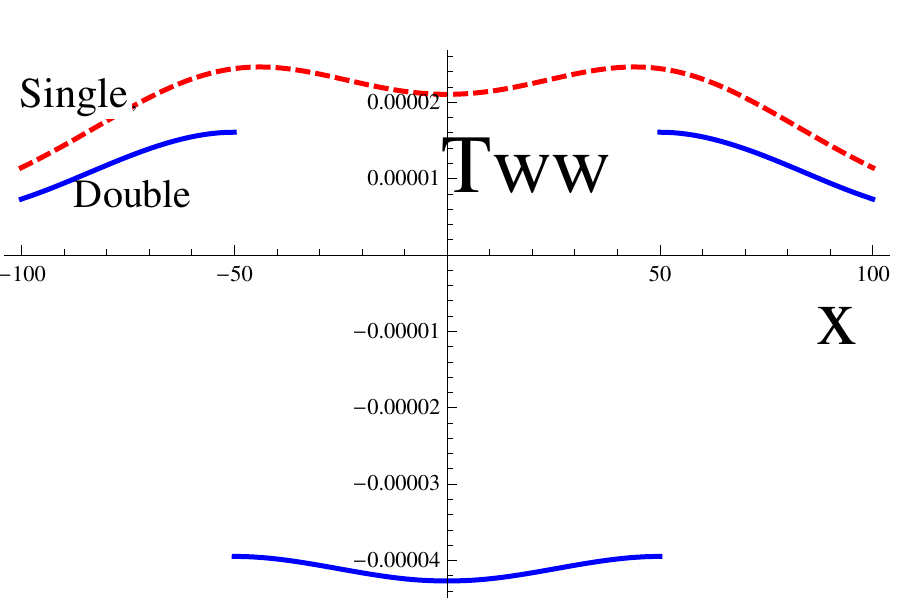}
    \includegraphics[width=5cm]{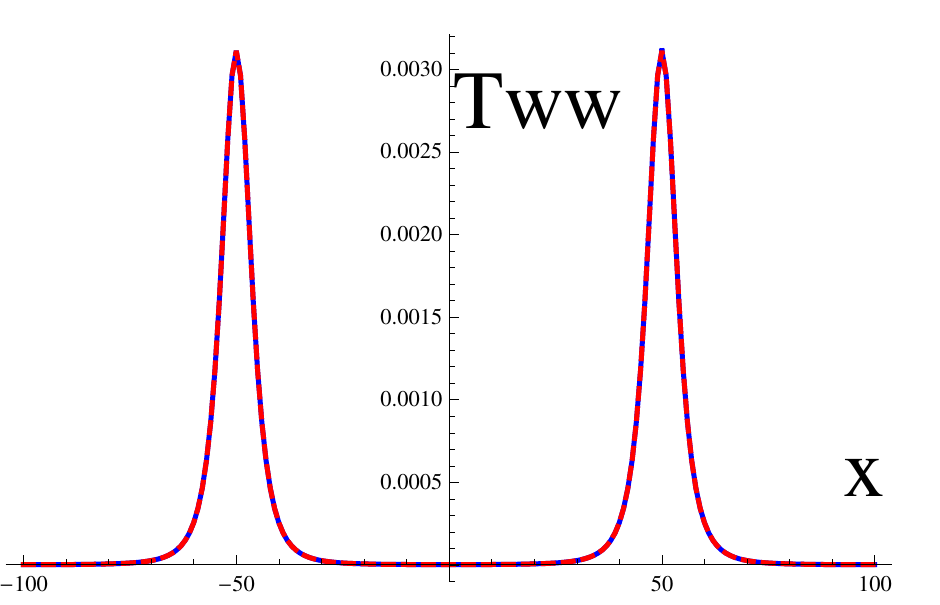}
\includegraphics[width=5cm]{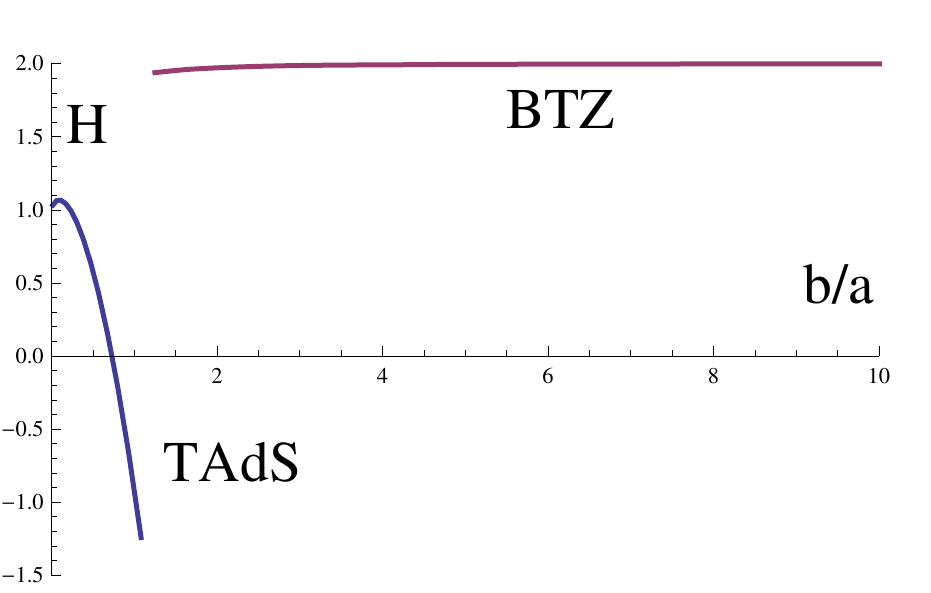}
	\caption{The behaviors of energy stress tensor $T_{ww}$ in a holographic CFT
under double splitting quenches at the time $t=0$. 
The left and middle graph describes the energy stress tensor 
as a function of the position $x$ with the quench parameter $a=6.3$ (BTZ phase) and $a=76.3$ (Thermal AdS phase), respectively (we always set $b=50$). The blue graph describes $T^D_{ww}$, while the red dotted one shows $T^{S(x=b)}_{ww}+T^{S(x=-b)}_{ww}$.
The right picture is a plot of the function $H(b/a)$ as a function of $b/a$. Note that there is a phase transition at $b/a=r_*\simeq 1.1064$. }\label{EMsp}
\end{figure}

The $T_{\nu\nu}$ is the energy stress tensor of a two dimensional CFT on an annulus 
with the flat metric. For a holographic CFT, the standard result tells us (in our normalization)
\ba
&& ({\rm I}) \mbox{Themal AdS phase}: T_{\nu\nu} = \frac{\pi^2c}{6s^2},\no
&&  ({\rm II}) \mbox{BTZ phase}: T_{\nu\nu} = -\frac{\pi^2c}{6}.
\ea
Finally, we obtain the behavior of energy stress tensor as plotted in figure \ref{EMsp}.
Note that in the thermal AdS phase, the energy density is discontinuous at the points 
$x=\pm b$ where the splitting quench is performed. We can confirm the inequality 
(\ref{ineqt}).

It is also useful to study the behavior of energy density in the far away limit $x\to\infty$.
We find the behavior 
\ba
T^{D}_{ww}\simeq H(b/a)\cdot \frac{ca^2}{8x^4}=H(b/a)\cdot T^{S}_{ww}, \label{hratg}
\ea
where $T^{D}_{ww}$ and $T^{S}_{ww}$ are the energy stress  tensor in the double and single splitting quench.
The function $H(b/a)$ is plotted in the right picture in figure \ref{EMsp}. Indeed, at the limit $b\to 0$, the double quench result coincides with that for the single quench $H(0)=1$. Also in the opposite 
limit $b\to \infty$, we find $H(\infty)=2$. Since $H$ always satisfies $H(b/a)\leq 2$, the inequality 
(\ref{ineqt}) is satisfied. Note that the property that  the asymptotic ratio $T^D/T^S$ reaches the value $H=2$ in the limit $b/a\to \infty$ is very special 
to the double splitting quenches, which is missing in joining quenches. As we will explain in subsection \ref{sec:bdyfsp} using the holography, this is
because the gravity dual includes two disconnected boundaries.

%%%%%%%%%%%%%%%%%%%%%%%%%%%%%
\subsection{Entanglement Entropy in Holographic CFTs}
%%%%%%%%%%%%%%%%%%%%%%%%%%%%%

Now we move on to the analysis of holographic entanglement entropy (HEE). 
We choose the subsystem $A$ to be an interval.
The holographic entanglement entropy is computed as $\min\{S^{con}_A,S^{dis}_A\}$.
As we will see below, in summary, we will be able to confirm (\ref{ineqJ}) for the connected geodesic contribution, 
which means that  the inequality for the full HEE (\ref{ineqJJ}) is again satisfied when $A$ is enough away from the quench points.

\begin{figure}
  \centering
 \includegraphics[width=6cm]{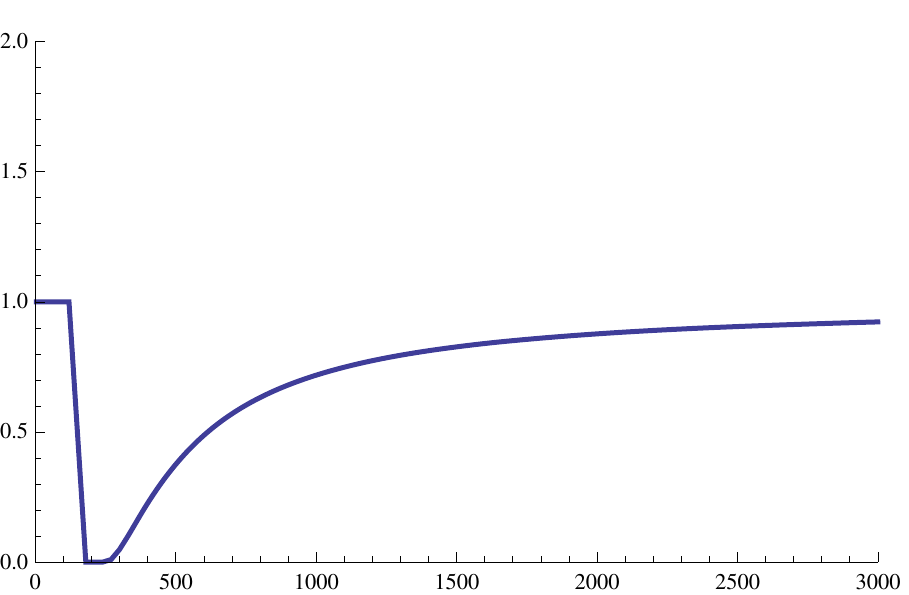}
  \caption{The ratio  $\frac{\Delta S_A^{D}}{\Delta S_A^{S(x=b)}
+\Delta S_A^{S(x=-b)}}$
under the time evolutions of HEE in the splitting quenches. We chose the subsystem (i) and set
$a=2$ and $b=50$.}
\label{fig:HEEcomparison}
\end{figure}

First, consider the HEE at $t=0$. If we choose $A$ to be an interval $[x-l/2,x+l/2]$ and
consider the value of HEE as a function of $l$, we find a qualitatively similar behavior as that the 
energy stress tensor $T^{D}_{ww}(x)$ discussed in the previous subsection. We can numerically confirm that the inequality (\ref{ineqJ}) is always satisfied (refer to figure \ref{fig:HEEcomparison} for an example of time evolution). In particular if we take the limit $l\to 0$, they are exactly proportional to each other via the first law (\ref{firstla}).

Similarly, the first law contribution gets dominant in the distant limit of subsystem $x\to \infty$ and the late time limit $t\to\infty$, where the connected geodesic is favored. In these limits we obtain:
\ba
&& \Delta S^{con, D}_A\simeq H(b/a)\cdot \frac{ca^2l^2}{24x^4}=H(b/a)\cdot \Delta S^{con, S}_A \ \ (x\to \infty),\no
&& \Delta S^{con, D}_A\simeq H(b/a)\cdot \frac{ca^2l^2}{24t^4}=H(b/a)\cdot \Delta S^{con, S}_A \ \ (t\to \infty),
\label{asymDSEE}
\ea
where $H$ is the function introduced in (\ref{hratg}). For these, clearly the inequality (\ref{ineqJ}) 
is satisfied.

Below we study the time evolution of HEE for the four different choices (i), (ii), (iii), (iv) of the subsystem $A$, depicted in figure \ref{fig:eesetup}. We parametrize 
the location of the subsystem $A$ as $A=[x_1,x_2]$.
 We define $\nu_{\mp,i}$ as the corresponding point of $w_\mp=x_i\mp t\ (i=1,2)$.
We will discuss the behavior of HEE separately in the thermal AdS and BTZ phase.

%%%%%%%%%%%%%%%%%%%%%%%%%%%%%
\subsubsection{HEE in Thermal AdS Phase ($s<1$)}
%%%%%%%%%%%%%%%%%%%%%%%%%%%%%
We write the distance between boundary points as $\Delta_{T,\mp}=(X_{T,1}\mp T_{L,T,1})-(X_{T,2}\mp T_{L,T,2})=-2\pi i(\nu_{\mp,1}-\nu_{\mp,2})=-2\pi i\Delta\nu_\mp$.

\begin{figure}[h!]
	\centering
	\includegraphics[width=15cm]{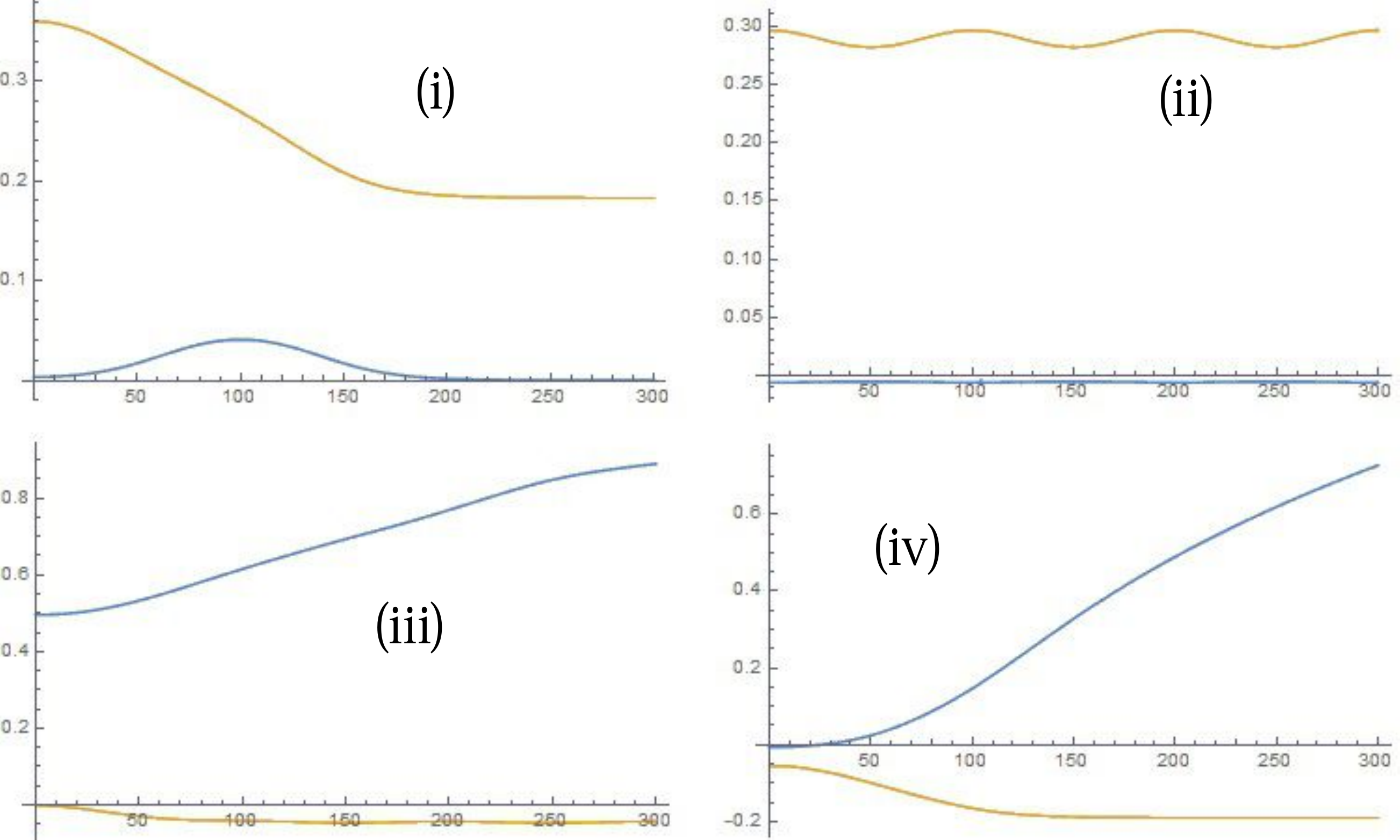}
	\caption{The time evolution of HEE in thermal AdS phase. The horizontal axis describes the time $t$, and the vertical axis describes $\Delta S_A$. We set $b=50$ and $a=50$ (this corresponds to $s= 0.945\ldots(<1)$), and the subsystem $A$ is chosen to be (i), (ii), (iii) and (iv). The upper left graph is for (i), the upper right is for (ii), the lower left is for (iii), and the lower right is for (iv). Blue and orange curves describe the HEE for connected and disconnected geodesics, respectively.}\label{DSQHEE_TAdS}
\end{figure}

Because the metric $ds_T^2$ asymptotically approaches to that of the Poincar\'{e} AdS, the UV cutoff $\epsilon$ in the  original $w$-coordinate is related to the UV cutoff $\delta_T$ in $(z_T,X_T,T_T)$-coordinate as
\begin{equation}
\epsilon = \left|\dfrac{dw}{d (X_T+iT_T)}\right| \delta_T = \frac{1}{2\pi} \left| \frac{dw}{d\nu} \right|\delta_T.
\end{equation}
The HEE for the connected geodesic is,
\begin{align}
S^{con}_A&=\frac{c}{12}\log \left[ \left(\frac{2s}{\delta_T}\right)^4\sin^2\left(\frac{\Delta_{T,-}}{2s} \right)\sin^2\left(\frac{\Delta_{T,+}}{2s} \right) \right]\\
&=\frac{c}{12}\log \left[ \left(\frac{s}{\pi\epsilon}\right)^4\frac{dw_{+,1}}{d\nu_{+,1}}\frac{dw_{-,1}}{d\nu_{-,1}}\frac{dw_{+,2}}{d\nu_{+,2}}\frac{dw_{-,2}}{d\nu_{-,2}}\sinh^2\left(\frac{\pi \Delta\nu_-}{s} \right)\sinh^2\left(\frac{\pi \Delta\nu_+}{s} \right) \right].
\end{align}
We can also calculate the HEE for the disconected geodesic.
\begin{equation}
S^{dis}_A=\min_{\sigma=\pm}\left(\frac{c}{12}\log \left[ \left(\frac{s}{\pi\epsilon}\right)^2\frac{dw_{+,1}}{d\nu_{+,1}}\frac{dw_{-,1}}{d\nu_{-,1}}\sinh^2\left(\frac{\pi L_{1,\sigma}}{s}\right)\right]\right)+(1\leftrightarrow 2)+2S_{bdy},
\end{equation}
where $L_{i,\pm}\ (i=1,2)$ is the distance between a twist operator and its ``mirror image'': 
\begin{equation}\label{defofL}
L_{i,\pm} = \nu_{-,i}-\nu_{+,i}\pm \frac{is}{2}.
\end{equation}
$S_{bdy}$ is the boundary entropy of $Q$. We assume that the tension is vanishing 
$T_{BCFT}=0$ in the AdS/BCFT and thus we have $S_{bdy}=0$.
We numerically computed the time evolution of HEE for the four difference choices of 
the subsystem $A$ in figure \ref{DSQHEE_TAdS}.
%%%%%%%%%%%%%%%%%%%%%%%%%%%%%
\subsubsection{HEE in BTZ phase ($s>1$)}
%%%%%%%%%%%%%%%%%%%%%%%%%%%%%
Here, we write the distance between boundary points as $\Delta_{B,\mp}=(X_{B,1}\mp T_{L,B,1})-(X_{B,2}\mp T_{L,B,2})=-2\pi i(\nu_{\mp,1}-\nu_{\mp,2})/s=-2\pi i\Delta\nu_\mp/s$.

\begin{figure}[h!]
	\centering
	\includegraphics[width=15cm]{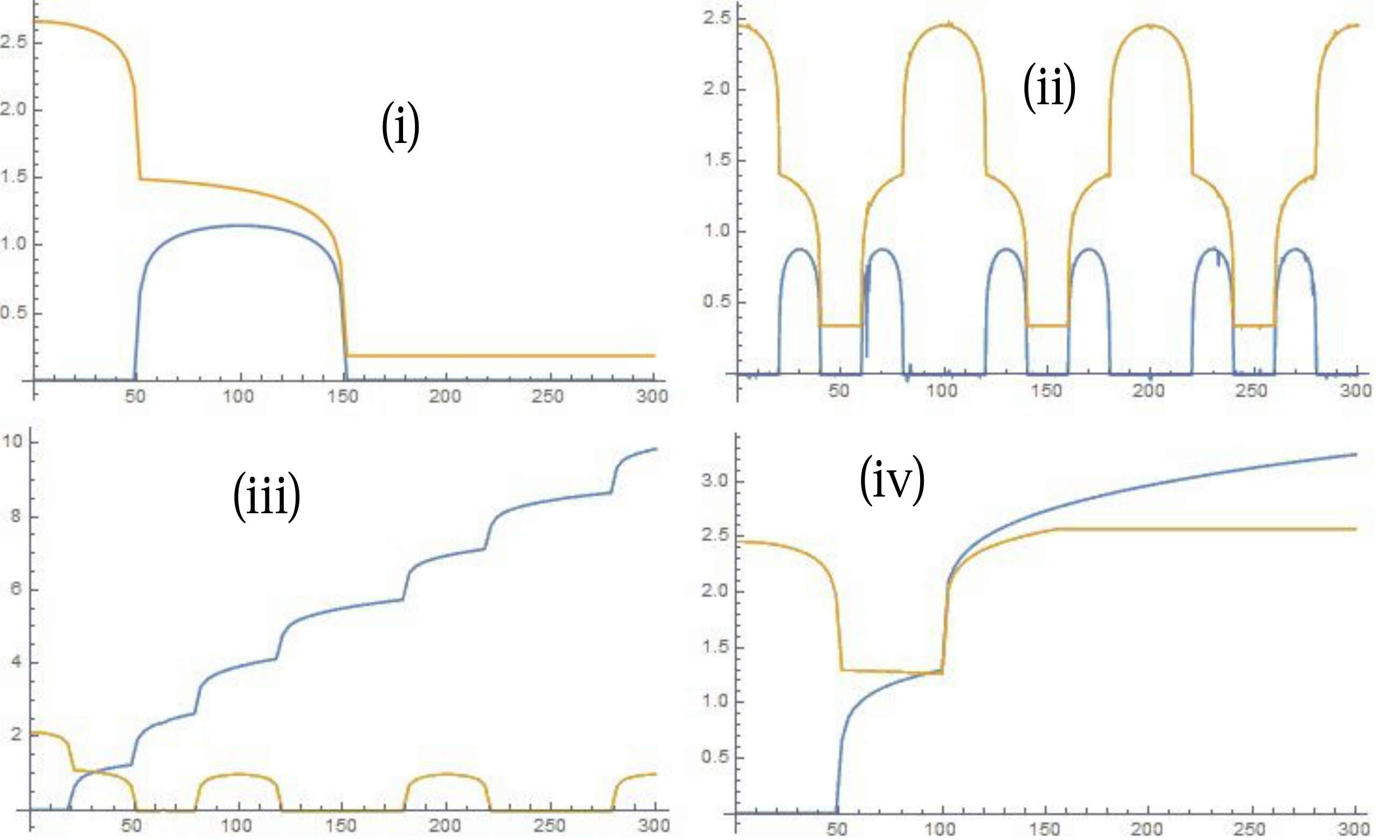}
	\caption{The time evolution of HEE in BTZ BH phase. The horizontal axis is for the time $t$, and the vertical axis is for $\Delta S_A$. We set $b=50$ and $a=0.05$ (corresponding to $s= 5.28\ldots(>1)$), and the subsystem $A$ is chosen to be (i), (ii), (iii) and (iv). The upper left graph is for (i), the upper right is for (ii), the lower left is for (iii), and the lower right is for (iv). Blue and orange curves describe the HEE for connected and disconnected geodesics, respectively.}\label{DSQHEE_BTZ}
\end{figure}

Because the metric is asymptotically the same as the Poincar\'{e} AdS, the UV cutoff $\epsilon$ of original $w$-coordinate is related to the UV cutoff $\delta_B$ of $(z_B,X_B,T_B)$-coordinate with
\begin{equation}
\epsilon = \left|\dfrac{dw}{d (X_B+iT_B)}\right| \delta_B = \frac{s}{2\pi} \left| \frac{dw}{d\nu} \right|\delta_B.
\end{equation}
The HEE for the connected geodesic is,
\begin{align}
S^{con}_A&=\frac{c}{12}\log \left[ \left(\frac{2}{s\delta_B}\right)^4\sinh^2\left(\frac{s\Delta_{B,-}}{2} \right)\sinh^2\left(\frac{s\Delta_{B,+}}{2} \right) \right]\\
&=\frac{c}{12}\log \left[ \left(\frac{1}{\pi\epsilon}\right)^4\frac{dw_{+,1}}{d\nu_{+,1}}\frac{dw_{-,1}}{d\nu_{-,1}}\frac{dw_{+,2}}{d\nu_{+,2}}\frac{dw_{-,2}}{d\nu_{-,2}}\sin^2\left(\pi \Delta\nu_- \right)\sin^2\left(\pi \Delta\nu_+ \right) \right].
\end{align}
We can also calculate the HEE for the disconected geodesic. However in BTZ black hole phase we have to be aware of homology constraint of geodesics, since there are two disconnected boundary surface $Q_1,Q_2$. Finally we get the formula of the HEE for disconected geodesic, 
\begin{equation}
\begin{split}
S^{dis}_A&=\min_{\sigma_1=\pm,\sigma_2=\pm}\left(\frac{c}{12}\log \left[ \left(\frac{s}{\pi\epsilon}\right)^2\frac{dw_{+,1}}{d\nu_{+,1}}\frac{dw_{-,1}}{d\nu_{-,1}}\sin^2\left(\frac{\pi L_{1,\sigma_1}}{s}\right)\right]+(1\leftrightarrow 2)\right.\\&~~~~\left.+\eta_{\sigma_1,\sigma_2}S_{BH}\right)
+S_{bdy,1}+S_{bdy,2},
\end{split}
\end{equation}
where $S_{BH}$ is given by  the black hole entropy  (\ref{eesplitv}) and 
 $L_i$ is defined by equation (\ref{defofL}). The signs $\sigma_{1}=\pm 1$ and $\sigma_2=\pm 1$ describe 
the two possible end points on the two disconnected boundary surfaces $Q_1$ (Im$\nu=s/2$) and 
$Q_2$ (Im$\nu=-s/2$) for each of the two disconnected geodesics. $\eta_{\sigma_1,\sigma_2}$ takes either $0$ or $1$, depending on whether the homology constraint of HEE requires us to include a horizon surface 
in addition to the two disconnected geodesics or not, which is explicitly given by
\begin{equation}
\eta_{\sigma_1,\sigma_2} = \frac{1-\sigma_1\sigma_2}{2}.
\end{equation}
$S_{bdy,1}$ and $S_{bdy,2}$ are the boundary entropies for the boundary surfaces $Q_1,Q_2$, which we set to zero again.
We numerically computed the time evolution of HEE for the four difference choices of 
the subsystem $A$ in figure \ref{DSQHEE_BTZ}.

%%%%%%%%%%%%%%%%%%%%%%%%%%%%%%%
\subsubsection{Interpretations of the time evolutions of HEE}
%%%%%%%%%%%%%%%%%%%%%%%%%%%%%%%

\begin{figure}
	\centering
	\includegraphics[width=12cm]{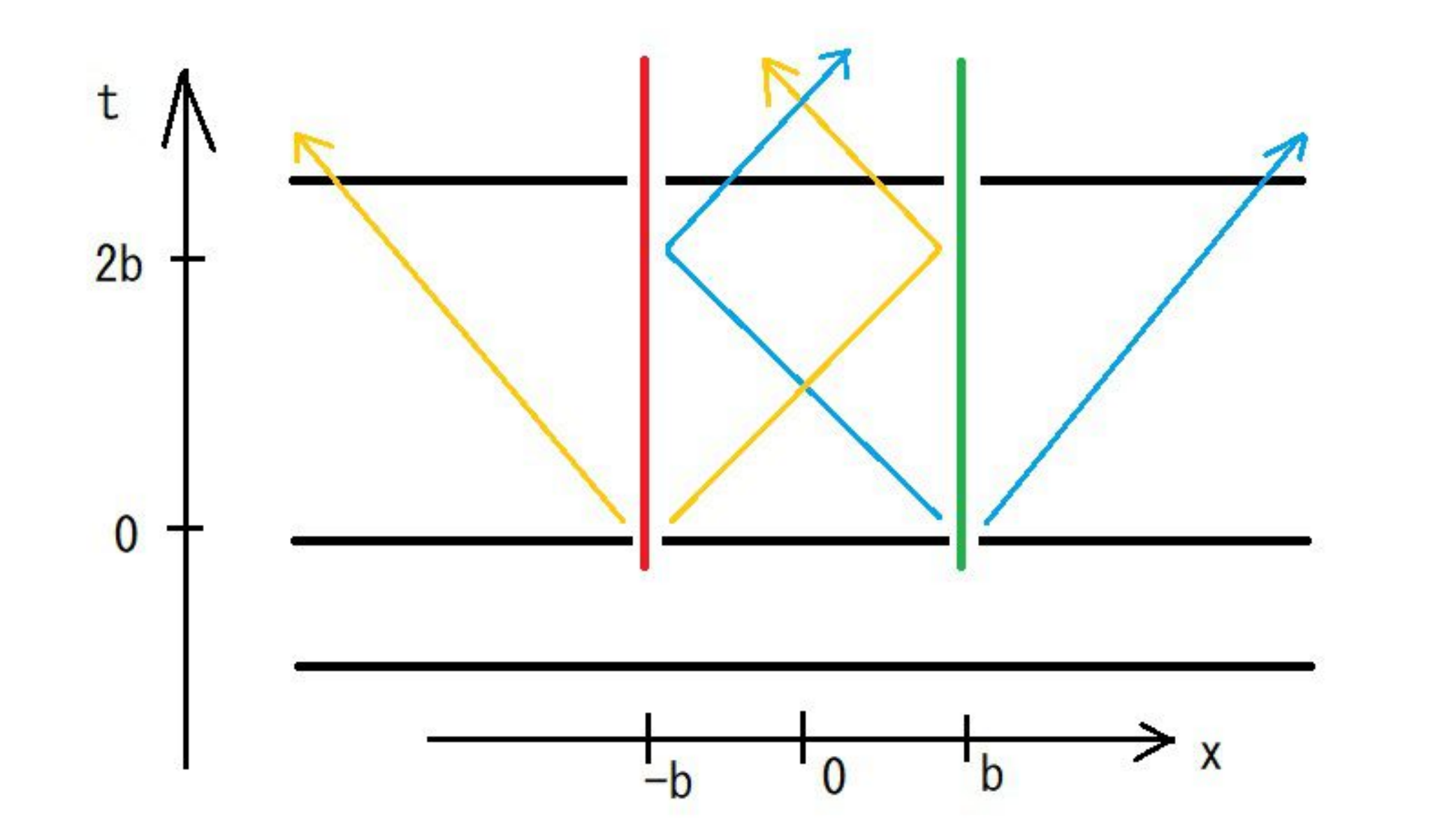}
	\caption{Sketch of a quasi-particle picture. Since single quench creates two quasi-particles, double splitting quench creates four quasi-particles, which is described by blue and yellow lines. These particles move at light speed. Two particles go away from quench point and the other two are trapped in the interval $[-b,b]$.}\label{DSQ_quasipp}
\end{figure}

Now we would like to give physical interpretations of the behaviors of HEE, computed before 
(refer to figure \ref{DSQHEE_TAdS} and \ref{DSQHEE_BTZ}). First we would like to note that we can understand most of qualitative features of the time evolutions from quasi-particle  propagations. Refer to figure \ref{DSQ_quasipp} for a sketch.
The quasi-particle picture is like this: when a simple splitting quench cuts a line into two semi-infinite lines,  entangled quasi-particles are created at the two end points, and they go away from the splitting point at the speed of light. In our double splitting quench case, this entangled pair creation occurs at both of $x=b$ and $x=-b$. Since the middle interval $[-b,b]$ is isolated after the quench, the entangled pairs created at the end points of this interval will confine inside it, bouncing due to 
the ``hard wall'' at $x=\pm b$.. Therefore we will observe 
oscillations for the excitation in $[-b,b]$. On the other hand, the other entangled pairs will propagate to $x\to\pm \infty$ under the time evolution.

Now let us examine the behaviors in more details for the four choices of subsystem $A=[x_1,x_2]$ 
of  figure \ref{fig:eesetup}.  Notice that the quasi-particle picture gets shaper for small values of $a$, as 
this quench cut off parameter $a$ essentially estimates the size of smearing length scale 
of the quench.  In the setup (i), we find that the connected geodesic is dominant. The HEE gets enhanced in the interval $[x_1-b,x_2-b]$, when the entangled pair enters into the subsystem $A$.

In the setup (ii), the connected contribution is again dominant. We observe the oscillations of entanglement pair propagations due to the bouncing at $x=\pm b$, mentioned in the above, though this effect is small for thermal AdS phase.

In the setup (iii), we only have the disconnected geodesic contribution in thermal AdS phase. 
In the BTZ phase, initially the disconnected geodesic is favored, while later the connected one 
dominates forever. In general, we expect that the HEE initially decreases as one part of the entangled pair escapes from the subsystem $A$ and that later the HEE oscillates as in the case of (ii).
We can estimate the final value of $\Delta S_A$ around which the HEE oscillates as follows:
\ba
\Delta S_A&=&\frac{1}{6}\log\frac{2(b-x_1)}{\ep}+\frac{1}{6}\log\frac{2(x_2-b)}{\ep}-\frac{1}{3}\log\frac{x_2-x_1}{\ep},\no
&=&\frac{1}{6}\log\frac{4(x_2-b)(b-x_1)}{(x_2-x_1)^2},
\ea
This can be evaluated as $\Delta S_A\simeq -0.034$ for our specific choice of (iii). 
This is because we can regard the system as the ground state of a holographic CFT on the three disconnected segments: $[-\infty,-b], [-b,b]$ and $[b,\infty]$.
Indeed, this estimation
agrees with our results in both thermal AdS and BTZ phase up to an error we expect.

In the setup of  (iv), the disconnected one always gets dominant in the thermal AdS phase.
In this case, the HEE decreases initially  because the entanglement parts get out of 
the subsystem $A$ and later approaches to a constant. This final value of $\Delta S_A$ is 
negative. On the other hand, in the BTZ phase, we find the initial increasing of  $\Delta S_A$ 
and it approaches to a positive constant. This is because there exists non-vanishing entanglement 
entropy between $[-b,b]$ and its complement, namely the black hole entropy (\ref{btzentp}) 
only in the BTZ phase. We can also estimate this final values as follows:
\ba
\Delta S_A&\simeq& \frac{1}{6}\log\frac{2(-x_1-b)}{\ep}+\frac{1}{6}\log\frac{2(x_2-b)}{\ep}-\frac{1}{3}\log\frac{x_2-x_1}{\ep}+\ti{\eta}\cdot \frac{\pi}{6}s\no
&=&\frac{1}{6}\log\frac{4(x_2-b)(x_1+b)}{(x_2-x_1)^2}+\ti{\eta}\cdot\frac{\pi}{6}s,
\ea
where $\ti{\eta}=0$ in thermal AdS phase and $\ti{\eta}=1$ in the BTZ phase.
This leads to the estimation $\Delta S_A\simeq -0.190$ in thermal AdS phase and 
$\Delta S_A\simeq 2.575$ in the BTZ phase and indeed these
agree with our results in both thermal AdS and BTZ phase up to an error we expect.

\begin{figure}[b!]
  \centering
 \includegraphics[width=10cm]{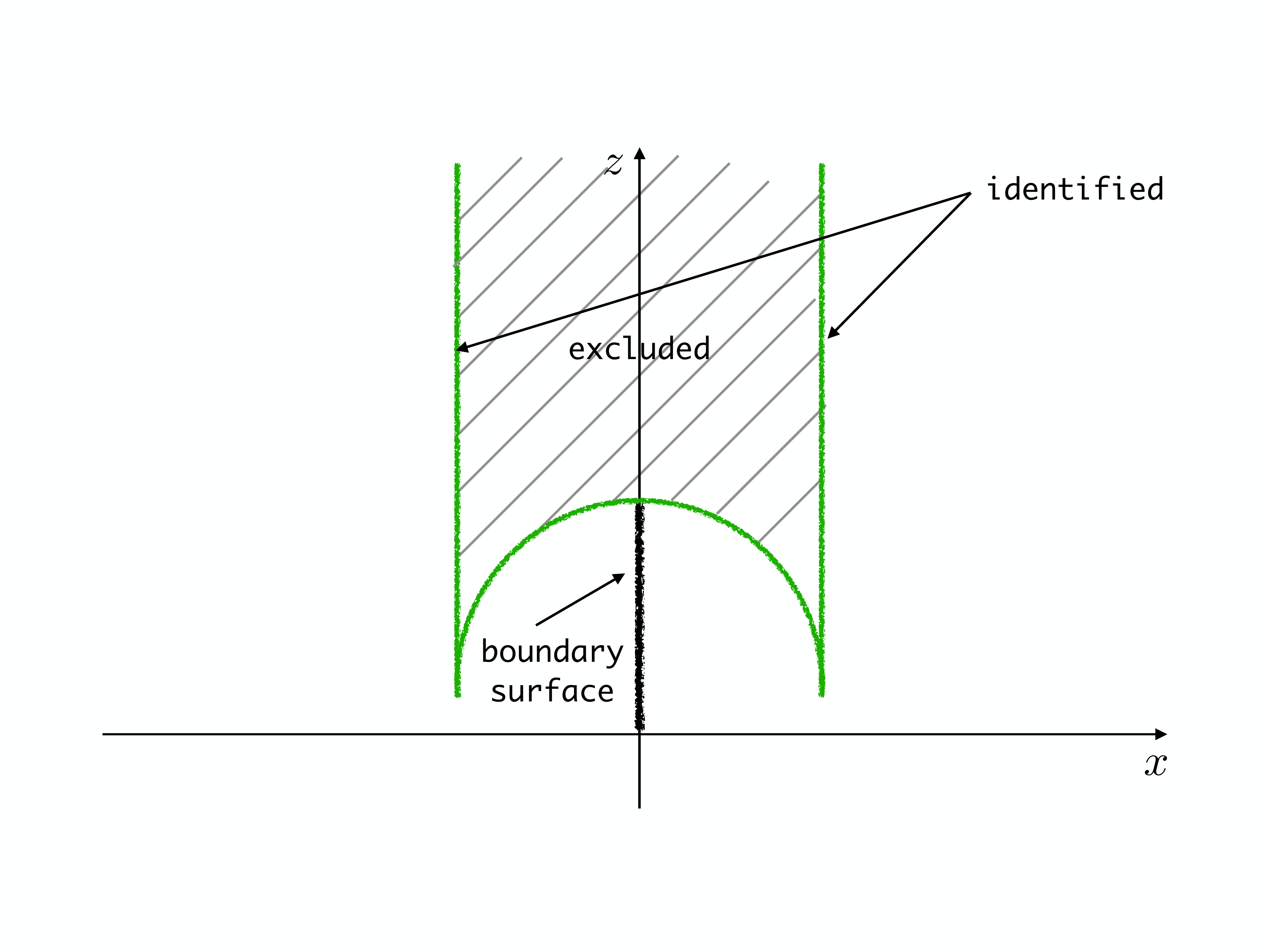}
  \caption{A time slice $t=$ const. in a single splitting quench. The green curve is an identification surface whose $+x$ part and $-x$ part should be identified to realize the true bulk geometry. The shaded part is excluded. The thick black line is the boundary surface, which extends from the quench point and falls towards $+z$ direction. Essentially, there is a falling boundary surface extending from the quench point and other parts of the bulk are connected.}
\label{fig:SSBS}
\end{figure}
%%%%%%%%%%%%%%%%%%%%%%%%%%%%%%%%%%%%%%%%%%%
\subsection{Boundary Surface in Holographic Double Splitting Local Quenches}\label{sec:bdyfsp}
%%%%%%%%%%%%%%%%%%%%%%%%%%%%%%%%%%%%%%%%%%%
Similarly to the single/double joining quench, there should also exist boundary surfaces in double splitting quench. Unfortunately, we cannot use technics introduced in section \ref{section:DJBS} to explicitly figure out the boundary surface this time. Nevertheless, we can still provide its intuitive description.

Let us start by reviewing the bulk geometry of the gravity dual of a single splitting quench at $x=0$ \cite{STW}. As figure \ref{fig:SSBS} shows, there is a boundary surface extending towards $+z$ direction. Besides, there exists a surface in the bulk given by equation (\ref{eq:SJBS}), which is realized as an identification surface. That is, we should identify $+x$ part and $-x$ part of this surface to get the true bulk geometry. The metric is given by (\ref{metads}). In a sentence, there is a boundary surface extending from the quench point to $+z$ direction, and other parts of the bulk are connected.   

\begin{figure}[h!]
  \centering
  \includegraphics[width=7.5cm]{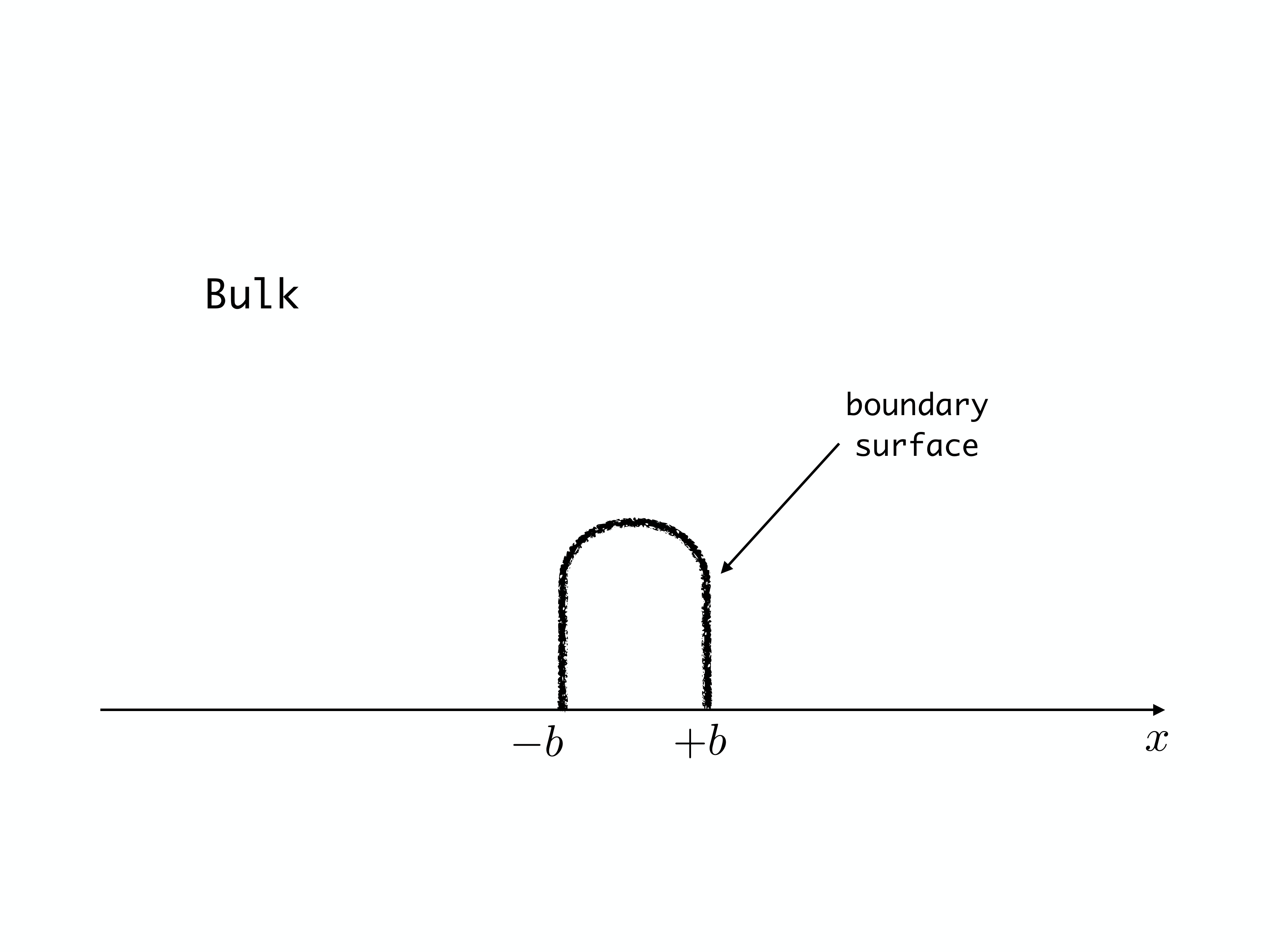}
  \includegraphics[width=7.5cm]{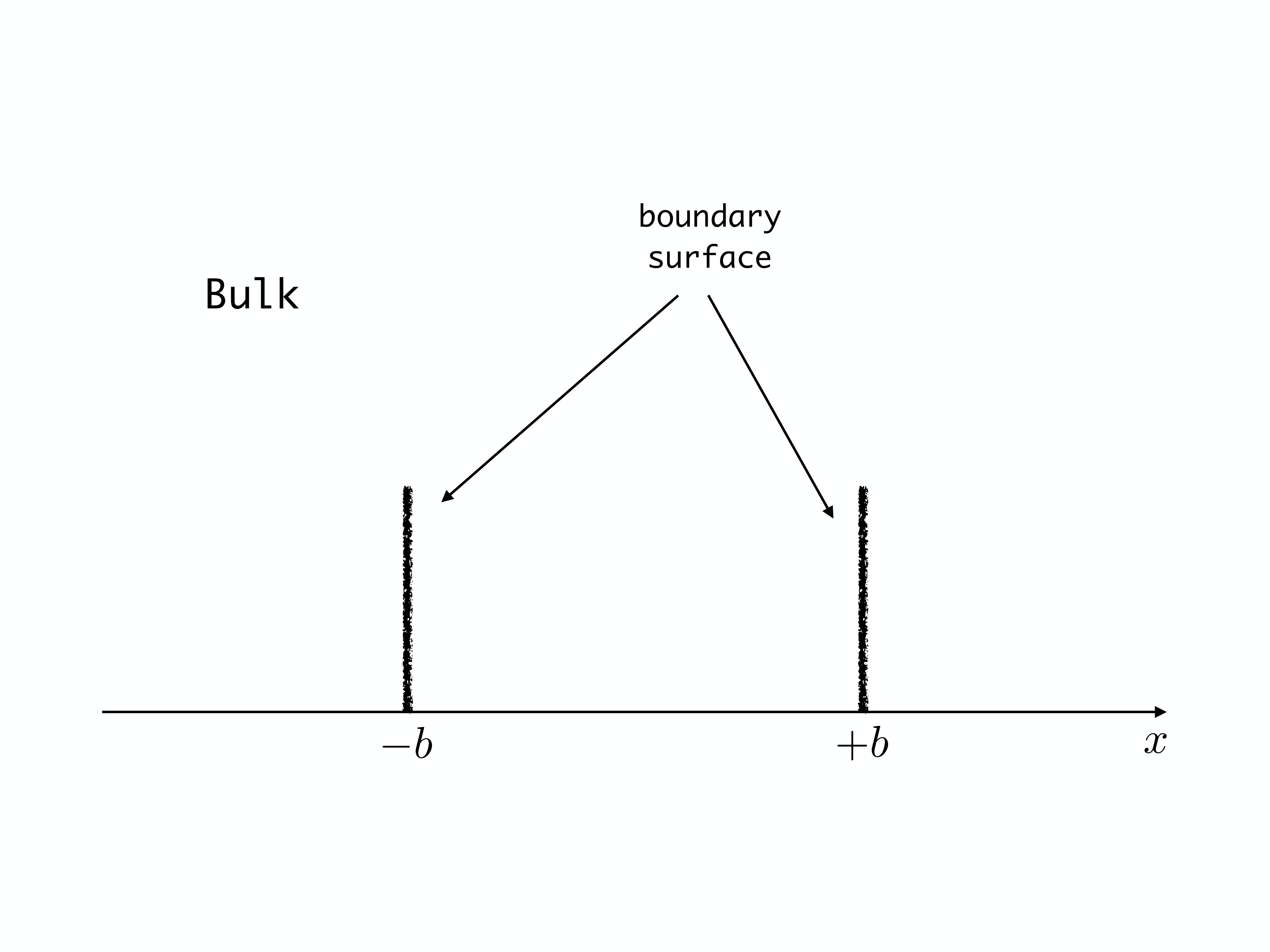}
  \caption{Time slice $t=0$ of the bulk in double splitting quench. The left figure shows case (I) thermal AdS phase and the right figure shows case (II) BTZ phase. This is the expected description of how the Hawking-Page transition should be realized in $(x,t,z)$ coordinate.}
\label{fig:DSBS}
\end{figure}

With the knowledge of the single joining quench, let us give an intuitive description of boundary surfaces in a double joining quench at $x=\pm b$. In (I) thermal AdS phase, $b/a<r_*$, we know that we have only one connected boundary surface in the Euclidean setup (c.f. figure \ref{TADSfig}). Therefore, the time slice $t=0$ is expected to be roughly given by the left figure of figure \ref{fig:DSBS}. The existence of this boundary surface separate the bulk into two part. This is reduced to the boundary surface of a single splitting quench at $b/a\rightarrow0$. On the other hand, in (II) BTZ phase, $b/a>r_*$, we know that we have two connected boundary surface in the Euclidean setup (c.f. figure \ref{BTZfig}).  Therefore, the time slice $t=0$ is expected to be roughly given by the right figure of figure \ref{fig:DSBS}. This is the expected description of how the Hawking-Page transition should be realized in $(x,t,z)$ coordinate. 

We can check that this expectation of boundary surfaces is consistent with the asymptotical behaviors of energy stress tensor (\ref{hratg}) and connected EE (\ref{asymDSEE}). At $b/a\rightarrow0$, the double splitting quench is simply reduced to a single splitting quench and thus we have $q^D/q^S = 1$. When $b/a$ is large, we have two heavy boundary surfaces. For an observer at distant $x$ or at distant $t$, the two boundary surfaces are almost at the same location. At the same time, there is two times of mass because the existence of two boundary surfaces. Thus we have $q^D/q^S = 2$. 

%%%%%%%%%%%%%%%%%%%%%%%%%%%%%%%%%%%%%%%%%%%
\subsection{Double Splitting Local Quenches in Massless Free Dirac Fermion CFT}
%%%%%%%%%%%%%%%%%%%%%%%%%%%%%%%%%%%%%%%%%%%

Now we move on to the double splitting quenches in the Dirac Fermion CFT.
Using the result in \cite{NSTW} of the two point function on a cylinder $(\nu,\bar{\nu})$,
we can calculate the EE under the double splitting local quenches as follows:
\ba
S_A &=&\frac{1}{12}\log\left[\frac{|\frac{dw_1}{d\nu_1}|^2|\frac{dw_2}{d\nu_2}|^2}{(2\pi\ep)^4}\right]\no
&+&\frac{1}{6}\log\left[\frac{\theta_1\left(\nu_1-\nu_2|is\right)\theta_1
\left(\bar{\nu}_1-\bar{\nu}_2|is\right)\theta_1\left(\nu_1-\bar{\nu}_1+\frac{is}{2}|is\right)\theta_1\left(\nu_2-\bar{\nu}_2+\frac{is}{2}|is\right)}{\eta(is)^6\cdot \theta_1\left(\nu_1-\bar{\nu}_2+\frac{is}{2}|is\right)\theta_1
\left(\nu_2-\bar{\nu}_1+\frac{is}{2}|is\right)}\right].\no
\ea

\begin{figure}[h!]
  \centering
 \includegraphics[width=6cm]{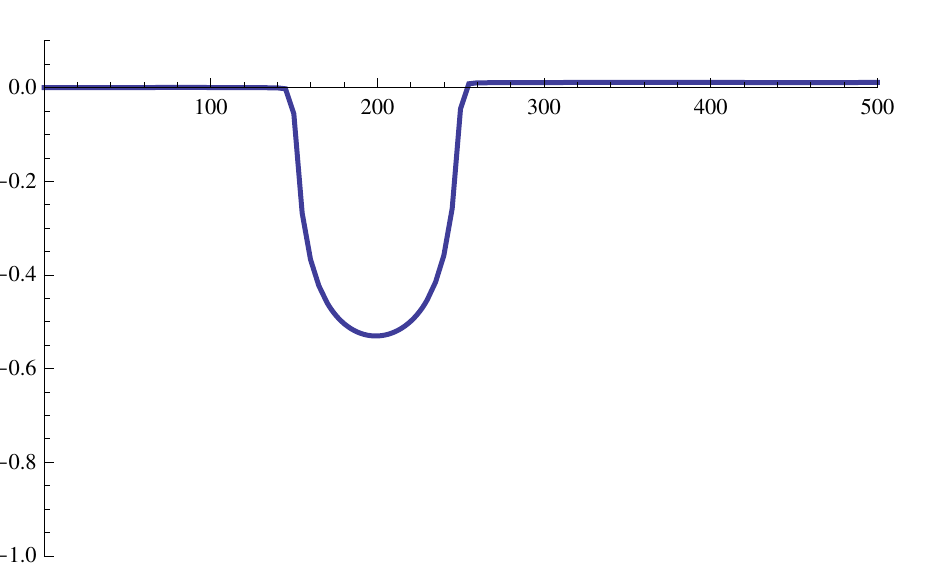}
  \caption{The ratio  
$\Delta S_A^{D}-(\Delta S_A^{S(x=b)}+\Delta S_A^{S(x=-b)})$
under the time evolutions of EE in the splitting quenches of Dirac fermion CFT. 
We chose the subsystem (i) and set
$a=2$ and $b=50$.}
\label{fig:Diraccomparioson}
\end{figure}

In the late time limit $t\to\infty$ or the distant limit of subsystem $x\to \infty$
with the subsystem size and quench parameters 
kept finite, we obtain the following results:
\ba
&& \Delta S^{D}_A\simeq \Delta S^{S}_A\simeq \frac{1}{6}\log\frac{4x_1x_2}{(x_1+x_2)^2}<0 \ \ (t\to \infty),\no
&& \Delta S^{D}_A\simeq 2\Delta S^{S}_A\simeq -\frac{a^4l^2}{12x^6} \ \ (x\to \infty).\label{Q6}
\ea
The result for $t\to\infty$ is easy to understand as it coincides with the entanglement entropy for vacuum state of a CFT defined on a half line.
The late time $t\to\infty$ result shows that the inequality  (\ref{ineqJ})  is violated. Refer also to 
figure \ref{fig:Diraccomparioson} for an explicit numerical computation.

%%%%%%%%%%%%%%%%%%%%%%%%%%%%%
\subsubsection{EE between two disconnected regions}
%%%%%%%%%%%%%%%%%%%%%%%%%%%%%

In the same way as in the holographic CFTs, we can regard the EE between the interval $[-b,b]$ and its complement  as the thermal entropy of the Dirac fermion CFT on a strip 
(in NS sector). The partition function at the inverse temperature $\beta$ is given by 
\ba
Z=e^{\frac{\beta}{24}}\prod_{m=1}^\infty (1+e^{-\beta(m-1/2)})^2=\frac{\theta_3(0,i\beta/2\pi)}{\eta(i\beta/2\pi)},
\ea
where we set the normalized inverse temperature (such that the width of strip is $\pi$)
\be
\beta=\frac{2\pi}{s}.
\ee
The entanglement entropy between the interval and its complement is computed as 
\ba
S_{th}&=&\beta^2\frac{\de}{\de\beta}\left[-\frac{\log Z}{\beta}\right] \no
&=&2\sum_{m=1}^\infty \log(1+e^{-\beta(m-1/2)})+2\beta\sum_{m=1}^\infty\frac{m-1/2}
{1+e^{\beta(m-1/2)}}.\label{entrp}
\ea
In the limit $b/a\to \infty$ (i.e. $s\to \infty$), we find $S_{th}\simeq \frac{\pi^2}{3\beta}$, which coincides with the black hole entropy $S_{BH}$ in the BTZ phase of holographic CFT. 

%%%%%%%%%%%%%%%%%%%%%%%%%%%%%
\subsubsection{Numerical Plots of Time Evolutions of EE}
%%%%%%%%%%%%%%%%%%%%%%%%%%%%%
\begin{figure}[t!]
  \centering
 \includegraphics[width=6cm]{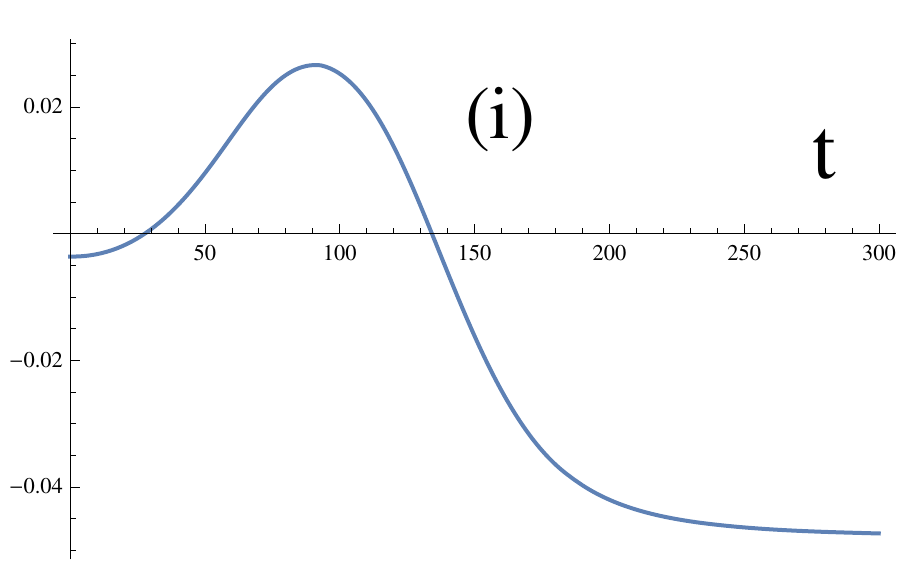}
 \includegraphics[width=6cm]{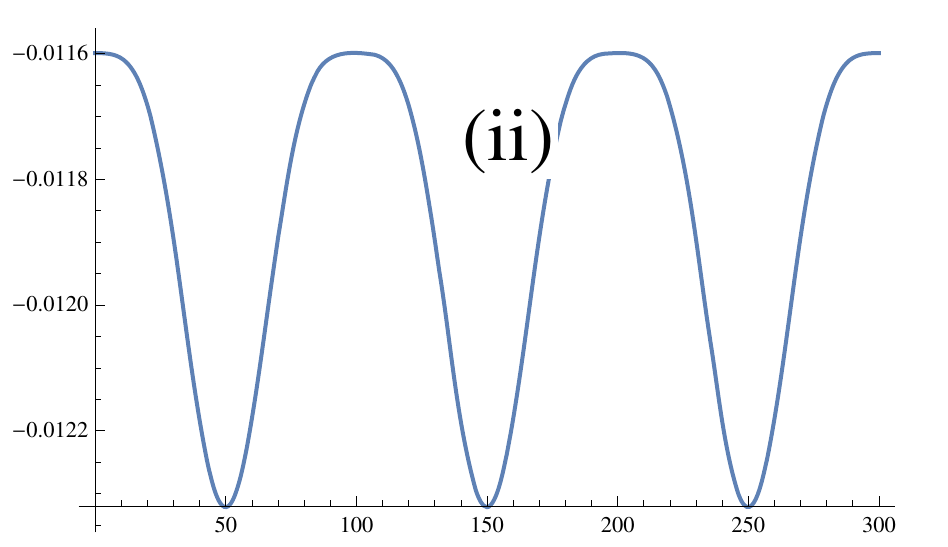}
\includegraphics[width=6cm]{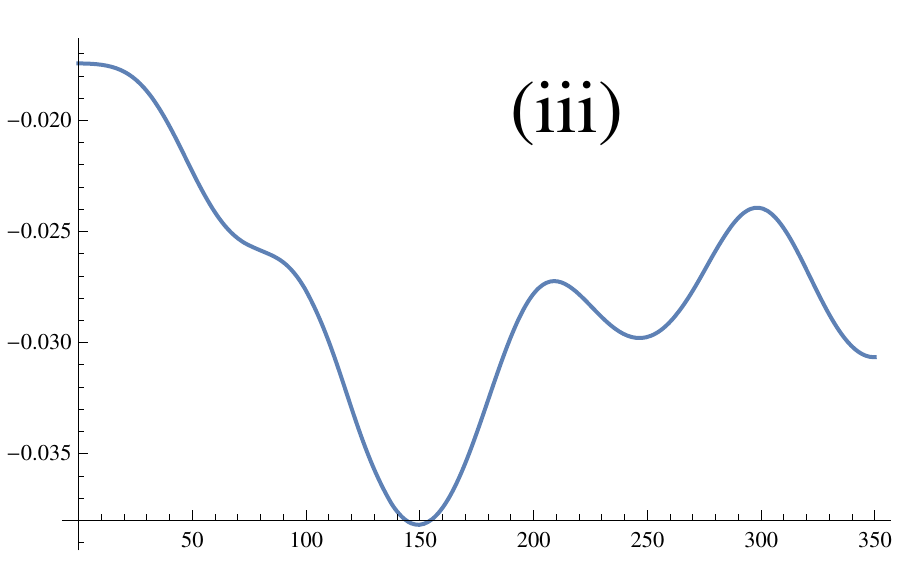}
 \includegraphics[width=6cm]{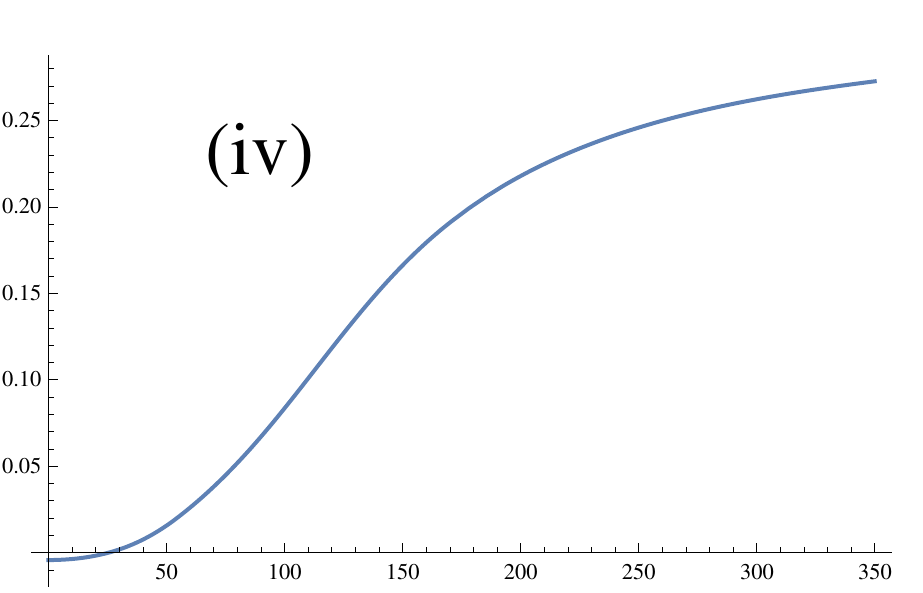}
 \caption{The time evolutions of EE in the Dirac Fermion CFT  for the choices (i), (ii), (iii), (iv)
of subsystem $A$. We chose $b=50$ and $a=50$  (this corresponds to $s= 0.945\ldots(<1)$). }
\label{fig:DiracSplit}
\end{figure}
\begin{figure}[h!]
  \centering
 \includegraphics[width=6cm]{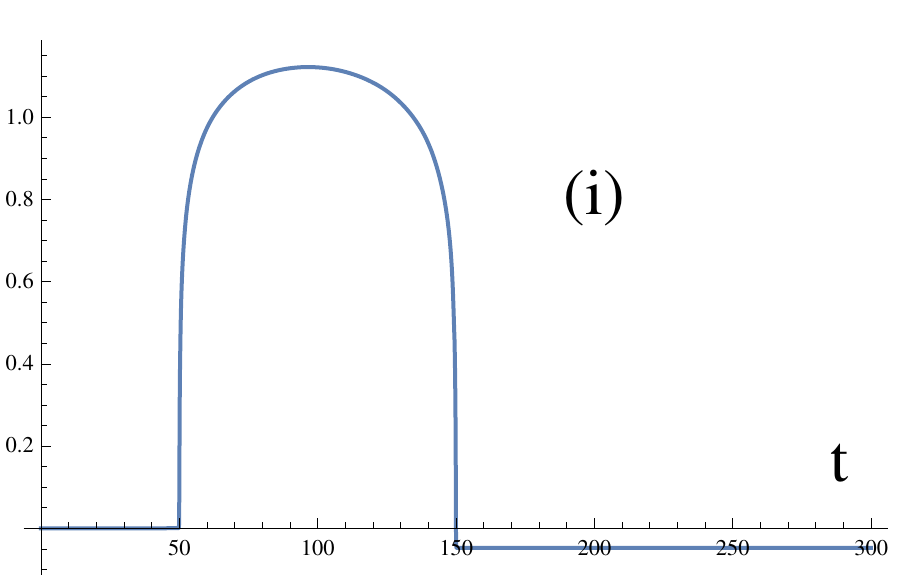}
 \includegraphics[width=6cm]{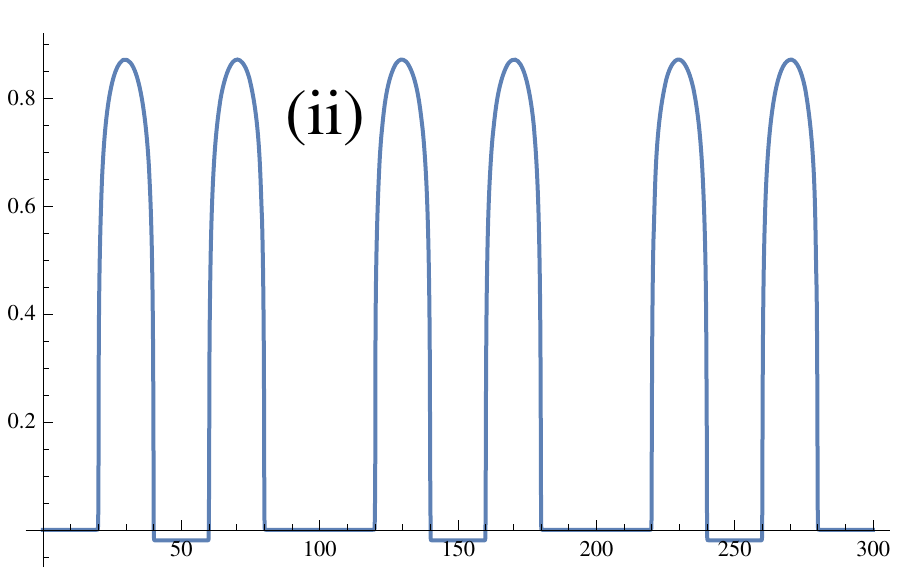}
\includegraphics[width=6cm]{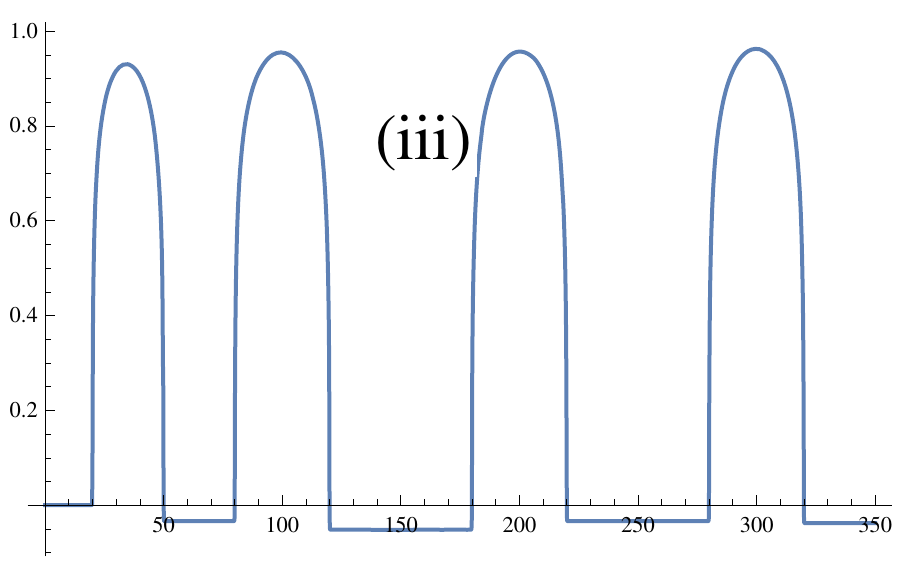}
 \includegraphics[width=6cm]{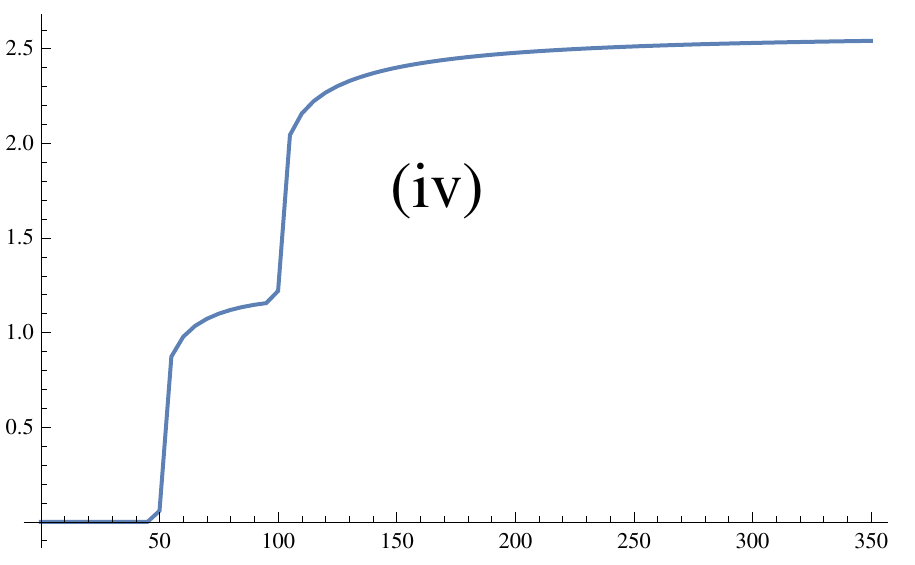}
 \caption{The time evolutions of EE in the Dirac Fermion CFT for the choices (i), (ii), (iii), (iv)
of subsystem $A$. We chose $b=50$ and $a=0.05$  (corresponding to $s= 5.28\ldots(>1)$). }
\label{fig:DiracSplitt}
\end{figure}

We plotted numerical computations of time evolutions of EE for the four different choices of the subsystem $A$ (see figure \ref{fig:eesetup})
in figure \ref{fig:DiracSplit} for $a=50$ and $a=5$ in figure \ref{fig:DiracSplitt}. We always chose $b=50$ for the quench parameter.
We find physical interpretations of these results in a way similar to those in holographic CFTs and 
thus we will not repeat them. 

%%%%%%%%%%%%%%%%%%%%%%%%%%%%%%%%%%%%%%%%%
%%%%%%%%%%%%%%%%%%%%%%%%%%%%%%%%%%%%%%%%%
\section{Double Operator Local Quenches}\label{sec:DOLQ}
%%%%%%%%%%%%%%%%%%%%%%%%%%%%%%%%%%%%%%%%%
%%%%%%%%%%%%%%%%%%%%%%%%%%%%%%%%%%%%%%%%%
A milder version of a local quench state can be modeled by inserting a local operator at some given position into the vacuum and performing the time evolution with original Hamiltonian \cite{NNT}. We call this an operator local quench. The distribution of the energy density in such state depends only on the conformal dimension of the local operator \cite{Caputa:2014vaa}. In particular, if it is a primary operator with the conformal dimension $\Delta_O=\frac{c}{16}$, then the energy stress tensor  of the operator local quench is 
identical to the joining \cite{CCL} and the splitting local quench \cite{STW} as pointed out in \cite{Asplund:2014coa}.
As we reviewed in section 2, it has been also known that the growth of EE
under the operator local quench scales as $\log t$ as is true for the joining quenches. 
 
In this section, we will analyze basic properties of a quantum state excited by inserting two local primary operators. In particular, we will focus on the energy density and compare it with the double joining quench results in the previous sections. We will also compute the difference between this energy density and the sum of energy densities from two ``independent" local quenches. We will find that this change in the energy density is negative supporting the interpretation in terms of the gravitational energy.

Generally, the evolution of entanglement entropy in double local operator quench requires more details from the underlying CFT. Nevertheless, in configurations for which we can apply the first-law \cite{Bhattacharya:2012mi}, we expect entanglement entropy to show similar features to the energy density itself. We will test this expectation by comparing our computations in the double local operator quench with double joining quench results from the previous section.

%%%%%%%%%%%%%%%%%%%%%%%%%%%%%%%%%%%%%%%
\subsection{Setup}
\begin{figure}[h!]
\begin{center}
 \includegraphics[width=10.0cm]{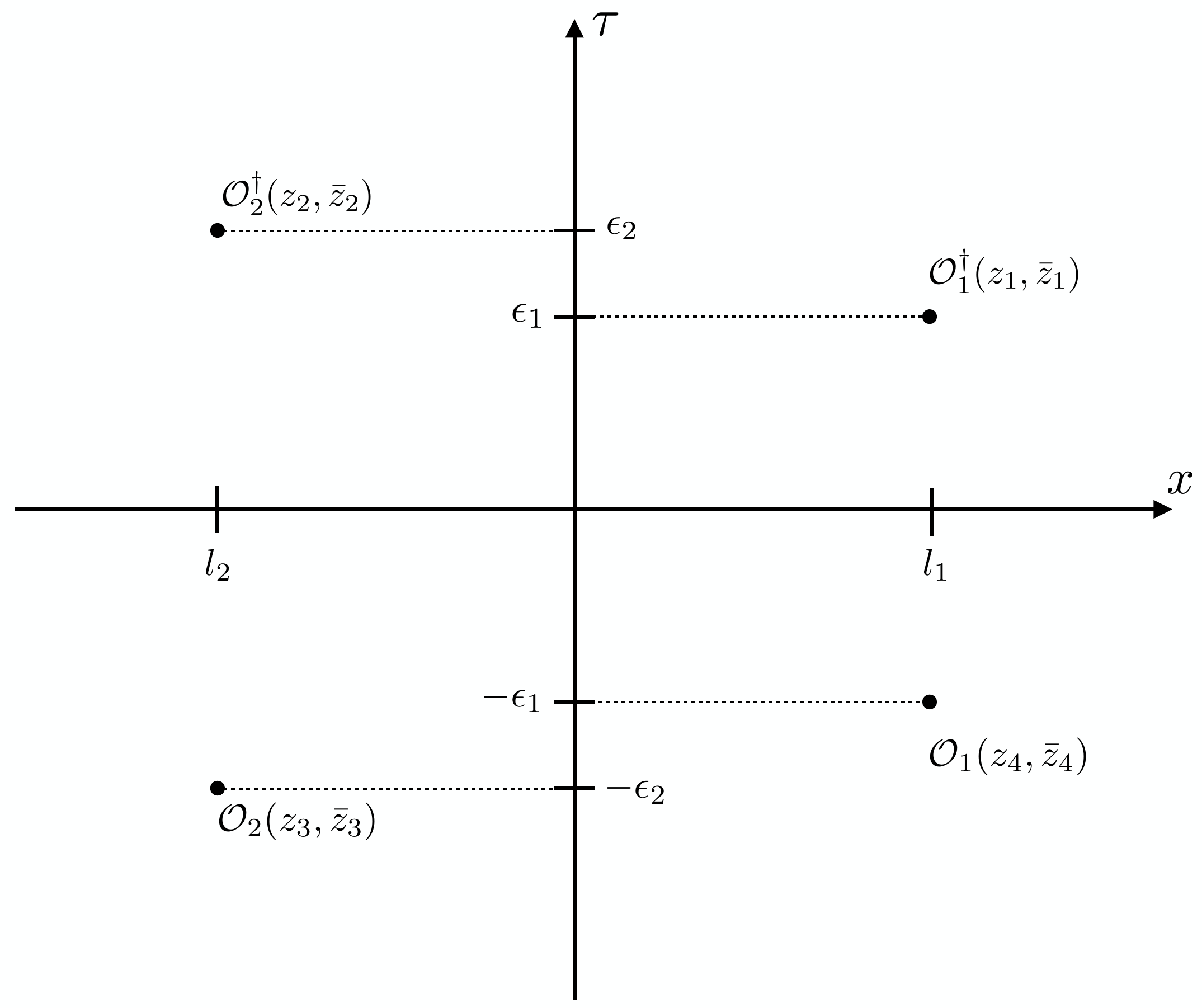}
 \end{center}
 \caption{Path integral representation of the density matrix of our setup.}
 \label{fig:SetupDP}
\end{figure}

%%%%%%%%%%%%%%%%%%%%%%%%%%%%%%%%%%%%%%%
Our setup consists of two local primary operators $\Op_1$  and $\Op_2$ of conformal dimensions $\Delta_1=2h_1=2\bar{h}_1$ and $\Delta_2=2h_2=2\bar{h}_2$. We insert them at positions $l_1$ and $l_2$ respectively, into the vacuum state of a CFT on the real line and then evolve such excited state with original CFT Hamiltonian (the state breaks translational invariance and has a non-trivial time evolution). We then define a double local operator quench state at time $t$ as
\be
\left|\Psi(t)\right>=e^{-iH t}\Op_2(l_2-i\epsilon_2)\Op_1(l_1-i\epsilon_1)\ket{0},
\ee
where in the above formula we ``smeared" the operators in Euclidean time in order to regulate the infinite energy from inserting them locally. More precisely we have
\be
\Op_i(l_i-i\epsilon_i)\equiv e^{-\epsilon_i H}\Op_i(l_i)e^{\epsilon_i H},
\ee
so that $\epsilon_i$ is a cut-off associated to the operator $\Op_i$.\\
The density matrix is then given by
\be
\rho(t)=\mathcal{N}e^{-iH t}\Op_2(l_2-i\epsilon_2)\Op_1(l_1-i\epsilon_1)\ket{0}\bra{0}\Op^\dagger_1(l_1+i\epsilon_1)\Op^\dagger_2(l_2+i\epsilon_2)e^{iH t},
\ee
and following the standard trick, we treat time $t$ as purely imaginary so that, after inserting the identity (exponents of $\pm iHt$) between the operators and the vacuum, we can write
\bea
\rho(t)&=&\mathcal{N}\Op_2(l_2-i(\epsilon_2+it))\Op_1(l_1-i(\epsilon_1+it))\ket{0}\bra{0}\Op^\dagger_1(l_1+i(\epsilon_1-it))\Op^\dagger_2(l_2+i(\epsilon_2-it))\nn
&\equiv&\mathcal{N}\Op_2(z_3,\bar{z}_3)\Op_1(z_4,\bar{z}_4)\ket{0}\bra{0}\Op^\dagger_1(z_1,\bar{z}_1)\Op^\dagger_2(z_2,\bar{z}_2).
\eea
The normalisation $\mathcal{N}$ of the density matrix assures $\Tr \rho(t)=1$ and is given by the inverse of the 4-point function that we denote by $C_4$
\be
\mathcal{N}^{-1}=\bra{0}\Op^\dagger_1(z_1,\bar{z}_1)\Op^\dagger_2(z_2,\bar{z}_2)\Op_2(z_3,\bar{z}_3)\Op_1(z_4,\bar{z}_4)\ket{0}\equiv C_4.
\ee
In the above, we also introduced complex coordinates $(z,\bar{z})=(x+i\tau,x-i\tau)$ such that
\bea
z_1&=&l_1+i(\epsilon_1-it),\qquad \bar{z}_1=l_1-i(\epsilon_1-it),\nn
z_2&=&l_2+i(\epsilon_2-it),\qquad \bar{z}_2=l_2-i(\epsilon_2-it),\nn
z_3&=&l_2-i(\epsilon_2+it),\qquad \bar{z}_3=l_2+i(\epsilon_2+it),\nn
z_4&=&l_1-i(\epsilon_1+it),\qquad \bar{z}_4=l_1+i(\epsilon_1+it).\label{Points}
\eea
The path integral representation of our density matrix is shown in figure \ref{fig:SetupDP}.
\\
In the following, we will use the four-point function in 2d CFTs defined by a general form
\bea
C_4&=&|z_{14}|^{-4h_1}|z_{23}|^{-4h_2}|1-z|^{4h_2}\bra{0}\Op^\dagger_1(0)\Op^\dagger_2(z,\bar{z})\Op_2(1)\Op_1(\infty)\ket{0},\nn
&\equiv& |z_{14}|^{-4h_1}|z_{23}|^{-4h_2} G(z,\bar{z}),\label{C4G}
\eea
where $\Op_1(\infty)\equiv\lim_{z_4,\bar{z}_4\to\infty}|z_4|^{4h_1}\Op_1(z_4,\bar{z}_4)$ and the conformal cross-ratios are
\be
z=\frac{z_{12}z_{34}}{z_{13}z_{24}},\qquad \bar{z}=\frac{\bar{z}_{12}\bar{z}_{34}}{\bar{z}_{13}\bar{z}_{24}}.
\ee
Finally, for later purpose, we can evaluate the cross-ratios for our complex insertion points. They are independent on time $t$ and equal to
\be
0<z=\bar{z}=\frac{(l_2-l_1)^2+(\epsilon_1-\epsilon_2)^2}{(l_2-l_1)^2+(\epsilon_1+\epsilon_2)^2}<1.
\ee
This way the geometry of our problem dictates particular limits of the CFT correlators.\\
Next, we will consider universal properties of the energy density in our double local quench states.

%%%%%%%%%%%%%%%%%%%%%%%%%%%%%%%%%%%%%%%
\subsection{Energy Density}
%%%%%%%%%%%%%%%%%%%%%%%%%%%%%%%%%%%%%%%
Now we consider the energy density in the double local operator excited state. We compute it by analytic continuation from the Euclidean regime where it is evaluated as
\be
T_{zz}(x)+\bar{T}_{\bar{z}\bar{z}}(x)\equiv\Tr\left[\rho(t)(T(x)+\bar{T}(x))\right],
\ee
where $T$ and $\bar{T}$ denotes chiral and anti-chiral components of the energy-momentum tensor.\\
In two-dimensions, the energy density is universally fixed by the OPE of the stress-tensor with the primary operator. Namely, we have the chiral one-point function of $T(x)$ \footnote{Anti-chiral part is obtained by complete analogy.} computed as
\be
T_{zz}(x)=\frac{\langle T(x)\Op^\dagger_{1}(z_1)\Op^\dagger_{2}(z_2)\Op_{2}(z_3)\Op_{1}(z_4)\rangle}{C_4}=\frac{1}{C_4}\sum^{4}_{i=1}\left(\frac{h_i}{(x-z_i)^2}+\frac{\partial_i}{x-z_i}\right)C_4.
\ee
where $h_4=h_1$ and $h_3=h_2$ and derivatives are $\partial_i=\partial_{z_i}$.\\
For example, in a free theory, our four-point function is given by
\be
C^F_4=|z_{14}|^{-4h_1}|z_{23}|^{-4h_2}
\ee
so $G(z,\bar{z})=1$ in conventions of \eqref{C4G}, such that the expectation value of the chiral stress tensor becomes
\be
T^F_{zz}(x)=\frac{h_{1}z^2_{14}}{(x-z_1)^2(x-z_4)^2}+\frac{h_{2}z^2_{23}}{(x-z_2)^2(x-z_3)^2}.
\ee
After analogous computation for the anti-chiral part and insertion of our points \eqref{Points}, we get the Lorentzian energy density\footnote{After analytic continuation, the Lorentzian energy is defined with an overall minus sign. With a slight abuse of notation, and consistently with previous sections, we still denote it by $T_{zz}(x)$ and similarly $\bar{T}_{\bar{z}\bar{z}}(x)$ for the anti chiral part.} 
\bea
T^F_{zz}(x)+\bar{T}^F_{\bar{z}\bar{z}}(x)&=&\frac{4h_1\epsilon^2_1}{((x-l_1-t)^2+\epsilon^2_1)^2}+\frac{4h_1\epsilon^2_1}{((x-l_1+t)^2+\epsilon^2_1)^2},\nn
&+&\frac{4h_2\epsilon^2_2}{((x-l_2-t)^2+\epsilon^2_2)^2}+\frac{4h_2\epsilon^2_2}{((x-l_2+t)^2+\epsilon^2_2)^2}.
\eea
At $t=0$, this energy describes two pulses (delta-function profiles regulated by $\epsilon_{i}$) at the insertion points of the operators. As time progresses, each pulse splits into left and right-moving parts that propagate from the insertion points with the speed of light (for us $c=1$). Without the loss of generality, we can assume $l_2<l_1$. Then the left-moving part from point $l_1$ scatters with the right-moving part from $l_2$ and, after this scattering process, both right-moving parts and left-moving parts continue propagation to the right and to the left respectively.\\
The total energy inserted to the system is a constant of motion and in free theory it becomes\footnote{We conveniently normalize it by $2\pi$.}
\be
E^F\equiv\int^\infty_{-\infty}(T^F_{zz}(x)+\bar{T}^F_{\bar{z}\bar{z}}(x))\,\frac{dx}{2\pi}=\frac{\Delta_1}{\epsilon_1}+\frac{\Delta_2}{\epsilon_2}.
\ee 
For interacting theories the four point function is more complicated and $G(z,\bar{z})\neq 1$. Then, the expectation value of the (chiral) energy momentum tensor becomes
\be
T_{zz}(x)=T^F_{zz}(x)+\sum^{4}_{i=1}\frac{\partial_i \log G(z,\bar{z})}{x-z_i}.
\ee
Since we will be interested in genuine interacting CFTs, it is natural to define the difference 
\be
\Delta T_{zz}(x)=T_{zz}(x)-T^F_{zz}(x)=\sum^{4}_{i=1}\frac{\partial_i \log G(z,\bar{z})}{x-z_i},\label{DTx}
\ee
which can be further simplified to
\be
\Delta T_{zz}(x)=\frac{z_{14}z_{23}}{\prod^4_{i=1}(x-z_i)}\,z\partial_z\log G(z,\bar{z}),
\ee
where we rewrote the derivatives w.r.t. $z_i$ in terms of $\partial_z$.
This way, the difference in the energy density in double local operator quench state between interacting and free theories can be computed as
\be
\Delta T_{zz}(x)+\Delta \bar{T}_{\bar{z}\bar{z}}(x)=\left[\frac{z_{14}z_{23}}{\prod^4_{i=1}(x-z_i)}\,z\partial_z+\frac{\bar{z}_{14}\bar{z}_{23}}{\prod^4_{i=1}(x-\bar{z}_i)}\,\bar{z}\partial_{\bar{z}}\right]\log G(z,\bar{z}).
\ee
Note that in our setup, the dependence on $t$ and $x$ is universal and identical for all 2d CFTs ($z=\bar{z}$ does not depend on $t$ or $x$). However, the sign and the ``magnitude" of the change in the energy depends on all the details of the interacting theory that are captured by $G(z,\bar{z})$.\\
More precisely, inserting points \eqref{Points} we have 
\be
\Delta T_{zz}(x)=\frac{4\epsilon_1\epsilon_2\,z\partial_z\log G(z,\bar{z})}{((x-l_1-t)^2+\epsilon^2_1)((x-l_2-t)^2+\epsilon^2_2)},%+\frac{2\epsilon_1\epsilon_2\,\bar{z}\partial_{\bar{z}}\log G(z,\bar{z})}{\pi((x-l_1+t)^2+\epsilon^2_1)((x-l_2+t)^2+\epsilon^2_2)}.
\ee
and similarly for $\Delta \bar{T}_{\bar{z}\bar{z}}(x)$ with replacing $t\to-t$.\\
Clearly, the denominators of the left- and the right-moving contributions are positive and the signs of the energy are determined by signs of derivatives of $G$ with respect to the real cross-ratios $0<z,\bar{z}<1$.\\
Moreover, in the limit of large distance or the late time, the energy vanishes with the fourth power of $x$ or $t$ respectively and with the same coefficient given by the derivatives of $\log G$, e.g.
\be
\Delta T_{zz}(x)+\Delta \bar{T}_{\bar{z}\bar{z}}(x)\simeq \frac{4\epsilon_1\epsilon_2\left(z\partial_z+\bar{z}\partial_{\bar{z}}\right)\log G(z,\bar{z})}{x^4},
\ee
and similarly for late $t$.\\
Finally, we can compute the total energy in the double local operator quench state in interacting CFTs (that is constant in time)
\be
E= \frac{\Delta_1}{\epsilon_1}+\frac{\Delta_2}{\epsilon_2}+\frac{2(\epsilon_2+\epsilon_1)(z\partial_z+\bar{z}\partial_{\bar{z}})\log G(z,\bar{z})}{(l_2-l_1)^2+(\epsilon_2+\epsilon_1)^2}.
\ee
Next, we will study a few canonical examples of the above energy density in rational (2d Ising model) and large-c holographic CFTs.

%%%%%%%%%%%%%%%%%%%%%%%%%%%%%%%%%%%%%%%
\subsection{Examples}
\begin{figure}[h!]
\begin{center}
 \includegraphics[width=11.0cm]{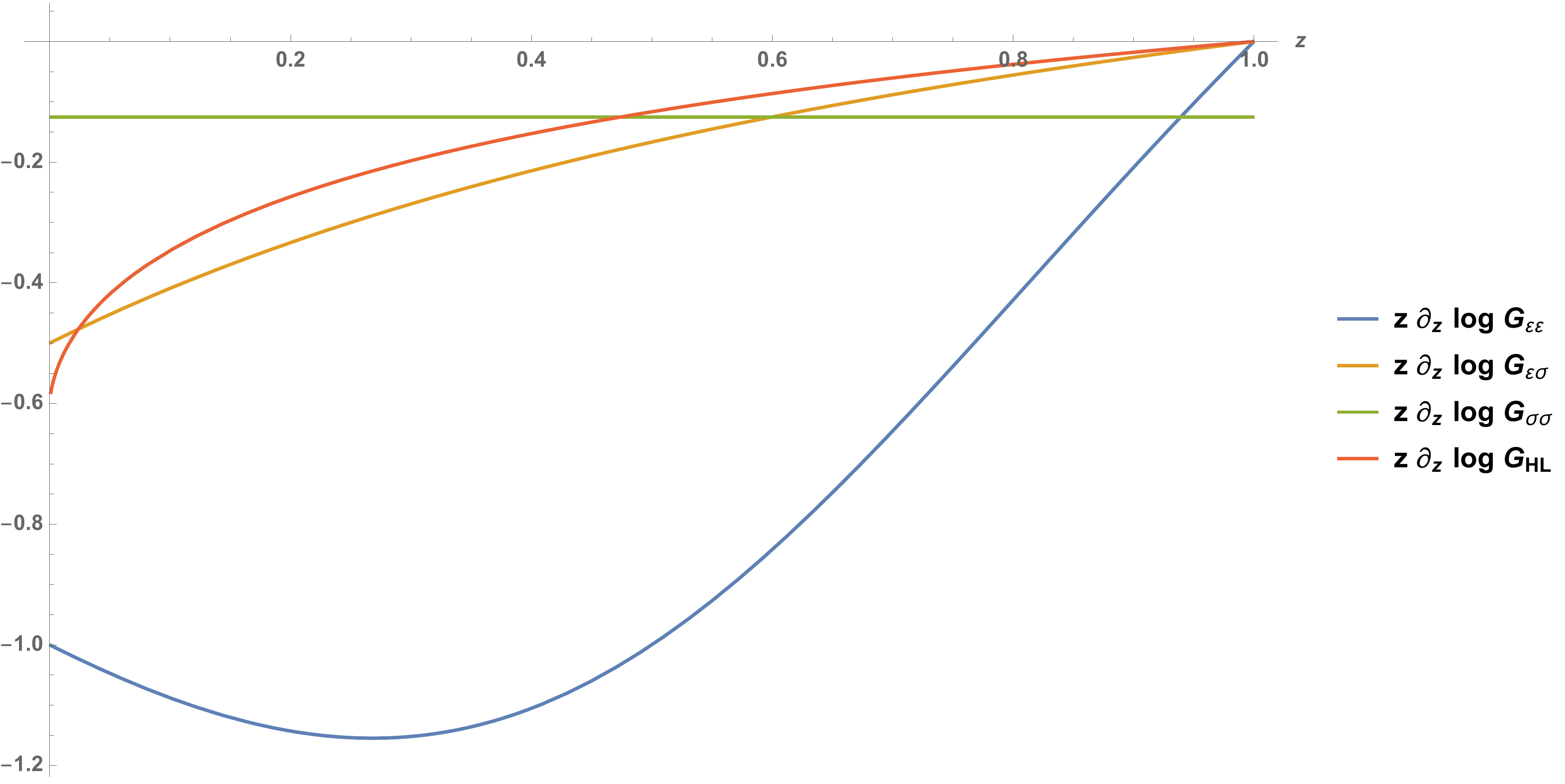}
 \end{center}
 \caption{Plots of $z\partial_z\log G$ that are all negative for our examples. We used $h_L=2$ and $\alpha=0.7$.}
 \label{fig:zDzG}
\end{figure}

%%%%%%%%%%%%%%%%%%%%%%%%%%%%%%%%%%%%%%%
In the 2d Ising Model we have the  energy operator $\varepsilon$ with dimensions $\Delta_\varepsilon=h_\varepsilon+\bar{h}_\varepsilon=1$ and  $\sigma$ operator $\Delta_\sigma=h_\sigma+\bar{h}_\sigma=\frac{1}{8}$. We can use the three relevant correlators\footnote{The subscripts of the operators denote their positions e.g. $\varepsilon_i=\varepsilon(z_i,\bar{z}_i)$.}
\bea
\langle \varepsilon_1\varepsilon_2\varepsilon_3\varepsilon_4\rangle&=&|z_{14}z_{23}|^{-2}\left|\frac{1-z+z^2}{z}\right|^2,\\
\langle\sigma_1\varepsilon_2\varepsilon_3\sigma_4\rangle&=&|z_{14}|^{-\frac{1}{4}}|z_{23}|^{-2}\left|\frac{1+z}{2\sqrt{z}}\right|^2\\
\langle \sigma_1\sigma_2\sigma_3\sigma_4\rangle&=&|z_{14}z_{23}|^{-\frac{1}{4}}\frac{|1+\sqrt{1-z}|+|1-\sqrt{1-z}|}{2|z|^{1/4}},
\eea
The first and second correlator get contributions only from a single conformal block since the fusion rules are $\varepsilon\times\varepsilon=1$ and $\sigma\times\varepsilon=\sigma$. However the third correlator gets contributions from two blocks since $\sigma\times \sigma=1+\varepsilon$.\\
On the other hand, as an example of a holographic correlator, we will take the Heavy-Light 4-point function given by
\be
\langle \Op^H_1 \Op^L_2 \Op^L_3 \Op^H_4\rangle=|z_{14}|^{-4h_H}|z_{23}|^{-4h_L}\left|\frac{z^{\frac{1-\alpha}{2}}(1-z^\alpha)}{\alpha (1-z)}\right|^{-4h_L}
\ee
with $0<\alpha=\sqrt{1-\frac{24h_H}{c}}\le1$.\\
We can show that for each of these correlators, the change in the energy density is negative with respect to the free theory 
\be
\Delta T_{zz}(x)\le 0.
\ee
This sign is determined by the sign of the derivative of the logarithm of $G(z,\bar{z})$ which for the above examples\footnote{The subscript of $G$ denotes the pair of operators used to excite the state.} becomes
\bea
z\partial_z\log G_{\varepsilon\varepsilon}&=&-\frac{(1-z)(1+z)}{1-z(1-z)},\\
z\partial_z\log G_{\varepsilon\sigma}&=&-\frac{(1-z)}{2(1+z)},\\
z\partial_z\log G_{\sigma\sigma}|_{\bar{z}=z}&=&-\frac{1}{8},\\
z\partial_z\log G_{\Op_H\Op_L}&=&-h_L\left[\frac{1+z}{1-z}-\alpha\frac{1+z^\alpha}{1-z^\alpha}\right].
\eea
Clearly, in each of the case above, for $0<z<1$ and $0<\alpha<1$, these expressions (as well as their anti-chiral counterparts) are negative hence the energy is smaller than in free theories (see Figure \ref{fig:zDzG}). Interestingly, for two $\sigma$ excitations, the derivative is independent on $z$ for $z=\bar{z}$.
\\
For the Heavy-Light operators, this is consistent with our observation that the attractive gravitational force is manifested in holographic CFTs by the decrease in the energy density relatively to free theories. On the other hand, results for the Ising model suggest that even in complicated quantum gravity theories, holographically dual to rational CFTs, gravitational force should remain attractive.
%%%%%%%%%%%%%%%%%%%%%%%%%%%%%%%%%%%%%%%
\subsection{Change in energy in holographic CFTs}
%%%%%%%%%%%%%%%%%%%%%%%%%%%%%%%%%%%%%%%
Let us now elaborate more on the holographic CFTs where the change in the energy and derivatives of $\log G(z,\bar{z})$ can be interpreted geometrically.\\
For example, if we focus on the Heavy-Light correlator, function $G(z,\bar{z})$ can be written as
\be
G_{\Op_H\Op_L}(z,\bar{z})=|1-z|^{4h_L}\langle \Op_H|\Op_L(z,\bar{z})\Op_L(1)|\Op_H\rangle\simeq V_0(z)\bar{V}_0(\bar{z}),\label{HolGV}
\ee
where the $V_0(z)\bar{V}_0(\bar{z})$ are the vacuum conformal blocks.\\\
First, in order to get a geometric picture behind our computations, recall that the two point function in the Heavy state (correlator in the middle) can be written in terms of a geodesic length between points $z$ and $1$ at the boundary of the AdS$_3$ conical singularity geometry dual to the Heavy state. More precisely, we have
\be
\langle \Op_H|\Op_L(z,\bar{z})\Op_L(1)|\Op_H\rangle\simeq \exp\left[-h_L( \mathcal{L}^\alpha_\gamma(z,1)+\mathcal{L}^\alpha_\gamma(\bar{z},1))\right],
\ee
where the geodesic length is given by
\be
\mathcal{L}^\alpha_\gamma(z,1)=2\log\left(\frac{z^{\frac{1-\alpha}{2}}(1-z^\alpha)}{\alpha\epsilon}\right).
\ee
Using this length, we can in fact rewrite the derivative of $\log G_{\Op_H\Op_L} $ as
\be
\partial_z \log G_{\Op_H\Op_L} =-h_L\, \partial_z \left(\mathcal{L}^\alpha_\gamma(z,1)-\mathcal{L}^1_\gamma(z,1)\right),
\ee
where the second term on the right is the geodesic length in the vacuum $\alpha=1$ and comes from the $|1-z|^{4h_L}$ pre-factor in \eqref{HolGV}. To write it as a geodesic length we added a cut-off term that can be freely included under the $z$ derivative. Analogous expression is obtained for $\bar{z}$'s.\\
We claim that, gravity and attractive gravitational force in particular, imply that the above derivative is negative. A physical interpretation of this claim is as follows. The derivative of the geodesic length with respect to $z\in(0,1)$ tells us how much the length decreases once we move the end-point of the geodesic closer to 1 (decrease the distance between the points). This decrease in the geodesic length is much bigger if the spacetime is empty or there is nothing  in spacetime that interacts gravitationally. On the other hand, if there is a massive object in the bulk of AdS$_3$ (as the conical singularity above) the decrease in the length is smaller and constrained by the gravitational force from the object. That is why the derivative of the difference between these lengths is always positive in holography or the derivative of the logarithm of $G$ is negative in holographic CFTs. 

\medskip
Let us see if/how these arguments could be generalized to arbitrary heavy correlators in holographic CFTs. One of the features of the so-called ``holographic CFTs" in two dimensions, that are expected to have at least a large central charge and a sparse spectrum, is that four-point correlators are dominated by the vacuum conformal block. In such theories, the 4-point correlators take the form
\be
C^{HOL}_4\simeq \left|z_{14}\right|^{-4h_{1}}\left|z_{23}\right|^{-4h_2}V_0(z)\bar{V}_0(\bar{z}),
\ee
where $G(z,\bar{z})\simeq V_0(z)\bar{V}_0(\bar{z})$ is approximated by the product of the vacuum blocks. At large central charge, blocks are represented by the exponential form 
\be
V_0(z)\simeq \exp\left(-\frac{c}{6}f(z)\right),
\ee
where function $f(z)$ depends on the cross-ratios as well as the dimensions of external operators. Function $f(z)$, as well as general conformal blocks, can in principle be determined using the Virasoro algebra, the monodromy method etc. (see e.g. \cite{Fitzpatrick:2014vua}) \\
In our setup, using this correlator we can again compute the expectation value \eqref{DTx}
\be
\Delta T_{zz}(x)=-\frac{c}{6}\sum^{4}_{i=1}\frac{\partial_i f(z)}{x-z_i}\equiv-\frac{c}{6}\sum^{4}_{i=1}\frac{c_i}{x-z_i},\label{DTMon}
\ee
where in the second equation we used the ``accessory parameters" defined as derivatives of $\partial_i f(z)=c_i$. The negativity of this expression  implies that, for our insertion points and $x$ on the real line, the sum with $c_i$ is always positive.\\
Next, from the behaviour of the one-point function of $T(x)$ at large $x$ (forth-order pole), we can get the following constraints between the parameters
\bea
\sum^{4}_{i=1}c_i=0,\qquad \sum^{4}_{i=1}\left(c_iz_i-\frac{6h_i}{c}\right)=0,\qquad \sum^{4}_{i=1}\left(c_iz^2_i-\frac{12h_i}{c}z_i\right)=0.
\eea
From which we can determine three parameters, say $c_1, c_3$ and $c_4$ such that, even for the simplest case of all equal operators, $\Delta \mathcal{T}(x)$ is expressed in terms of the dimensions $h_1/c$ and one function $c_2$. Hence, it is still a complicated expression that makes it not entirely obvious why the change in the energy should be negative. To make further progress along this route, one would have to impose further monodromy constraints what is beyond the scope of this work.\\
On the other hand given the simplicity of \eqref{DTMon} we believe that there may exist a simple physical argument to prove that this quantity is negative and we leave it as an interesting open problem.

%%%%%%%%%%%%%%%%%%%%%%%%%%%%%%%%%%%%%%%
\subsection{Comparison with double joining quenches}
\begin{figure}[h!]
\begin{center}
 \includegraphics[width=11.0cm]{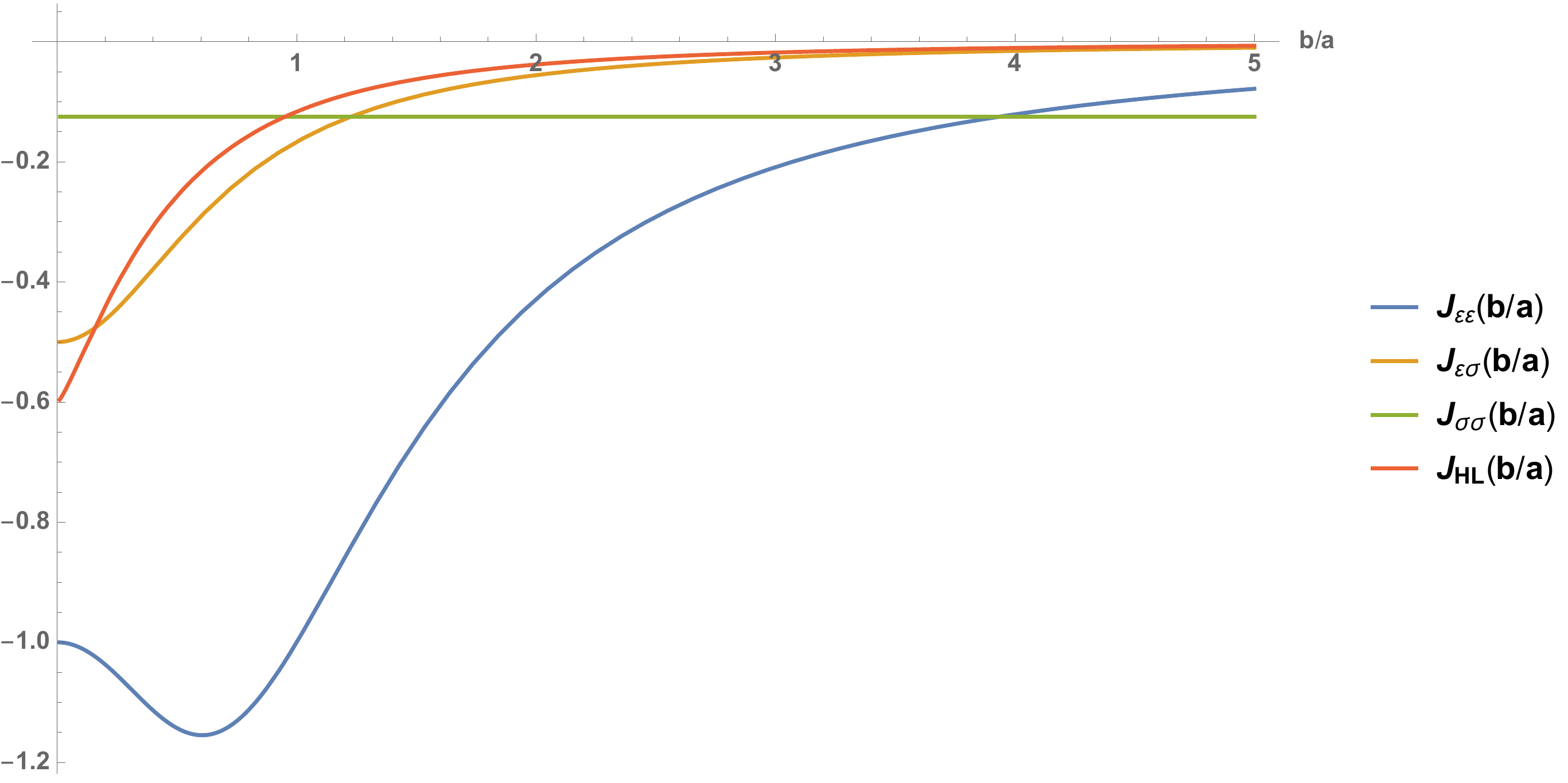}
 \end{center}
 \caption{Plots of $J(b/a)$ in our examples. We used $h_L=2$ and $\alpha=0.7$.}
 \label{fig:Jba}
\end{figure}

%%%%%%%%%%%%%%%%%%%%%%%%%%%%%%%%%%%%%%%
Let us finally compare the results with the double joining quench and rewrite more general computations of the previous section in terms of parameters used before. Connecting to parameters of the double joining quench, for $l_2=-l_1=b$ and $\epsilon_2=\epsilon_1=a$ the cross-ratios become
\be
z=\bar{z}=\frac{(b/a)^2}{1+(b/a)^2}.
\ee
For the operator quench in free theories we have the energy density given as the sum of two independent quenches
\be 
T_{zz}^{D,F}=T^S_{zz}(l_1)+T^S_{zz}(-l_1).
\ee
If we use parameters $l_2=-l_1=b$ and $\epsilon_2=\epsilon_1=a$ then explicitly
\bea
T^F_{zz}(x)+ \bar{T}^F_{\bar{z}\bar{z}}(x)&=&\frac{4h_1a^2}{((x-b-t)^2+a^2)^2}+\frac{4h_1a^2}{((x-b+t)^2+a^2)^2}\nn
&+&\frac{4h_2a^2}{((x+b-t)^2+a^2)^2}+\frac{4h_2a^2}{((x+b+t)^2+a^2)^2}.
\eea
%At $t=0$ we have
%\be
%T^F_{zz}(x)+\bar{T}^F_{\bar{z}\bar{z}}(x)=\frac{8h_1a^2}{((x-b)^2+a^2)^2}+\frac{8h_2a^2}{((x+b)^2+a^2)^2},
%\ee
%and 
In particular, at $t=0$ and for $b=0$, when we bring operators together, the energy looks like a single quench at $t=0$ with the effective energy $h_1+h_2$
\be
T^F_{zz}(x)=\frac{4(h_1+h_2)a^2}{(x^2+a^2)^2}.\label{ETxF}
\ee
This is analogous to what we observed in joining quenches.\\
In interacting theories, we have generally
\be
T_{zz}(x)=T^F_{zz}(x)+\Delta T_{zz}(x),\label{ETx}
\ee
where
\be
\Delta T_{zz}(x)=\frac{4a^2 z\partial_z\log G}{\left((x-b+t)^2+a^2\right)\left((x+b+t)^2+a^2\right)}%+\frac{2a^2 \bar{z}\partial_{\bar{z}}\log G}{\pi\left((x-b-t)^2+a^2\right)\left((x+b-t)^2+a^2\right)}
\ee
and similarly for the anti-chiral part with $t\to-t$.\\
At $t=0$, the change in the energy density becomes
\be
\Delta T_{zz}(x)=\frac{4a^2 z\partial_z\log G(z,\bar{z})}{((x-b)^2+a^2)((x+b)^2+a^2)}.
\ee
Finally, for large $x$ we can compute the total (chiral) energy density at $t=0$
\be
T_{zz}(x)\simeq\frac{4a^2(h_1+h_2+z\partial_z\log G)}{ x^4}.
\ee
This way, we can analytically compute the energy in double operator quench as a function of $b/a$ at $t=0$ in our examples. Namely, we define at large $x$ 
\be
\Delta T^D_{zz}(x)=\frac{4a^2z\partial_z\log G}{x^4}\equiv \frac{4a^2}{x^4}J(b/a).   \label{Q8}
\ee
Now, since we have\footnote{in all cases $(z\partial_z+\bar{z}\partial_{\bar{z}})\log G=2z\partial_z\log G=2\bar{z}\partial_{\bar{z}}\log G$ since $z=\bar{z}$.}
\bea
J_{\varepsilon\varepsilon}\left(b/a\right)&=&-\frac{1+2\frac{b^2}{a^2}}{1+\frac{b^2}{a^2}+\frac{b^4}{a^4}},\\
J_{\varepsilon\sigma}\left(b/a\right)&=&-\frac{1}{2(1+2\frac{b^2}{a^2})},\\
J_{\sigma\sigma}\left(b/a\right)|_{\bar{z}=z}&=&-\frac{1}{8},\\
J_{\Op_H\Op_L}\left(b/a\right)&=&-h_L\left[1+2\frac{b^2}{a^2}+\alpha\frac{1+\left(\frac{1+\frac{b^2}{a^2}}{\frac{b^2}{a^2}}\right)^\alpha}{1-\left(\frac{1+\frac{b^2}{a^2}}{\frac{b^2}{a^2}}\right)^\alpha}\right],
\eea
we can plot functions $J(b/a)$ in the above examples in figure \ref{fig:Jba}.

\begin{figure}[h!]
\begin{center}
\includegraphics[width=11.0cm]{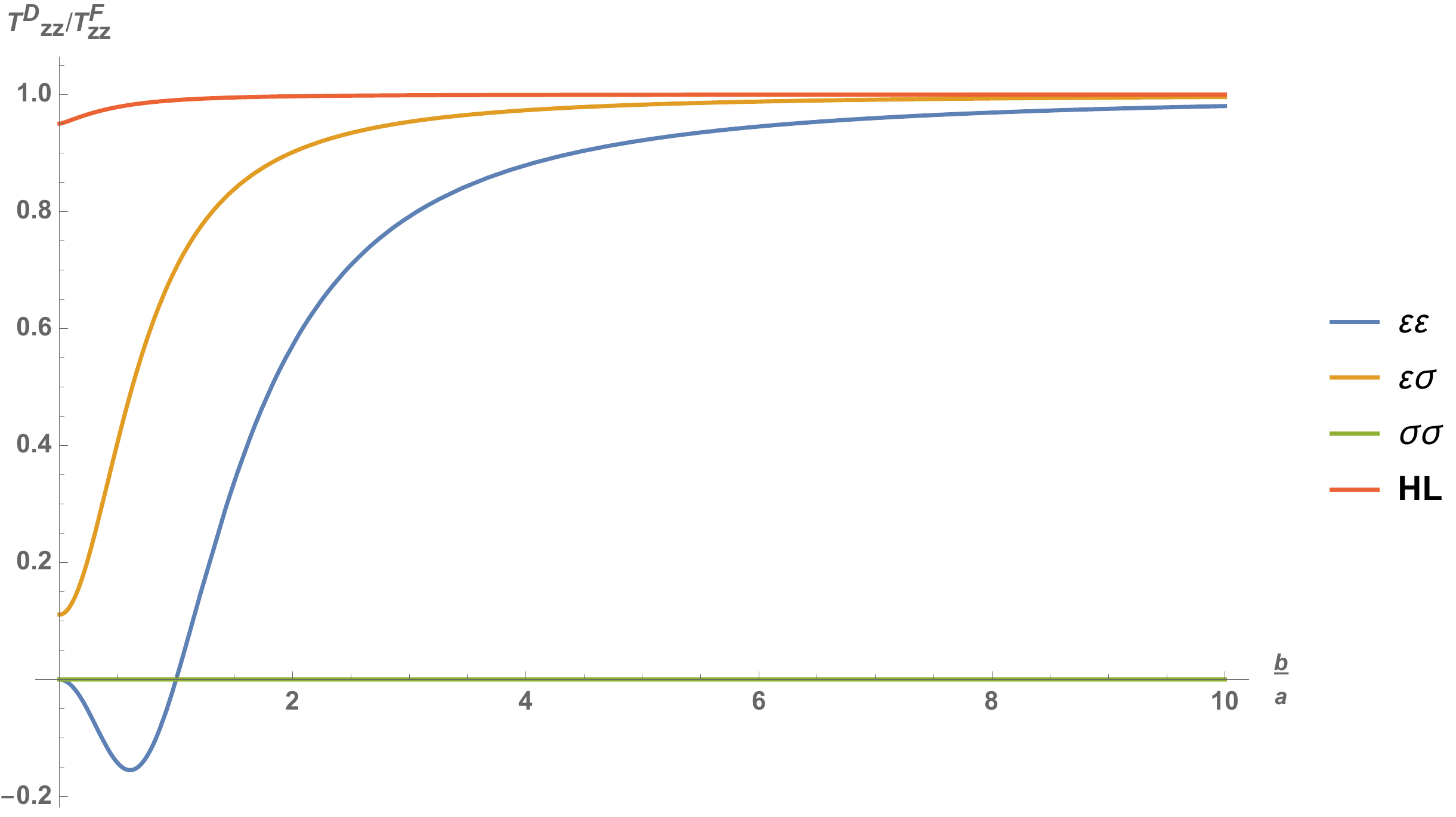}
\end{center}
\caption{Plots of the ratio \eqref{RatTTf}. We used $h_L=2$ and $\alpha
=0.7$ and $h_H=10$.}
\label{fig:RatioTTf}
\end{figure}

Clearly, the changes in the energy density are always negative. They start from the dimension of the operator $-2h_i$ for the same operators and $-h_\varepsilon$ for the mixed correlator as well as $-h_L(1-\alpha)$ for Heavy-Light at $b/a=0$ and decrease with the separation distance $b/a$.\\
It is also interesting to consider the total energy density \eqref{ETx} divided by the total energy density in free theories \eqref{ETxF} in the limit of large $x$ (at $t=0$)
\be
\frac{T^D_{zz}(x)}{T^F_{zz}(x)}\simeq \left(1+\frac{z\partial_z \log G}{
h_1+h_2}\right)=\left(1+\frac{J(b/a)}{h_1+h_2}\right).\label{RatTTf}
\ee
We plot this ratio for our examples in figure \ref{fig:RatioTTf}.
\\
For $\sigma$'s this ratio is identically zero. At small separations this 
ratio vanishes for $\varepsilon$ and is equal to $1/9$ for the mixed $\varepsilon$-$\sigma$ case. For the Heavy-Light correlator we have the 
initial value of the ratio given by $1-\frac{h_L(1-\alpha)}{h_L+h_H}$. 
At large $b/a$ the energy density approaches the free theory result and 
the ratio is given by $1$. We can also see that the $\varepsilon$ and 
mixed $\varepsilon$-$\sigma$ examples have a minimum at $b/a=\sqrt{\frac
{1}{2}\left(\sqrt{3}-1\right)}$ and $b/a=0$ respectively. \\
From this behaviour we can see that from a large distance $x$ the two 
local excitations look independent when separated by large $b/a$. On the 
other hand, when they are close from each other, the energy is always 
decreased due to interactions which result in ``screening" of the actual 
content of the two operators. These features are indeed similar to what 
we observed for both the double joining quench (refer to the middle picture 
of  Fig.\ref{fig:ratio}) and the double splitting quench (refer to the right picture of Fig.\ref{EMsp}).
This monotonically behavior of the ratio common to all three local quenches when $b$ is large, can be understood  from perturbative gravitational 
attractive force. However we would like to notice a difference that when the two quenches 
occur at the same location i.e. $b=0$, the ratio $T^D/T^S$ goes back to $1$ in the double joining and splitting quench (\ref{bc}), 
while in the operator local quench the ratio is still monotonically decreasing and is smaller than $1$ as $b$ gets smaller.

%%%%%%%%%%%%%%%%%%%%%%%%%%%%%%%%%%%%%%%%%
%%%%%%%%%%%%%%%%%%%%%%%%%%%%%%%%%%%%%%%%%
\section{Conclusions and Discussions}
%%%%%%%%%%%%%%%%%%%%%%%%%%%%%%%%%%%%%%%%%
%%%%%%%%%%%%%%%%%%%%%%%%%%%%%%%%%%%%%%%%%

In this work, we studied the dynamics of double local quenches in two dimensional CFTs. 
We discussed three different types of double local quenches: (a) joining, (b) splitting, and (c) operator. 

 In section \ref{sec:DJLQ} and \ref{sec:DSLQ}, we analyzed the first two types of double quenches (a) and (b), respectively.
 We showed that we can calculate the energy density via conformal transformations into an upper half plane (case (a))
and a cylinder (case (b)), respectively. Thus the setups of (a) and (b) can be treated as examples of boundary conformal field theories (BCFTs).  
For holographic CFTs, we constructed their gravity duals by using the framework of AdS/BCFT and this allows us to calculate the entanglement entropy (EE). Moreover, we worked out how the boundary surface $Q$ looks like in the original coordinate for the joining quench by directly performing the coordinate transformation. We also gave a qulitative discussion about the the boundary surface $Q$ in splitting quench.
For a Dirac fermion CFT, we can explicitly write the twist operator by the bosonization procedure and thus we can calculate the EE exactly by the conformal maps.  

For the double operator local quench (c), we manage to calculate the values of energy stress tensor using known expressions of conformal blocks in section \ref{sec:DOLQ}. Besides free CFTs, certain examples in the Ising model and holographic CFTs are examined. 

Throughout our discussions in this paper, we focus on the inequality (\ref{ineq}): $q^{D(x=\pm b)}-q^{S(x=b)}-q^{S(x=-b)}\leq 0$, where $q$ is a quantity which measures local excitation due to initial local quenches. This inequality means that $q$ for the double quench is upper bounded by that for the simple sum of two single local quenches and it is inspired by the attractive nature of the gravitational force.

Below let us give a more detailed summary of our new results in this paper.
From our explicit calculations, we find that the energy stress tensor for the double quench is always upper bounded by that for the simple sum of two single local quenches, i.e. the inequality 
(\ref{ineqt}). This is true for any 2d CFTs in the case of (a) joining (\ref{Q1})
and (b) splitting quench (\ref{hratg}). In the case of (c) operator quench, we were able to confirm this inequality analytically for the free, Ising as well as holographic CFT where one of the operator is heavy and the other is light. For the free CFT in the case (c), we found that the inequality is always saturated, while not in the other cases, for which refer to (\ref{Q8}).

It will be an intriguing future problem to confirm that this inequality of energy stress tensor is true for any double operator quenches. For holographic CFTs, we argued that this inequality is due to the attractive nature of gravitational forces, because they will make two heavy objects dual to the double quench get closer and because this will make the back reactions to a far away observer smaller. 
 
Moreover, we analyzed the time evolutions of EE under double (a) joining and (b) splitting local quenches. In the former (a), we were able to interpret the evolutions of the EE in terms of relativistic particle propagations both for the holographic CFT (refer to Fig.17 and 19) and Dirac fermion CFTs (refer to Fig.28) except the logarithmic growth $\Delta S_A\sim \frac{c}{3}\log t$ when the subsystem $A$ is semi-infinite. Interestingly this growth is a half of the simple sum of two single joining local quenches (\ref{timehalf}). In the latter case (b), again we find that the qualitative behaviors can be explained by the particle picture and there is no logarithmic growth found as was so in the single quench case. However, only for the holographic CFTs, there is a phase transition dual to the 
Hawking-Page transition, when we change the value of the quench parameters $(a,b)$, where $a$ is the regularization parameter of the quench and $2b$ is equal to the distance of the two quench points. 

Moreover, the difference between the holographic CFTs and free Dirac fermion CFT becomes 
sharp when we consider the inequality (\ref{ineqJJ}).
For the holographic entanglement entropy (HEE), this inequality is always satisfied 
if the subsystem $A$ is enough separated from the quench points. Refer to (\ref{Q2}) and (\ref{Q3})
for (a) joining case and to (\ref{asymDSEE}) for (b) splitting case.
However,  the inequality (\ref{ineqJJ}) is violated in the free Dirac fermion CFT 
even in such a case as in (\ref{Q4}) for (a) joining case and (\ref{Q6}) for (b) splitting case.
 Note also that the inequality (\ref{ineqJ}) is always true for the connected geodesic. 
This inequality for the HEE can be regarded as a non-linear extension of that (\ref{ineqt}) for 
the energy stress tensor. Indeed if we take the limit of vanishing size of the subsystem $A$, the first law of EE tells us that  (\ref{ineqJJ}) is reduced to  (\ref{ineqt}).

There will be several future directions to study the double local quenches from different 
viewpoints. For example, it will be interesting to probe the double quenches 
by other quantum information theoretic quantities such as
the computational complexity, information metric and related quantities
\cite{Stanford:2014jda,MIyaji:2015mia,Brown:2015bva,Lehner:2016vdi,Caputa:2017yrh,Belin:2018bpg} 
(see \cite{Ageev:2019fxn} for such calculations under a single holographic local quench).
Another direction will involve higher dimensional generalizations, where the AdS/BCFT construction 
will play an important role because the CFT analysis will become very difficult. 

Finally, we have to admit that our discussions about the relation between quantities in CFT and gravitational force in its AdS gravity dual are not quantitative enough. This is mainly because in 3D gravity there is no dynamically propagating gravitons in the bulk \cite{1984DJtH} and thus there is no gravitational force 
which obeys a standard Newton's law. Instead, what we have in mind as gravitational force in this paper
is that due to so called boundary gravitons or Brown-Henneaux boundary excitations \cite{1986BH, 2007MW}, which is expected to lead to attractive back-reactions (see e.g \cite{Fitzpatrick:2014vua}).
To make a quantitative comparison between CFT quantities and gravitational force, it will be an interesting 
direction to try to repeat a similar analysis of energy density and entanglement entropy in higher dimensional CFTs.

%%%%%%%%%%%%%%%%%%%%%%%%%%%%%%%%%%%%%%%%%
%%%%%%%%%%%%%%%%%%%%%%%%%%%%%%%%%%%%%%%%%
\section*{ Acknowledgements}
%%%%%%%%%%%%%%%%%%%%%%%%%%%%%%%%%%%%%%%%%
%%%%%%%%%%%%%%%%%%%%%%%%%%%%%%%%%%%%%%%%%
We are grateful to Alvaro Veliz-Osorio for participation in the initial stages of this project, discussions and comments on the draft and Yuya Kusuki and Masamichi Miyaji for useful correspondences. PC, TN and TT are supported by the Simons Foundation through the ``It from Qubit'' collaboration. TT would like to thank
Matthew Headrick, Robert Myers and Mark Van Raamsdonk for useful conversations. We are also grateful to the long term workshop ''Quantum Information and String Theory'' (YITP-T-19-03) held at Yukawa Institute for Theoretical Physics, Kyoto University and participants for useful discussions. TT is supported by JSPS Grant-in-Aid for Scientific Research (A) No.16H02182 and JSPS Grant-in-Aid for Challenging Research (Exploratory) 18K18766. TT is also supported by World Premier International Research Center Initiative (WPI Initiative) from the Japan Ministry of Education, Culture, Sports, Science and Technology (MEXT).

%%%%%%%%%%%%%%%%%%%%%%%%%%%%%%%%%%%%%%%%%
%%%%%%%%%%%%%%%%%%%%%%%%%%%%%%%%%%%%%%%%%
\appendix
%%%%%%%%%%%%%%%%%%%%%%%%%%%%%%%%%%%%%%%%%
%%%%%%%%%%%%%%%%%%%%%%%%%%%%%%%%%%%%%%%%%

%%%%%%%%%%%%%%%%%%%%%%%%%%%%%%%%%%%%%%%%%
%%%%%%%%%%%%%%%%%%%%%%%%%%%%%%%%%%%%%%%%%
\section{Analytical Expressions of EE in Single Joining/Splitting Local Quench}\label{EESQ}
%%%%%%%%%%%%%%%%%%%%%%%%%%%%%%%%%%%%%%%%%
%%%%%%%%%%%%%%%%%%%%%%%%%%%%%%%%%%%%%%%%%
In this appendix, we summarize all the analytical expressions for the EE after a single joining/splitting local quench, in both holographic CFT and Dirac free fermion CFT. These are worked out under $a\rightarrow 0$ and $\ep\rightarrow0$. The subsystem $A$ is chosen to be $A = [x_1, x_2]$ where $0<|x_1|<x_2$. This is a general choice due to the symmetry.
\subsection{Single Joining Local Quench}
{\bf EE in holographic CFT}

The connected EE $S^{con}_A(t)$ and the disconnected EE $S^{dis}_A(t)$ are shown below. At $0<t<|x_2|$, 
\ba
 S^{con}_A&=&\left\{\begin{aligned}\frac{c}{3}\log (x_2-x_1)/\ep, \quad\qquad\qquad (x_1>0),\\ \frac{c}{6}\log \frac{4(x_1^2-t^2)(x_2^2-t^2)}{a^2\ep^2}, \ \ \ (x_1<0)\end{aligned}\right.\no 
 S^{dis}_A&=&\frac{c}{6}\log \left(4x_2|x_1|/\ep^2\right)+2S_{bdy}. \label{hppa}
\ea
At $|x_1|<t<x_2$, 
\ba
S^{con}_A=\frac{c}{6}\log \frac{2(x_2-x_1)(t-x_1)(x_2-t)}{a\ep^2},\ \ \
 S^{dis}_A=\frac{c}{6}\log \frac{4x_2(t^2-x_1^2)}{a\ep^2}+2S_{bdy}. \label{hppb}
\ea
At late time $t>x_2$,
\ba
S^{con}_A=\frac{c}{3}\log (x_2-x_1)/\ep, \ \ \
 S^{dis}_A=\frac{c}{6}\log \frac{4(t^2-x_2^2)(t^2-x_1^2)}{a^2\ep^2}+2S_{bdy}.  \label{hppc}
\ea
\vspace{5mm} 
%%%%%%%%%%%%%%%%%%
{\bf EE in Dirac free fermion CFT}
%%%%%%%%%%%%%%%%%%

The EE $S_A(t)$ are shown below. At $0<t<|x_1|$,
\ba
S_A=\frac{1}{6}\log \frac{4|x_1|x_2(x_2-x_1)^2}{(x_2+|x_1|)^2\ep^2}, \label{psqa}
\ea
At $|x_1|<t<x_2$,
\ba
S_A=\frac{1}{6}\log \frac{4(x_2-x_1)x_2(x_2-t)(t^2-x_1^2)}{a(x_1+x_2)(x_2+t)\ep^2},\ \ \ \label{psqb}
\ea
At $t>x_2$,
\ba
S_A=\frac{1}{3}\log (x_2-x_1)/\ep.  \label{psqc}
\ea
%%%%%%%%%%%%%%%%%%%
\subsection{Single Splitting Quench}
%%%%%%%%%%%%%%%%%%%
\vspace{5mm}
{\bf EE in holographic CFT}

The connected EE $S^{con}_A(t)$ and the disconnected EE $S^{dis}_A(t)$ are shown below. At $0<t<|x_1|$, 
\ba
 S^{con}_A&=&\frac{c}{3}\log (x_2-x_1)/\epsilon, \ \ \ \no
 S^{dis}_A&=&\frac{c}{6}\log \frac{4(x_1^2-t^2)(x_2^2-t^2)}{a^2\epsilon^2}+2S_{bdy}. \label{hpsa}
\ea
At $|x_1|<t<x_2$, 
\ba
S^{con}_A&=&\left\{\begin{aligned}\frac{c}{6}\log \frac{2(x_2-x_1)(t-x_1)(x_2-t)}{\ap\ep^2},\ \ (x_1>0)\\
\frac{c}{6}\log \frac{2(x_2-x_1)(t+x_1)(x_2+t)}{\ap\epsilon^2},\ \ (x_1<0)\end{aligned}\right. \no
 S^{dis}_A&=&\frac{c}{6}\log \frac{4|x_1|(x_2^2-t^2)}{a\ep^2}+2S_{bdy}.  \label{hpsb}
\ea
At $t>x_2$, we have
\ba
S^{con}_A&=&\left\{\begin{aligned}\frac{c}{3}\log (x_2-x_1)/\ep, \quad\qquad\qquad  (x_1>0)\\
\frac{c}{6}\log \frac{4(t^2-x_1^2)(t^2-x_2^2)}{a^2\ep^2},\ \ (x_1<0)\end{aligned}\right. \no
S^{dis}_A&=&\frac{c}{6}\log \frac{4|x_1|x_2}{\ep^2}+2S_{bdy}. \label{hpsc}
\ea
%%%%%%%%%%%%%%%%%%%
\vspace{5mm}
{\bf EE in Dirac free fermion CFT}
%%%%%%%%%%%%%%%%%%%

The EE $S_A(t)$ are shown below. At $0<t<|x_1|$, 
\ba
 S_A&=&\frac{1}{3}\log (x_2-x_1)/\ep.  \label{saa}
\ea
At $|x_1|<t<x_2$, 
\ba
S_A=\frac{1}{6}\log \frac{4|x_1|(x_2-x_1)(t-|x_1|)(x_2^2-t^2)}{(x_1+x_2)(t+|x_1|)a\ep^2}. \label{sab}
\ea
At $t>x_2$, 
\ba
S_A=\frac{1}{6}\log \frac{4|x_1|x_2(x_2-|x_1|)^2}{(x_2+x_1)^2\ep^2}. \label{sac}
\ea

%%%%%%%%%%%%%%%%%%%%%%%%%%%%%%%%%%%%%%%%%
%%%%%%%%%%%%%%%%%%%%%%%%%%%%%%%%%%%%%%%%%
\section{Time evolution of connected EE in single joining quench from the length of geodesic}
\label{SJCEE}
%%%%%%%%%%%%%%%%%%%%%%%%%%%%%%%%%%%%%%%%%
%%%%%%%%%%%%%%%%%%%%%%%%%%%%%%%%%%%%%%%%%

\begin{figure}
  \centering
   \includegraphics[width=7cm]{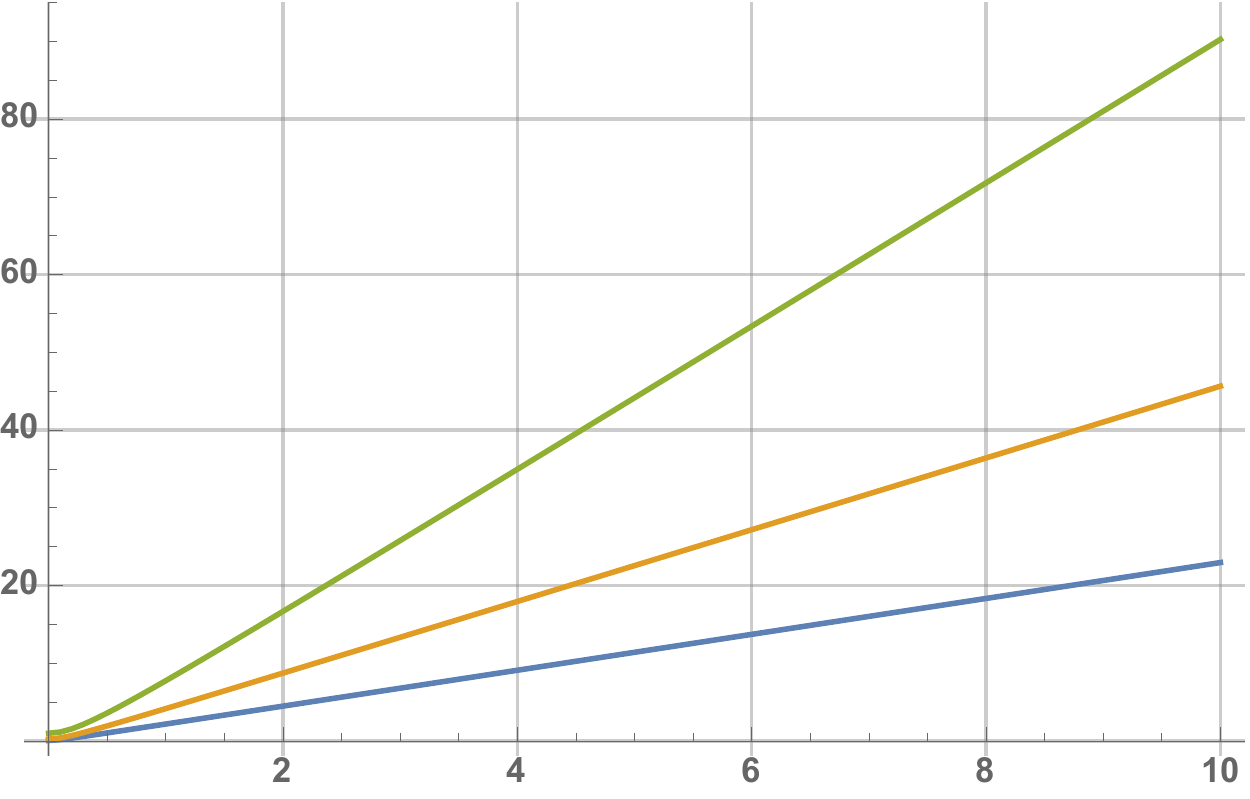}
   \includegraphics[width=7cm]{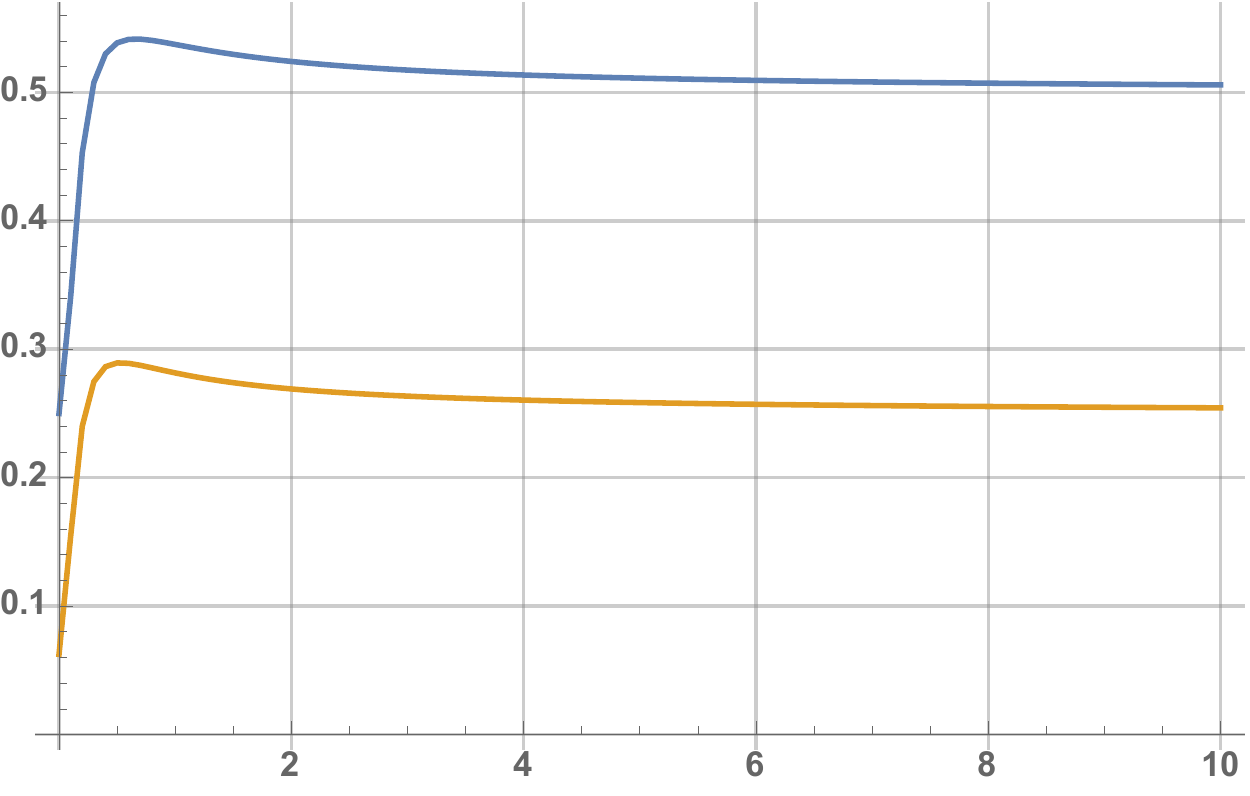}
 \caption{Numerical evidence for that $z$ coordinate of the tip is proportional to $\sqrt{at}$ at $t\gg a$. (Left figure) Vertical axis: $z^2$, horizontal axis: $t$. Green line: $a=1$, yellow line: $a=1/2$, blue line: $a=1/4$. We can see that the curve becomes linear at late time. (Right figure) Vertical axis: $z^2$ ratio for different $a$, horizontal axis: $t$. Blue line: ($z^2$ at $a=1/2$)/ ($z^2$ at $a=1$), yellow line: ($z^2$ at $a=1/4$)/ ($z^2$ at $a=1$). We can see that at late time they goes to 1/2 and 1/4 respectively.}
\label{fig:tip}
\end{figure}

Take a time slice $t=\rm{const.}$ of the boundary surface (figure \ref{fig:SJBSTE}) and then let us focus on the point who gives the minimal $z$. We call this a ``tip" and clearly we can see that, at $t>0$, there is exactly one tip on $\pm x$ side respectively. Figure {\ref{fig:tip}} shows numerical evidence for that the $z$ coordinate of plus side tip is proportional to $\sqrt{at}$ at $t\gg a$.

Let us consider a large subregion $A=[l_1,l_2]$ where $0<l_1\ll l_2$ and focus on the connected entanglement entropy at $l_1\ll t \ll l_2$: (refer to equation (\ref{SJEEcon}))
\ba
S^{con}_A - S^{(0)}_A\simeq \frac{c}{6}\log\frac{t}{a} + ...
\label{SJconEE}
\ea
\begin{figure}
  \centering
   \includegraphics[width=12cm]{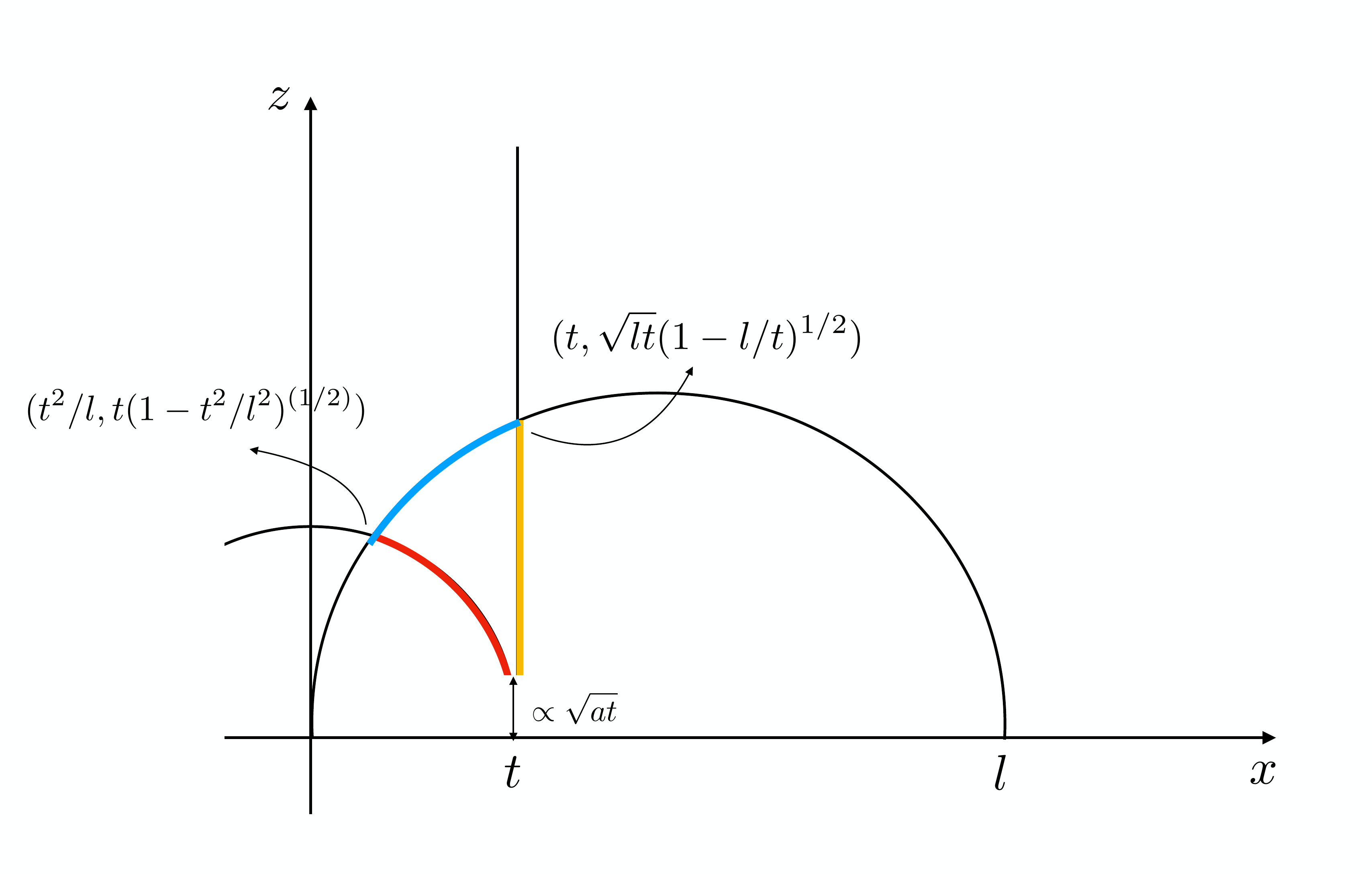}
 \caption{Evaluating the contribution of making a detour at the sharp corner of the boundary surface on a time slice. We approximately regard the boundary surface as a half circle $x^2+z^2=t^2$ and a vertical line $x=t$ with a cut off $\sqrt{at}$ at the tip.}
\label{fig:SJgeodesic}
\end{figure}

This contribution is expected to be given by making a detour at the tip of the boundary surface. For simplicity, let us set $l_1 = 0$, $l_2 = l$, and evaluate the effect by making a detour on a time slice $t=\rm{const.}$ (figure {\ref{fig:SJgeodesic}). We approximately regard the boundary surface as a half circle $x^2+z^2=t^2$ and a vertical line $x=t$ with a cut off at the tip. Besides, we approximately regard the metric as Poincar\'{e} metric. The variation of the geodesic is expected to be roughly given by $(red~line + yellow~line - blue~line)$. Noting that the length of a quarter circle with radius $R$ and cut off $\epsilon$ is given by
\ba
\log(\frac{R+\sqrt{R^2-\epsilon^2}}{\epsilon}),
\ea
we can evaluate the length of the three lines:
\ba
(length~of~red~line) \simeq  \log \left(\frac{t}{\sqrt{a t}}\right) - \log \left(\frac{t+\sqrt{t^2-t^2+t^4/l^2}}{\sqrt{t^2-t^4/l^2}}\right) \simeq \log\sqrt{{t}/{a}},
\ea
\ba
(length~of~yellow~line) \simeq \log\left(\frac{\sqrt{lt}(1-t/l)^{1/2}}{\sqrt{at}}\right) \simeq \log\sqrt{l/a},
\ea
\ba
&(length~of~blue~line) \simeq \log\left(\frac{l/2+\sqrt{l^2/4-t^2+t^4/l^2}}{\sqrt{t^2-t^4/l^2}}\right) - \log\left(\frac{l/2+\sqrt{l^2/4-lt+t^2}}{\sqrt{lt-t^2}}\right)\nonumber\\
&\simeq \log\sqrt{l/t}.
\ea
Thus, $(red~line + yellow~line - blue~line)$ gives a $(1/2)\log(t/a\times l/a\times t/l)=\log(t/a)$ length variation, which exactly matches the connected entanglement entropy variation (\ref{SJconEE}).

\section{A derivation of some asymptotic behaviors of entanglement entropy in joining quenches in Dirac Fermion CFT}\label{derdirac}
In this section, we derive analytically $\Delta S_A = -\f{l^2}{24 x^2} + \cdots$ for $A = [x-\f{l}{2},x+\f{l}{2}]$ in the $x \to \infty$ limit with fixed $l$ at $t = 0$ for both of single joining local quenches and double joining local quenches in Dirac fermion CFT.
The entanglement entropy is given by 
\be
S^{Dirac}_A=\frac{1}{6}\log\left(\frac{|\xi_1-\xi_2|^2|\xi_1-\bar{\xi}_1||\xi_2-\bar{\xi}_2||g'(\xi_1)||g'(\xi_2)|}{\ep^2(\xi_1-\bar{\xi}_2)(\xi_2-\bar{\xi}_1)}\right).
\ee
In both cases, $\xi$ takes the value on the unit circle and we can write as $\xi = e^{i\theta}$.
In single joining local quenches, the conformal map to the UHP is 
\be
w = g(\xi )= ia\f{\xi^2 + 1}{\xi^2 - 1}.
\ee
Using $\xi = e^{i\theta}$, we can write as 
\ba
w &=&  a \cot \theta, \qquad  \theta = \arctan \f{a}{w}, \notag \\
g'(e^{i\theta}) &=&  \f{i a e^{-i\theta}}{\sin ^2 \theta} \label{eq:inverses}
\ea 
Entanglement entropy in $\theta$ variable becomes
\be
S^{Dirac}_A=\frac{1}{6}\log\left(\frac{4a^2 \sin^2 \f{\theta_1 - \theta_2}{2} }{\ep^2\sin^2 \f{\theta_1 + \theta_2}{2} \sin \theta_1 \sin \theta_2}\right).
\ee
For calculating the $x^{-2}$ term in entanglement entropy, we need up to  next next leading term in (\ref{eq:inverses}):
\be
\theta = \f{a}{w}  - \f{a^3}{3 w^3}+ \cdots = \f{a}{w} \Big( 1 - \f{a^2}{3 w^2} \Big) + \cdots.
\ee
In our setting, $w_1 = \bar{w}_1 = x- \f{l}{2}$ and $w_2 = \bar{w}_2 = x + \f{l}{2}$. 
Therefore, we can evaluate entanglement entropy as 
\ba
S_A ^{Dirac} &\sim&   \f{1}{6} \log \Big[   \f{l^2}{\epsilon^2} \Big( 1 - \f{l^4}{4x^2} \Big) \Big] \sim \f{1}{3} \log \f{l}{\epsilon} - \f{l^2}{24 x^2}. 
\ea
This show that the subleading part of entanglement entropy is given by $-\f{l^2}{24 x^2}$ and therefore $\Delta S_A \sim -\f{l^2}{24x^2}$.

Similarly, we can compute the $x^{-2}$ term in double joining local quenches.
The conformal map from double local quenches to the UHP is 
\be
w = g(\xi) = i \f{a}{a_0(\alpha)} \Big( \f{1}{2} \sin^2 \alpha  \log (- i \xi) + \f{\xi ^2 + 1 }{2(\xi^2 -1)} \cos ^2 \alpha \Big),
\ee
where $a_0(\alpha) = \f{1}{2} \sin^2 \alpha \log (\cot \f{\alpha}{2}) + \f{1}{2} \cos \alpha$.
In $\xi = e^{i\theta}$ coordinate, the map becomes
\ba
w &=& g(\theta) =  \f{a}{a_0(\alpha)} \Big( \f{1}{2} \cos^2 \alpha  \cot \theta + \Big( \f{\pi}{2} - \theta \Big)\f{1}{2} \sin ^2 \alpha \Big),  \notag \\
g'(e^{i\theta })&=&  i \f{a}{2 a_0(\alpha)} e^{-i\theta} \Big( \sin^2 \alpha + \f{\cos^2 \alpha}{\sin ^2 \theta} \Big).
\label{eq:dlmapap}
\ea
and entanglement entropy is written as 
\be
S_A ^{Dirac} = \f{1}{6} \log\left(\f{ a^2 \sin^2 \alpha}{a_0(\alpha)^2} \frac{ \sin^2 \f{\theta_1 - \theta_2}{2}  \sin \theta_1 \sin \theta_2}{\ep^2  \sin^2 \f{\theta_1 + \theta_2}{2}} \Big(\sin^2 \alpha + \f{\cos^2 \alpha}{\sin ^2 \theta_1}\Big)\Big(\sin^2 \alpha + \f{\cos^2 \alpha}{\sin ^2 \theta_2} \Big) \right). \label{eq:dleetheta}
\ee
What we need is to invert (\ref{eq:dlmapap}) analytically when $w$ is large ($\theta$ is small).
Generically, we can determine the inverse of $\delta = f(\epsilon)$ order by order.
When $\delta$ and $\epsilon$ are small and $f$ has an expansion $f(\epsilon) = f_1\epsilon + f_2 \epsilon^2 + f_3 \epsilon^3  + \cdots$, the inverse function is expanded as 
\be
\epsilon = \f{1}{f_1} \delta - \f{f_2}{(f_1) ^3}\delta^2 + \Big(\f{2 (f_2)^2}{(f_1)^5} - \f{f_3}{(f_1)^4} \Big)\delta^3 + \cdots. \label{eq:invertgen}
\ee
To apply this technique, it is convenient to rewrite the map (\ref{eq:dlmapap}) as 
\be
\f{a}{2a_0(\alpha)} \f{\cos^2 \alpha}{w - b} = \f{ \tan \theta}{1 - (\tan^2 \alpha )\theta \tan \theta } = \theta + \Big( \f{1}{3} + \tan^2 \alpha  \Big) \theta ^3 +  \cdots.
\ee
where $b = a \f{b_0(\alpha)}{a_0(\alpha)}$ and $b_0(\alpha) = \f{\pi}{4} \sin^2 \alpha$.
Then, when $w$ is large we can invert this map using (\ref{eq:invertgen}) as 
\be
\theta =  \Big(\f{a}{2a_0(\alpha)} \f{\cos^2 \alpha}{w - b}\Big) - \Big(\f{1}{3} + \tan ^2 \alpha \Big)\Big(\f{a}{2a_0(\alpha)} \f{\cos^2 \alpha}{w - b}\Big)^3 + \cdots.
\ee
Using this expansion and expression for entanglement entropy (\ref{eq:dleetheta}) with $w_1 = x - \f{l}{2}$ and $w_2 = x + \f{l}{2}$, we can read off the $x^{-2}$ term as 
\be
S_A^{Dirac} = \f{1}{3} \log \f{l}{\epsilon} - \f{l^2}{24 x^2} + \cdots.
\ee
This show that the subleading part of entanglement entropy is given by $-\f{l^2}{24 x^2}$ even in double joining local quenches and therefore $\Delta S_A \sim -\f{l^2}{24x^2}$.

In the same manner, we can derive the $t\to \infty $ limit of entanglement entropy for $A = [ x - \f{l}{2}, x+ \f{l}{2}]$ with fixed $x$ and $l$ in single joining local quenches.
In time dependent cases, $\xi \neq \bar{\xi}$ and therefore $\theta \neq \bar{\theta}$.
Entanglement entropy in single joining local quenches is 
\be
S_A ^{Dirac} = \f{1}{6} \log \Bigg( \f{4a^2 \sin \f{\theta_1 - \theta_2}{2}\sin \f{\bar{\theta_1} - \bar{\theta_2}}{2}  \sin \f{\theta_1 + \bar{\theta}_1}{2} \sin \f{\theta_2 + \bar{\theta}_2}{2} }{\sin \f{\theta_1 + \bar{\theta}_2}{2} \sin \f{\theta_2 + \bar{\theta}_1}{2} \sin  \theta_1 \sin  \bar{\theta}_1\sin  \theta_2 \sin  \bar{\theta}_2} \Bigg). \label{eq:eeDStimeap}
\ee 
The conformal map (\ref{eq:inverses}) is expanded as 
\be
\theta = \f{a}{w}  - \f{a^3}{3 w^3} + \f{a^5}{5 w^5} - \f{a^7}{7 w^7} + \mathcal{O}(w^{-8}).
\ee
In our choice of the branch, the expansion in large negative real $w$ becomes
\be
\theta =\pi +  \f{a}{w}  - \f{a^3}{3 w^3} + \f{a^5}{5 w^5} - \f{a^7}{7 w^7} + \mathcal{O}(w^{-8}).
\ee
Using this expansion and expression for entanglement entropy (\ref{eq:dleetheta}) with $w_1 = x -t -  \f{l}{2}$, $\bar{w}_1 = x + t -  \f{l}{2}$, $w_2 = x - t + \f{l}{2}$ and $w_2 = x + t + \f{l}{2}$ , we can read off the $t^{-6}$ term as 
\be
S_A^{Dirac} = \f{1}{3} \log \f{l}{\epsilon} + \f{a^2 l^2 x^2}{3 t^6} + \cdots.
\ee
Similarly, we can derive the $t\to \infty $ limit of entanglement entropy for $A = [ x - \f{l}{2}, x+ \f{l}{2}]$ with fixed $x$ and $l$ in double joining local quenches.
Entanglement entropy in double joining local quenches becomes 
\ba
S_A ^{Dirac} &=& \f{1}{6} \log \Bigg( \f{a^2}{a_0(\alpha)^2} \f{ \sin \f{\theta_1 - \theta_2}{2}\sin \f{\bar{\theta_1} - \bar{\theta_2}}{2}  \sin \f{\theta_1 + \bar{\theta}_1}{2} \sin \f{\theta_2 + \bar{\theta}_2}{2} }{\sin \f{\theta_1 + \bar{\theta}_2}{2} \sin \f{\theta_2 + \bar{\theta}_1}{2}}  \notag \\
&& \qquad  \times\s{\sin^2 \alpha + \f{\cos^2 \alpha}{\sin^2  \theta_1 }} \s{\sin^2 \alpha + \f{\cos^2 \alpha}{\sin^2  \bar{\theta}_1 }}\s{\sin^2 \alpha + \f{\cos^2 \alpha}{\sin^2  \theta_2 }} \s{\sin^2 \alpha + \f{\cos^2 \alpha}{\sin^2  \bar{\theta}_2 }}
\Bigg) . \notag \\ \label{eq:eeDDtimeap}
\ea 
The conformal map for double joining local quenches is expanded up to the fifth order of $\theta$ as
\be
\f{a}{2a_0(\alpha)} \f{\cos^2 \alpha}{w - b} = \f{ \tan \theta}{1 - (\tan^2 \alpha )\theta \tan \theta } = \theta + \Big( \f{1}{3} + \tan^2 \alpha  \Big) \theta ^3 +\Big( \f{2}{15} + \f{2}{3}\tan^2 \alpha + \tan^4 \alpha  \Big) \theta ^5 + \cdots.
\ee
The inverse of $\delta = f(\epsilon)$ with the expansion $f(\epsilon) = f_1\epsilon + f_2 \epsilon^2 + f_3 \epsilon^3 +f_4 \epsilon^4 + f_5 \epsilon^5   + \cdots$ is given by
\ba
\epsilon &=& \f{1}{f_1} \delta - \f{f_2}{(f_1) ^3}\delta^2 + \Big(\f{2 (f_2)^2}{(f_1)^5} - \f{f_3}{(f_1)^4} \Big)\delta^3  + \Big(-\f{5(f_2)^3}{(f_1)^7} + \f{5f_2f_3}{(f_1)^6}- \f{f_4}{(f_1)^5}\Big) \delta^4 \notag \\
&& \qquad  \qquad + \Big(\f{14 (f_2) ^4}{(f_1)^9} -  \f{21 (f_2) ^2 f_3}{(f_1)^8} + \f{3 (f_3) ^2 }{(f_1)^7} +  \f{6 f_2  f_4}{(f_1)^7} - \f{f_5}{(f_1)^6} \Big)\delta^5 + \mathcal{O}(\delta^6).
\ea
Using this expansion, we can invert the map (\ref{eq:dlmapap}) up to the order of $w^{-5}$ as 
\ba
\theta &=& \Big(\f{a}{2a_0(\alpha)} \f{\cos^2 \alpha}{w - b}\Big) - \Big(\f{1}{3} + \tan ^2 \alpha \Big)\Big(\f{a}{2a_0(\alpha)} \f{\cos^2 \alpha}{w - b}\Big)^3 \notag \\
&& \qquad \qquad + \Big(\f{1}{5} + \f{4}{3} \tan^2 \alpha + 2 \tan^4 \alpha \Big) \Big(\f{a}{2a_0(\alpha)} \f{\cos^2 \alpha}{w - b}\Big)^5 + \mathcal{O}(w^{-6}).
\ea
When $w$ is large in negative real direction, we get the following expansion in our choice of the branch:
\ba
\theta &=&\pi + \Big(\f{a}{2a_0(\alpha)} \f{\cos^2 \alpha}{w + b}\Big) - \Big(\f{1}{3} + \tan ^2 \alpha \Big)\Big(\f{a}{2a_0(\alpha)} \f{\cos^2 \alpha}{w + b}\Big)^3 \notag \\
&& \qquad + \Big(\f{1}{5} + \f{4}{3} \tan^2 \alpha + 2 \tan^4 \alpha \Big) \Big(\f{a}{2a_0(\alpha)} \f{\cos^2 \alpha}{w + b}\Big)^5 + \mathcal{O}(w^{-6})
\ea
From these expansions and (\ref{eq:eeDDtimeap}), we obtain the $t\to \infty$ behavior in double local quenches as 
\be\label{derGba}
S_A ^{Dirac} = \f{1}{3} \log \f{l}{\epsilon} + \f{\cos^2\alpha \sin ^2 \alpha}{a_0(\alpha)^2} \f{a^2 l^2}{24 t^4} + \cdots. 
\ee
Through the relation $\f{b}{a} = \f{b_0(\alpha)}{a_0(\alpha)}$, the coefficient of the subleading term $\f{\cos^2 \alpha \sin ^2 \alpha}{a_0(\alpha)^2}$ becomes the function on $\f{b}{a}$.
This function $G(b/a) = \f{\cos^2 \alpha \sin ^2 \alpha}{a_0(\alpha)^2}$ is the one that is given in the main part of this paper.

%%%%%%%%%%%%%%%%%%%%%%%%%%%%%%%%%%%%%%%%%
%%%%%%%%%%%%%%%%%%%%%%%%%%%%%%%%%%%%%%%%%

\end{document}